\documentclass{article}

\usepackage[utf8]{inputenc}
\usepackage{amsmath,amssymb,amsfonts,stmaryrd}
\usepackage{physics}
\usepackage{dsfont}
\usepackage{graphicx}
\usepackage{xcolor}
\usepackage{graphicx}

\usepackage{cancel}
\usepackage{color}
\usepackage{tikz}
\usepackage{mathrsfs}
\usepackage{relsize}
\usepackage[linktocpage]{hyperref}
\usepackage[all]{xy}
\hypersetup{colorlinks=true,citecolor=blue,linkcolor=blue, urlcolor=blue, breaklinks=true}

\usepackage{bm} 
\usepackage{subfigure} 
\usepackage{environ}
\usepackage{url}
\usepackage[margin=0.75in]{geometry}
\usepackage[export]{adjustbox}
\usepackage{float}

\usepackage{mathtools}

\newcommand{\ZZ}{{\mathbb Z}}

\newcommand{\g}{{\gamma}}

\newcommand{\Z}{{\mathbb Z}}
\newcommand{\Sq}{{\mathrm{Sq}}}
\newcommand{\U}{{\mathrm{U}(1)}}

\newcommand{\ra}{\rightarrow}

\newcommand{\lr}[1]{ \langle {#1} \rangle}

\NewEnviron{eqs}{%
\begin{equation}\begin{split}
    \BODY
\end{split}\end{equation}}

\title{Codimension-2 defects and higher symmetries \\ in (3+1)D topological phases}

\author{Maissam Barkeshli$^{1}$, Yu-An Chen$^{1,2}$, Sheng-Jie Huang$^{1}$, \\
Ryohei Kobayashi$^{1}$, Nathanan Tantivasadakarn$^{3,4}$, and Guanyu Zhu$^5$}

\date{
    $^1$ Department of Physics, Condensed Matter Theory Center, and Joint Quantum Institute, University of Maryland, College Park, Maryland 20742, USA\\
    $^2$Joint Center for Quantum Information and Computer Science, University of Maryland, College Park, Maryland 20742, USA\\
    $^3$ Walter Burke Institute for Theoretical Physics and Department of Physics, California Institute of Technology, Pasadena, CA 91125, USA\\
    $^4$ Department of Physics, Harvard University, Cambridge, MA 02138, USA\\
    $^5$ IBM Quantum, IBM T.J. Watson Research Center, Yorktown Heights, NY 10598 USA\\
}

\begin{document}

\maketitle

\begin{abstract}
(3+1)D topological phases of matter can host a broad class of non-trivial topological defects of codimension-1, 2, and 3, of which the well-known point charges and flux loops are special cases. The complete algebraic structure of these defects defines a higher category, and can be viewed as an emergent higher symmetry. This plays a crucial role both in the classification of phases of matter and the possible fault-tolerant logical operations in topological quantum error-correcting codes. In this paper, we study several examples of such higher codimension defects from distinct perspectives. We mainly study a class of invertible codimension-2 topological defects, which we refer to as twist strings. We provide a number of general constructions for twist strings, in terms of gauging lower dimensional invertible phases, layer constructions, and condensation defects. We study some special examples in the context of $\mathbb{Z}_2$ gauge theory with fermionic charges, in $\mathbb{Z}_2 \times \mathbb{Z}_2$ gauge theory with bosonic charges, and also in non-Abelian discrete gauge theories based on dihedral ($D_n$) and alternating ($A_6$) groups. The intersection between twist strings and Abelian flux loops sources Abelian point charges, which defines an $H^4$ cohomology class that characterizes part of an underlying 3-group symmetry of the topological order. The equations involving background gauge fields for the 3-group symmetry have been explicitly written down for various cases. We also study examples of twist strings interacting with non-Abelian flux loops (defining part of a non-invertible higher symmetry), examples of non-invertible codimension-2 defects, and examples of the interplay of codimension-2 defects with codimension-1 defects. We also find an example of geometric, not fully topological, twist strings in (3+1)D $A_6$ gauge theory. 
\end{abstract}

\tableofcontents

\section{Introduction}

It is well-known that topologically ordered phases of matter are characterized by the existence of topological  excitations with certain fusion and braiding properties. In (2+1)D, these correspond to anyons, which are point-like quasi-particles whose fusion and braiding properties are described mathematically by a unitary modular fusion category \cite{nayak2008,wang2008}. In (3+1)D, topologically ordered phases possess point and loop excitations, with non-trivial mutual braiding statistics. In particular, for discrete gauge theory with gauge group $G$, which is believed to describe all (3+1)D topological phases \cite{lan2018,lan2019}, the point and loop excitations correspond to gauge charges and flux loops.

Over the last decade, it has been understood that there is significantly more to the story, in the sense that there is an additional structure in the universal properties of topological phases of matter. This structure corresponds to the distinct kinds of extrinsic topological defects of varying codimension that can be supported \cite{bombin2010,kapustin2010,kapustin2011,KitaevKong12,barkeshli2012a,clarke2013,lindner2012,cheng2012,barkeshli2013genon,you2012,barkeshli2013defect,barkeshli2013defect2}. Understanding the properties of these defects is crucial to fully understand the types of physical phenomena that can occur, while also providing the foundation for describing symmetry in topological phases of matter \cite{barkeshli2019}.
For example, in (2+1)D, in addition to anyons, topologically ordered phases of matter can support topologically non-trivial line defects, which are codimension-1 defects. In some cases, anyons near a line defect can be annihilated or created by a local operator; in other cases, an anyon may be converted to a topologically distinct anyon upon crossing the line defect. In general, these line defects can be thought of as topologically distinct classes of gapped interfaces between the given topological order and itself. The codimension-2 junction between two distinct line defects can localize exotic zero modes and give rise to topologically protected degeneracies, thus forming a non-Abelian defect (Ref.~\cite{barkeshli2010} Sec.~V) \cite{bombin2010,barkeshli2012a,clarke2013,lindner2012,cheng2012,barkeshli2013genon,you2012,barkeshli2013defect,barkeshli2013defect2}. The anyons themselves can be viewed as special kinds of codimension-2 defects. 

Since the line and point defects can be fused together in myriad ways, it is expected that the combined algebraic structure of line and point defects in (2+1)D topological phases of matter is described mathematically by a unitary fusion 2-category \cite{kapustin2010,douglas2018}. While this is understood at a somewhat abstract mathematical level \cite{douglas2018}, it is ongoing work to fully understand this structure in concrete terms amenable to calculations in physical models \cite{barter2019,bridgeman2019,bridgeman2020}. The properties of these line and point defects are crucial to understanding symmetry-enriched topological phases of matter and their modern characterization in terms of $G$-crossed braided tensor categories \cite{barkeshli2019,barkeshli2022invertible,bulmash2022,aasen2021}. 

Similarly, in (3+1)D, the point-like gauge charges and flux loops constitute only part of the story. (3+1)D topological phases can support distinct types of codimension-1, 2, and 3 defects. The codimension-1 defects correspond to gapped interfaces between the topological order and itself; the codimension-2 defects correspond to gapped interfaces between distinct codimension-1 defects, and so on. The conventional gauge charges and flux strings constitute just a special case of the more general kinds of codimension-3 and codimension-2 defects, respectively. 
Extrapolating from (2+1)D, one expects that there exists a mathematical structure corresponding to a unitary fusion 3-category, which would describe the combined algebraic structure of codimension-1,2, and 3 defects in (3+1)D topologically ordered phases of matter. However, the mathematics of fusion $n$-categories for $n > 2$ is much less well-developed. 

From a contemporary perspective, the entire structure of these topological defects of varying codimension can be thought of as a ``higher symmetry" \cite{gaiotto2014,benini2019,cordova2022,mcgreevy2022, bhardwaj2022universal}. For example, in a $(d+1)$-dimensional topological phase, one can think of implementing a non-trivial symmetry operation on a $(d-k+1)$-dimensional subspace of the system by sweeping a codimension-$k$ defect along some closed trajectory in time.
A special class of defects that are invertible, as we will define more precisely below, define a higher group symmetry. Invertible codimension-$k$ defects define a $(k-1)$-form symmetry \cite{gaiotto2014}. In a $(d+1)$-dimensional topological phase, this leads to a series of groups $K_k$, for $k = 1,\dots, d$. $K_k$ is an Abelian group if $k > 1$. 

The non-trivial interaction between invertible defects of varying codimension implies a higher group structure. In (2+1)D, the invertible defects are known mathematically to define a categorical 2-group symmetry (see Ref.~\cite{ENO2010} and Appendix D of Ref.~\cite{barkeshli2019}). Extrapolating to (3+1)D, one expects the invertible defects to define a categorical 3-group symmetry.\footnote{A ``categorical" $n$-group can also be thought of as an $(n+1)$-group.} Some aspects of this 3-group symmetry in (3+1)D topological phases were studied in Refs.~\cite{Hsin2020liquid,kapustinThorngren2017FermionSPT}.

The purpose of this paper is to make progress in further understanding higher codimension defects in (3+1)D topological phases of matter. A long-term goal is to develop a complete understanding of the fusion 3-category of defects in (3+1)D topological phases of matter. In the short term, a satisfactory understanding of the categorical 3-group of invertible defects may be within reach. 

Most of the focus in this paper is on studying a class of invertible codimension-2 defects, which we refer to as ``twist strings." These can be thought of as loop-like defects in (3+1)D topological phases, but which do not correspond to the conventional flux strings. Nevertheless, the twist strings have several striking properties. For example, a twist string that crosses a flux string can source a non-trivial point charge; state differently, the linking between twist strings and flux strings can induce point charges in the system.

\subsection{Overview of results}

\begin{figure}
    \centering
    \includegraphics[width=1\textwidth]{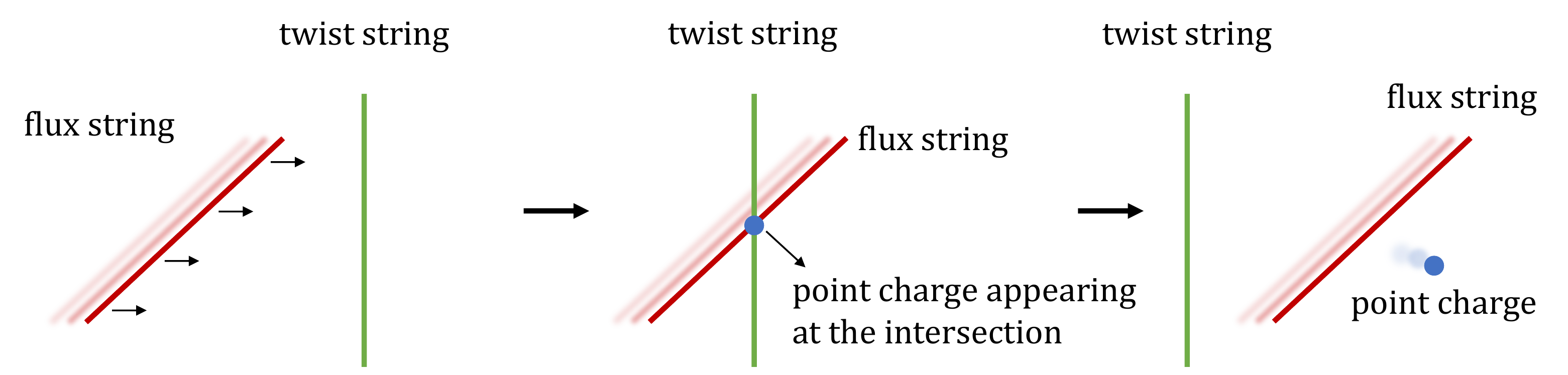}
\caption{The vertical (green) line represents the topological twist string (codimension-2 defect), and the horizontal (red) line is the flux string. When the flux string is moved across the twist string, a point charge appears from their intersection.}
\label{fig: loop defect intersection produces charge}
\end{figure}

\begin{figure}
    \centering
    \includegraphics[width=0.8\textwidth]{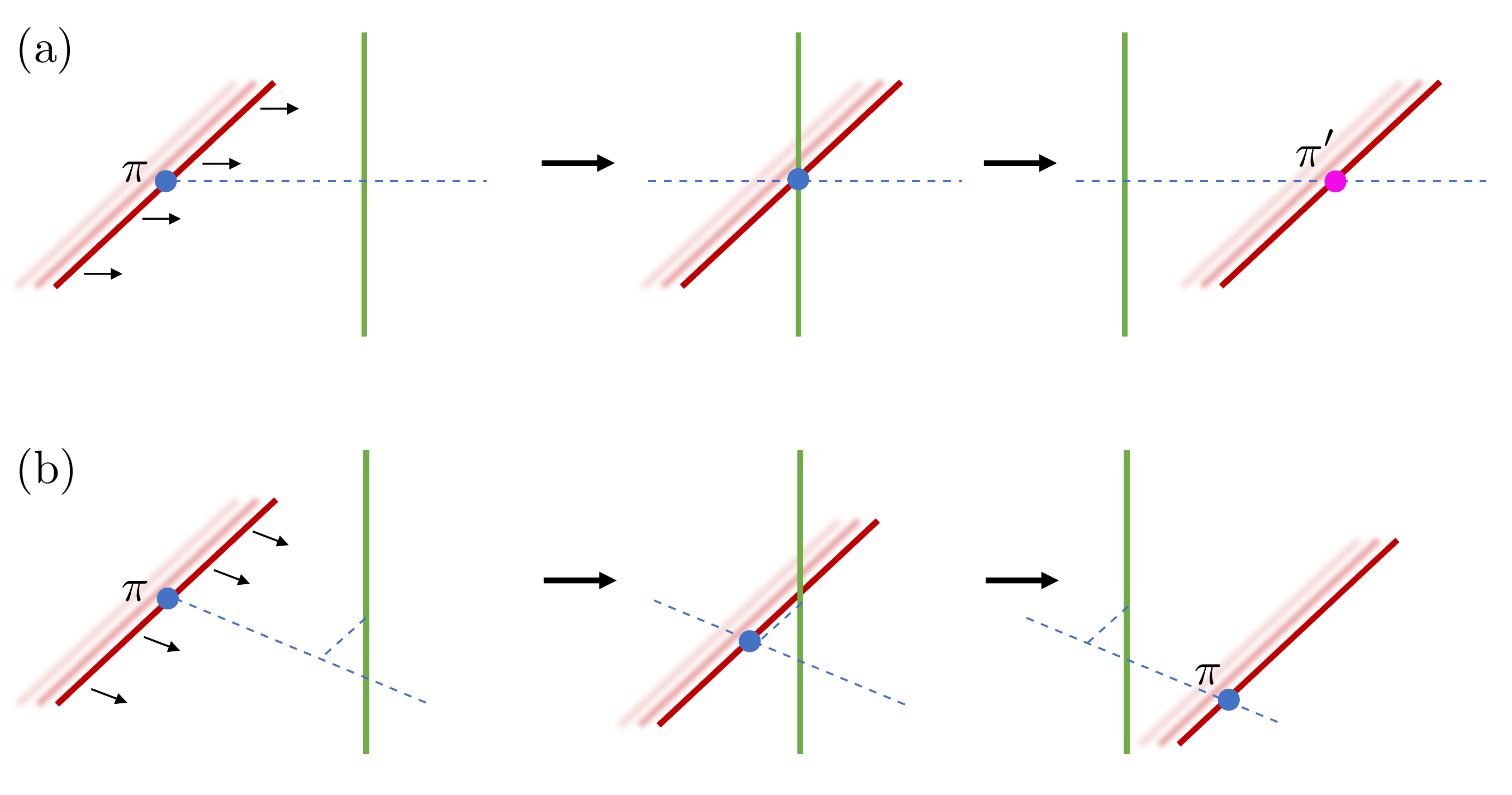}
\caption{Illustration of the generic properties of a non-topological twist string (green line).  There exist a certain flux string with an attached charge $\pi$, such that the transformation of this charge depends on the detailed geometric trajectory of the flux string and the charge.  (a)  If the worldline of the attached charge $\pi$ (blue) intersects with the geometric twist string, the attached charge will be transformed into $\pi'$ (purple).  (b) If the worldline of the attached charge avoids the twist string, the attached charge remains the same.}
\label{fig:geometric_defect_intersection}
\end{figure}

Here we provide a brief overview of our main results. 

In Sec.~\ref{sec:defect}, we provide a rough definition of the notion of codimension-$k$ defects and some of their basic properties. The definition is analogous to the definition of a gapped phase of matter, except applied to a subspace of the total space. We also briefly explain how codimension-$k$ defects correspond to higher ($(k-1)$-form) symmetries.

We study a class of invertible codimension-2 topological defects of (3+1)D discrete $G$ gauge theory, which we refer to as twist strings. One way to construct these twist strings is as follows. First, we prepare a (3+1)D trivial invertible topological phase with $G$ symmetry, which can be either bosonic or fermionic. Then, we decorate the codimension-2 surface embedded in the (3+1)D spacetime with a (1+1)D invertible topological phase with the same symmetry $G$. We then gauge the $G$ symmetry of the whole spacetime. The resulting theory is given by the (3+1)D $G$ gauge theory, where the decorated (1+1)D invertible phase now defines an invertible codimension-2 defect of the (3+1)D $G$ gauge theory. Other constructions of twist strings are given by layer constructions (Sec.~\ref{sec:layer}) or the notion of condensation defects (Sec.~\ref{sec:highergauge}). 

The main defining property of the twist strings is that they have a nontrivial interplay with the point charges and flux loops of the (3+1)D topological order (see Fig.~\ref{fig: loop defect intersection produces charge}). Namely, the crossing between a twist string and a flux string sources a point charge. This relationship between the twist string and the flux string in (3+1)D is reminiscent of the codimension-1 defects in (2+1)D topological orders that can permute anyons, which can also be regarded as attaching an additional anyon to the anyon crossing the defect. In (3+1)D, a codimension-2 defect cannot permute the label of the flux strings, but it can still act on the flux string by attaching a point charge upon linking:
\begin{eqs}
    \includegraphics[width=0.65\textwidth, valign=c ]{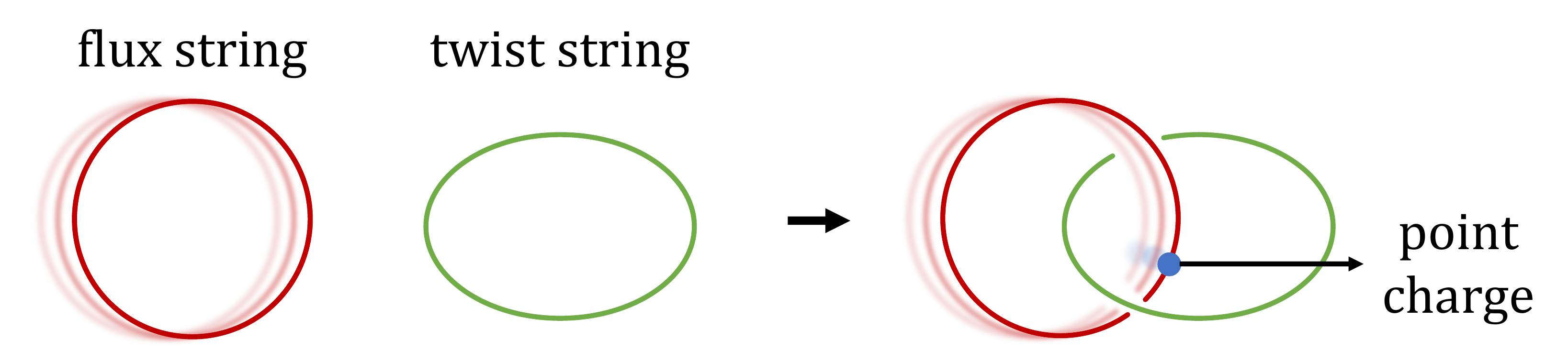}
\end{eqs}

In the most general situation, the flux string can be non-Abelian and is labeled by a conjugacy class $[g]$ of the gauge group $G$. In these cases, one can consider charged flux loops, where the charge is labeled by an irreducible representation (irrep) $\pi_g$ of the centralizer of $g$.\footnote{These charges can be distinguished by a two-loop braiding process. } When the flux string crosses the twist string, a point charge $\pi_g$ can appear on the string, whose worldline emanates from the space-time crossing point between the twist string and the flux string. $\pi_g$ must, on general grounds, be a one-dimensional irrep. In many simple examples, the point charge is a \it deconfined \rm Abelian point charge, although there are also examples where the charge is confined to the flux string. The above process implies that the total charge of the system is directly related to the linking number between twist strings and magnetic flux strings.

In Sec.~\ref{sec: Exactly solvable models for defects in (2+1)D toric codes}, we warm up by showing how some familiar invertible codimension-1 defects (which we also refer to here as twist strings) in (2+1)D topological phases can be understood in terms of gauging an invertible phase with a non-trivial (1+1)D invertible phase decorated on the line defect. This includes in particular the $e \leftrightarrow m$ twist string, permuting $e$ and $m$ anyons in $\Z_2$ gauge theory, which arises from decorating a codimension-$1$ line with a gauged Kitaev chain, where the procedure to gauge fermion parity on the 2d square lattice is reviewed in Appendix.~\ref{sec: review of 2d bosonization}. We show how this understanding can be used to provide exactly solvable models with codimension-1 defects in the ground state, specifically focusing on the example of $\Z_2$ gauge theory and $\Z_2 \times \Z_2$ gauge theory. A codimension-1 defect in the $\Z_2 \times \Z_2$ gauge theory comes from gauging the (1+1)D $\ZZ_2 \times \ZZ_2$ cluster state. In Appendix~\ref{sec: Z2 x Z2 automorphism}, it is shown that 1d Kitaev chains and 1d cluster states have exhausted all invertible codimension-1 defects in the (2+1)D $\ZZ_2 \times \ZZ_2$ toric code. The same argument holds for the (2+1)D $\ZZ_2^N$ toric code.

We then generalize the (2+1)D analysis in Sec.~\ref{sec:exact3d} to construct exactly solvable Hamiltonian models of (3+1)D topologically ordered phases that host the twist strings. This is done by performing the gauging procedure on a 3d cubic lattice for a given trivial (3+1)D invertible phase, in the presence of the (1+1)D invertible phase decorated on the 1d defect line embedded in the 3d cubic lattice. Concretely, we describe the twist string realized in an exactly solvable model of a (3+1)D $\Z_2\times\Z_2$ toric code with bosonic particles, and a (3+1)D $\Z_2$ toric code with a fermionic particle. The twist string in the (3+1)D $\Z_2\times\Z_2$ toric code corresponds to the 1d cluster state with $\Z_2\times\Z_2$ symmetry before gauging, and the twist string in a (3+1)D $\Z_2$ toric code with a fermionic particle is obtained by starting with a (3+1)D trivial fermionic invertible phase with a decoration of the 1d Kitaev chain, and then gauging the $\Z_2^f$ fermion parity symmetry by performing the bosonization valid for three spatial dimensions. The review of the procedure to gauge fermion parity on the 3d cubic lattice is found in Appendix~\ref{sec: review of 3d bosonization}.

\begin{figure}
    \centering
    \includegraphics[width=1\textwidth]{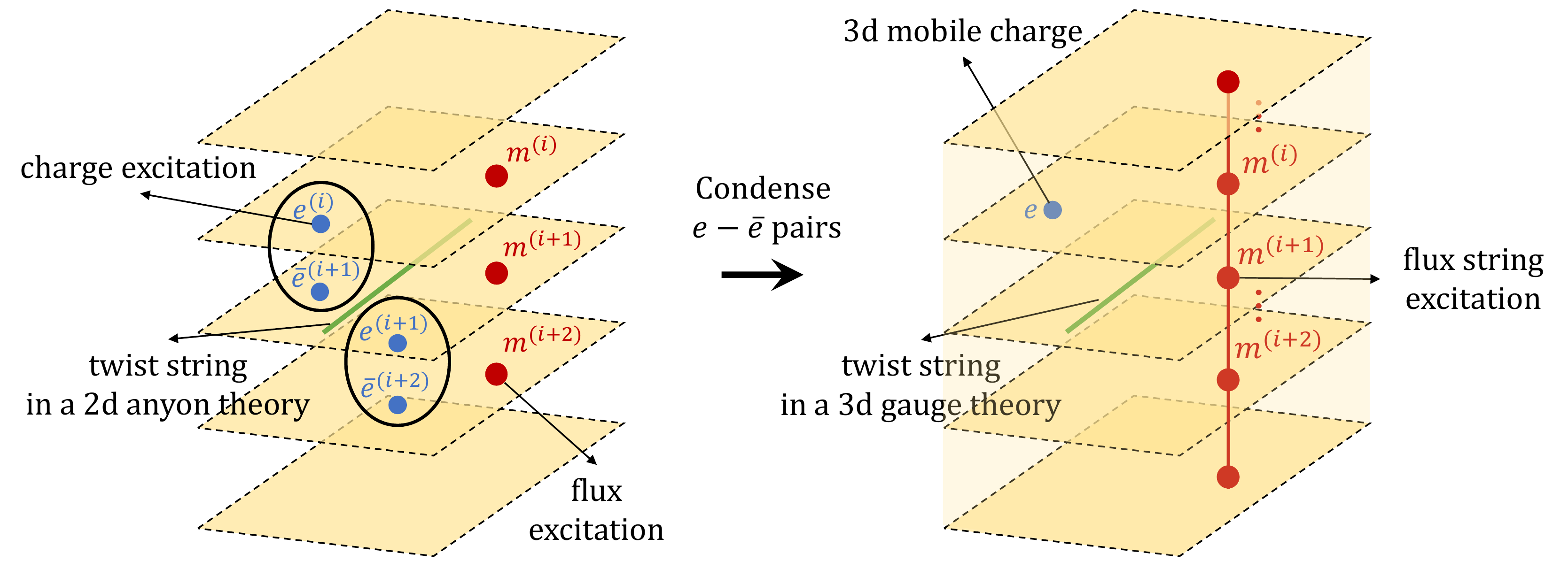}
\caption{Layers of (2+1)D anyon theories are prepared, and then pairs of particle excitations in adjacent layers are condensed.}
\label{fig: introduction to layer construction}
\end{figure}

 In addition to the above perspective, we show in Sec.~\ref{sec:layer} that the twist strings can also be understood from a layer construction, starting with layers of (2+1)D topological phases. Any discrete $G$ gauge theory in (3+1)D can arise from layering (2+1)D discrete $G$ gauge theories, and then condensing pairs of charges from neighboring layers; we present a layer construction for general non-Abelian $G$ in Sec.~\ref{sec:non-Abelian_layer_construction}, which is a technical result that may be of independent interest. The condensation causes the point charges to be fully mobile in 3d, and confines the fluxes from different layers into flux strings. Invertible codimension-1 defects in (2+1)D topological phases correspond to anyon permutation symmetries, such that anyons get permuted into other anyons with identical braiding and fusion rules as they cross the defect. We show that only certain invertible codimension-1 defects in a given layer are compatible with the layer construction. After condensation of charges between layers, a compatible invertible codimension-1 defect in (2+1)D becomes a twist string (invertible codimension-2 defect) in the (3+1)D topological phase. (see Fig.~\ref{fig: introduction to layer construction}). The compatibility conditions are that the anyon permutations must (i) satisfy a flux-preserving condition in order to give rise to a twist string in (3+1)D. However, this alone is not enough to guarantee that the properties of the twist string are fully topological in (3+1)D. We find that a second, stronger condition (ii) is necessary: a charge-independence condition that implies the anyon permutation only depends on the flux label. 
 
 Armed with the above perspective on twist strings coming from the layer construction, we use it to construct a non-trivial twist string in non-Abelian $D_4$ and $A_6$ gauge theories. These examples give novel codimension-2 defects that lead to a rich structure of the global symmetry involving the mixture of the non-invertible 1-form symmetry generated by magnetic strings, invertible 1-form symmetry generated by the twist string, and the 2-form symmetry generated by electric particles.
 
 Remarkably, we find that there are examples, e.g., in $A_6$ gauge theory, in which condition (ii) above is violated while (i) is satisfied. In such cases, the twist strings are \it non-topological\rm: their properties have some geometric dependence as illustrated in Fig.~\ref{fig:geometric_defect_intersection}, reminiscent of fracton physics \cite{NH19, PCY20}.

Next, we provide a general description of twist strings from the perspective of gauging lower dimensional SPT phases. In Sec.~\ref{sec:wrapping}, we generally derive the action of the twist strings on the magnetic strings when the twist string corresponds to (1+1)D bosonic symmetry-protected-topological (SPT) phase with $G$ symmetry. We explain how a 1d $G$-SPT, when dimensionally reduced on a circle with $G$ flux, has a $G$ charge in its ground state. This dimensional reduction argument can be used to understand in general why the decorated 1d SPT acts as a twist string that attaches charge to a magnetic string crossing it. In Sec.~\ref{sec:fermiongauge}, we further extend this argument to derive the action of the twist string when it corresponds to a fermionic invertible phase with $\Z_2^f$ or, more generally, a fermionic $G_f$ symmetry. We show explicitly in Sec.~\ref{nonAbelianSPT} that the construction of twist strings from gauging (1+1)D SPTs can always be understood from the layer construction perspective. In Sec.~\ref{subsec:sigmamodularinvariant}, we provide evidence, but not a complete proof, that topological twist strings from the layer construction can always be obtained by gauging (1+1)D invertible phases. 

Yet another perspective on the twist strings, given in Sec.~\ref{sec:highergauge},  is in terms of a so-called ``condensation defect'' of a (3+1)D gauge theory. In general, a condensation defect in codimension-$k$ is defined in Ref.~\cite{Konstantinos2022Higher} as a defect obtained by summing over insertions of defects with codimension higher than $k$, supported on the codimension-$k$ surface. 
In the case of (3+1)D $\Z_2$ and $\Z_2\times\Z_2$ gauge theory, we explicitly derive that the twist string can be described as the condensation of the electric Wilson lines supported on the codimension-2 surface. 

In Sec.~\ref{sec:3group}, we explain how the structure of the global symmetries generated by twist strings, flux strings, and point charges of the (3+1)D discrete Abelian gauge theory together form a 3-group. In particular, the 3-group is partially characterized by a 4-cocycle $H^4(B^2K_1, K_2)$, where $K_1$ corresponds to the 1-form symmetry group generated by the worldsheets of the twist string and magnetic string, and $K_2$ corresponds to the 2-form symmetry generated by the Wilson line of the electric particle and $B^2G$ is the Eilenberg-Maclane space $K(G,2)$. The non-trivial 4-cocycle mathematically expresses the fact that the intersection of the twist string and magnetic string produces a point charge. By utilizing the expression of the twist strings as condensation defects, we show how this 4-cocycle can be computed mathematically. This can be viewed as a higher dimensional version of the symmetry-localization obstructions discussed in the condensed matter physics literature \cite{barkeshli2019,barkeshli2018,FidkowskiVishwanath17}, which corresponds to an $H^3$ class, and which determines an underlying 2-group symmetry of the corresponding (2+1)D topological quantum field theory.

Furthermore, in (3+1)D $(\Z_2)^3$ gauge theory given by three copies of the $\Z_2$ toric code, it is known that there is an interesting $\Z_2$ 0-form symmetry that acts by permuting the magnetic and twist strings in a nontrivial way~\cite{Yoshida15, WebsterBartlett18}. In this case, we derive a rich structure of the 3-group that involves the 0-form, 1-form, and 2-form symmetries realized in the (3+1)D $(\Z_2)^3$ gauge theory. We explicitly present the equations satisfied for closed background gauge fields of this 3-group. 

Finally, an important aspect of the twist strings is that they are breakable, and can exist on finite open segments. In Sec.~\ref{sec:codim3}, we discuss the codimension-3 defects that arise from considering open segments of the twist strings. These codimension-3 defects are non-Abelian defects, in the sense that they give rise to a topologically protected degeneracy that essentially corresponds to the edge zero modes of a gauged (1+1)D invertible phase embedded in the (3+1)D topological order. This gives a mechanism, for example, to have Majorana zero modes dangling at certain points in a 3d space. We define logical operations for topological quantum computation by explicitly constructing membrane operators on the lattice enclosing distinct sets of codimension-3 defects.

We note that non-trivial codimension-1 and codimension-2 defects were also studied in some examples in Ref.~\cite{mesaros2013}. There, the notion of a particular kind of twist defect in (2+1)D, referred to as a genon \cite{barkeshli2010,barkeshli2012a,barkeshli2013genon}, was generalized to (3+1)D. The example of Ref.~\cite{mesaros2013} is properly thought of as a non-trivial codimension-1 defect, whose boundary gives a non-trivial codimension-2 twist defect. In contrast, the codimension-2 defects studied in this work are purely codimension-2, in the sense that they are not the boundary of a non-trivial codimension-1 defect. 

More closely related to the present work are results from the quantum information literature \cite{WebsterBartlett18,Yoshida15,Y16,Yoshida17, Webster:2020, zhu2021topological}. In particular, Refs.~\cite{Yoshida15,Y16,Yoshida17} pointed out the possibility of codimension-$k$ defects decorated by SPT states before gauging and the connection to transversal logical gates.
In the context of the codimension-2 defects decorated by SPTs, which correspond to our twist strings, Refs.~\cite{Yoshida17} emphasized that these strings are unbreakable; however, we show explicitly that these twist strings are indeed breakable (see Sec.~\ref{sec:defect} and \ref{sec:codim3}). Ref.~\cite{WebsterBartlett18} also discussed some examples of twist strings in (3+1)D $\Z_2$ and $\Z_2 \times \Z_2$ gauge theory, and pointed out the appearance of a point charge upon crossing with a flux string in these models. Our work further develops these insights by providing a variety of general constructions and computations valid for arbitrary gauge groups, examples of twist strings involving non-Abelian gauge theories, a wide class of exactly solvable lattice models in the presence of open or closed twist strings, and developing the relationship to higher group symmetries and field theory.

Finally, we note that a special class of codimension-$2$ defects, referred to as ``Cheshire charges," were studied in Ref.~\cite{else2017}. Cheshire charges are examples of non-invertible defects, which arise from breaking the $G$ gauge symmetry down to a subgroup $H$ along a loop. In Appendix \ref{app:cheshire}, we provide a field theoretic description of Cheshire charge in (3+1)D $\Z_2$ gauge theory using the ideas of gauging a lower dimensional spontaneous symmetry breaking phase and a condensation defect. 

\subsection{Remark on terminology}

In this paper, the flux loop excitation in (3+1)D discrete gauge theory is sometimes referred to as a flux string, magnetic string, magnetic loop excitation, and so on. When we refer to a codimension-2 surface operator that corresponds to the worldsheet of the flux loop excitation, we sometimes call it a membrane operator, magnetic surface operator, Wilson surface operator, and so on. Similarly, a line operator that corresponds to the worldline of an electric particle is referred to as a Wilson line operator.

\section{Codimension-k defects: basic concepts and definitions}
\label{sec:defect}
\subsection{Basic definitions}

Let us begin by providing a rough definition of the notion of a codimension-$k$ defect in a topological phase of matter, and some basic properties like invertibility.

Our setup is a quantum many-body system on a Hilbert space $\mathcal{H}$ that decomposes as a tensor product of local Hilbert spaces $\mathcal{H} = \otimes_i \mathcal{H}_i$, corresponding to a discretization of some $d$-dimensional spatial manifold $M^d$. The system is described by a geometrically local Hamiltonian $H$, with a bulk energy gap in the thermodynamic limit. 

For each topologically ordered phase of matter, let us consider a translationally invariant, gapped, local Hamiltonian $H_{0}$ on a $d$-dimensional torus, $T^d$. Next, let us consider a local gapped Hamiltonian $H_{k} = H_{0} + V_{k}$, where $V_{k}$ is a local potential with support on a $T^{d-k} \subset T^d$ submanifold. We require that $V_{k}$ be translationally invariant in the $(d-k)$-dimensions along $T^{d-k}$.\footnote{The requirement of translational invariance on $H_0$ and $V_{k}$ in principle can be replaced with a weaker notion of homogeneity; however, a proper definition of homogeneity is beyond the scope of this work.} The ground states of $H_{k}$ host a codimension-$k$ defect along the support of $V_{k}$. 

Next, we may group codimension-$k$ defects with the same support into topological equivalence classes. Let us consider two different Hamiltonians
$H_{k} = H_{0} + V_{k}$ and $H_{k}' = H_{0} + V_{k}'$, where $V_{k}'$ also has support along the same submanifold $T^{d-k}$ and is translationally invariant. The ground states of $H_{k}$ and $H_{k}'$ host topologically equivalent codimension-$k$ defects if there is an adiabatic path $V_{k}(t)$ without closing the energy gap such that $V_{k}(0) = V_{k}$, $V_{k}(1) = V_{k}'$, and $V_{k}(t)$ only has non-trivial support on the same $T^{d-k}$ submanifold for all $0 \leq t \leq 1$. Otherwise, the codimension-$k$ defects are topologically distinct. 

Alternatively, we may instead adopt a definition based on ground states and finite depth local quantum circuits \cite{zeng2015book}. For example, ground states $|\Psi_{k}\rangle$ and $|\Psi_{k}'\rangle$ of $H_{k}$ and $H_{k}'$ host equivalent defects if there is a local constant depth circuit, with support only on the defect, that approximately converts $|\Psi_{k}'\rangle$ to $|\Psi_{k}\rangle$.
If no ground state of $H_{k}$ can be approximately converted to a ground state of $H_{k}'$ by a local constant depth circuit, then the codimension-$k$ defects are topologically distinct.

We have for convenience defined the codimension-$k$ defects to lie along a $T^{d-k}$ submanifold and to correspond to a ground state of $H_k$ with a translationally invariant potential $V_k$. In general, one can consider the codimension-$k$ defects to lie along any codimension-$k$ submanifold, and one can also weaken the translation invariance condition. We will adopt this more general perspective in this work, although giving a proper definition of topological equivalence classes in such cases is more complicated and beyond the scope of this work.

The codimension-$k$ defects we have defined so far are fully supported on a torus $T^{d-k}$ in a translationally invariant way. We can further consider codimension-$k$ domain walls between distinct codimension-$(k-1)$ defects, and similarly, we can group these domain walls into topological equivalence classes. We will refer to a codimension-$k$ defect as `pure' if it is not a domain wall between non-trivial codimension-$(k-1)$ defects. Alternatively, a 'pure' codimension-$k$ defect can be thought of as a domain wall between \it trivial \rm codimension-$(k-1)$ defects. 

Flux strings and point charges in (3+1)D topological phases provide simple examples of topologically non-trivial pure codimension-2 and codimension-3 defects, respectively. In order to create a flux string along some loop $\gamma$ out of the vacuum, for example, one must apply a membrane operator to a sheet $D^2$ such that $\gamma = \partial D^2$. Similarly, local operators cannot create the non-trivial point charges, and can only be created at the boundaries of Wilson line operators. However, there are in general many more topologically non-trivial defects, beyond simply the well-known point charges and flux strings. 

Finally, we will say that a codimension-$k$ defect in the topological equivalence class $A$ is \it invertible \rm if there exists another codimension-$k$ defect in an equivalence class $\bar{A}$, such that if the two codimension-$k$ defects are near each other, they are topologically equivalent to the trivial codimension-$k$ defect. 

Not all defects are invertible. Simple examples of non-invertible point defects are gauge charges corresponding to higher dimensional irreducible representations of the gauge group $G$. Non-invertible codimension-$2$ defects include flux strings with non-Abelian flux, and `Cheshire' charges (which we discuss in Appendix~\ref{app:cheshire}). Invertible defects include Abelian point charges (corresponding to 1-dimensional irreducible representations of the gauge group), Abelian flux loops, and the twist strings studied in this paper.

In general, the set of topologically distinct codimension-$k$ defects is infinite when $d - k > 2$, since one can always decorate any $(d-k)$-dimensional topologically ordered phase of matter on a $(d-k)$-dimensional submanifold. Nevertheless, the \it invertible \rm codimension-$k$ defects for a given topological order may be finitely generated, similar to the case of invertible topological phases of matter in a given dimension, and thus amenable to a complete description. As a step towards fully describing the categorical 3-group of invertible codimension-$k$ defects, in this paper, we focus mainly on invertible codimension-$2$ defects in (3+1)D topological phases of matter. 

We note that while the familiar flux loops provide a special example of codimension-2 defects in (3+1)D topological phases, they also have a special status among codimension-2 defects. Namely, flux loops are distinguished by the conservation of flux and therefore it is not possible to have a segment of flux terminating on the vacuum. On the other hand, the pure codimension-2 defects studied in this work can be supported on a line segment and terminate on the vacuum.\footnote{Mathematically the domain walls between codimension-2 defects form the 2-morphisms of a fusion 3-category. The above statement can be rephrased as saying that there is no 2-morphism between the flux loops and the trivial loop, in contrast to the twist strings studied in this work.} This property has implications for the dynamics and stability of defects, as discussed below. 

\begin{figure}
    \centering
    \includegraphics[width=0.8\textwidth]{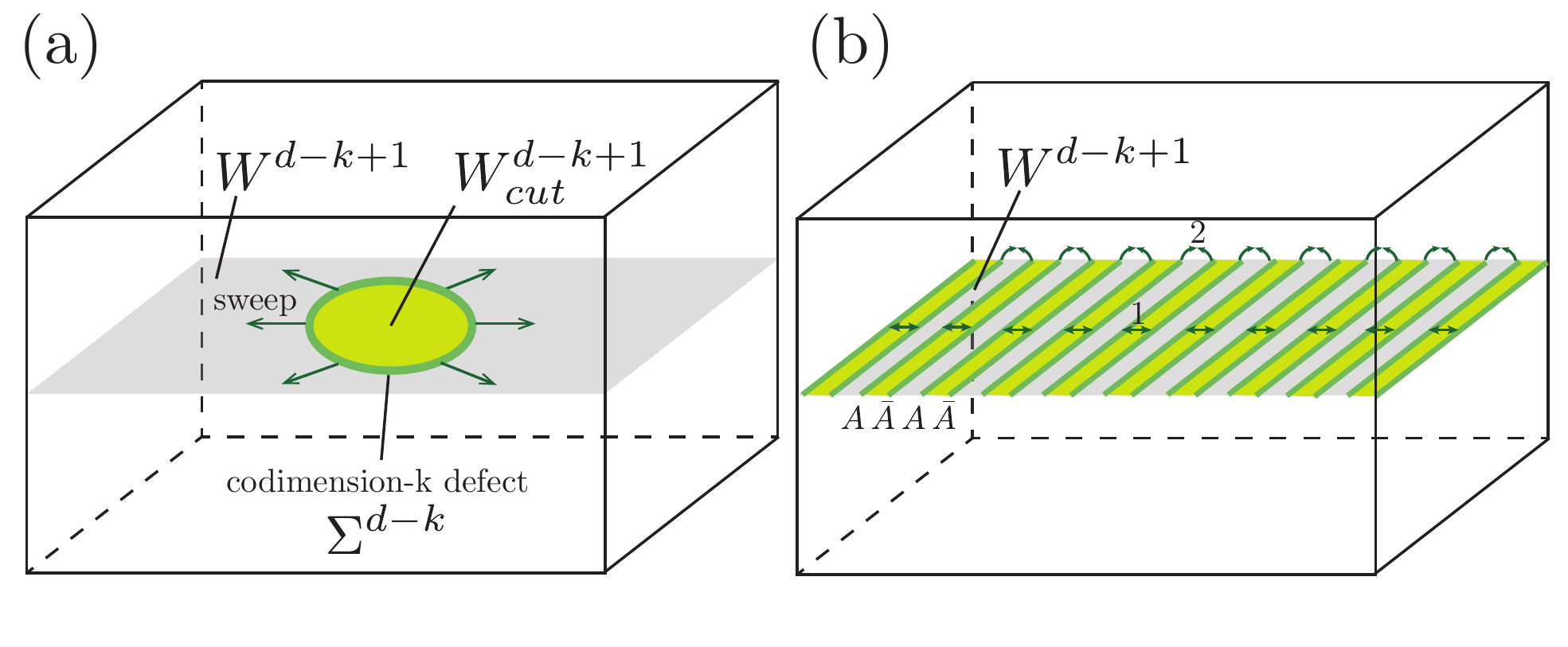}
\caption{(a)  Sweeping the codimension-$k$ defect  on a codimension-$(k-1)$ submanifold (grey sheet)  $W^{d-k+1}$ gives rise to an emergent $(k-1)$-form symmetry.  The swept region is an open manifold $W^{d-k+1}_{cut}$ (light green), while the codimension-$k$ defect (green) is located on its boundary $\Sigma^{d-k} = \partial W_{cut}^{d-k+1}$.  (b) For an invertible defect, the corresponding $(k-1)$-form symmetry operator is a constant-depth local quantum circuit, since the sweeping of the codimension-$k$ invertible defect $A$ can be done in two steps in $O(1)$ time:  (1) Parallel creation of a defect $A$ and its inverse $\bar{A}$ separated with $O(1)$ distance.  (2)  Parallel annihilation of the defect $A$ with its neighboring inverse $\bar{A}$ on the other side. Note that the box in the figure represents a 3-torus with periodic boundary conditions.}
\label{fig:domain_wall_sweeping}
\end{figure}

We have discussed topological equivalence classes for codimension-$k$ defects; however, there is another notion of a topological condition. One can ask whether, in the topological quantum field theory (TQFT) description of the system, the given codimension-$k$ defect is fully topological in the sense that the path integral in the presence of the codimension-$k$ defect only depends on topology. It is possible to consider defects that are topologically non-trivial, yet not fully topological in this way. For example, it is known that the TQFT path integrals for (2+1)D chiral topological orders (with non-zero chiral central charge) are not fully topological and depend on a choice of framing \cite{witten1989}. Similarly, one can envision codimension-$1$ defects in (3+1)D topological orders corresponding to decorating with a chiral (2+1)D topological phase, in which case the corresponding path integral of the TQFT would not be fully topological in the presence of the defect, and would depend on additional geometric structure. It is unclear how to define this field theoretic notion of a topological defect entirely from the perspective of a quantum many-body system. We will in Sec.~\ref{sec:layer} encounter some examples of invertible codimension-$2$ defects (twist strings) which are not fully topological in this field theoretic sense. Unless otherwise stated, the codimension-$k$ defects that we consider in this paper are fully topological in the TQFT sense.

\subsection{Defects, emergent higher symmetries, and logical gates}

There is a correspondence between codimension-$k$ topological defects and ``higher symmetries." An $n$-form symmetry is an example of a higher symmetry. It is an operator supported on a codimension-$n$ subspace, which commutes with the Hamiltonian. By viewing the ground state space as the code space of a topological quantum error correcting code, higher-form symmetry operators correspond to fault-tolerant logical operations on the quantum code \cite{zhu2021topological}. Below we elaborate on the connection between topological defects and higher-form symmetries.

Given a codimension-$k$ defect with support on a null bordant submanifold one can consider nucleating it from the vacuum, sweeping the defect through a closed codimension-$(k-1)$ submanifold $W^{d-k+1}$ of the system\footnote{Given a triangulation of the manifold, defect configurations can be viewed as cocycles. Pachner moves on this triangulation are allowed, which means that defects can be re-connected in arbitrary ways (subject to some self-consistency conditions).}, and then shrinking it back to zero, see e.g. Fig.~\ref{fig:domain_wall_sweeping}(a).
This then gives an operator that maps the ground state subspace of the clean Hamiltonian $H_0$ back to itself. Since the ground state subspace is left invariant, this operation can be thought of as an emergent symmetry with support on $W^{d-k+1}$. 

If the codimension-$k$ defect is invertible, the corresponding symmetry operator is invertible, and therefore defines a group structure. In this case, the invertible codimension-$k$ defect defines a $(k-1)$-form symmetry of the topological quantum field theory. Moreover, in the invertible case this $(k-1)$-form symmetry, or equivalently the logical operation on the code space, can be implemented by a \textit{constant-depth local quantum circuit} on the corresponding lattice model [see Fig.~\ref{fig:domain_wall_sweeping}(b)] (or a \textit{locality-preserving unitary} which can also be defined in a continuum system).  Such locality-preserving unitary maps a local operator to a local operator in the $O(1)$ neighborhood of its support, and its connection to the fault-tolerant logical gate has been previously studied in the context of (2+1)D TQFT \cite{Beverland:2016bi}. Such logical gates are fault-tolerant since errors can only propagate linearly with time according to a `light cone' \cite{Bravyi:2013dx}.  A more restrictive class of such constant-depth local quantum circuits or locality-preserving unitaries, which correspond to \textit{on-site}   
 symmetries, are the \textit{transversal logical gates} \cite{ WebsterBartlett18,Yoshida15,Y16,Yoshida17, Zhu:2017tr, zhu2021topological}, which can be expressed as a product of local unitaries, i.e., $\otimes_j U_j$, and hence  correspond to depth-1 circuits.  Such logical gates are even more desirable since they do not spread errors within each code block. Certain examples of higher-form symmetries discussed in this paper belong to this class.   

On the other hand, given a symmetry operator with support on a closed submanifold $W^{d-k+1}$, one can truncate it to an open submanifold $W_{cut}^{d-k+1}$. The resulting state is then locally in the ground state everywhere except for the boundary $\Sigma^{d-k} = \partial W_{cut}^{d-k+1}$. In this way, emergent symmetry operators with support on codimension-$(k-1)$ manifolds in space define codimension-$k$ defects.

If we only focus on the invertible defects, the higher symmetry that they define forms the structure of a higher group. If we instead consider non-invertible defects in a $(d+1)$D topological phase as well, then we expect to obtain the structure of a fusion $d$-category. The codimension-$k$ defects are expected to correspond to $(k-1)$-morphisms of the $d$-category. 

\subsection{Dynamics and energetics}

Let us consider a Hamiltonian $H_0$ whose gapped ground state has no defects. Consider the average energy $E_k = \langle \psi_k | H_0 | \psi_k \rangle$ for a state $|\psi_k\rangle$ that has a pure codimension-$k$ defect. On general grounds we expect that $E_k$ grows with the volume of the defect, because $|\psi_k\rangle$ is the ground state of a Hamiltonian that differs from $H_0$ by a defect potential $V_k$ which has support on the defect. This means that for a defect with characteristic linear size $L$, the codimension-$k$ defect will generically have an energy cost $\propto L^{3-k}$. From this perspective, the energetic cost of all loop-like defects in (3+1)D topological phases is linear in the length of the loop, similar to the conventional flux loops. 

However, there is an important distinction among defects, which is that of stability. Flux loops are topologically stable, in the sense that if a flux loop is created and then evolved according to the Hamiltonian, conservation of flux requires that the loop cannot break apart into small open line segments. Nevertheless, generically a flux loop can decay into distinct loops, potentially carrying different fluxes, as long as total flux is conserved. Similarly, some point-like particles may be fully stable, and others may be unstable and decay into more stable point-like particles. Whether these decays occur and which excitations are stable depends on local energetics that is not universal and depends on details of the underlying Hamiltonian. 

The kinds of codimension-2 defects that we study may be even less stable than the flux loops, since there is no flux conservation that forbids the loop from breaking up into distinct smaller line segments that can propagate individually. Whether the loop indeed breaks up or is stable depends again on local energetics and non-universal microscopic details of the underlying Hamiltonian. 
It is therefore important to note that the notion of a non-trivial codimension-$k$ defect is somewhat distinct from the question of its dynamical stability, which is an interesting question to study further in specific models. 

\section{Exactly solvable models for twist strings in (2+1)D toric codes} \label{sec: Exactly solvable models for defects in (2+1)D toric codes}

In this section, we provide lattice constructions for twist strings (invertible codimension-1 defects) in (2+1)D toric codes. Our constructions are closely related but not identical to lattice models previously provided in the literature (see e.g. \cite{bombin2010,KitaevKong12,you2012,you2013,Yoshida15,Y16,Kesselring18}). In particular, we construct all twist strings for both $\ZZ_2$ toric code and $\ZZ_2 \times \ZZ_2$ toric code from the perspective of gauging (1+1)D bosonic and fermionic SPT phases that are decorated along a codimension-1 submanifold \cite{Yoshida15,Y16}.\footnote{Note that gauging (1+1)D bosonic and fermionic SPT phases do not produce all defects for $\mathbb{Z}_N$ toric codes. For example, the $e \leftrightarrow m$ twist string in the $\mathbb{Z}_3$ toric code does not arise from gauging bosonic or fermionic SPT phases. It would require $\mathbb{Z}_3$ parafermions; however, the bosonization of parafermions in (2+1)D hasn't been developed yet.}

\subsection{Twist strings in the $\ZZ_2 \times \ZZ_2$ toric code}

\subsubsection{Review of gauging 0-form symmetries}

\begin{figure}[htb]
    \centering
    \includegraphics[width=0.8\textwidth]{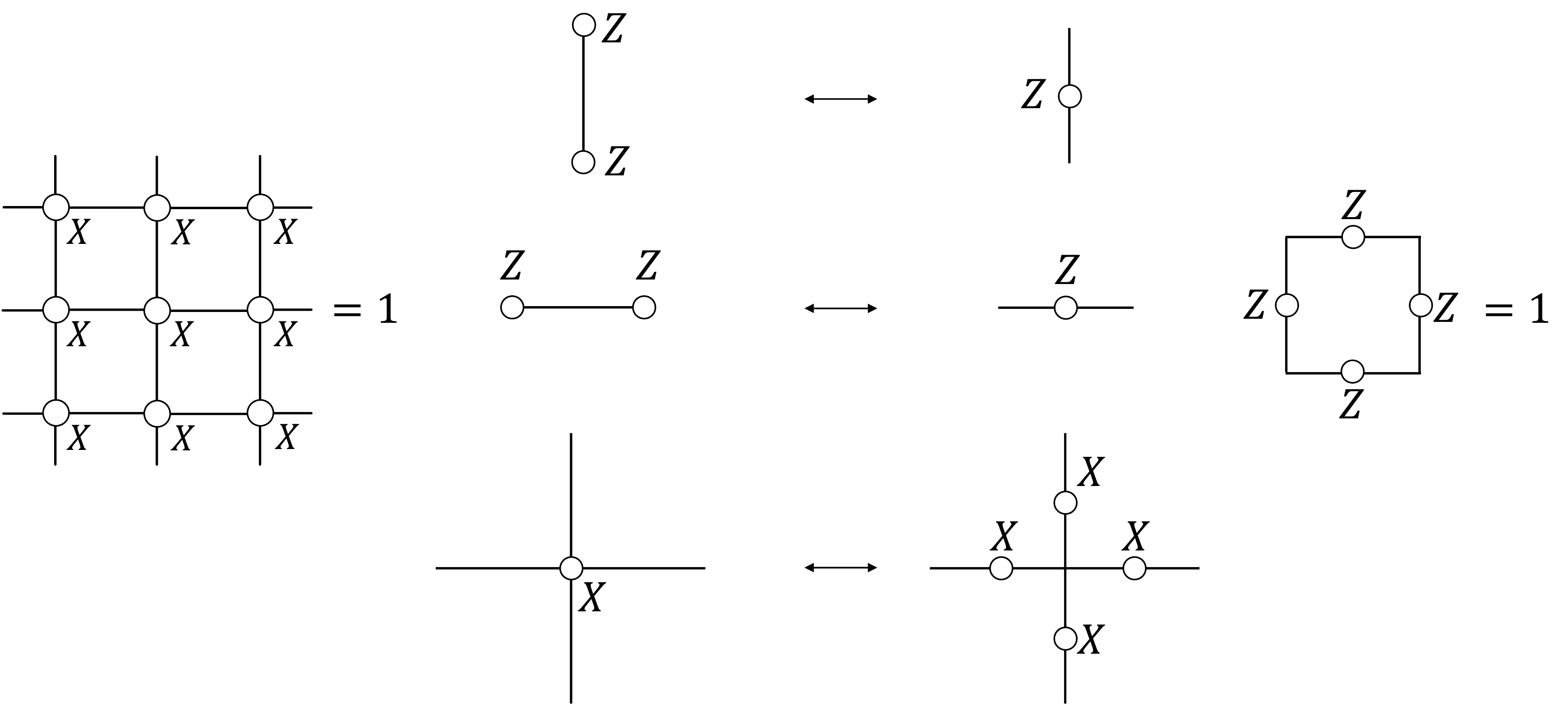}
\caption{Gauging $\ZZ_2$ symmetry \cite{Y16}. Left: each dot represents a qubit on a vertex. We consider the symmetric sector of the Hilbert space: $\prod_v X_v = 1$. Symmetric operators are generated by the single $X_v$ and the product of adjacent $Z_v$. Right: The Hilbert space contains qubits at all edges, with the gauge constraint $\prod_{e\subset f} Z_e = 1$ for each face $f$. For non-simply connected manifolds, there are additional constraints that the product of $Z_e$ along any cycle equals $+1$.}
\label{fig: 2d gauging}
\end{figure}

We first review the procedure of gauging 0-form $\ZZ_2$ symmetries in Ref.~\cite{Y16}, which is analogous to the Kramers-Wannier duality \cite{Kogut1979}. First, we start with a Hamiltonian respecting a global $\ZZ_2$ symmetry $\prod_v X_v$ (the Hilbert space is formed by qubits at vertices). The terms in the Hamiltonian are generated by $X_v$ and $Z_{v} Z_{v'}$. Then, we reformulate the $\Z_2$ symmetric subspace of the full Hilbert space and the symmetric operators in terms of new degrees of freedom, summarized in Fig.~\ref{fig: 2d gauging}. Before gauging the $\ZZ_2$ symmetry (left side of Fig.~\ref{fig: 2d gauging}), the symmetric subspace of the full Hilbert space consists of a tensor product of qubits placed at vertices and a global $\ZZ_2$ symmetry constraint $\prod_v X_v = 1$. After gauging (right side of Fig.~\ref{fig: 2d gauging}), the dual Hilbert space has qubits on the edges and local gauge constraints $\prod_{e \subset f} Z_e = 1$ in addition to one-form symmetry constraints $\prod_{e \subset \gamma} Z_e = 1$ for all closed loops $ \gamma \in Z_1(M,\ZZ_2)$ (where $Z_1$ denotes closed 1-cycles). If we start with the paramagnetic fixed point of the transverse field Ising model $H = - \sum_v X_v$, the dual Hamiltonian realizes the $\ZZ_2$ toric code model after $\ZZ_2$ symmetry.\footnote{We impose the gauge constraint $\prod_{e \subset f} Z_e = 1$ energetically, which realizes the $Z$-plaquette term in the gauged Hamiltonian.}

\subsubsection{1d cluster state and the $m_1 \ra m_1 e_2, ~ m_2 \ra m_2 e_1$ twist string}
\label{subsubsec:clusterdefect}

\begin{figure}[htb]
    \centering
    \includegraphics[width=0.95\textwidth]{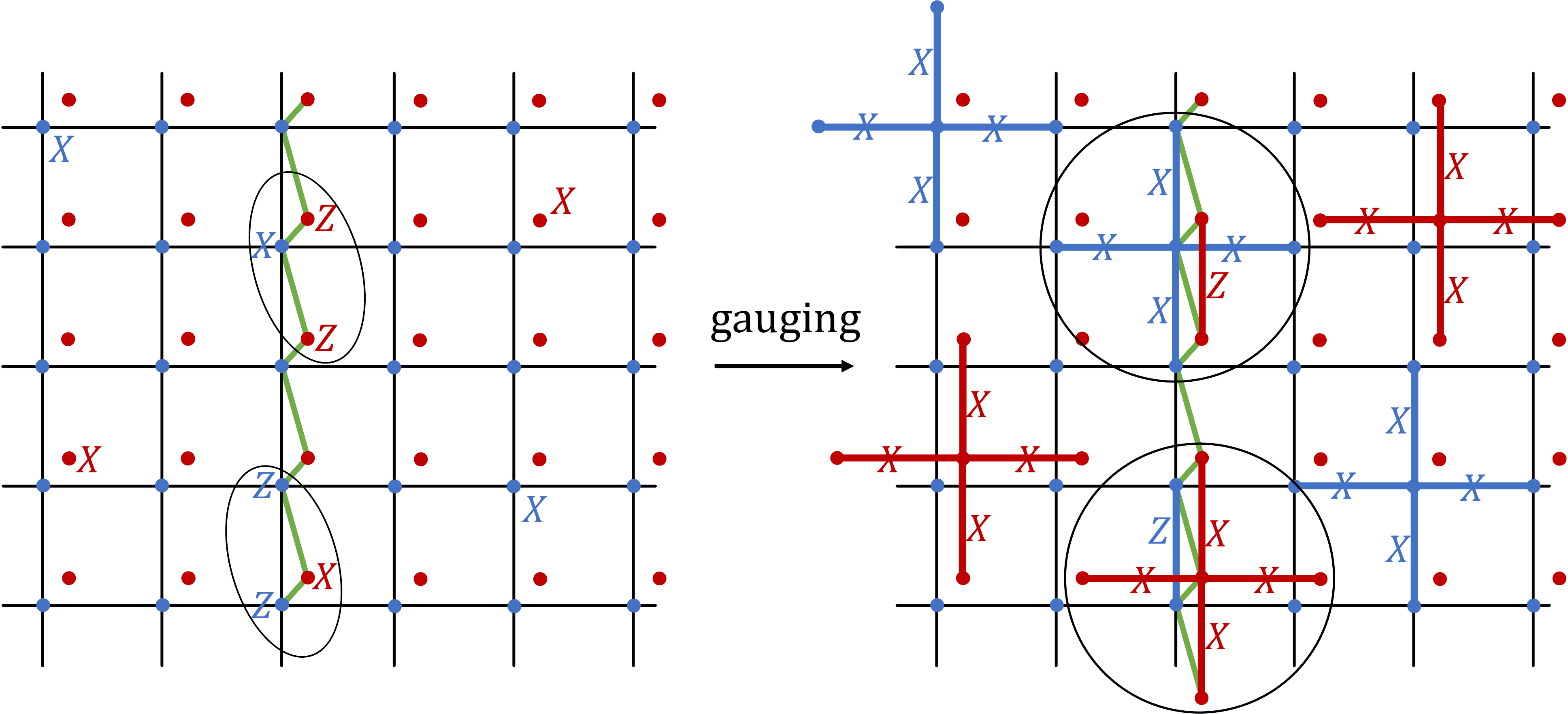}
\caption{Left: each blue or red dot represents a qubit. The cluster state is decorated on the defect (green line), which makes the Hamiltonian having Pauli $X$ coupled to adjacent Pauli $Z$; otherwise, each dot gives Pauli $X$ to the Hamiltonian. Right: after gauging $\ZZ_2 \times \ZZ_2$ symmetry, the singe Pauli $X$ away from the defect becomes the $X$-star term. On the twist string, the $X$-star term is dressed with an additional $Z$ of a different color due to the cluster state Hamiltonian.}
\label{fig: 2d gauging cluster state}
\end{figure}

This example was first shown in Ref.~\cite{Y16}, which utilizes a 3-colorable 2d triangulation. Instead of the triangular lattice, we use the square lattice for convenience and give a direct generalization to the codimension-2 defect on the 3d cubic lattice in Sec.~\ref{subsec:3dZ2Z2exact}.

\begin{figure}[htb]
    \centering
    \includegraphics[width=0.8\textwidth]{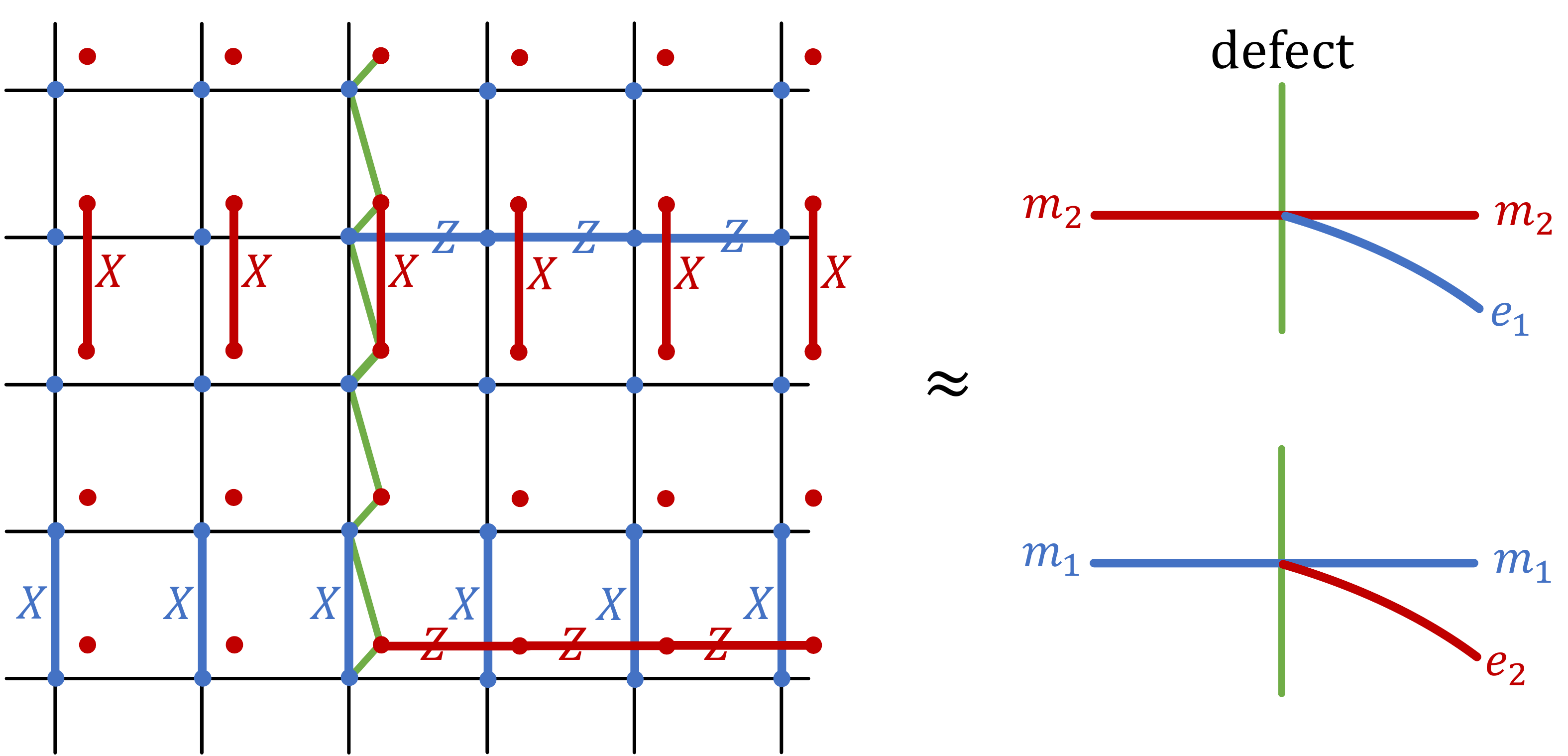}
\caption{The string operators that commute with all terms except near endpoints. When the flux $m_1$ ($m_2$) crosses the twist string, the charge $e_2$ ($e_1$) appears.
}
\label{fig: string operator m1-m1e2 defect}
\end{figure}
We begin with two copies of the paramagnetic fixed point of the 2d Ising model $H_0 = -\sum_v X_v$. Then, we decorate a closed loop with the 1d cluster state. Namely, we modify the product state by acting with a local unitary $U$ creating the cluster state \textit{only} along this closed loop. This means that the state is the ground state of a modified Hamiltonian $UH_0U^\dagger$, which only differs from the original Hamiltonian near the closed loop. The stabilizer terms of this modified Hamiltonian are shown on the left of Fig.~\ref{fig: 2d gauging cluster state} (the blue and red dots are referred to as species 1 and 2, respectively). After gauging $\ZZ_2 \times \ZZ_2$ symmetry, the terms in the dual Hamiltonian are drawn in the right of Fig.~\ref{fig: 2d gauging cluster state} where the stabilizers are given by those of the usual toric code away from the closed loop but differ along the loop (green line). This defect in the toric code is what we refer to as the twist string. We note that the $Z$-plaquette terms for red and blue qubits, which realizes the $\ZZ_2$ gauge constraints are unmodified throughout. This twist string has a direct consequence on how anyons -- which are codimension-2 defects -- move. The $e$-anyon generated by the red or blue string of $Z$ operators can freely pass the twist string as in the usual toric code. However, for $m$-anyons generated by a string of red or blue $X$, as it passes through the twist string, it will violate the $X$-vertex term for the other color. Therefore, passing an $m$-string through the twist string results in the accompaniment of an $e$-string of the other color coming out from the intersection of the string and the twist string. The explicit string operators of this process are shown in Fig.~\ref{fig: string operator m1-m1e2 defect}. To conclude, the anyons transform (i.e. get permuted) as they travel through the twist string:
\begin{eqs}
    e_1 &\ra e_1, \\
    e_2 &\ra e_2, \\
    m_1 &\ra m_1 e_2,\\
    m_2 &\ra m_2 e_1.
\label{eq: 2d m1-m1e2 defect}
\end{eqs}

Let us also present a complimentary perspective where we derive the above permutation using quantum circuits. To demonstrate the circuit for the cluster state and its gauged version, we consider a small loop of the cluster state generated by a product $CZ$ gates:
\begin{eqs}
    \includegraphics[width=0.6\textwidth, valign=c ]{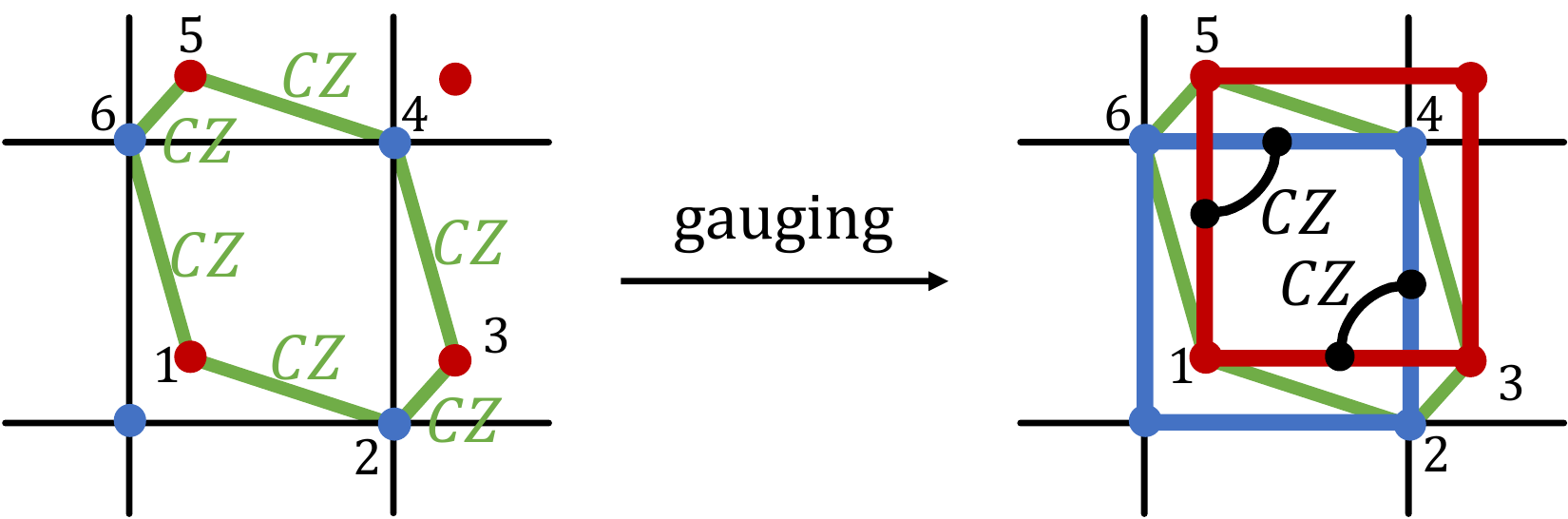}
    \quad ,
\label{eq: gauging CZ loop}
\end{eqs}
To unpack the above equation, we first recall that the exact unitary used to create the cluster state is given by $U=\prod CZ$, consisting of a product of Controlled-Z gates around the loop. In the computational basis $\ket{a_1,a_2}$ where $a_1,a_2=0,1$, $CZ=(-1)^{a_1a_2}$. Therefore, around a closed loop consisting of sites $1$ to $6$ as shown above, we have that the phase factor assigned to this closed loop is $$a_1a_2 + a_2a_3 + a_3a_4 +a_4 a_5 + a_5 a_6 + a_6 a_1 = (a_1 + a_3) (a_2 + a_4) + (a_1 + a_5) (a_4 + a_6) \ \text{(mod 2)}.$$ 
The latter expression is composed of brackets that are invariant under the $\ZZ_2\times \ZZ_2$ symmetry, and therefore allows us to directly gauge the unitary $U$. The sign assigned to the dual unitary is therefore
$$ (-1)^{a_{13}a_{24} + a_{15}a_{46}} = CZ_{13,24}CZ_{15,46} $$

We can now use this to create the cluster state along the vertical line in Fig.~\ref{fig: 2d gauging cluster state} by applying the small $CZ$ loops on all faces in the right half plane. After gauging $\ZZ_2 \times \ZZ_2$ symmetry, the original $m_1$/$m_2$ (blue/red) excitations generated by $X_1$/$X_2$-string operators are now conjugated by RHS of Eq.~\eqref{eq: gauging CZ loop} on the half plane, which will induce a $Z_2$/$Z_1$ string. Hence, the intersection of the $m_1$/$m_2$ string with the twist string must result in an extra $e_2$/$e_1$ string coming out of the intersection.

We remark that in the $\ZZ_2 \times \ZZ_2$ toric code, the twist string corresponding to the cluster state along with the $e_1 \leftrightarrow m_1$ and $e_2 \leftrightarrow m_2$ twist strings generate all possible automorphism of anyons, which is proven in Appendix~\ref{sec: Z2 x Z2 automorphism}. Concretely, the group of all possible automorphisms of anyons in the $\ZZ_2 \times \ZZ_2$ toric code is isomorphic to the real Clifford group on two qubits, which is generated by the Hadamard gates $H_1$, $H_2$ and the controlled-Z gate $CZ_{12}$~\cite{Yoshida15,Kesselring18}.\footnote{In fact, this correspondence is exact at the level of logical qubits in two copies of the surface code.}
In the next subsection, we will now discuss the $e \leftrightarrow m$ twist string in the $\ZZ_2$ toric code, which will complete the discussion of lattice constructions for invertible defects in layers of $\ZZ_2$ toric codes.

\subsection{Twist strings in the $\ZZ_2$ toric code}\label{sec: 2d defect}

\begin{figure}[htb]
    \centering
    \includegraphics[width=0.95\textwidth]{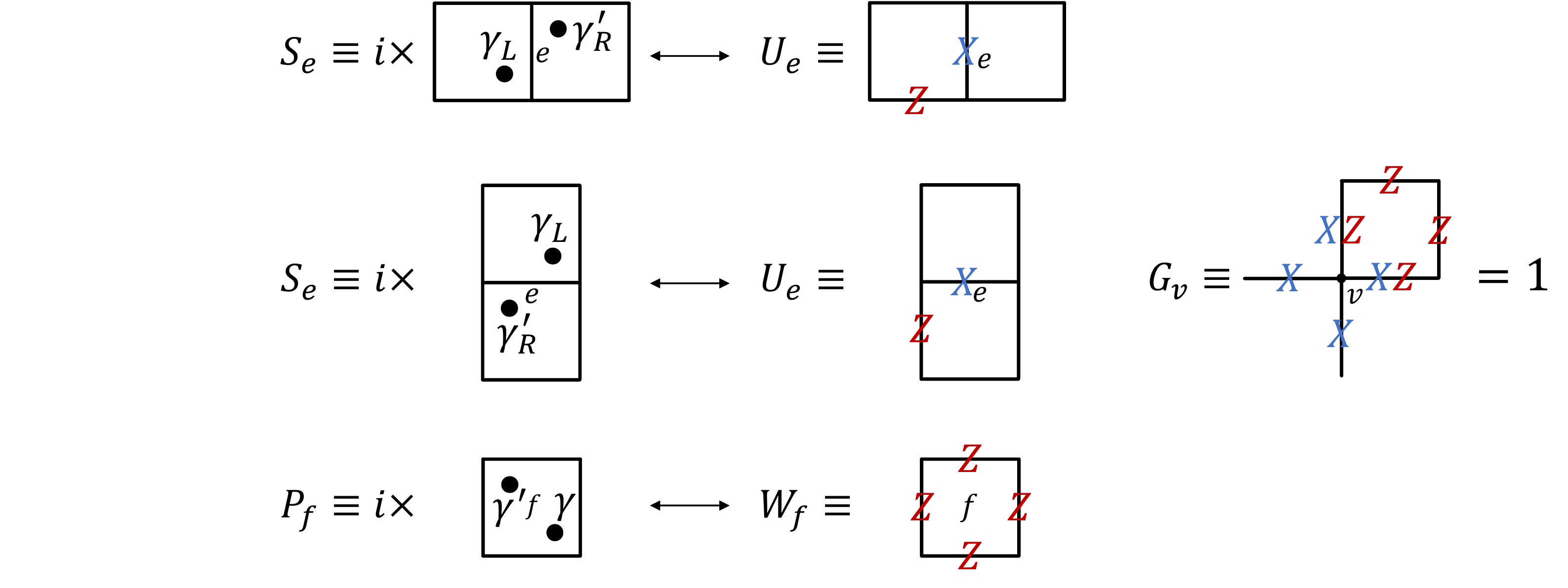}
\caption{Gauging $\ZZ_2^f$ ($=\prod_{f} (i \g_f^\prime \g_f)$) symmetry. This is the summary of 2d bosonization on the square lattice. There are two Majorana fermions $\g_f$, $\g_f^\prime$ at each vertex $f$ on the left-hand side, and a qubit on each edge $e$ on the right-hand side, described by the Pauli matrices $X_e$, $Y_e$, $Z_e$. The Majorana hopping operator $S_e = i \g_{L} \g'_{R}$ ($L$/$R$ is the face left/right to $e$)
and the on-site fermion parity operator $P_f = i \g'_f \g_f$ defined in the figure generate all $\ZZ_2^f$ symmetric operators.
$S_e$ and $P_f$ are mapped to Pauli operators $U_e$ and $W_f$ shown in the figure. Note that the bosonic side is required to satisfy the gauge constraint $G_v$ on each face. The fermionic side requires even fermion parity: $\ZZ_2^f = 1$.}
\label{fig: 2d bosonization}
\end{figure}

We give the lattice construction of the $e \leftrightarrow m$ twist string in the 2d toric code by gauging fermion parity of the Kitaev chain decorated along a 1-cycle of a 2d trivial fermionic theory (an atomic insulator). Gauging fermion parity symmetry on the lattice is described by 2d bosonization \cite{CKR18, CX22}. The isomorphism of local operators is summarized in Fig.~\ref{fig: 2d bosonization}, and is reviewed in Appendix~\ref{sec: review of 2d bosonization}.  It is worth noting the similarities to gauging a bosonic $\ZZ_2$ symmetry in Fig.~\ref{fig: 2d gauging}.

\begin{figure}[htb]
    \centering
    \includegraphics[width=0.9\textwidth]{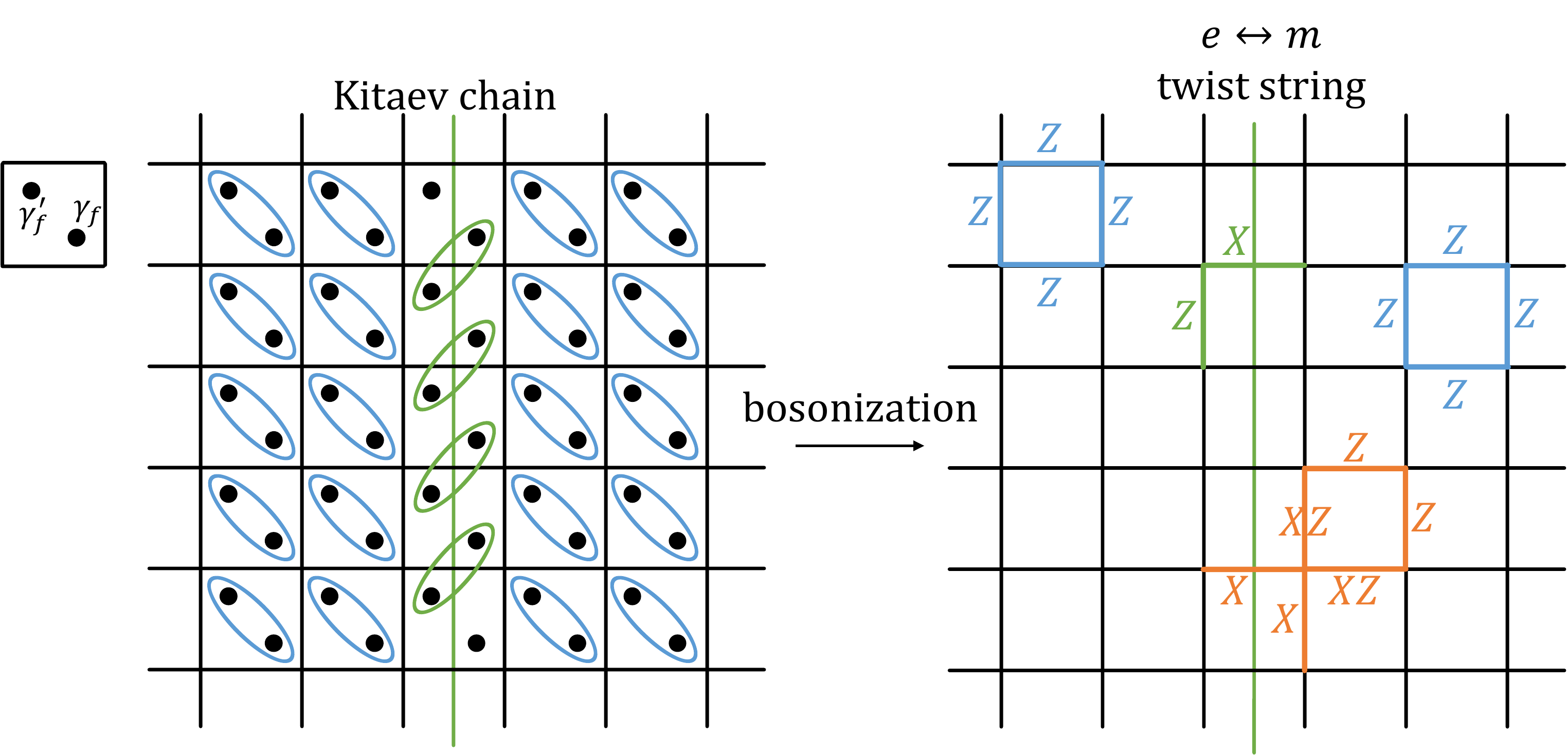}
\caption{Left: each dot represents a Majorana fermion. The Kitaev chain is decorated on the green line, which makes the Majorana fermion paired with the other one on its adjacent face; otherwise, Majorana fermions are paired within each face and form an atomic insulator. Right: after bosonization, the pairing term on the Kitaev chain becomes the hopping operator $U_e$ on the twist string, while the pairing term in each face not on the Kitaev chain becomes the $Z$ fluxes. Note that the gauge constraint $G_v = 1$ holds at all vertices.}
\label{fig: Kitaev chain and e-m defect}
\end{figure}

Now, we set up the construction of the defect on a 2d square lattice. We place a complex fermion on each face, which can be decomposed into two Majorana fermions $\g_f, \g^\prime_f$ as shown in Fig.~\ref{fig: Kitaev chain and e-m defect}. The stabilizers of the atomic insulators are given by $P_f = i  \g^\prime_f \g_f$, which pair up the Majorana fermions within the square. However, along the specified 1-cycle (green line), we decorate a Kitaev chain by instead pairing the Majorana fermion across the edges. This is represented by the operator $S_e = i \g_{L(e)} \g^\prime_{R(e)}$, where $\g_{L(e)}$ and $\g^\prime_{R(e)}$ correspond to Majorana fermions in faces left to and right to the edge $e$.
We then perform 2d bosonization (gauging fermion parity described in Fig.~\ref{fig: 2d bosonization}) on this system.
The $P_f$ term away from the green line is mapped to the flux term $W_f = \prod_{e \subset f} Z_e$, and the $S_e$ term on the green line is mapped to the hopping operator $ U_e = X_e Z_{e'}$ ($e'$ is southwest to $e$), as shown in Fig.~\ref{fig: Kitaev chain and e-m defect}. Note that bosonization will require the gauge constraints $G_v = 1$ (defined in Fig.~\ref{fig: 2d bosonization} or Eq.~\eqref{eq:gauge constraint at vertex}) at all vertices. Next, we are going the study the property of this defect line and show that it corresponds to the $e \leftrightarrow m$ twist string.

\begin{figure}[H]
    \centering
    \includegraphics[width=0.8\textwidth]{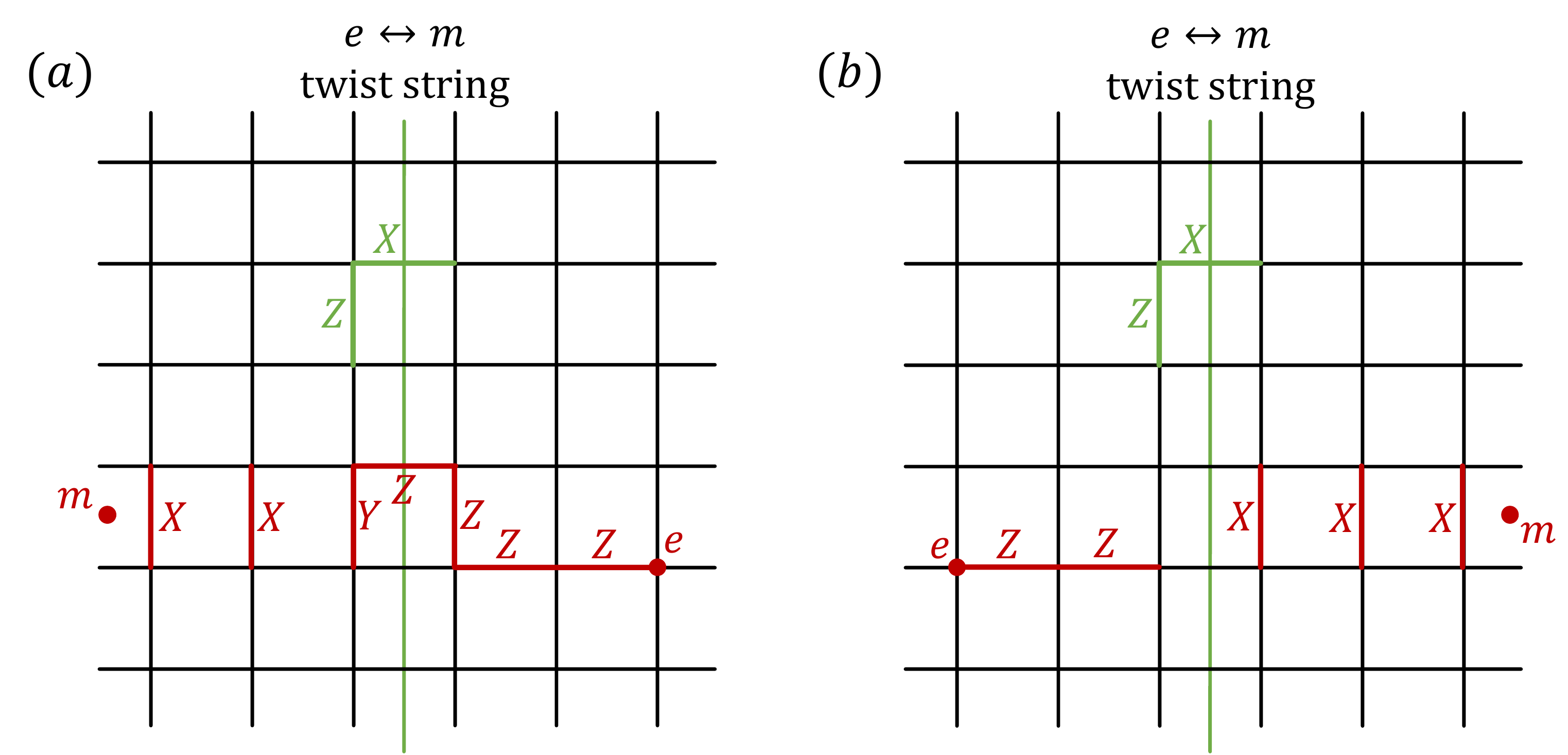}
\caption{The string operators that commute with all terms except near endpoints. We can see that $e$ and $m$ excitations are exchanged when it crosses the twist string.}
\label{fig: e-m defect string operator}
\end{figure}

\begin{figure}[H]
    \centering
    \includegraphics[width=0.5\textwidth]{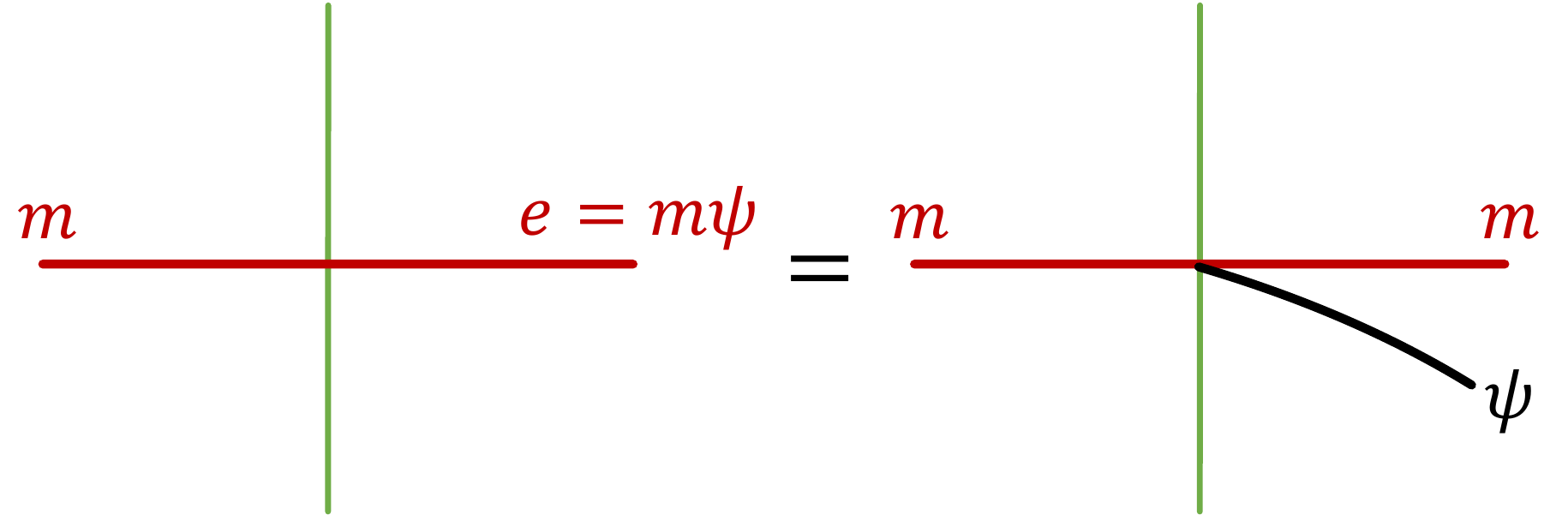}
\caption{We can think of this $e \leftrightarrow m$ twist string as the $\psi$ line coming out from the intersection of $m$-line and the twist string.}
\label{fig: 2d defect}
\end{figure}

Outside the defect, the gauge constraints and the flux terms project the ground state to be the toric code ground state, i.e., the star term $\prod_{e \supset v} X_e$ obtained from the product of $G_v$ and $W_f$. Within each toric code region, we have standard $e$ and $m$ excitations, violating the star term and the flux term, respectively. On the other hand, we can also consider string operators across the defect, shown in Fig.~\ref{fig: e-m defect string operator}. Note that the string operators only violate a finite number of terms near its two endpoints, and commute with all terms in its middle. In particular, the string operators are designed for commuting with
\vspace{-0.5cm}
\begin{align}
    U_e=
    \raisebox{0.3cm}{\includegraphics[width=0.1\textwidth, valign=c]{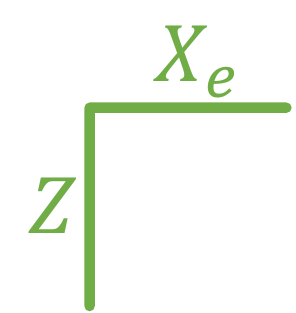}}
\end{align}
on the defect. An important property of these string operators is that the $e$ excitation on one side becomes the $m$ excitation on the other side. Therefore, this defect is the $e \leftrightarrow m$ twist string that permutes the anyons in the toric code. The $e \leftrightarrow m$ twist string can be thought of as the fermion line coming out from the intersection of the defect and the $m$ line, shown in Fig.~\ref{fig: 2d defect}.
\begin{figure}[H]
    \centering
    \includegraphics[width=1.0\textwidth]{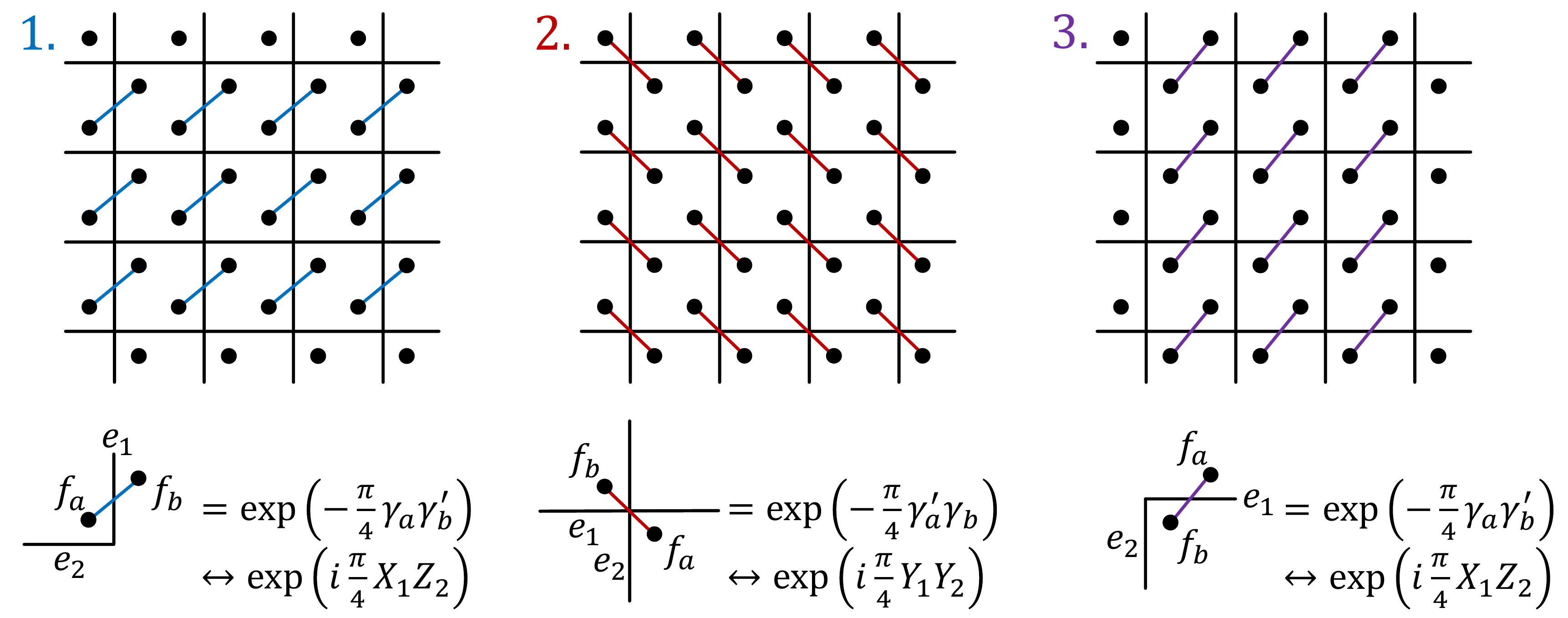}
\caption{The operator pumping the Kitaev chain or the $e \leftrightarrow m$ twist string. The circuit has three steps, and each step is a product of swap operations of Majorana fermions.}
\label{fig: large em defect}
\end{figure}
Similar to the previous $\ZZ_2 \times \ZZ_2$ cluster state case, this defect can be generated by the product of local operators (however, note that they do not give a transversal logical gate). The operators consist of three steps, shown explicitly in Fig.~\ref{fig: large em defect}. Consider applying the above  three steps in a small region
\begin{eqs}
    \includegraphics[scale=0.6]{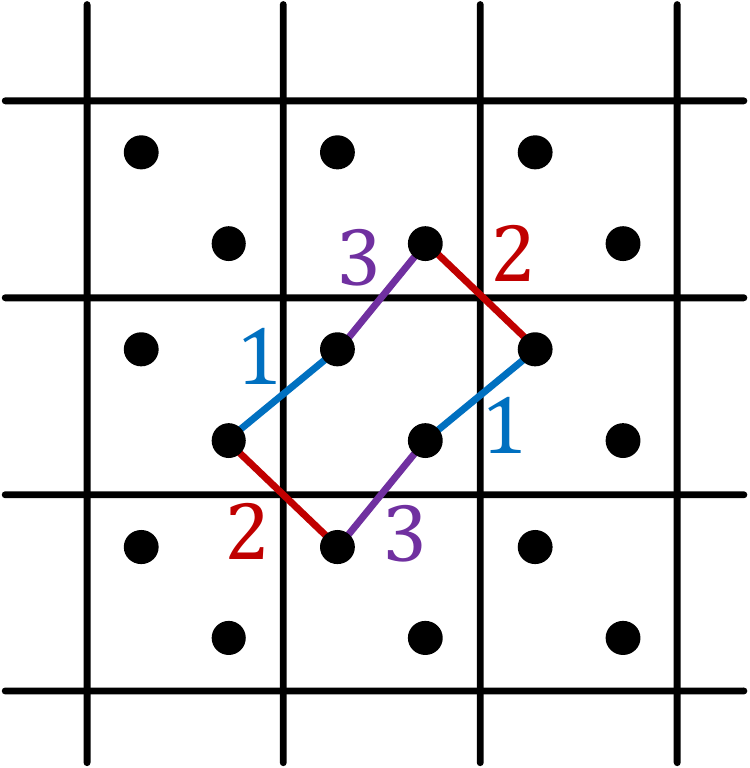}
\end{eqs}
and observe that its boundary hosts the Kitaev chain (before bosonization):
\begin{eqs}
    \includegraphics[scale=0.5]{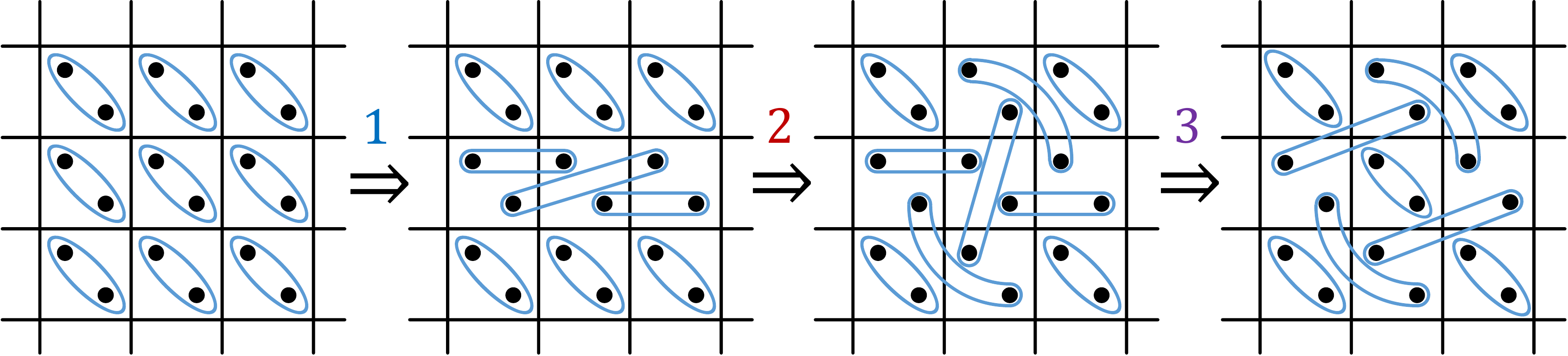} ~.
\end{eqs}
As shown explicitly in the figure, the action of the operator pumps the Kitaev chain to the boundary of the region. After gauging, we therefore pump the $e \leftrightarrow m$ twist string to the boundary. The unitary used to perform the pumping is closely related to the Floquet unitary discussed in Refs.~\cite{Po2017, FPPV19,PotterMorimoto16}.

\section{Exactly solvable models for twist strings in (3+1)D toric codes}
\label{sec:exact3d}

\subsection{Twist string in the (3+1)D $\ZZ_2 \times \ZZ_2$ toric code}
\label{subsec:3dZ2Z2exact}

\begin{figure}[ht]
    \centering
    \includegraphics[width=0.95\textwidth]{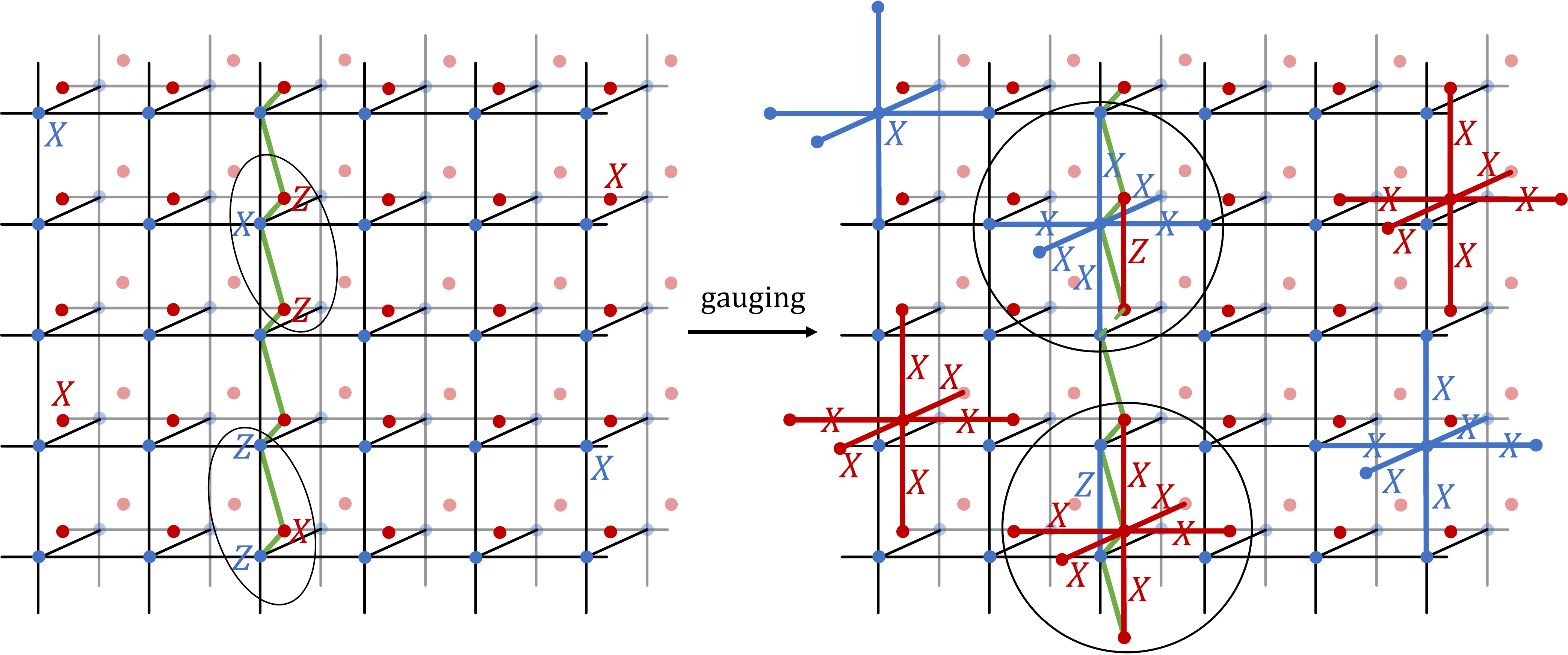}
\caption{Left: each blue or red dot represents a qubit. The cluster state is decorated on the green line, which makes the Hamiltonian have Pauli $X$ coupled to adjacent Pauli $Z$; otherwise, each dot gives Pauli $X$ to the Hamiltonian. Right: after gauging $\ZZ_2 \times \ZZ_2$ symmetry, the singe Pauli $X$ becomes the $X$-star term. On the green line, the $X$-star term is dressed with an additional $Z$ of a different color due to the cluster state Hamiltonian.}
\label{fig: 3d cluster state to codim 2 defect}
\end{figure}

In this section, we construct the twist string (invertible codimension-2 defect) in the 3d $\ZZ_2 \times \ZZ_2$ toric code by gauging the 1d $\ZZ_2 \times \ZZ_2$ cluster state. Gauging global $\ZZ_2$ symmetry in the 3d cubic lattice is similar to Fig.~\ref{fig: 2d gauging}, where the only differences are that the star term contains Pauli $X$ on edges in the third direction, and the zero-flux condition holds for all faces in three directions.

Before gauging $\ZZ_2 \times \ZZ_2$ symmetry, we decorate the 1d cluster state in a cubic lattice, shown as Fig.~\ref{fig: 3d cluster state to codim 2 defect}. At each vertex, there is a blue dot and a red dot, representing two qubits. Away from the defect (green line), the Hamiltonian contains a single Pauli $X$ on each blue or red qubit.
On the defect line, the Hamiltonian term on each qubit $i$ becomes $Z_{i-1} X_i Z_{i+1}$
where $i-1$ and $i+1$ represent the adjacent qubits on the defect.
This Hamiltonian has a global $\ZZ_2 \times \ZZ_2$ symmetry, corresponding to the product of $X_i$ on all blue qubits or all red qubits respectively.
This Hamiltonian is obtained by conjugation of a unitary operator $U \equiv \prod_{i \in \text{defect}} CZ_{i, i+1}$ on the trivial Hamiltonian $H_0 = - \sum_{i} X_i$ with $i$ summed over all qubits.

\begin{figure}[H]
    \centering
    \includegraphics[width=0.8\textwidth]{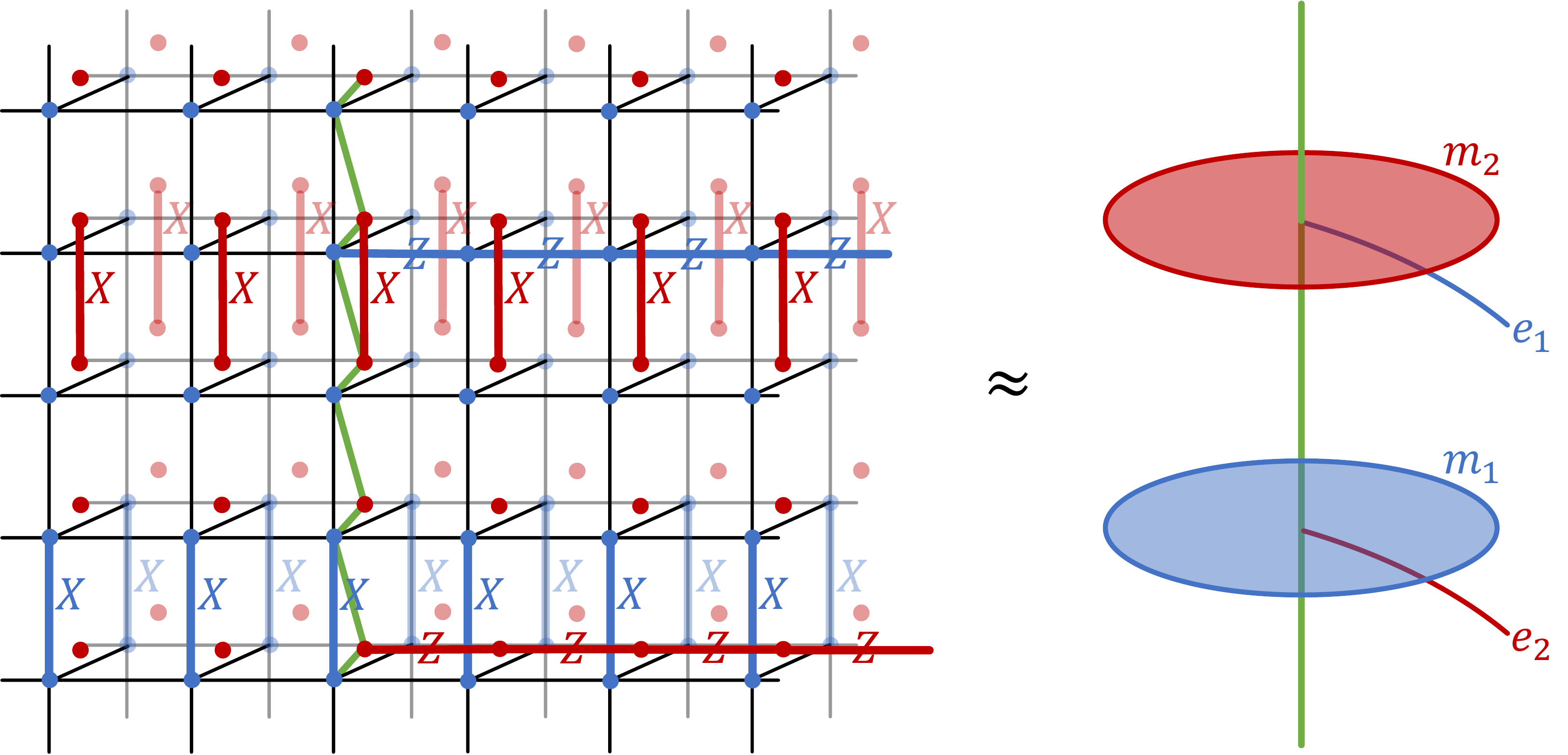}
\caption{The membrane operators that commute with all terms except near boundaries. When the $m_1$ ($m_2$) membrane operator intersects with the twist string, the string operator of charge $e_2$ ($e_1$) comes out from the intersection point.
}
\label{fig: membrane operator m1-m1e2 defect}
\end{figure}

Next, we can gauge the $\ZZ_2 \times \ZZ_2$ symmetry of this Hamiltonian, and the defect becomes a twist string in (3+1)D $\ZZ_2 \times \ZZ_2$ gauge theory. Away from the twist string, the Pauli $X_i$ becomes the $X$-star term, shown in Fig.~\ref{fig: 3d cluster state to codim 2 defect}. Blue and red qubits are decoupled in this region as expected. On the other hand, on the twist string, $Z_{i-1} X_i Z_{i+1}$ becomes an $X$-star term dressed with an additional $Z$ of the other color, shown in Fig.~\ref{fig: 3d cluster state to codim 2 defect}, which is simply the 3d generalization of Fig.~\ref{fig: string operator m1-m1e2 defect}. Given the lattice Hamiltonian, we can study its excitations. The special property of this twist string is that when a $m_1$-loop crosses the twist string, a charge $e_2$ comes out. The explicit membrane operator is provided in Fig.~\ref{fig: membrane operator m1-m1e2 defect}. The membrane operator only anti-commutes with Hamiltonian terms near its boundary, and commutes with all terms in its interior, including terms at the twist string. To achieve this, an $e_2$-line must appear from the intersection of the $m_1$-membrane and the defect. Similarly, an $e_1$-line comes from the intersection of the $m_2$-membrane and the defect. This is the 3d version of the 2d anyon permutation Eq.~\eqref{eq: 2d m1-m1e2 defect}.

\subsection{Twist string in the (3+1)D toric code with a fermionic charge}
\label{subsec:3dZ2exact}

Similar to the 2d case, we decorate the 1d Kitaev chain along a 1-cycle in a trivial fermionic system (atomic insulator), and gauge fermion parity symmetry of the whole system. Gauging fermion parity symmetry on the cubic lattice is described by 3d bosonization \cite{CK19, C20}. The isomorphism of local operators is summarized in Fig.~\ref{fig: 3d bosonization}, and is reviewed in Appendix~\ref{sec: review of 3d bosonization}.
Gauging a Kitaev chain on a codimension-2 defect is shown in Fig.~\ref{fig: 3d kitaev chain to codim-2 defect}.

Next, we want to construct a membrane operator which commutes with all terms in its bulk and only violates specific terms near its boundary. If we only apply the operator for the $m$-loop excitation, which is the product of Pauli $X$ on edges perpendicular to the membrane, it will anti-commute with a single term on the Kitaev chain. To fix this issue, we introduce another string operator for the $\psi$ particle, which also anti-commutes with the single term on the Kitaev chain. Therefore, the membrane operator for a $m$-loop excitation together with a $\psi$-line will commute with all terms except their boundaries. To sum up, whenever the $m$-loop is linked with the twist string, a fermion line must come out from this linking.

\begin{figure}[H]
    \centering
    \includegraphics[width=0.8\textwidth]{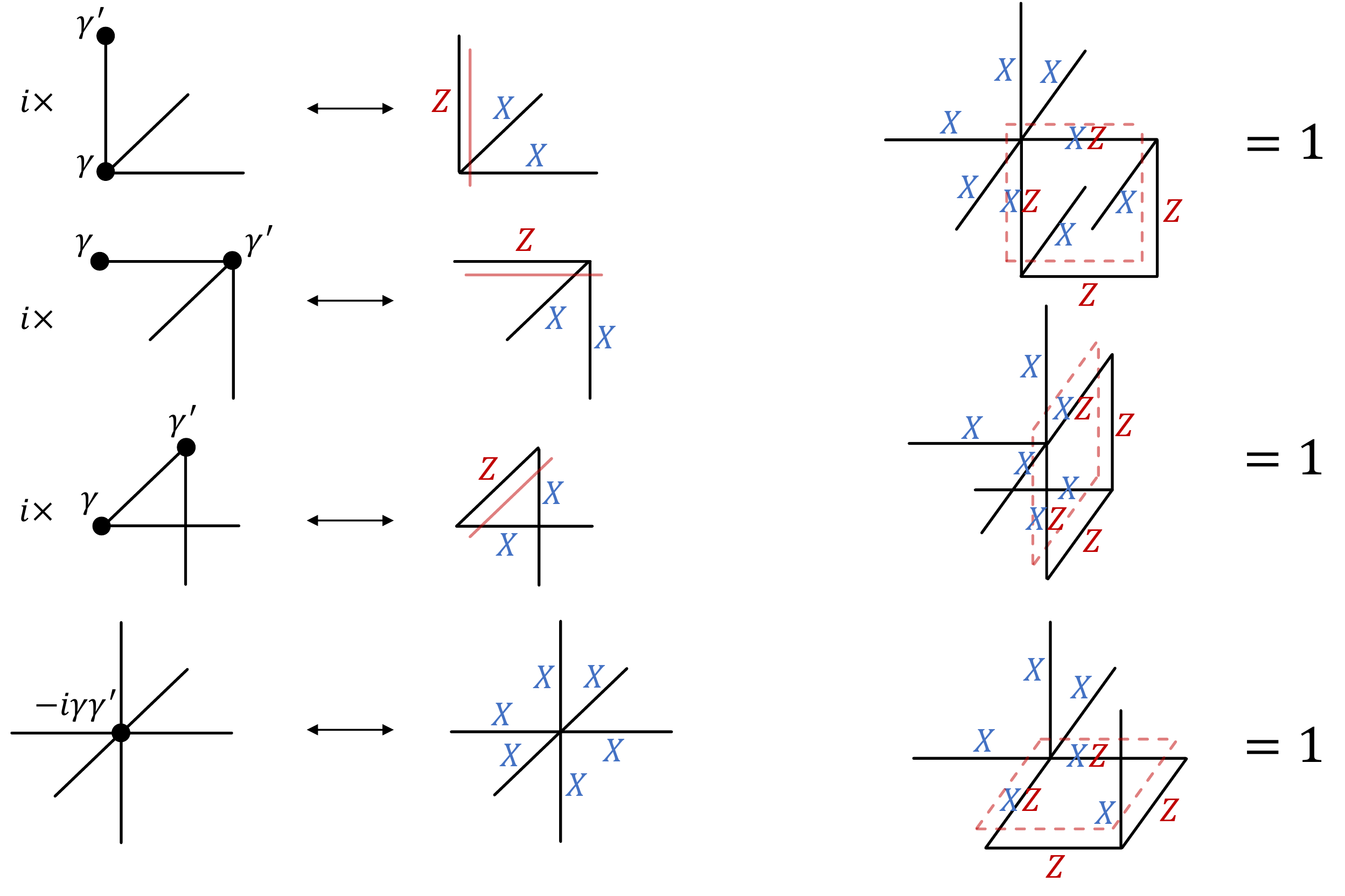}
\caption{This is the summary of 3d bosonization on the cubic lattice. There are two Majorana fermions $\g_v$, $\g_v^\prime$ at each vertex $v$, and a qubit on each edge $e$, described by the Pauli matrices $X_e$, $Y_e$, $Z_e$. The mapping between the Majorana operators and Pauli operators is shown above.
The fermionic hopping operators $i \g_i \g^\prime_j$ along the edge $e=\lr{ij}$ are mapped to the Pauli hopping operators: Pauli $Z_e$ dressed with other Pauli $X$ matrices on edges crossing the ``framing'' of edge $e$. The on-site fermion parity $P_v = -i \g_v \g_v^\prime$ at the vertex $v$ is mapped to the $X$-star term. On the fermionic side, we focus on the even subspace, i.e., $\prod_v P_v = +1$. On the other hand, the bosonic side is restricted to the subspace satisfying the gauge constraints.
}
\label{fig: 3d bosonization}
\end{figure}

\begin{figure}[H]
    \centering
    \includegraphics[width=1\textwidth]{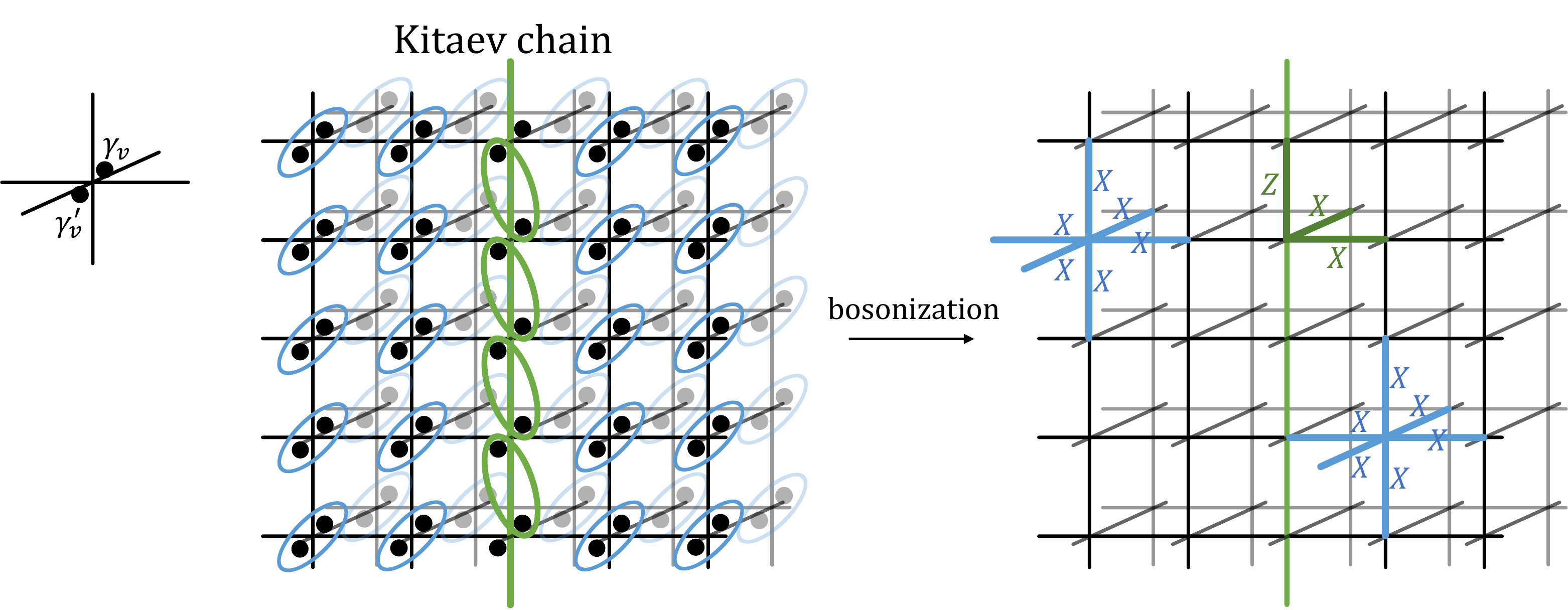}
\caption{Left: each dot present a Majorana fermion. The Kitaev chain is decorated on the defect (green line), which makes the Majorana fermion paired with the other one on its adjacent vertex; otherwise, Majorana fermions are paired within each vertex. Right: after bosonization, the pairing term on the Kitaev chain becomes the Pauli hopping operator on the twist string, while the pairing term in each face not on the Kitaev chain becomes the $X$ star term. The gauge constraints in Fig.~\ref{fig: 3d bosonization} hold at all faces.}
\label{fig: 3d kitaev chain to codim-2 defect}
\end{figure}

\begin{figure}[H]
    \centering
    \includegraphics[width=0.8\textwidth]{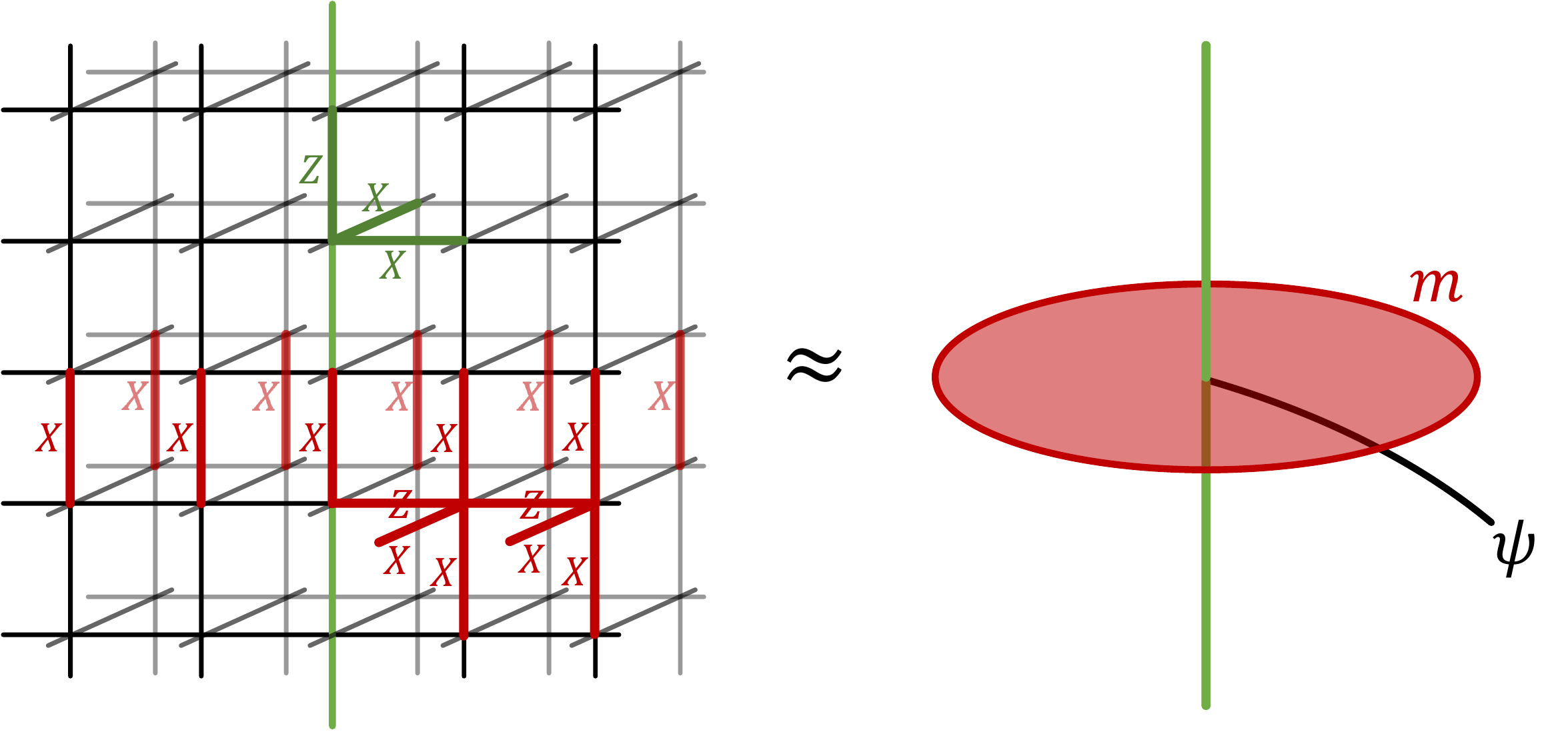}
\caption{The membrane operator which commutes will all terms in its bulk. This operator can be thought of as the $\psi$-line coming out from the intersection of the twist string and the membrane operator of the $m$-loop excitation.}
\label{fig: 3d defect membrane operator}
\end{figure}

\section{Layer construction for twist strings in general topological states: from (2+1)D to (3+1)D}
\label{sec:layer}

In this section we describe a general method, using the layer construction, to construct twist strings in general (3+1)D topological phases. Our approach is inspired by Refs.~\cite{WangSenthil13, JianQi14, TJV21-1, TJV21-2}, and in our new setting we can deal with defects in the (3+1)D topological order. Not only does this provide a simple way to understand twist strings in (3+1)D from simpler properties of defects in (2+1)D, we will also understand how to construct twist strings in non-Abelian (3+1)D topological phases, which have non-trivial interaction with non-Abelian flux loops. 

\begin{figure}[htb]
    \centering
    \includegraphics[width=1\textwidth]{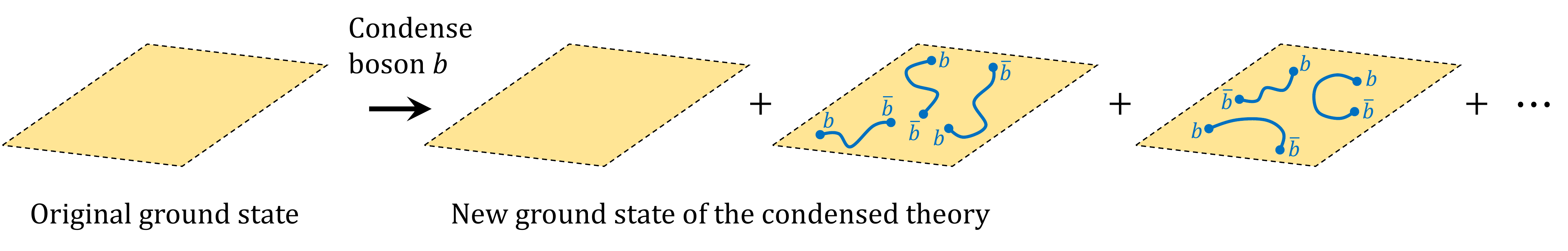}
\caption{Condensation of the boson $b$. Given an anyon theory $\mathcal{A}$, we can condense a boson $b \in \mathcal{A}$ by proliferating the particles $b$ in the space. The blue lines are short string operators creating the boson $b$ and its anti-particle $\bar b$. In this manuscript, we will condense the boson $b$ that is abelian (quantum dimension $1$) and has its inverse $\bar b$ ($b \times \bar b = 1$).}
\label{fig: condense_boson_b}
\end{figure}

\subsection{Abelian theory}

In this section, we describe the layer construction approach starting from 2d Abelian anyon theories. We will demonstrate two main examples: the $\ZZ_2 \times \ZZ_2$ toric codes with two bosonic charges, and the $\ZZ_2$ toric code with a fermionic charge. First, we illustrate how 2d toric codes can be stacked into a 3d toric code with a bosonic charge or a fermionic charge, depending on which anyons are condensed. Second, we insert a twist string in one layer of 2d toric codes, which needs to be compatible with the condensed anyons. The precise criteria for allowed twist strings are that the condensed anyons must be invariant when they cross the twist strings. We will show that this procedure can recover the twist strings constructed in Sec.~\ref{sec:exact3d}.

\subsubsection{Layer construction for the (3+1)D toric code}

We start with the simplest example of constructing a 3d $\ZZ_2$ toric code via coupling layers of 2d $\ZZ_2$ toric codes stacked in the $z$-direction. To do this, we condense pairs of $e$ charge  in the neighboring layers, i.e., the anyons  in the form of $e^{(j)}e^{(j+1)}$, where $j$  are the layer labels $1\le j\le L$ and we have chosen the periodic boundary condition $e^{(L+1)}=e^{(1)}$. The condensation of a boson $b$ is described by Fig.~\ref{fig: condense_boson_b}, where the new ground state is proliferation of this boson. Obviously, all such anyons $e^{(j)}e^{(j+1)}$ can be condensed at the same time since the pair  $e^{(j)}e^{(j+1)}$ is a boson and has trivial mutual braiding statistics with any other condensed anyons. After condensing these anyons, the $e$ charge in each layer $j$, denoted by $e^{(j)}$, becomes a deconfined excitation, since it braids trivially with the condensed pair $e^{(j)}e^{(j+1)}$.  Note that a single charge $e^{(j)}$ can freely tunnel to the next layer and identified as $e^{(j+1)}$ through the fusion with the condensed pair $e^{(j)}e^{(j+1)}$, i.e., $e^{(j)} \times e^{(j)}e^{(j+1)} =  e^{(j+1)}$.

\begin{figure}[htb]
    \centering
    \includegraphics[width=0.75\textwidth]{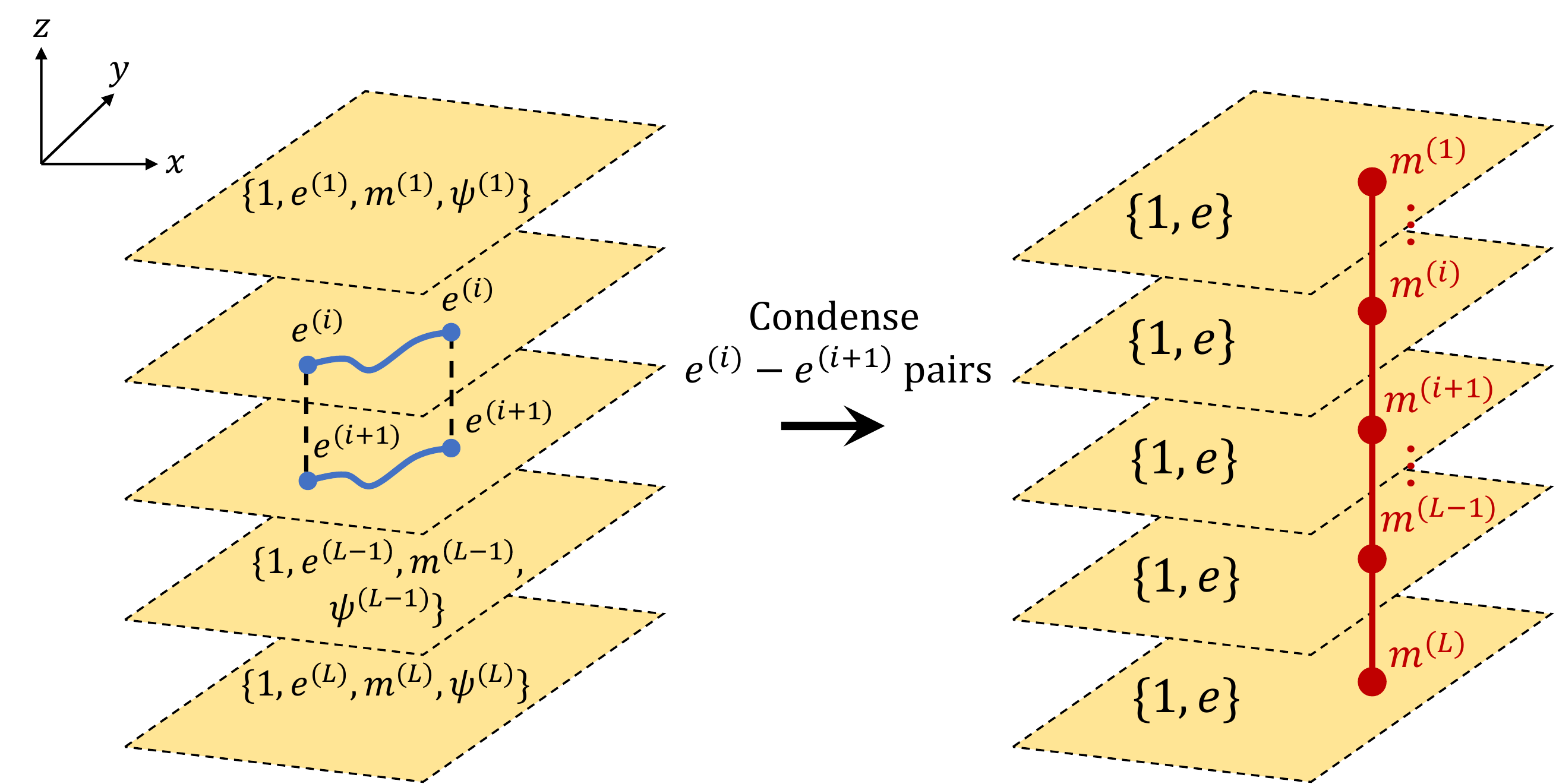}
\caption{Condensation of the pair of $e$-charges $e^{(j)}e^{(j+1)}$ in the $L$ layers of 2d $\Z_2$ toric codes. The deconfined excitations after the condensation is given by a single $e$-charge $e^{(j)}$, and the $m$-flux string given by an array of $m$-anyons in all layers.}
\label{fig: layer_e_e_condense}
\end{figure}

For the $m$-type excitations, a single $m$-anyon in layer $j$, denoted by  $m^{(j)}$, must be confined since it braids  non-trivially with $e^{(j)}e^{(j+1)}$ and $e^{(j-1)}e^{(j)}$. The formation of the condensate forces the $m$-anyon in each layer to be bound together to form a loop-like $m$-flux, which is a deconfined excitation, as shown in Fig.~\ref{fig: layer_e_e_condense}.  Such a $m$-flux string is expressed as $\prod_{j=1}^{L} m^{(j)}$, i.e., the array of $m$-anyon in each layer in a straight vertical string (loop) along the  $z$-direction (with periodic boundary condition). Note that this $m$-flux loop  has trivial mutual braiding statistics with the condensed pairs of $e$-charges  $e^{(j)}e^{(j+1)}$ since they overlap in two consecutive layers and the $\pi$ braiding phases cancel with each other. 
The braiding statistics between the two remaining deconfined excitations, i.e,  $e$-charge and  $m$-flux string, is given by the mutual braiding phase between $e$ and $m$ anyons in the 2d case, i.e., $M_{e,m\text{-flux}}= -1$.  

\begin{figure}[htb]
    \centering
    \includegraphics[width=0.5\textwidth]{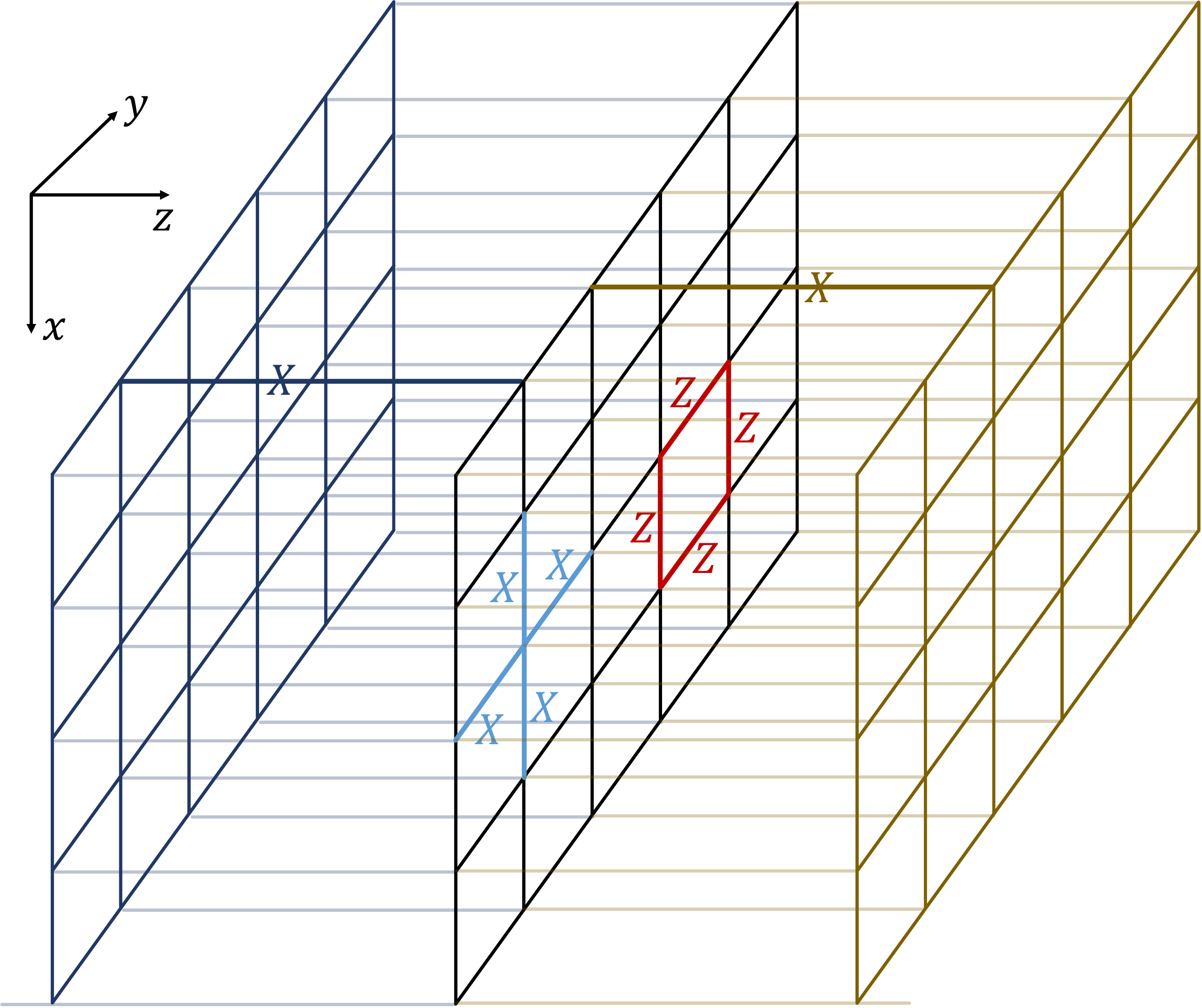}
\caption{We prepare layers of 2d toric codes first. On the inter-layer edges, we put ancilla qubits in $X_e = +1$ states. These ancilla qubits will become the background gauge field when we condense bosons (gauge the 1-form symmetry).}
\label{fig: layers of toric codes}
\end{figure}

The lattice Hamiltonian can be constructed directly from this procedure. We first prepare layers of 2d toric codes with inter-layer ancilla qubits, shown as Fig.~\ref{fig: layers of toric codes}. The initial Hamiltonian is
\begin{eqs}
    H_0 = -\sum
    \includegraphics[width=0.08\textwidth, valign=c]{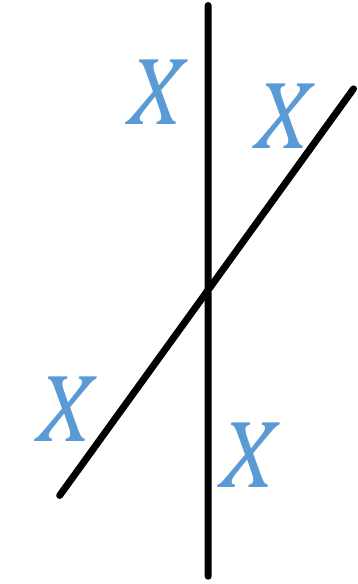}
    -\sum
    \includegraphics[width=0.09\textwidth, valign=c]{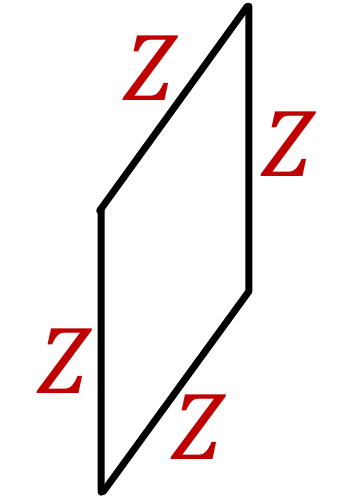}
    -\sum_{\text{inter-layer } e} X_e.
\label{eq: initial Hamiltonian}
\end{eqs}
Next, we condense adjacent $e^{(j)}e^{(j+1)}$, which enforce $Z$ operators in two layers are coupled together with the background field:
\begin{eqs}
    \includegraphics[width=0.25\textwidth]{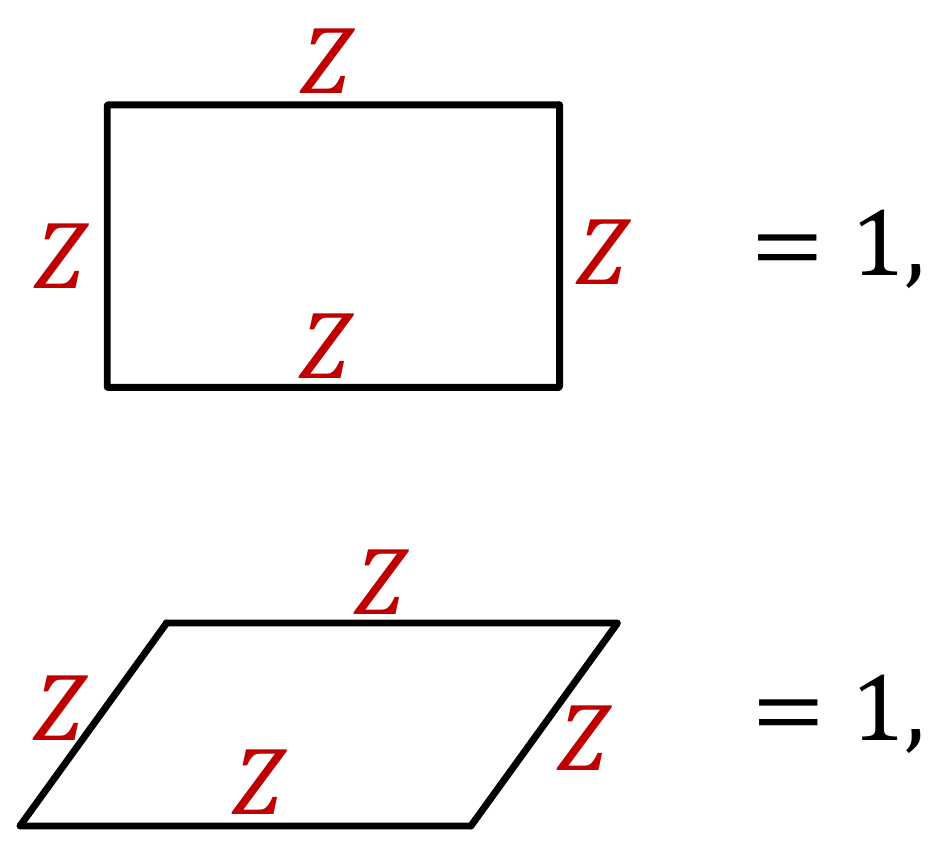}
\label{eq: condense e e}
\end{eqs}
which can be treated as the gauge constraints when gauging 1-form symmetry generated by $e^{(j)}e^{(j+1)}$. Since we impose the above condition, the initial Hamiltonian is no longer valid and we should only keep terms that commute with Eq.~\ref{eq: condense e e}. It is straightforward to write down the remaining terms:
\begin{eqs}
    \includegraphics[width=0.22\textwidth, valign=c]{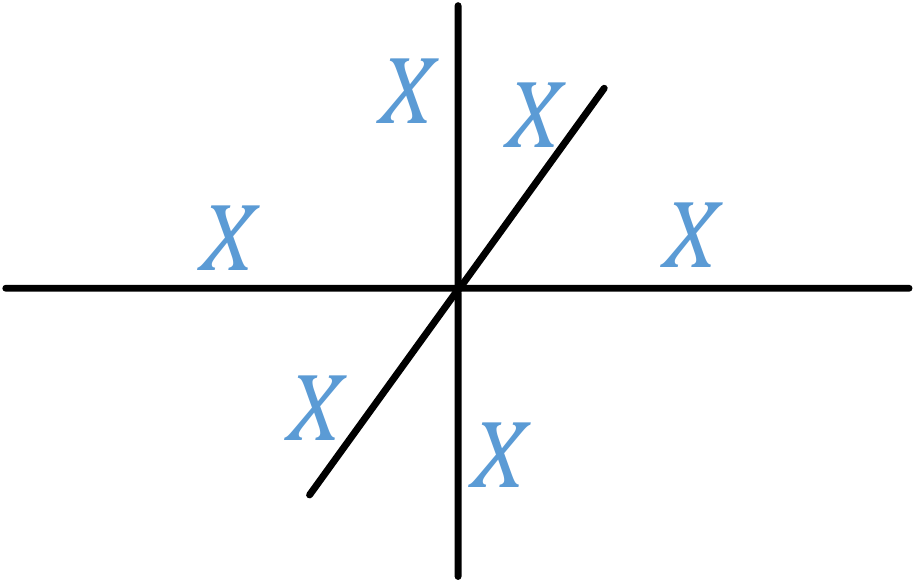}
    \quad \text{and} \quad
    \includegraphics[width=0.09\textwidth, valign=c]{Z_plaquette_term.pdf}
    \quad.
\label{eq: 3d X and Z terms}
\end{eqs}
If we impose the gauge constraints Eq.~\eqref{eq: condense e e} energetically, the condensed Hamiltonian as obtained as:
\begin{eqs}
    H_{\text{condensed}}= -\sum_v
    \includegraphics[width=0.15\textwidth, valign=c]{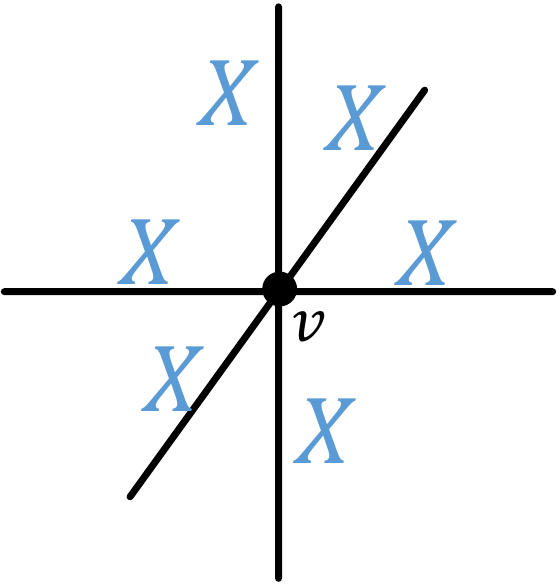}
    - \sum_{f} \includegraphics[width=0.12\textwidth, valign=c]{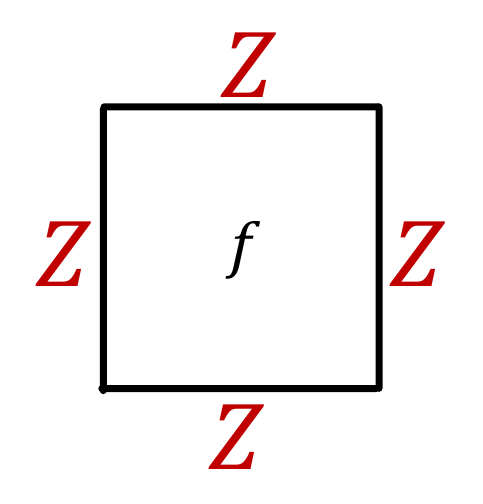}
    - \sum_{f} \includegraphics[width=0.09\textwidth, valign=c]{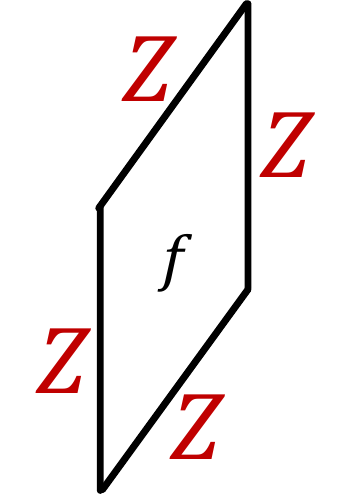}
    - \sum_{f} \includegraphics[width=0.12\textwidth, valign=c]{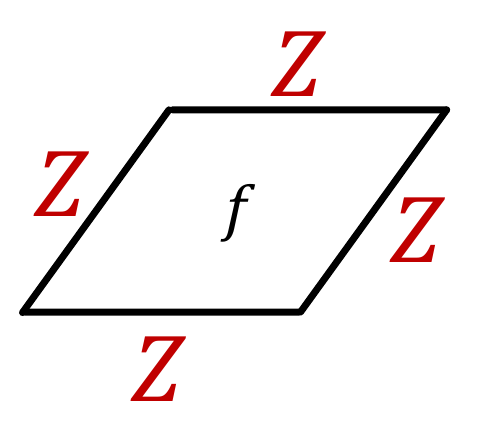},
\end{eqs}
which is exactly the 3d toric code (with a bosonic charge).

\subsubsection{Layer construction for the (3+1)D toric code with a fermionic charge}
\label{subsubsec:layerfermion}

\begin{figure}
    \centering
    \includegraphics[width=0.75\textwidth]{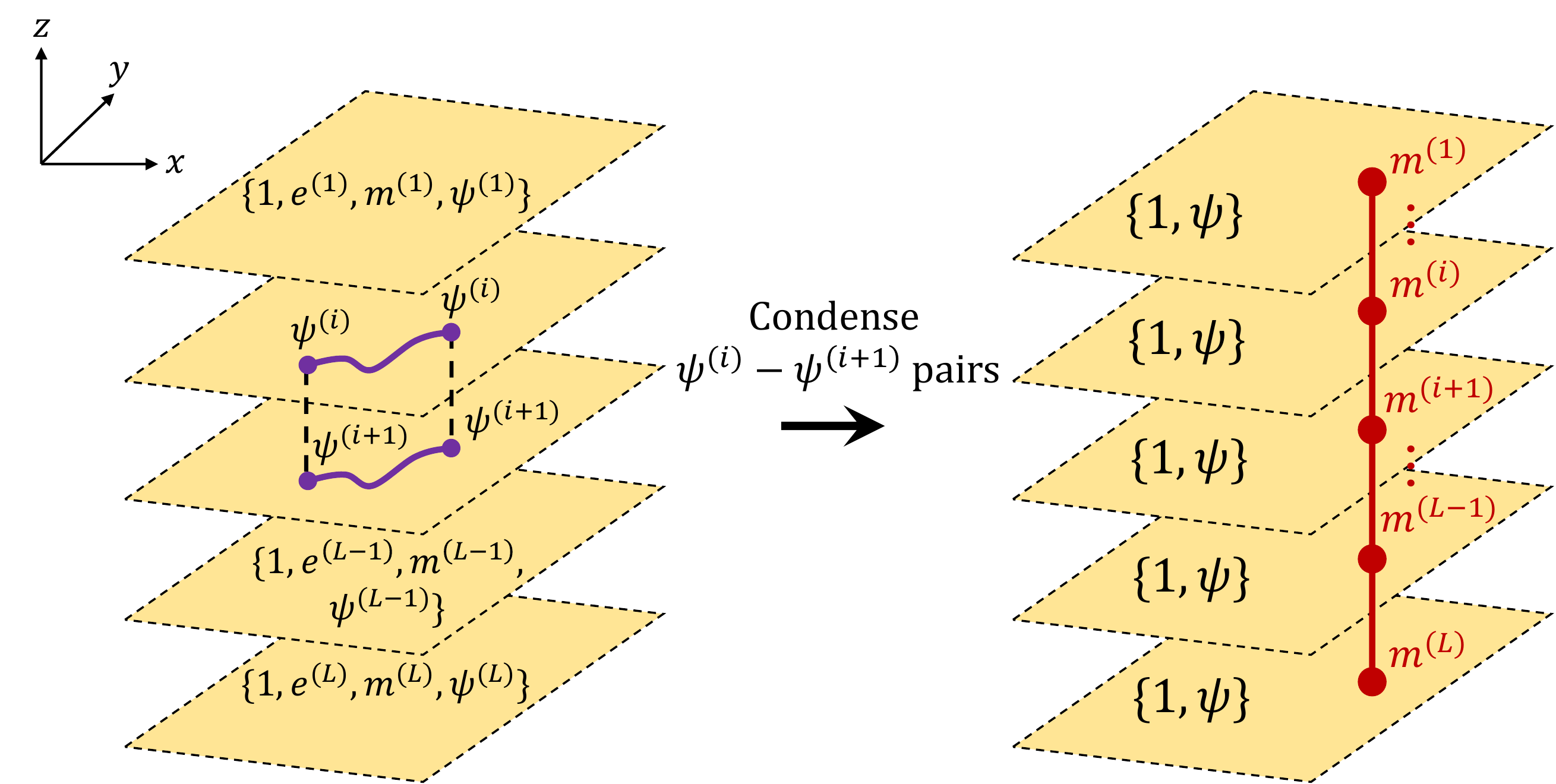}
\caption{Condensation of the pair of $\psi$-charges $\psi^{(j)}\psi^{(j+1)}$ in the $L$ layers of $\Z_2$ toric codes. The deconfined excitations after the condensation is given by a single $\psi$-charge $\psi^{(j)}$, and the $m$-flux string given by an array of $m$-anyons in all layers.}
\label{fig: layer_psi_psi_condense}
\end{figure}

Here we describe the procedure to obtain the (3+1)D $\Z_2$ toric code with a fermionic particle via the layer construction, starting from the layers of (2+1)D $\Z_2$ toric codes. First, we prepare $L$ layers of (2+1)D $\Z_2$ toric codes, whose anyons are labelled by $\{ 1, e^{(j)}, m^{(j)}, \psi^{j}\}$ for $j=1,2,\dots, L$. We then condense each pair of fermions in the adjacent layers $\psi^{(j)} \psi^{(j+1)}$ for $1\le j \le L$, as described in Fig.~\ref{fig: layer_psi_psi_condense}. After we condense the $\psi^{(j)} \psi^{(j+1)}$ pairs in all adjacent layers, the single fermion $\psi^{(j)}$ in each layer becomes a deconfined particle. The $\psi$ particles in different layers are identified by the fusion of condensed anyons $\psi^{(j)} \psi^{(j+1)}$. Therefore, the fermion $\psi$ becomes a particle excitation in the resulting (3+1)D topological order. Meanwhile, the anyon $m^{(j)}$ braids non-trivially with $\psi^{(j)} \psi^{(j+1)}$ and $\psi^{(j-1)} \psi^{(j)}$, so a single $m$ particle is confined. The deconfined excitation is an array of $m$ anyons in $L$ layers given by  $\prod_{j=1}^{L} m^{(j)}$, which describes a $m$-flux string excitation in the resulting (3+1)D topological order. After all, we get a (3+1)D $\Z_2$ toric code with a fermionic particle $\psi$ and a $m$ magnetic string excitation.

The lattice Hamiltonian can be constructed in a similar way as before. We prepare copies of toric codes with an ancilla qubit on each inter-layer edge $e$ into the $X_e=+1$ state, as shown in Fig.~\ref{fig: layers of toric codes}. The initial Hamiltonian is still Eq.~\eqref{eq: initial Hamiltonian}. These ancillas will serve as the background gauge field when we gauge the 1-form symmetry generated by $\psi^{(j)} \psi^{(j+1)}$.
Next, we condense $\psi-\psi$ in the adjacent layers. There are different conventions to perform this condensation on the lattice. We choose the following way:
\begin{eqs}
    \includegraphics[width=0.2\textwidth]{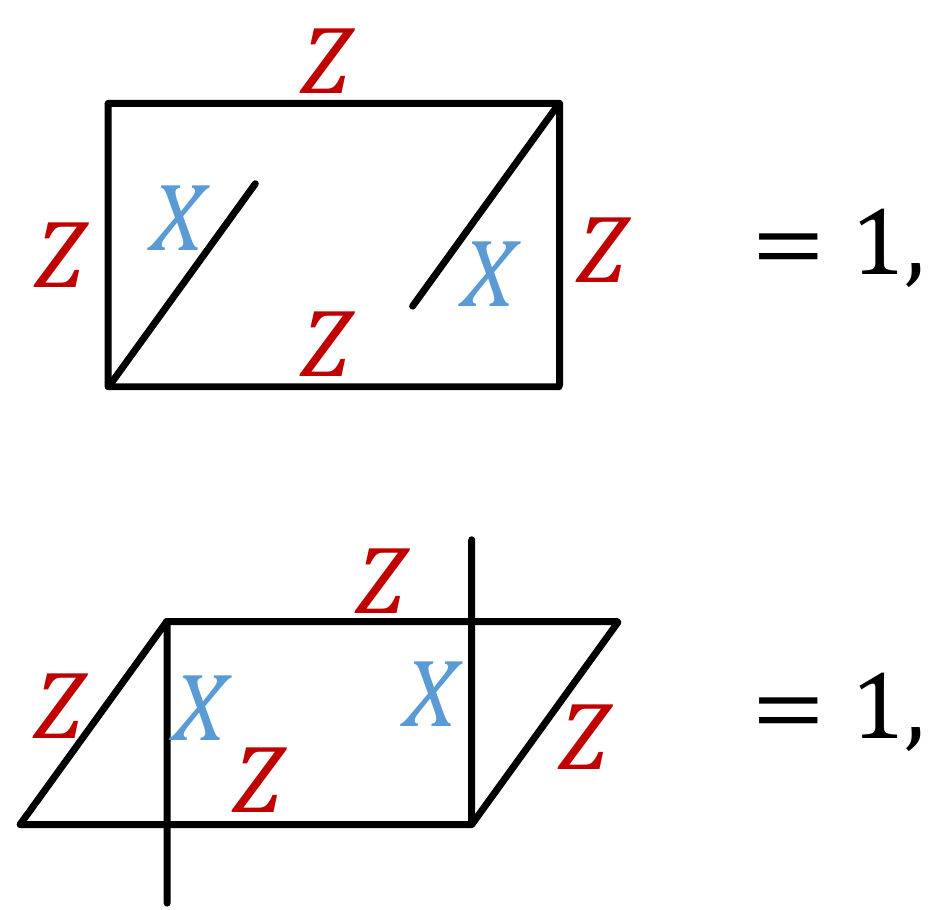}
\label{eq: condense psi psi}
\end{eqs}
such that these $\psi-\psi$ condensation terms in the inter-layers commute with each other. Since we couple the $\psi-\psi$ hopping operator to the background field, we also need to modify the star term and plaquette term to commute with Eq.~\eqref{eq: condense psi psi}. In other words, some Hamiltonian terms in Eq.~\eqref{eq: initial Hamiltonian} will be dropped, and only the combinations of them that commute with Eq.~\eqref{eq: condense psi psi} will be kept during this condensation process:
\begin{eqs}
    \includegraphics[width=0.2\textwidth, valign=c]{modified_X_star_term.pdf}
    \quad \text{and} \quad
    \includegraphics[width=0.2\textwidth, valign=c]{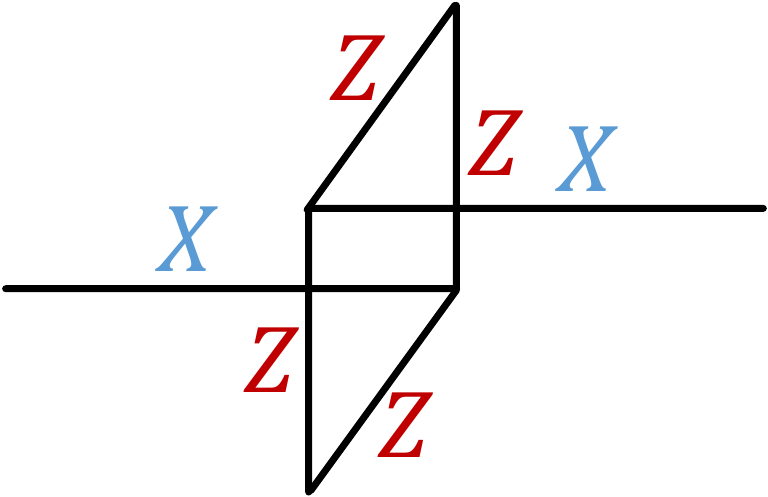}
    \quad.
\label{eq: modified X and Z terms}
\end{eqs}
Focusing on the ground state, we impose Eqs.~\eqref{eq: condense psi psi} and \eqref{eq: modified X and Z terms} energetically as
\begin{eqs}
    H_{\text{condensed}} = -\sum_v \includegraphics[width=0.15\textwidth, valign=c]{Av.pdf}
    - \sum_{f \in yz} \includegraphics[width=0.12\textwidth, valign=c]{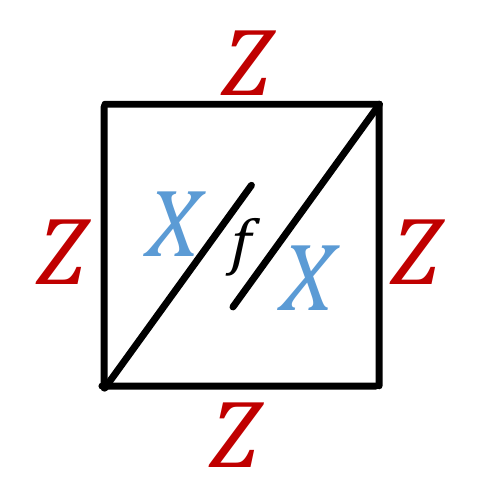}
    - \sum_{f \in xz} \includegraphics[width=0.10\textwidth, valign=c]{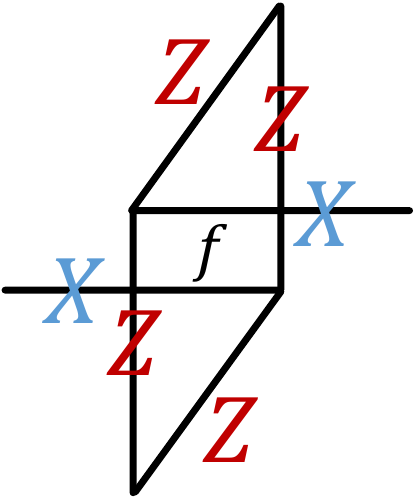}
    - \sum_{f \in xy} \includegraphics[width=0.12\textwidth, valign=c]{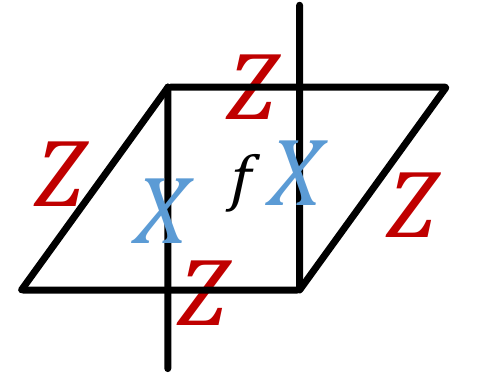}.
\end{eqs}
We can see that the last three terms are equivalent to the gauge constraints of the 3d bosonization in Fig.~\ref{fig: 3d bosonization} up to multiplying the first term. Therefore, the Hamiltonian $H_{\text{condensed}}$ has the same ground state as the 3d toric code with a fermionic charge.

\begin{figure}
    \centering
    \includegraphics[width=0.6\textwidth]{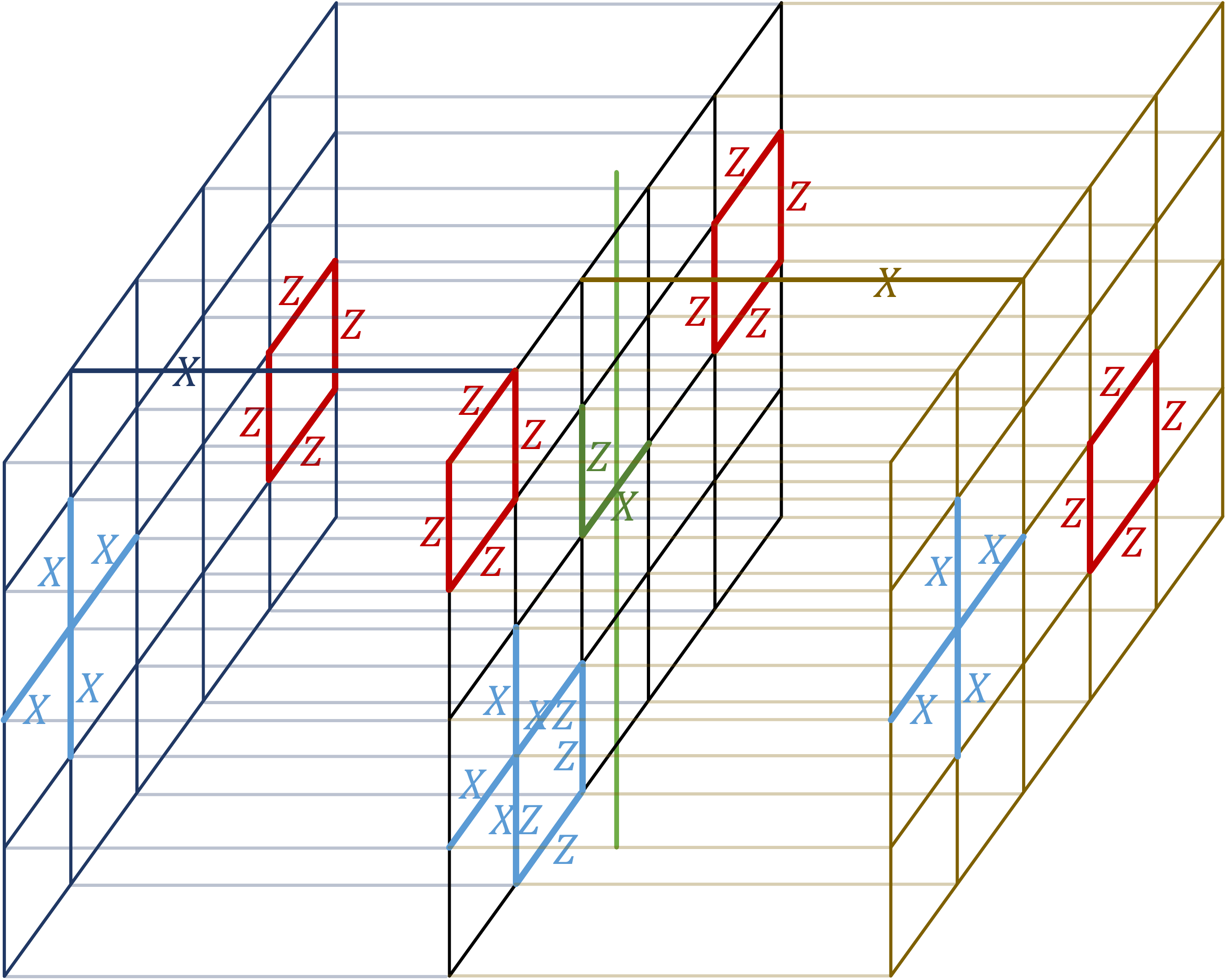}
\caption{We prepare layers of 2d toric codes, and we put the $e \leftrightarrow m$ twist string in one of them. On the inter-layer edges, ancilla qubits are in $X_e = +1$ states.}
\label{fig: layers of toric codes with 1 defect}
\end{figure}

\subsubsection{Inserting twist strings in the (3+1)D toric code with a fermionic charge}
\label{sec:twistdefect3Dtcfromlayer}

We have constructed 3d toric codes with bosonic and fermionic charges from layers of 2d toric codes. The next step is to introduce twist strings in these systems.

First, we demonstrate the twist string (invertible codimension-2 defect) in the 3d toric code with a fermionic charge. As shown in Fig.~\ref{fig: layers of toric codes with 1 defect}, we introduce an $e \leftrightarrow m$ twist string in only one layer of the 2d toric codes. To condense $\psi-\psi$ pairs, we choose a slightly different convention from Eq.~\eqref{eq: condense psi psi}:
\begin{eqs}
    \includegraphics[width=0.3\textwidth]{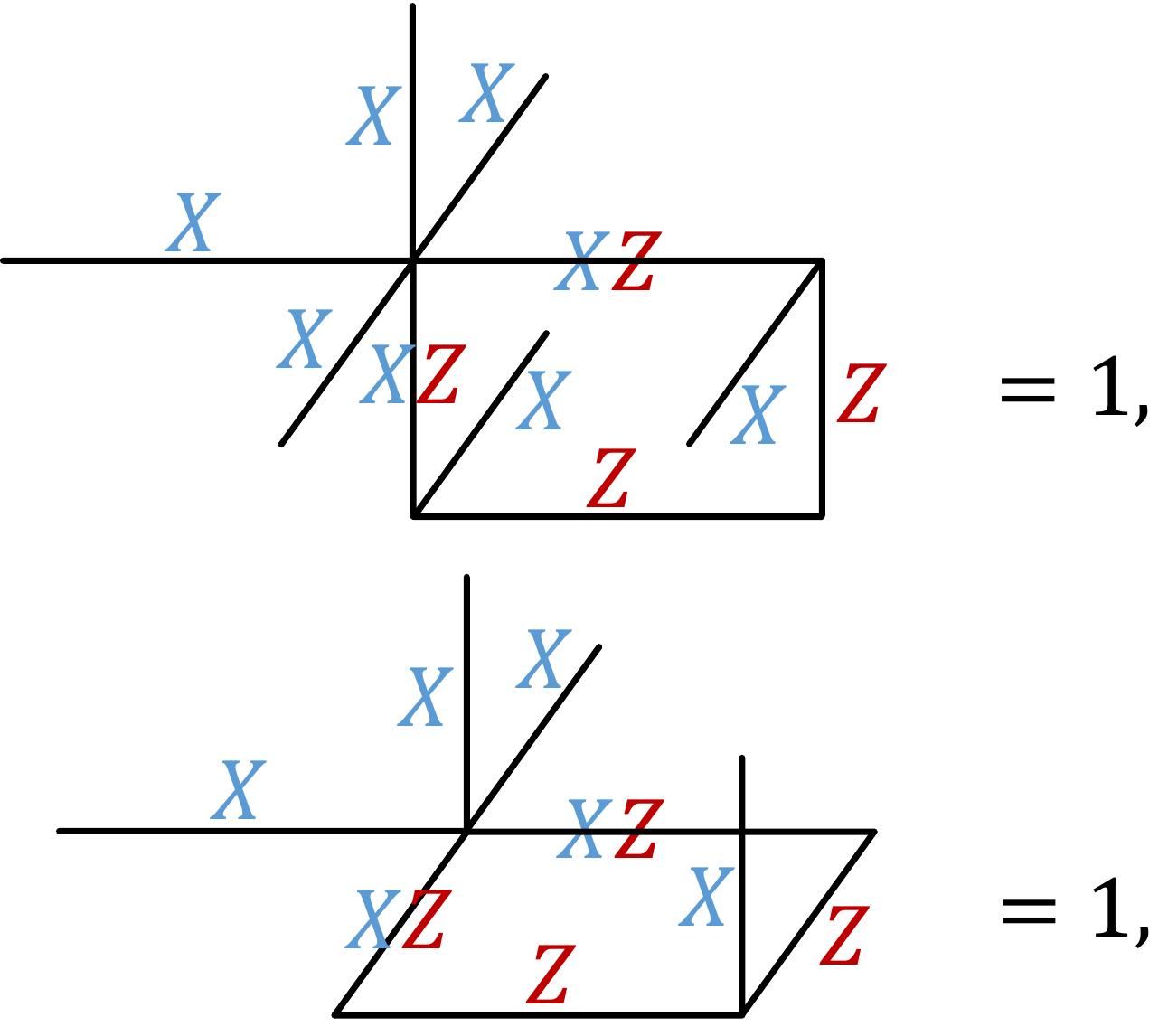}
\label{eq: condense psi psi 2}
\end{eqs}
where we have multiplied the $X$-star term on the left layer of toric codes and dress appropriate $X$ on the inter-layer edges to make them commute. Away from the twist string, this condensation is equivalent to Eq.~\eqref{eq: condense psi psi} by multiplying a 3d $X$-star term. The above convention is chosen to be compatible with the twist string in the 2d layer.
We modified the terms on each layer by multiplying the inter-layer $X_e$:
\begin{eqs}
    \includegraphics[width=0.35\textwidth, valign=c]{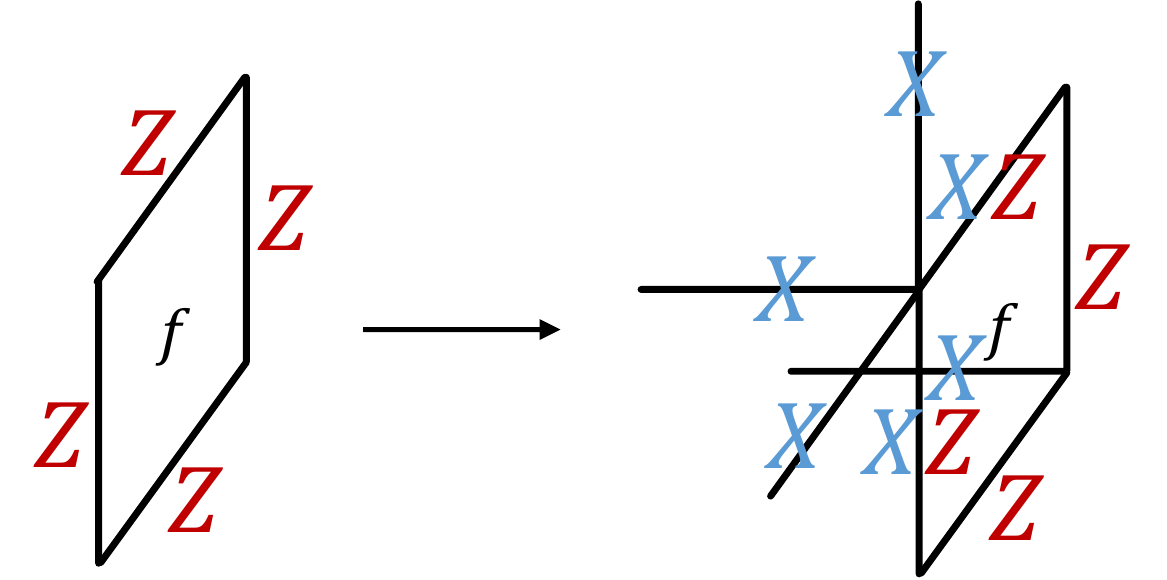}
    \quad \quad
    \text{and}
    \quad \quad
    \includegraphics[width=0.35\textwidth, valign=c]{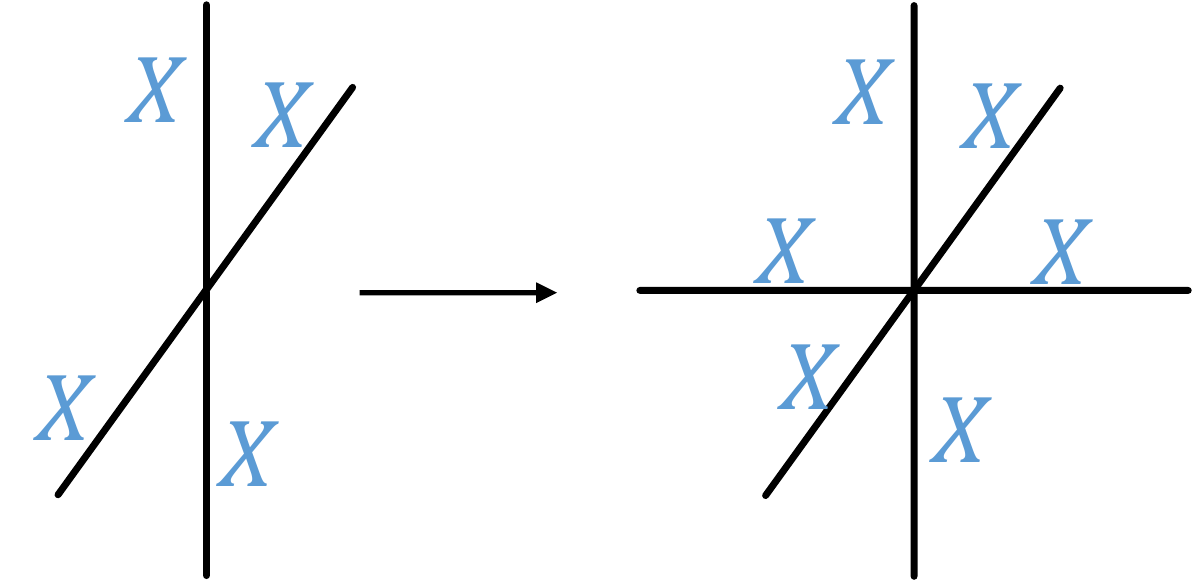}
    \quad.
\end{eqs}
On the twist string, the term becomes
\begin{eqs}
    \includegraphics[width=0.28\textwidth, valign=c ]{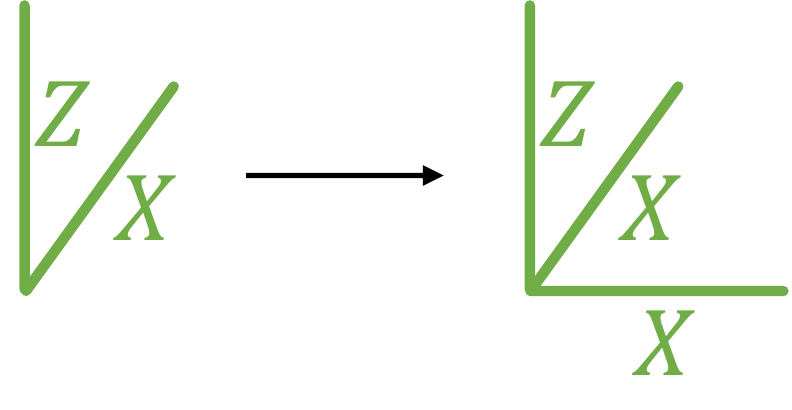}
    \quad .
\end{eqs}
Therefore, we recover the construction of the twist string in Fig.~\ref{fig: 3d defect membrane operator}.

\subsubsection{General layer construction with (2+1)D Abelian anyon theory}
\label{sec:layerfromabelian}
Now we describe a general recipe to construct a (3+1)D topological order from the layer construction for (2+1)D Abelian anyon theory. The (2+1)D Abelian anyon theory on each layer is described by a modular tensor category, whose set of anyons is denoted as $\mathcal{A}$. 
The first step is to identify a subset $\mathcal{M} \subset \mathcal{A}$ satisfying
\begin{enumerate}
    \item $\mathcal{M}$ contains only bosons and fermions which mutually braid trivially with each other.
    \item For any anyon $a \notin \mathcal{M}$, there exist an anyon $b \in \mathcal{M}$ such that $a$ and $b$ have non-trivial mutual braiding, $M_{a,b}\neq 1$.
\end{enumerate}

From these two conditions, $\mathcal{M}$ forms a group under fusion since if $a, b \in \mathcal{M}$, $a \times b$ braids trivially with elements in $\mathcal{M}$, which implies $a \times b \in \mathcal{M}$. 

Next, we prepare $L$ layers of (2+1)D anyon theories $\mathcal{A}$, and, for each $a \in \mathcal{M}$, we condense the pair $a^{(j)}\overline{a}^{(j+1)}$ in the adjacent $j$-th and $(j+1)$-th layers. After condensation, anyons in $\mathcal{M}$ are deconfined in the (3+1)D theory and become the particle excitations. As such, we may refer to the particles in $\mathcal{M}$ as electric charges. Other anyons $b\in\mathcal{A}$ not contained in $\mathcal{M}$ are confined, though the array of $b$ particles in the layers given by $\prod_{j=1}^L b^{(j)}$ becomes deconfined and form loop-like flux excitations. As such, we refer to equivalence classes in $\mathcal{A}/\mathcal{M}$ as fluxes. After all, the resulting theory realizes a (3+1)D topological order where the particle excitations have the group-like fusion rule $\mathcal{M}$, while the loop excitations in the form of $\prod_{j=1}^L b^{(j)}$ have the group-like fusion rule given by $\mathcal{A}/\mathcal{M}$.

\subsubsection{Inserting twist strings for general layer construction with (2+1)D Abelian anyon theory}
\label{sec:twiststringabelianlayer}

Here we construct a twist string in the (3+1)D theory, which is obtained from the layer construction with the (2+1)D Abelian anyon theory. As we have done in Sec.~\ref{sec:twistdefect3Dtcfromlayer}, we introduce the twist string by inserting a codimension-1 defect $\mathcal{D}$ of the (2+1)D Abelian anyon theory in the single layer of the layered system, and then performing the layer construction in the presence of the twist string. We want the inserted codimension-1 defect $\mathcal{D}$ in the (2+1)D theory to behave as a codimension-2 defect of the resulting (3+1)D theory after the layer construction. 

Suppose that we are inserting the defect $\mathcal{D}$ at the $k$-th layer in the $L$ layers of the Abelian anyon theories.
Obviously, an insertion of the defect $\mathcal{D}$ in the single layer defines an automorphism $\rho_{\mathcal{D}}$ that acts on $L$ copies of the (2+1)D Abelian anyon theory before layer construction. 
For the codimension-1 defect $\mathcal{D}$ to define the twist string after the layer construction, we require that $\rho_{\mathcal{D}}$ acts on the loop excitations formed by any $b\in\mathcal{A}$ as
\begin{align}
    \rho_{\mathcal{D}}:\quad  \left(\prod_{j=1}^L b^{(j)}\right)\  \to \ \left(\prod_{j=1}^L b^{(j)} \right)\times a^{(k)},  \quad  \text{for some $a\in\mathcal{M}$} .
\end{align}
so that it preserves the set of deconfined particles after the anyon condensation.
It follows that $\rho_{\mathcal{D}}$ acts on the anyons $b\in\mathcal{A}$ of the $k$-th layer by attaching an anyon in $\mathcal{M}$,
\begin{align}
    \rho_{\mathcal{D}}: b\to b\times a, \quad \text{for some $a\in\mathcal{M}$}.
    \label{eq:fluxpreservingabelian}
\end{align}
We call the automorphism with this property a ``\textit{flux-preserving automorphism},'' 
since the label of the fluxes after the layer construction is given by $\mathcal{A}/\mathcal{M}$, and the above action leaves the elements of $\mathcal{A}/\mathcal{M}$ invariant.
The flux-preserving condition in Eq.~\eqref{eq:fluxpreservingabelian} is equivalent to the condition that $\rho_{\mathcal{D}}$ leaves the anyons in the set $\mathcal{M}$ invariant,
\begin{align}
    \rho_{\mathcal{D}}: a\to a, \quad \text{for all $a\in\mathcal{M}$.}
    \label{eq:chargeconserveabelian}
\end{align}
To see that Eq.~\eqref{eq:fluxpreservingabelian} implies Eq.~\eqref{eq:chargeconserveabelian}, suppose that $\rho_{\mathcal{D}}$ acts on $a\in\mathcal{M}$ as $\rho_{\mathcal{D}}: a\to a'$ with some $a'\in\mathcal{M}$. Since the mutual braiding is invariant under the action of the automorphism, we have $M_{a,b}=M_{a',b}$ for any $b\in\mathcal{A}$. This implies that $a\times \overline{a'}$ is a transparent particle in the modular tensor category, which means $a=a'$, hence Eq.~\eqref{eq:chargeconserveabelian} follows. 
Conversely, to see that Eq.~\eqref{eq:chargeconserveabelian} implies Eq.~\eqref{eq:fluxpreservingabelian}, suppose that $\rho_{\mathcal{D}}$ acts as $\rho_{\mathcal{D}}:b\to c$ for $b,c\in\mathcal{A}$. According to the invariance of the mutual braiding, we have $M_{a,b}=M_{a,c}$ for $a\in\mathcal{M}$. This implies that $b\times \overline{c}\in\mathcal{M}$, so Eq.~\eqref{eq:fluxpreservingabelian} follows. Physically, Eq.~\eqref{eq:chargeconserveabelian} guarantees that the twist string after the layer construction preserves the labels of particle excitations. 

In summary, an insertion of the defect $\mathcal{D}$ in a single layer with the property Eq.~\eqref{eq:fluxpreservingabelian}, or equivalently Eq.~\eqref{eq:chargeconserveabelian}, realizes the codimension-2 twist string of the resulting (3+1)D theory that gives the action on the flux loops illustrated in Fig.~\ref{fig: loop defect intersection produces charge}.

Here, let us describe the example of the twist string of (3+1)D $\Z_2$ toric code with a fermionic particle constructed in Sec.~\ref{sec:twistdefect3Dtcfromlayer}. In that case, the (2+1)D Abelian theory is taken as the 2d $\Z_2$ toric code $\mathcal{A}=\{1,e,m,\psi\}$, and we take $\mathcal{M}=\{1,\psi\}$ so that the particle excitation becomes a fermion after the layer construction.
The twist string is then described via the $e \leftrightarrow m$ twist string, which obviously satisfies the flux-preserving condition, $e\times\psi=m$. Accordingly, the fermionic particle $\psi$ is left invariant under the action of the defect.

\subsection{Non-Abelian theory}

\subsubsection{General layer construction with non-Abelian discrete  $G$-gauge theory}\label{sec:non-Abelian_layer_construction}

Here we describe the layer construction to obtain a (3+1)D discrete $G$-gauge theory when $G$ is possibly non-Abelian, from the layer construction of (2+1)D discrete $G$-gauge theory. Our analysis below mainly restricted to the case where the (3+1)D discrete gauge theory only has bosonic point charges, although we comment on the generalization to the case of fermionic charges as well. We also restrict to the case where the (3+1)D and (2+1)D discrete gauge theories are `untwisted,' meaning they have trivial $\mathcal{H}^4(G, \U)$ and $\mathcal{H}^3(G, \U)$ cocycles respectively. 

The underlying algebraic structure of the (2+1)D discrete $G$-gauge theory is associated with the quantum double $D(G)$.  The anyons in the quantum double model $D(G)$ can be expressed in the form of  $a=([g], \pi_g)$.  Here, $g \in G$ is the group element and $[g]$ denotes the conjugacy class of $G$ that contains $g$, which corresponds to the magnetic flux carried by the anyon. The electric charge  $\pi_g\in\mathrm{Rep}(C_g)$ corresponds to the irreducible representation of the centralizer of $g$, which is denoted by  $C_g$. 
We call $([g], \mathbf{1})$, which has a trivial representation $\mathbf{1}$, a  pure magnetic flux. On the other hand, we call $([1], \pi)$, which is in the trivial conjugacy class $[1]$, a  pure electric charge. Here, $\pi\equiv \pi_1$ is the irreducible representation of the entire symmetry group $G$. Finally, we call $([g], \pi_g)$ with nontrivial conjugacy class $[g]$ and representation  $\pi_g$ a dyon.

\begin{table*}
    \centering
    \begin{tabular}{l l}
    \hline
    Conjugacy class &  Centralizer \\
    \hline
    $[{(0,0)}]=\{(0,0)\}$ & $C_{(0,0)}=D_n$ \\
    $[(\frac{n}{2},0)]=\{(\frac{n}{2},0)\}$ & $C_{(\frac{n}{2},0)}=D_n$ \\
    $[(r,0)] = \{(r,0),(n-r,0)\}$ for $r\neq 0,\frac{n}{2}$ & $C_{(r,0)}=\Z_{n}= \{(r,0)\}_{r=0}^{n-1}$ \\
    $[(0,1)] = \{(2r,1)\}_{r=0}^{\frac{n}{2}}$ & $C_{(0,1)}=D_2 = \{(\frac{nj}{2}, k)\}_{j,k\in\{0,1\}}$ \\
    $[(1,1)] = \{(2r+1,1)\}_{r=0}^{\frac{n}{2}}$ & $C_{(1,1)}=D_2= \{(\frac{nj}{2}+k, k)\}_{j,k\in\{0,1\}}$ \\
    \hline
\end{tabular}
    \caption{Conjugacy classes of the dihedral group $D_{n}$ for even $n$ and their centralizers. }
    \label{tab:conjcent}
\end{table*}

As a concrete example, we consider the quantum double $D(D_{4})$ of the dihedral group $D_{4}$ of order $8$.
The group element of a dihedral group $D_n$ is labeled by $(r,s)\in\Z_n \rtimes \Z_2 \cong D_n$, with the group multiplication law given by
\begin{align} \label{eq: group law of Dn}
    (r_1,s_1)\cdot (r_2, s_2) = (r_1+r_2(1-2s_1), s_1+s_2),
\end{align}
where the first entry is modulo $n$ and the second entry is modulo $2$.
Since this will be our main example in the following sections, here we explicitly list all the anyons. We present the conjugacy classes of $D_{n}$ with even $n$ and their centralizers in Table ~\ref{tab:conjcent}. Restricting to $n=4$ is straightforward. We will also need the irreducible representations of the centralizers. We denote the irreducible representations of the centralizers by using the notation in Table ~\ref{tb:D4} when it is isomorphic to $D_{4}$, and denote the one-dimensional irreducible representations of the centralizer $C_{(1,0)} = \mathbb{Z}_{4}$ by $\beta_{l} = q^{l}$, where $q = \exp{i2\pi/4}$ and $l=0,1,2,3$. When the centralizer is isomorphic to $D_{2}$, we use the notation in Table ~\ref{tb:D2}. In total, there are $22$ anyons. The quantum dimension of the anyon is given by the product of the number of elements in the conjugacy class and the dimension of the representation of the corresponding centralizer, which is summarized in Table.~\ref{tb:DD4}.

\begin{table*}
\begin{center}
 \begin{tabular}{ c | c c c c c } 
 $D_{4}$ & $[(0,0)]$ & $[(2,0)]$ & $[(1,0)]$ & $[(0,1)]$ & $[(1,1)]$ \\
 \hline
 $J_{0}$ & $1$ & $1$ & $1$ & $1$ & $1$
 \\
 
 $J_{1}$ & $1$ & $1$ & $1$ & $-1$ & $-1$
 \\
 $J_{2}$ & $1$ & $1$ & $-1$ & $1$ & $-1$
 \\
 $J_{3}$ & $1$ & $1$ & $-1$ & $-1$ & $1$
 \\
 $\alpha$ & $2$ & $-2$ & $0$ & $0$ & $0$
 \end{tabular}
\end{center}
\caption{The character table of $D_{4}$, where $q = \exp{i2\pi/4}$.}
\label{tb:D4}
\end{table*}

\begin{table*}
\begin{center}
 \begin{tabular}{ c | c c c c } 
 $D_{2}$ & $[(0,0)]$ & $[(1,0)]$ & $[(0,1)]$ & $[(1,1)]$ \\
 \hline
 $A_{0}$ & $1$ & $1$ & $1$ & $1$
 \\
 
 $A_{1}$ & $1$ & $1$ & $-1$ & $-1$
 \\
 $A_{2}$ & $1$ & $-1$ & $1$ & $-1$
 \\
 $A_{3}$ & $1$ & $-1$ & $-1$ & $1$
 \end{tabular}
\end{center}
\caption{The character table of $D_{2}$.}
\label{tb:D2}
\end{table*}

\begin{table*}
\begin{center}
 \begin{tabular}{ c | c c c c c c c} 
 $D(D_{4})$ & $([(0,0)],J_{i})$ & $([(0,0)],\alpha)$ & $([(2,0)],J_{i})$ & $([(2,0)],\alpha)$ & $([(1,0)],\beta_{l})$ & $([(0,1)],A_{i})$ &
 $([(1,1)],A_{i})$
 \\
 \hline
 $d_{([(r,s)],\pi_{(r,s)})}$ & 1 & 2 & 1 & 2 & 2 & 2 & 2
 \end{tabular}
\end{center}
\caption{Irreducible representations of $D(D_{4})$ and their dimensions.}
\label{tb:DD4}
\end{table*}

We discuss the layer construction for a (3+1)D $G$-gauge theory, starting with layers of (2+1)D $G$-gauge theory described by the (untwisted) quantum double $D(G)$. The idea for the layer construction is largely the same as the Abelian case. 
That is, we again think of condensing a pair of pure electric charges in the form of $a^{(j)}\overline{a}^{(j+1)}$ in the neighboring $j$-th and $(j+1)$-th layer for each $j$, where $a \equiv ([1], \pi)$ represents a pure charge and  $\overline{a}\equiv ([1], \overline{\pi})$ represents the anti-particle of $a$ (where $\overline{\pi}$ represents complex conjugation of $\pi$). Due to this process, we expect that the electric charge in each layer gets identified, and represents a single electric charge of a resulting (3+1)D theory after condensation. Meanwhile, a single pure flux string $([g],\mathbf{1})^{(j)}$ in each layer gets confined due to non-trivial braiding statistics between the condensed anyon pair $a^{(j)}\overline{a}^{(j+1)}$, since the condensed particles should not be detectable by any deconfined excitation. The  formation of the condensate hence forces the pure flux string in each layer to be bound together into a flux string $\prod_{j=1}^{L} ([g],\mathbf{1})^{(j)}$, such that the Aharonov-Bohm phase between the pure flux $([g], \mathbf{1})^{(j)}$ and the pure charge $a^{(j)}$ cancels with the Aharonov-Bohm between the pure flux $([g], \mathbf{1})^{(j)}$ and the pure anti-charge $\bar{a}^{(j+1)}$ in the neighboring layer. Therefore, there is only trivial braiding statistics between the flux string and the pure charge-anticharge pair $a^{(j)}\bar{a}^{(j+1)}$. The flux string then becomes the deconfined magnetic excitation of the resulting (3+1)D theory after condensation. After all, we expect to get a (3+1)D $G$-gauge theory with electric particles and flux strings.

To phrase the above layer construction in a precise way, we need to employ an algebraic description of anyon condensation valid for a non-Abelian topological phase in (2+1)D, given by $L$ copies of $G$-gauge theory in our case. In general, anyon condensation in a topological order $\mathcal{C}$ corresponds to a gapped interface between $\mathcal{C}$ and the other topological order $\mathcal{C}'$ obtained after anyon condensation. This is equivalent to a gapped boundary of a topological order $\mathcal{C}\boxtimes\overline{\mathcal{C}}'$ after folding the picture along the interface. The gapped boundary of a (2+1)D topological order $\mathcal{C}\boxtimes\overline{\mathcal{C}}'$ is algebraically formulated in terms of Lagrangian algebra anyon of $\mathcal{C}\boxtimes\overline{\mathcal{C}}'$. 
In our case, we start with the $L$ copies of the $G$-gauge theory in (2+1)D given by $\mathcal{C}=(D(G))^L$ ($L$-th power means stacking $D(G)\boxtimes D(G)\boxtimes\dots\boxtimes  D(G)$ of $L$ layers), and we expect to obtain a (2+1)D $G$-gauge theory $\mathcal{C}'=D(G)$ after proper anyon condensation. This (2+1)D theory $\mathcal{C}'$ is physically regarded as a theory obtained by compactifying an effective (3+1)D theory for the layered system obtained by anyon condensation, where the string-like object in the form of $\prod_{j=1}^L ([g],\mathbf{1})^{(j)}$ is treated as a magnetic particle in $\mathcal{C}'$ after compactification.

Hence, the layer construction is described by the Lagrangian algebra anyon of the modular category given by $(D(G))^L\boxtimes \overline{D(G)}$. Here we propose an explicit form of the Lagrangian algebra anyon that corresponds to the layer construction sketched above.
The Lagrangian algebra anyon physically represents condensed anyons of the gapped boundary, which has the form of
\begin{align}
    \mathcal{L} = \sum_{a\in (D(G))^L\boxtimes \overline{D(G)}} Z_{0a} a,
\end{align}
where $a$ is the anyon condensed on the boundary, and $Z_{0a}$ is the non-negative integer.

We express the anyon of $(D(G))^L\boxtimes \overline{D(G)}$ in the form of $[\prod_{j=1}^L a^{(j)}, b]$ for $a^{(j)}\in D(G)^{(j)}$, $b\in\overline{D(G)}$. 
We then propose that the Lagrangian algebra anyon of $(D(G))^L\boxtimes \overline{D(G)}$ for the layer construction is given by 
\begin{align}
    \mathcal{L}=\bigoplus_{[g]\in\mathrm{Conj}(G)}\bigoplus_{\pi_g\in\mathrm{Rep}(C_g)}\left[\left(\prod_{j=1}^L([g],\mathbf{1})^{(j)}\right)\times ([1],\pi_g)^{(1)}\times \left(\prod_{j=1}^{L-1}\bigoplus_{\pi'_g\in\mathrm{Rep}(C_g)}\left(([1],\pi'_g)^{(j)}\times ([1], \overline{\pi'_g})^{(j+1)}\right)\right), \quad ([g],\overline{\pi_g})\right]
    \label{eq:lagrangianeq}
\end{align}
where the object in the form of $([1],\pi'_g)$ with $\pi'_g\in \mathrm{Rep}(C_g)$ is an abuse of notation, since it does not correspond to an anyon of $D(G)$ by itself. This defines an anyon after fusing with a magnetic flux $([g], \pi_g)$, where we define the fusion as
\begin{align}
    ([g], \pi_g)\times ([1], \pi'_g) := \bigoplus_{\epsilon_g} N_{\epsilon}([g], \epsilon_g)
\end{align}
with $\epsilon_g\in\mathrm{Rep}(C_g)$, where the decomposition into irreducible representation is given by $\pi_g\otimes \pi'_g=\bigoplus_{\epsilon_g}N_{\epsilon} \epsilon_g$. 
Let us explain the physical intuitions behind the form of the Lagrangian algebra anyon in Eq.~\eqref{eq:lagrangianeq}. First, it has the effect of identifying the array of magnetic fluxes in $D(G)$ in the form of $\prod_{j=1}^L([g],\mathbf{1})^{(j)}$ as the magnetic particle $([g],\mathbf{1})$ of $D(G)$ in the condensed theory. Also, when an electric charge $([1], \pi_g)^{(1)}$ is attached to a magnetic flux in a single (first) layer, it becomes identified with the dyon $([g],\pi_g)$ of the condensed theory. Finally, we are condensing the pair of electric particles in the neighboring layer by the term $\prod_{j=1}^{L-1}\bigoplus_{\pi'_g\in\mathrm{Rep}(C_g)}\left(([1],\pi'_g)^{(j)}\times ([1], \overline{\pi'_g})^{(j+1)}\right)$. In particular, when $g=1$, this term has the effect of identifying the pair of electric particles in the form of $([1],\pi')^{(j)}\times ([1], \overline{\pi'})^{(j+1)}$ for $\pi\in \mathrm{Rep}(G)$ as a trivial anyon of the condensed theory. This means that we are condensing these pairs of electric particles. 

We can check the Lagrangian property is indeed satisfied for $\mathcal{L}$ in Eq.~\ref{eq:lagrangianeq}. That is, the quantum dimension of the Lagrangian algebra anyon $\mathcal{L}$ defined as $\mathrm{dim}(\mathcal{L}):=\sum_{a} Z_{0a}d_a$ must be identical to the total quantum dimension of the modular tensor category $\mathcal{C}$,
\begin{align}
    \mathrm{dim}(\mathcal{L}) = \sqrt{\sum_{a\in\mathcal{C}}d_a^2}.
    \label{eq:lagrangianproperty}
\end{align}
To see this, let us explicitly compute the quantum dimension of $\mathcal{L}$. In the expression of Eq.~\eqref{eq:lagrangianeq}, each anyon in the big parenthesis $[]$ contains $L$ pairs of electric particles in the form of $\{([1],\pi_g), ([1],\overline{\pi_g})\}$, and by summing over $\pi_g$ for each pair, each contributes as $\sum_{\pi_g}\mathrm{dim}(\pi_g)^2= |C_g|$ to the quantum dimension. So, after summing over the labels of electric particles, the contribution of the electric particles is evaluated as $|C_g|^L$ for $L$ pairs. Hence, the quantum dimension of $\mathcal{L}$ is rewritten as
\begin{align}
\begin{split}
        \mathrm{dim}(\mathcal{L})&=\sum_{[g]\in\mathrm{Conj}(G)}\mathrm{dim}\left(\left[\prod_{j=1}^L([g],\mathbf{1})^{(j)}, \quad ([g],\mathbf{1})\right]\right)\cdot |C_g|^L \\
        &= \sum_{[g]\in\mathrm{Conj}(G)}|[g]|^{L+1}\cdot |C_g|^L \\
        &= \sum_{[g]\in\mathrm{Conj}(G)} |G|^L\cdot |[g]| \\
        &= |G|^{L+1}.
\end{split}
\end{align}
Meanwhile, $\mathrm{dim}(D(G)^L\boxtimes\overline{D(G)})= \mathrm{dim}(D(G))^{L+1}=|G|^{L+1}$.
We hence have the property Eq.~\eqref{eq:lagrangianproperty}.

The Lagrangian algebra anyon must also satisfy that the each anyon in the summand of $\mathcal{L}$ carries the trivial spin, and trivial $F$ and $R$ symbols,
\begin{align}
    (F^{\mathcal{L} \mathcal{L} \mathcal{L}}_{\mathcal{L}})_{\mathcal{L} \mathcal{L}} = 1,\quad R^{\mathcal{L} \mathcal{L}}_{\mathcal{L}}=1,
\end{align}
which means the $F$ and $R$ symbols are 1 for all possible fusion vertices that represent $\mathrm{Hom}(\mathcal{L}\otimes\mathcal{L},\mathcal{L})$. An explicit proof of these properties for the anyon in Eq.~\eqref{eq:lagrangianeq} is left for future work. 

We remark that the way of performing layer construction for $D(G)^L$ to obtain $D(G)$ is not unique in general, and Eq.~\eqref{eq:lagrangianeq} is not the most general form of the Lagrangian algebra anyon for possible layer construction. For instance, Eq.~\eqref{eq:lagrangianeq} does not capture the example of condensing a pair of fermions from neighboring layers in the $\Z_2$ toric code described in Sec.~\ref{subsubsec:layerfermion}, since Eq.~\eqref{eq:lagrangianeq} corresponds to condensing a pair of bosonic particles. If we instead consider discrete gauge theory based on a fermionic symmetry group $G_f$, then a similar analysis as above should go through. We leave a systematic analysis of this case for future work.

\subsubsection{Twist strings in  non-Abelian discrete $G$-gauge theory}
\label{sec:twiststringnonabelianlayer}

Here, we generalize the construction of twist strings in the Abelian case  discussed in Sec.~\ref{sec:twiststringabelianlayer} to the case of non-Abelian discrete $G$-gauge theory in the framework of layer construction. That is, we construct a codimension-2 twist string in the (3+1)D $G$-gauge theory by inserting a codimension-1 defect $\mathcal{D}$ of the (2+1)D $G$-gauge theory $D(G)$ in a single layer, and perform the layer construction in the presence of the single insertion of the defect.
Here we introduce a condition required for $\mathcal{D}$, described in Eq.~\eqref{eq:fluxpreserving} or Eq.~\eqref{eq:twiststringtopological} below. 
The first condition Eq.~\eqref{eq:fluxpreserving} is required for the twist string to be compatible with the condensation procedure that determines the layer construction, The second condition Eq.~\eqref{eq:twiststringtopological} is strictly stronger than Eq.~\eqref{eq:fluxpreserving}, and further guarantees that the twist string is a topological defect in the (3+1)D $G$-gauge theory after the layer construction.
Let us explain these conditions by steps.

Analogously to the discussion in Sec.~\ref{sec:twiststringabelianlayer}, we require the twist string to induce a map between the set of deconfined excitations in the (3+1)D $G$-gauge theory after the layer construction, and such a map must obviously preserve the labels of all flux strings (loops).  Therefore, the corresponding domain wall $\mathcal{D}$ in the (2+1)D $G$-gauge theory in the context of layer construction must induce a {flux-preserving automorphism} 
$\rho_{\mathcal{D}}\in \mathrm{Aut}(D(G))$, i.e.,
\begin{equation}
\rho_{\mathcal{D}} : ([g], \pi_g) \rightarrow ([g], \pi'_g) \quad \text{for all anyons $([g],\pi_g)\in D(G)$}.
\label{eq:fluxpreserving}
\end{equation}
The above equation means that the automorphism associated with the domain wall $\mathcal{D}$ preserves the conjugacy class $[g]$, i.e., the magnetic flux, while in general could permute the representation, i.e., the electric charge, from $\pi_g$ to $\pi'_g$. 

We expect that an insertion of the defect satisfying Eq.~\eqref{eq:fluxpreserving} defines a codimension-2 defect of the (3+1)D theory after the layer construction, but it turns out that the resulting codimension-2 defect is not necessarily ``topological''; the action of the defect on the flux string can depend on the detail of where the point-like irrep $\pi_g$ is attached along the extended flux string. Also, the action can depend on where the defect acts on the flux string, since the defect acts at the specific point on the extended flux string by crossing. If we want the twist string to be topological, we want the action of the defect to be independent of the detail of the configuration of the flux string and irreps attached to it. For this purpose, we require a further condition that the action of $\rho_{\mathcal{D}}$ only depends on the flux label $[g]$ of the anyon $([g],\pi_g)$, i.e.,
\begin{align}
    \rho_{\mathcal{D}}: ([g],\pi_g) \to ([g],\pi_g\times \sigma_{[g]}), \quad \text{for all anyons $([g],\pi_g)\in D(G)$}.
    \label{eq:twiststringtopological}
\end{align}
where $\sigma_{[g]}\in \mathrm{Rep}(C_g)$ only depends on the flux label of the anyon $[g]$, and independent of the irreps $\pi_g$ attached to it. This condition guarantees that the action of the twist string on the flux string does not depend on which layer the point-like irrep $\pi_g$ is attached to the flux string. For $\rho_{\mathcal{D}}$ to be an automorphism, we require $\mathrm{dim(}\sigma_{[g]})=1$ so that the quantum dimension of the anyons are preserved under $\rho_{\mathcal{D}}$. The irreps $\sigma_{[g]}$ can then be regarded as an element $\sigma_{[g]}\in \mathcal{H}^1(C_g,\U)$.

Summarizing, given an automorphism with the flux-preserving property Eq.~\eqref{eq:fluxpreserving} and the corresponding domain wall $\mathcal{D}$ in the (2+1)D $G$-gauge theory, one can construct the corresponding twist string in (3+1)D $G$-gauge theory through the layer construction, i.e., by condensing all the pairs of pure electric charges and anti-charges $a_j \bar{a}_{j+1}$ with $a \equiv ([1], \pi)$ as discussed in Sec.~\ref{sec:non-Abelian_layer_construction}, in the presence of the domain wall $\mathcal{D}$ inserted in a single layer. If we want the twist string to be topological, we further require the condition Eq.~\eqref{eq:twiststringtopological}. 

For an Abelian $G$ gauge theory (a special case of the description in Sec.~\ref{sec:layerfromabelian}), we note that the flux-preserving condition Eq.~\eqref{eq:fluxpreserving} implies Eq.~\eqref{eq:twiststringtopological}. This is because the flux-preserving condition is equivalent to the property that the pure charges $(1,\pi)$ are invariant under the action of $\rho_{\mathcal{D}}$, so the action of $\rho_{\mathcal{D}}$ on dyons $(g,\pi)$ are automatically determined as Eq.~\eqref{eq:twiststringtopological} once we define $\sigma$ by the the action of $\rho_{\mathcal{D}}$ on the pure flux as $\rho_{\mathcal{D}}: (g,1)\to (g,\sigma_{g})$. 

However, in a non-Abelian $G$ gauge theory, Eq.~\eqref{eq:fluxpreserving} does not necessarily imply Eq.~\eqref{eq:twiststringtopological}. In Sec.~\ref{subsubsec:A6}, we indeed find an example of an automorphism that satisfies Eq.~\eqref{eq:fluxpreserving} while violating Eq.~\eqref{eq:twiststringtopological}, in the (2+1)D $A_6$ gauge theory $D(A_6)$. This automorphism leads to a non-topological codimension-2 defect of (3+1)D $A_6$ gauge theory. Meanwhile, we can also find an automorphism satisfying Eq.~\eqref{eq:twiststringtopological} in $D(A_6)$, and it leads to a topological twist string of (3+1)D $A_6$ gauge theory.

In non-Abelian $G$ gauge theory, we remark that the automorphism in Eq.~\eqref{eq:fluxpreserving} in general cannot be realized by fusing the pure electric charge in the form of $([1],\pi)$ to the dyon $([g], \pi_g)$. For example, the $A_6$ gauge theory $D(A_6)$ considered in Sec.~\ref{subsubsec:A6} does not contain any Abelian pure electric charge, reflecting that $A_6$ is a perfect group and does not admit an Abelian irreducible representation. The fusion of a pure electric charge in $D(A_6)$ hence cannot give an automorphism.

Let us investigate the properties of the flux-preserving automorphism $\rho_{\mathcal{D}}$ satisfying Eq.~\eqref{eq:fluxpreserving}, and show some fundamental constraints that $\rho_{\mathcal{D}}$ needs to satisfy. For this purpose, we study the modular $S$ and $T$ matrices, which are left invariant under the automorphism $\rho_{\mathcal{D}}$.  The modular matrices are given by the following formulae~\cite{beigi2011}: 
\begin{equation}
    {S}_{([g],\pi_g)([g'],\pi_{g'})} = \frac{1}{\abs{C_g}\abs{C_{g'}}} \sum_{h:hg'h^{-1} \in C_g} \text{tr}_{\pi_g}(hg'^{-1}h^{-1})\text{tr}_{\pi_{g'}}(hg^{-1}h^{-1}), \label{eq:modular_S}
\end{equation}
\begin{equation}
    T_{([g],\pi_g)([g],\pi_g)}=\theta_{([g],\pi_g)}=\frac{\text{tr}_{\pi_g}(g)}{\text{tr}_{\pi_g}(1)},    \label{eq:modular_T}
\end{equation}
where $\text{tr}_{\pi_g}(\cdot)$ represents the character of the representation $\pi_g$.

One can show that the conjugacy class preserving condition in Eq.~\eqref{eq:fluxpreserving} for the automorphism $\rho_{\mathcal{D}}$ is equivalent to the condition that $\rho_{\mathcal{D}}$ preserves all the pure charges in the form of $([1],\pi)$,
\begin{align}
    \rho_{\mathcal{D}}: ([1],\pi)\to ([1],\pi) \quad \text{for all the irreps $\pi$ of $G$}.
    \label{eq:purechargeconserved}
\end{align}
To see the equivalence, we firstly derive the property in Eq.~\eqref{eq:purechargeconserved} from Eq.~\eqref{eq:fluxpreserving}. This can be immediately shown by checking the invariance of the modular $S$ matrix. Suppose that $\rho_{\mathcal{D}}$ transforms the anyons as $\rho_{\mathcal{D}}: ([1],\pi)\to ([1],\pi')$ and $\rho_{\mathcal{D}}: ([g],\mathbf{1})\to ([g],\pi_g'')$, where $\mathrm{dim}(\pi''_g)=1$ since the quantum dimension is preserved under $\rho_{\mathcal{D}}$. We then must have $S_{([1],\pi),([g],\mathbf{1})} = S_{([1],\pi'),([g],\pi_g'')}$. This condition can be expressed as 
\begin{align}
    \mathrm{tr}_{\pi}(g^{-1}) = \mathrm{tr}_{\pi'}(g^{-1}).
\end{align}
Since this is valid for any $g\in G$, the irreps $\pi,\pi'$ must be equivalent, so we have Eq.~\eqref{eq:purechargeconserved}.

Conversely, one can also show the flux-preserving property Eq.~\eqref{eq:fluxpreserving} from Eq.~\eqref{eq:purechargeconserved}, by utilizing the invariance of the modular $S$ matrix. Suppose that $\rho_{\mathcal{D}}$ transforms the anyons as $\rho_{\mathcal{D}}: ([1],\pi'')\to ([1],\pi'')$ and $\rho_{\mathcal{D}}: ([g],\pi_g)\to ([g'],\pi'_{g'})$. We then have $S_{([1],\pi''),([g],\pi_g)} = S_{([1],\pi''),([g],\pi'_{g'})}$, which is expressed as
\begin{align}
\begin{split}
    &\frac{\text{tr}_{\pi''}(g^{-1}) \dim(\pi_g) }{\abs{C_g}} =\frac{\text{tr}_{\pi''}(g{'}^{-1}) \dim(\pi'_{g'}) }{\abs{C_{g'}}}
    \end{split}
\end{align}
Using the invariance of the quantum dimensions under $\rho_{\mathcal{D}}$ given by $|[g]|\mathrm{dim}(\pi_g)=|[g']|\mathrm{dim}(\pi'_{g'})$, one can simply rewrite the condition as
\begin{align}
    \text{tr}_{\pi''}(g^{-1}) =  \text{tr}_{\pi''}(g'^{-1}).
\end{align}
Since this is valid for any irreps $\pi''$ of $G$, the conjugacy class has to be preserved: $[g]=[g']$, hence Eq.~\eqref{eq:fluxpreserving}.

Below we discuss several examples of flux-preserving automorphism. 

\subsubsection{Example: Twist string in $D_{4}$ gauge theory}

The first example is the quantum double $D(D_{4})$. There exists a flux-preserving automorphism which we summarize below:
\begin{align}
    ([2,0],J_{0}) &\leftrightarrow ([2,0],J_{1}), \  ([0,1],A_{0}) \leftrightarrow ([0,1],A_{2}), \ 
    ([1,1],A_{0}) \leftrightarrow ([1,1],A_{3}),
    \\
    ([2,0],J_{2}) &\leftrightarrow ([2,0],J_{3}), \
    ([0,1],A_{1}) \leftrightarrow ([0,1],A_{3}), \ 
    ([1,1],A_{1}) \leftrightarrow ([1,1],A_{2}).
    \label{eq:D4automorphism}
\end{align}
In Appendix ~\ref{app:D4ST}, we check that the automorphism leaves the $S$ and $T$ matrices invariant. In Sec.~\ref{sec:wrapping}, we show that the twist string that realizes this automorphism can be obtained by gauging the non-trivial $(1+1)$D SPT state protected by the $D_{4}$ symmetry.

\subsubsection{Example: Twist strings in $A_6$ gauge theory}
\label{subsubsec:A6}
Another example is given by the quantum double model $D(A_6)$. The corresponding group $G=A_6$ is the alternating group of order six, i.e., the group of even permutations of six objects: $\{1,2, ..., 6\}$.  We consider a particular group element composed of two elementary permutations, i.e., $\alpha=(1,2)(3,4)$. The corresponding conjugacy class is the set of permutations with form of a product of two elementary permutations, i.e., $(i_1,i_2)(i_3,i_4)$. The centralizer  group of $\alpha$ is then $C_\alpha=\{1,\alpha,\beta_1,\beta_2,\beta_3,\beta_4,\gamma_1,\gamma_2\}$, with $\beta_1=(1,2)(5,6), \beta_2=(3,4)(5,6), \beta_3=(1,3)(2,4), \beta_4=(1,4)(2,3), \gamma_1=(1,3,2,4)(5,6)$ and $\gamma_2=(1,4,2,3)(5,6)$. The conjugacy classes of the centralizer subgroup $C_\alpha$ are $\{1\}$, $\{\alpha\}$, $\{\beta_1,\beta_2\}$, $\{\beta_3,\beta_4\}$, and $\{\gamma_1,\gamma_2\}$. The corresponding character table is listed in Table \ref{tb:A6}. There exist two types of flux-preserving automorphisms $\rho_1,\rho_2\in\mathrm{Aut}(D(A_6))$ along with the corresponding domain walls, which has been studied in Ref.~\cite{beigi2011} in the (2+1)D theory. The nontrivial action of each automorphism on anyons is given by:
\begin{align}
    \rho_1 : ([\alpha], \pi_1) &\leftrightarrow ([\alpha], \pi_2) \\
    \rho_2 : ([\alpha], \pi_3) &\leftrightarrow ([\alpha], \pi_4) = ([\alpha], \pi_3 \times \pi_2)
    \label{eq:automorphismA6}
\end{align}
and leaves other anyons invariant. Here, $\pi_1$ denotes the trivial representation $\pi_1=\mathbf{1}$.
One can check that both types of automorphism leave the modular $S$, $T$ matrices invariant according to Eq.~\eqref{eq:modular_S} and Eq.~\eqref{eq:modular_T}~\cite{beigi2011}. These two examples provide new twist strings in $(3+1)$D non-Abelian gauge theory by using the layer construction.

\begin{figure}[t]
    \centering
    \includegraphics[width=0.8\textwidth]{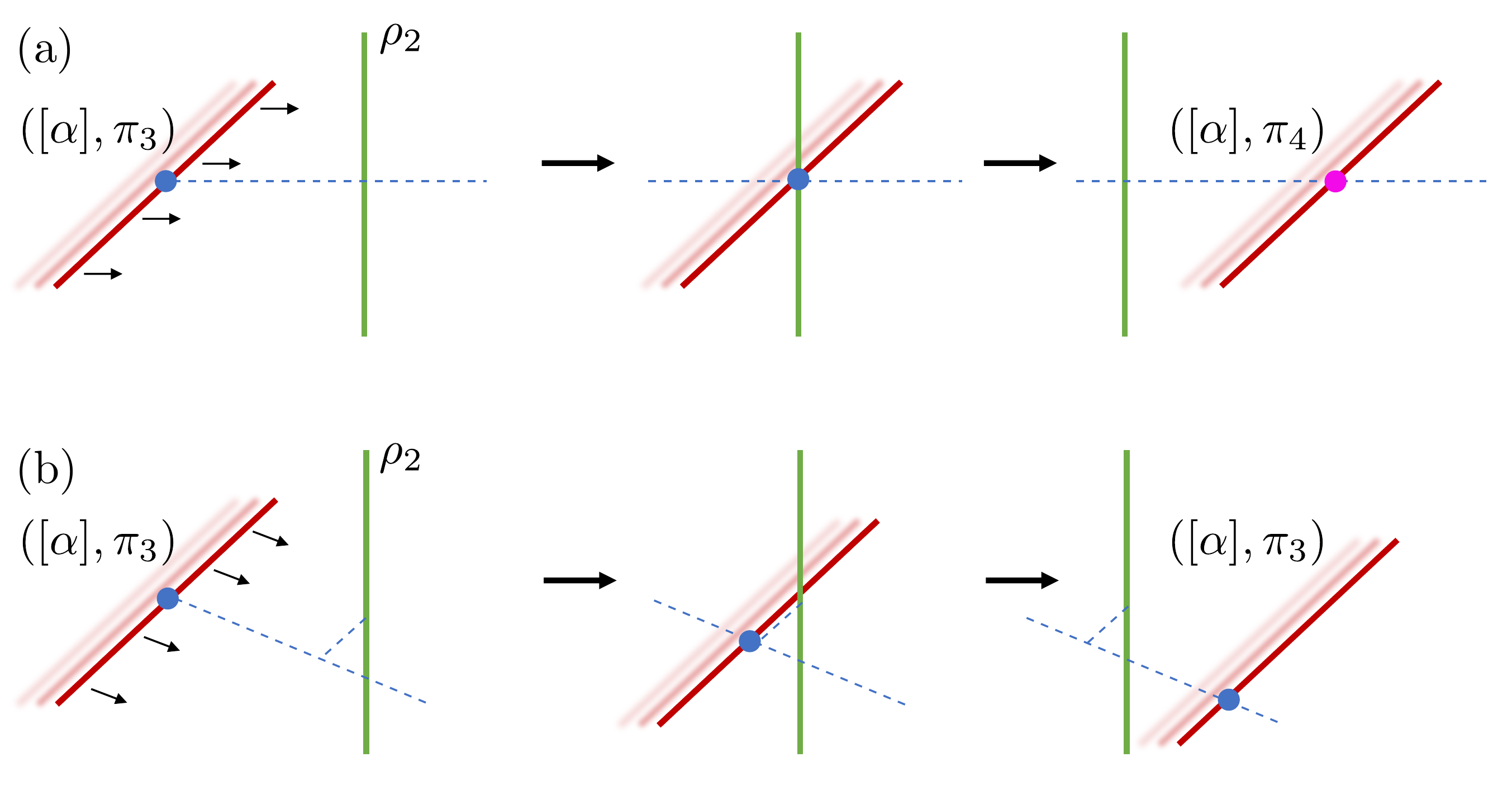}
\caption{Illustration of the properties of geometric twist string $\rho_2$ (green line).  Consider the flux string with an attached charge labeled as $([\alpha],\pi_3)$, the permutation this attached charge depends on the detailed geometric trajectory of the flux string and the charge.  (a)  If the worldline of the attached charge $\pi_3$ (blue) intersects with the geometric twist string, the attached charge will be transformed into $\pi_4$ (purple).  (b) If the worldline of the attached charge avoids the twist string, the attached charge remains the same.}
\label{fig:geometric_defect_A6}
\end{figure}

Note that both of the automorphisms $\rho_1$, $\rho_2$ obviously violate the condition Eq.~\eqref{eq:twiststringtopological}, since their actions on the anyons explicitly depend on the label of the irrep $\pi_i$ attached to the flux. Therefore, the twist strings constructed out of $\rho_1$, $\rho_2$ give examples of non-topological twist strings.  We call such defects \textit{geometric twist strings}, since the permutation of charges depends on the detailed geometry of the trajectory.   For example, when considering the $\rho_2$ twist string, a flux string attached with an additional charge labeled as $([\alpha], \pi_3)$ only gets a permutation on the attached charge, i.e., transformed to $([\alpha], \pi_4)$ when the worldline of the attached charge intersects with the twist string, as illustrated in Fig.~\ref{fig:geometric_defect_A6}(a).   On the other hand, if the worldline of the charge is slightly perturbed away from intersecting with the twist string, the charge label remains to be $\pi_3$, as illustrated in Fig.~\ref{fig:geometric_defect_A6}(b).  As one can see, the charge permutation phenomenon is not topologically robust in this case.   The lack of topologically robust intersection between the worldline of attached charge and the twist string can be understood via their intersection dimension  $d_\text{int}=1+1-3=-1<0$.~\footnote{The intersection dimension between an $n$-dimensional submanifold with another $m$-dimensional submanifold in a $k$-dimensional manifold is a topological invariant given by the following formula: $d_\text{int}=n+m-k$.}
In contrast, the intersection dimension between the worldsheet of a flux string and the twist string is $d_\text{int}=2+1-3=0$, indicating the robust intersection at a single point (0-dimensional).  

Now we consider the more subtle case of twist string $\rho_1$. If a pure flux string $([\alpha],\pi_1) \equiv ([\alpha],\bf{1})$ passes through twist string $\rho_1$, it will gets attached by an additional charge $\pi_2$, i.e., the label is changed to $([\alpha],\pi_2)$. This phenomenon is topologically robust. However, if the flux string is attached with a charge and labeled as $([\alpha], \pi_3)$, this attached charge $\pi_3$ will remain there when the worldline of the charge intersects with the twist string and hence no $\pi_2$ charge is produced in this case. If the worldline is slightly perturbed away from intersecting with the twist string, a $\pi_2$ charge will still be produced, and the total charge becomes $\pi_3 \times \pi_2 =\pi_4$ instead of $\pi_3$.   Therefore, the total charge after the transformation depends on the detailed geometry of the trajectory, and the twist string $\rho_1$ is also not fully topological.  

Meanwhile, the composite automorphism given by $\rho_1\circ\rho_2$ gives rise to the following automorphism, 
\begin{equation}
\rho_1\circ\rho_2 : ([\alpha], \pi_i) \rightarrow ([\alpha], \pi_i\times \pi_2), 
\end{equation}
where $\pi_i$ refers to any irreducible representation of $C_\alpha$. This satisfies Eq.~\eqref{eq:twiststringtopological}, so the corresponding twist string becomes topological.
In this case, independent of the charge attached to the initial flux string, the total charge after the transformation always gets multiplied by $\pi_2$.

Also, as mentioned in Sec.~\ref{sec:non-Abelian_layer_construction}, the actions of $\rho_1$, $\rho_2$ and $\rho_1\circ\rho_2$ cannot be understood in terms of attaching a pure charge $(1,\pi)$ to anyons. It follows that in this example, the twist string does not attach a deconfined point charge to the flux string that crosses it; rather there is an irreducible representation $\pi_i$ on the flux string that gets permuted upon crossing the twist string.

\begin{table*}
\begin{center}
 \begin{tabular}{ c | c c c c c } 
 $C_\alpha$ & $1$ & $\alpha$ & $\beta_1, \beta_2$ & $\beta_3, \beta_4$ & $\gamma_1, \gamma_2$ \\
 \hline
 $\pi_1$ & $1$ & $1$ & $1$ & $1$ & $1$
 \\
 
 $\pi_2$ & $1$ & $1$ & $-1$ & $-1$ & $1$
 \\
 $\pi_3$ & $1$ & $1$ & $1$ & $-1$ & $-1$
 \\
 $\pi_4$ & $1$ & $1$ & $-1$ & $1$ & $-1$
 \\
 $\pi_5$ & $2$ & $-2$ & $0$ & $0$ & $0$
 \end{tabular}
\end{center}
\caption{The character table of the centralizer subgroup $C_\alpha$.}
\label{tb:A6}
\end{table*}

\subsubsection{Relation between twist strings from layer construction and gauging (1+1)D SPT phases}
\label{subsec:sigmamodularinvariant}

Here we discuss a mathematical consequence of the topological twist strings constructed in Sec.~\ref{sec:twiststringnonabelianlayer}. Specifically, we show below that the automorphism $\rho_{\mathcal{D}}$ satisfying the property Eq.~\eqref{eq:twiststringtopological} defines a modular invariant $Z_{(g,h)}$, which can be interpreted as a partition function of a (1+1)D bosonic $G$-SPT phase on a torus. This provides a partial correspondence between the topological twist strings constructed using the layer construction Sec.~\ref{sec:twiststringnonabelianlayer} and those constructed by gauging (1+1)D bosonic SPT phases.

We have seen in Sec.~\ref{sec:twiststringnonabelianlayer} that the topological twist string in (3+1)D $G$-gauge theory is obtained from an automorphism $\rho_{\mathcal{D}}$ of the (2+1)D $G$-gauge theory satisfying the condition Eq.~\eqref{eq:twiststringtopological}.
To make contact with the (1+1)D SPT phase, for a given automorphism $\rho_{\mathcal{D}}$ satisfying Eq.~\eqref{eq:twiststringtopological} let us define
\begin{align}
    Z_{(g,h)}:= \sigma_{[g]}(h)
\end{align}
for a commuting pair $g,h\in G$ satisfying $[g,h]=1$. Here $\sigma_{[g]}\in \mathcal{H}^1(C_g,\U)$ characterizes the permutation action of $\rho_{\mathcal{D}}$ on anyons in Eq.~\eqref{eq:twiststringtopological}.
One can regard $Z_{(g,h)}$ as a function of the flat $G$-gauge field on a torus $T^2$ with holonomy $(g,h)$ on fundamental cycles. 
Later, we conjecture that the phase $Z_{(g,h)}$ is always understood as a partition function of a (1+1)D bosonic $G$-SPT phase on a torus.

One can show that $Z_{(g,h)}$ defines a modular invariant on a torus. To show this, we utilize the invariance of the modular data of $D(G)$ under the action of $\rho_{\mathcal{D}}$. Since the $T$ matrix Eq.~\eqref{eq:modular_T} is invariant under the action of $\rho_{\mathcal{D}}$ on anyons, the spins of the anyons are preserved under the action of $\rho_{\mathcal{D}}$. This gives $\theta_{([g],\mathbf{1})} = \theta_{([g],\sigma_{[g]})}$, so we have $\sigma_{[g]}(g) = 1$. We then get $\frac{Z_{(g,gh)}}{Z_{(g,h)}} = \sigma_{[g]}(g) = 1$, so the $T$ invariance follows
\begin{align}
    Z_{(g,gh)} = Z_{(g,h)}.
\end{align}
 Also, according to the invariance of the $S$ matrix element Eq.~\eqref{eq:modular_S} under the action of $\rho_{\mathcal{D}}$, we have $S_{([g],1),([h],1)} = S_{([g],\sigma_{[g]}),([h],\sigma_{[h]})}$. This equation is rewritten as $1 = \sigma_{[g]}(h^{-1})\sigma_{[h]}(g^{-1})$, where we used $\mathrm{Tr}_{\sigma_{[g]}}(h^{-1})=\sigma_{[g]}(h^{-1})$.
 which gives $\frac{Z_{(h^{-1},g)}}{Z_{g,h}} = \frac{\sigma_{[h^{-1}]}(g)}{\sigma_{[g]}(h)} = \sigma_{[h^{-1}]}(g)\sigma_{[g^{-1}]}(h) = 1$. The $S$ invariance then follows
\begin{align}
    Z_{(h^{-1},g)} = Z_{(g,h)}.
\end{align}
Further, since $\sigma_{[g]}\in \mathcal{H}^1(C_g,\U)$, we obviously have
\begin{align}
    Z_{(g,h)}Z_{(g,k)} = Z_{(g,hk)} \quad \text{for $h,k\in C_g$.}
    \label{eq:Zrep}
\end{align}

Conversely, for a given torus modular invariant $Z_{(g,h)}$ for a commuting pair $g,h\in G$ and satisfying $Z_{(g,h)}Z_{(g,k)} = Z_{(g,hk)}$, one can define $\sigma_{[g]}(h):= Z_{(g,h)}$ and the permutation action of $\rho_{\mathcal{D}}$ as $([g],\pi_g)\to ([g],\pi_{g}\times\sigma_{[g]})$ leaving $S,T$ invariant. Therefore, there is one-to-one correspondence between a torus modular invariant $Z_{(g,h)}$ satisfying Eq.~\eqref{eq:Zrep}, and a permutation action of $\rho_{\mathcal{D}}$ on anyons leaving the full $S,T$ matrices invariant and satisfying Eq.~\eqref{eq:twiststringtopological}.

In the following section, we construct topological twist strings by gauging a $G$ symmetry after decorating a string with a (1+1)D SPT. The (1+1)D SPT in this case is characterized by an element $\omega \in \mathcal{H}^2(G, \U)$. We would like to prove an equivalence between the gauged (1+1)D SPT construction and the layer construction discussed in Sec.~\ref{sec:twiststringnonabelianlayer}. To do so, we need to show that an automorphism $\rho_{\mathcal{D}}$ satisfying Eq.~\eqref{eq:twiststringtopological} defines the same twist string as specifying some element of $\mathcal{H}^2(G, \U)$, and vice versa. In Sec.~\ref{sec:wrapping}, we will show how an element of $\mathcal{H}^2(G, \U)$ indeed specifies an automorphism $\rho_D$ satisfying Eq.~\eqref{eq:twiststringtopological}. So we need to consider the opposite direction, of how to go from $\rho_{\mathcal{D}}$ satisfying Eq.~\eqref{eq:twiststringtopological} to an element of $\mathcal{H}^2(G, \U)$. 

Above, we have shown that we can go from $\rho_{\mathcal{D}}$ to a torus modular invariant $Z_{g,h} \in \U$. This takes us part of the way to specifying an element of $\mathcal{H}^2(G, \U)$. However, we saw that the torus modular invariant $Z_{g,h}$ only specifies the permutation action of $\rho_{\mathcal{D}}$. To specify the full automorphism of the modular category, we also need information about how $\rho_{\mathcal{D}}$ acts on anyon fusion / splitting vertices \cite{barkeshli2019}. It is unlikely that this information is contained in $Z_{g,h}$, since $Z_{g,h}$ only guarantees invariance of the modular $S, T$ matrices, and not the ``beyond modular data" of the (2+1)D TQFT~\cite{mignard2021beyondmodular, bonderson2019beyondmodular}. It thus seems unlikely that $Z_{g,h}$ is sufficient to fully define an element of $\mathcal{H}^2(G,\U)$. Therefore, to extract an element of $\mathcal{H}^2(G, \U)$ from $\rho_{\mathcal{D}}$, we presumably need to do more than to simply extract the torus modular invariant $Z_{g,h}$, and also study how the action of $\rho_{\mathcal{D}}$ on anyon fusion / splitting vertices generically defines an element $\omega \in \mathcal{H}^2(G, \U)$. We leave a detailed study of this for future work.

\section{Twist strings via gauging (1+1)D bosonic SPT phases}
\label{sec:wrapping}
\begin{figure}
    \centering
    \includegraphics[width=0.75\textwidth]{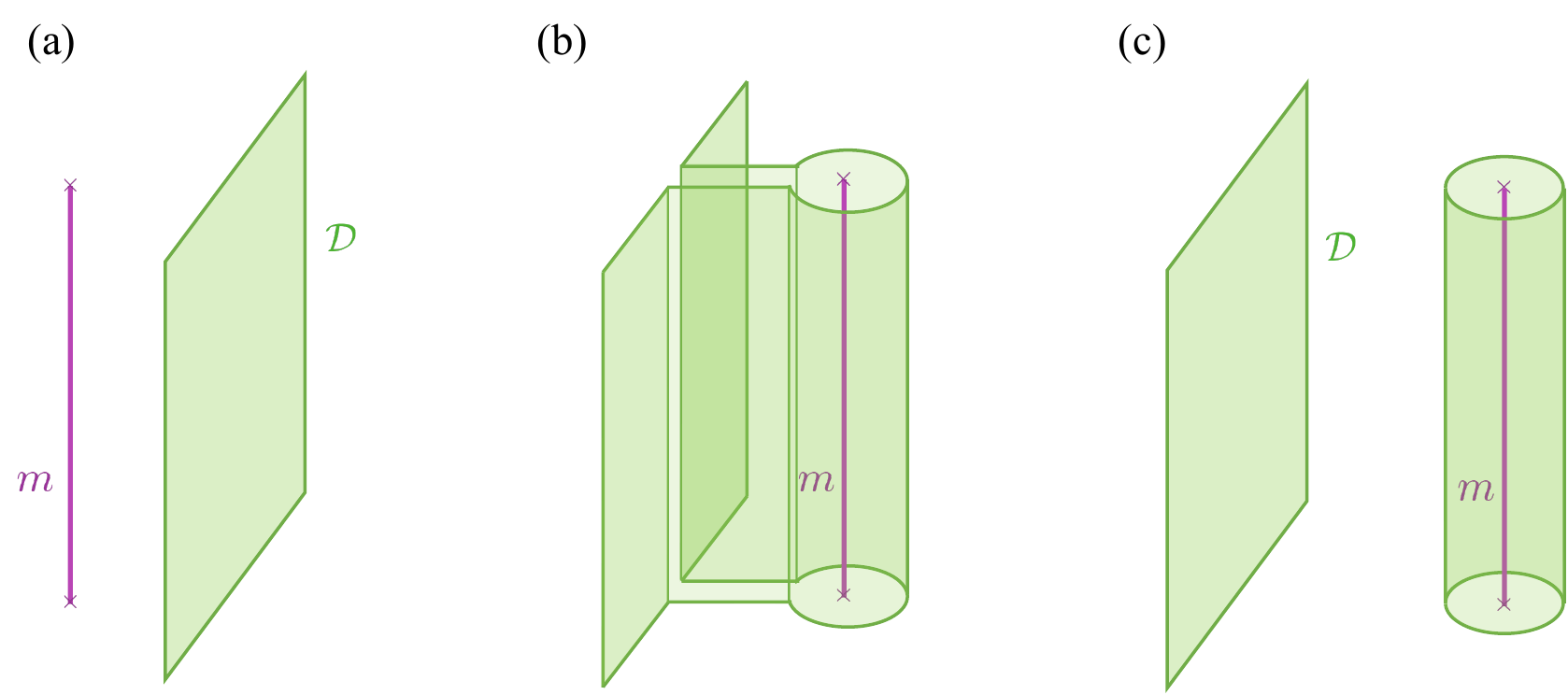}
    \label{fig:wrap}
\caption{The process of a magnetic flux passing through a codimension-1 invertible twist string in a $(2+1)$D theory. The figure shows the (2+1)D spacetime. In this figure, the magnetic flux $m$ is a line operator (purple line) extended in the vertical direction.
Meanwhile, the codimension-1 twist string defect $\mathcal{D}$ is described as a 2D surface in the spacetime. As a magnetic flux passes through the defect, the defect will wrap around it as shown in (b). One can shrink the wrapping cylinder of the defect so that the magnetic flux is dressed by a dimensionally-reduced twist defect as shown in (c).
Note that the whole process leaves the correlation function invariant, due to bordism invariance of the partition function of the (1+1)D SPT phase that corresponds to the defect $\mathcal{D}$. 
}

\label{fig:2dwrap}
\end{figure}

Here we give a systematic construction of twist strings in discrete $G$-gauge theory from the perspective of gauging lower dimensional SPT phases. Though the logic in this section is applicable for discrete gauge theory in generic spacetime dimensions, we start with the case of (2+1)D $G$-gauge theory as a warm-up.

We want to construct codimension-1 twist strings of a $(2+1)$D $G$-gauge theory. For this purpose, we start with a trivial bosonic $G$-SPT phase in (2+1)D, and then decorate a codimension-1 submanifold with a $(1+1)$D bosonic SPT phase with $G$ symmetry. After gauging the $G$ symmetry, it defines a codimension-1 invertible defect of a (2+1)D $G$-gauge theory, with its fusion rule given by the stacking rule of the (1+1)D SPT phase. A general picture on how the twist string acts on anyons can be understood by the following argument. As shown in Fig.~\ref{fig:2dwrap}, when a magnetic excitation of the $G$-gauge theory passes through the defect, the defect will wrap around the magnetic flux excitation. By shrinking the wrapping part of the defect, we find that the magnetic flux is dressed with a (0+1)D object obtained by ``dimensional reduction'' of the invertible defect. 

\subsection{Abelian $G$}

Now we consider a twist string built by decorating a $(1+1)$D bosonic SPT state labeled by $\omega \in \mathcal{H}^{2}(G,\mathrm{U}(1))$ in the $(2+1)$D untwisted gauge theory. To illustrate the main idea in a simple setup, let us assume that the gauge group is Abelian. The case that $G$ is non-Abelian is studied later.
We think of this $(2+1)$D gauge theory as a result of gauging a trivial (2+1)D invertible phase with $G$ symmetry. The  magnetic flux is labeled by $g \in G$, while the electric particle is labeled by an irreducible $G$-representation in $ \mathrm{Rep}({G})\cong \mathcal{H}^1(G,\mathrm{U}(1))$. Before we gauge the symmetry, we consider a background $g$ magnetic flux and bring it through the defect. When a $g$ flux passes through the defect, by the wrapping argument, we find that the $g$ flux is dressed by a dimensional reduction of a $(1+1)$D bosonic SPT state with a $g$ flux insertion. The result of the dimensional reduction is a $(0+1)$D invertible phase whose ground state carries electric charge ${\pi}_{g}\in \mathcal{H}^1(G,\mathrm{U}(1))$ given by the slant product $i_{g}$: 
\begin{equation}
    {\pi}_{g}(h) = i_{g}\omega(h) = \frac{\omega(g,h)}{\omega(h,g)}.
\label{eq:attcharge}
\end{equation}
 When we gauge the $G$ symmetry, the  magnetic flux becomes dynamical, and regarded as a line operator that corresponds to the worldline of the magnetic particle. We therefore conclude that a magnetic flux is attached to the Wilson line for the electric charge ${\pi}_{g}$ when passing through the defect,
 \begin{align}
     \xi_{g}(\gamma)\to \xi_g(\gamma)\cdot \eta_{\pi_g}(\gamma),
 \end{align}
 where $\xi_{g}(\gamma)$ and $\eta_{\pi_g}(\gamma)$ are line operators supported on a closed curve $\gamma$, which correspond to the magnetic particle $(g,1)$ and electric particle $(1,\pi_g)$ in the $G$-gauge theory in (2+1)D respectively. That is, the twist string acts on the anyons of the (2+1)D discrete gauge theory by permutation 
 \begin{align}
     (g,1)\to (g, \pi_g).
     \label{eq:twiststringaction}
     \end{align}
We see that the twist string implements a flux-preserving automorphism.

The action of the twist string in Eq.~\eqref{eq:twiststringaction} can be directly generalized to discrete $G$-theory in generic $(d+1)$ spacetime dimensions with $d>2$. 
In that case, the magnetic surface $\xi_g$ is a codimension-2 operator supported on $(d-1)$D submanifold $M^{d-1}$ of the spacetime, while the codimension-$(d-1)$ twist string $\mathcal{D}$ is defined as a (1+1)D $G$-SPT action evaluated on a 2D surface. The twist string then acts on the magnetic surface $m_g$ by attaching the Wilson line at a closed loop $\gamma$ embedded in $M^{d-1}$. To see this, we prepare the twist string $\mathcal{D}$ on the cylinder $S^1\times \gamma$, where $S^1$ is linking with $M^{d-1}$. By shrinking the $S^1$ into a point, the twist string acts on $m_g$ at the loop $\gamma$ by attaching the Wilson line $\eta_{\pi_g}(\gamma)$ on the magnetic surface. So, the action of the twist string is expressed as
\begin{align}
    \mathcal{D}(S^1\times \gamma)\cdot \xi_g(M^{d-1}) = \eta_{\pi_g}(\gamma)\cdot \xi_g(M^{d-1})
    \label{eq:twistactiongenericd}
\end{align}
in generic spacetime dimensions. For the case of $d>2$, note that the twist string cannot permute the label of the surface operator supported on $(d-1)$-manifold, due to the difference in dimensions between the magnetic surfaces and the electric Wilson lines. However, it can still act on the magnetic surfaces by choosing a closed loop embedded in $M^{d-1}$ and attaching the Wilson line $\eta_{\pi_g}$ on it. In the case of (3+1)D discrete gauge theory with $d=3$, the action of the defect $\mathcal{D}$ in Eq.~\eqref{eq:twistactiongenericd} gives a general description for the action of the twist string in (3+1)D constructed in Sec.~\ref{sec:exact3d} and Sec.~\ref{sec:layer} where the electric particle emits from the intersection between the magnetic surface and the twist string in the (3+1)D spacetime, for example described in Fig.~\ref{fig: 3d defect membrane operator}. To clarify the relation between the attachment of the Wilson line Eq.~\eqref{eq:twistactiongenericd} and the intersection effect in e.g., Fig.~\ref{fig: 3d defect membrane operator}, it is convenient to think about the intermediate process of shrinking a cylinder of the the defect $\mathcal{D}$ for only a half-region of the cylinder. Then there should be the intersection between the defect $\mathcal{D}$ and the magnetic surface, where the Wilson line comes out of the intersection to attach the closed Wilson line after shrinking the cylinder everywhere. This makes the connection between the two effects of the twist string on the magnetic surface.

\subsection{Example: twist string of $\mathbb{Z}_{2} \times \mathbb{Z}_{2}$ gauge theory in $(2+1)$D and $(3+1)$D }
\label{subsec:wrap3dZ2Z2}

Let us consider a simple example where $G = \mathbb{Z}_{2} \times \mathbb{Z}_{2}$. The (2+1)D $\Z_2\times \Z_2$ gauge theory corresponds to two copies of the $\Z_2$ toric code studied in Sec.~\ref{subsec:3dZ2Z2exact}. We label the magnetic particles as $m_1, m_2$ and the electric particles as $e_1,e_2$.
The non-trivial $(1+1)$D $\Z_2\times \Z_2$ SPT is the cluster state characterized by $\omega = \frac{1}{2} a\cup a'\in \mathcal{H}^2(G,\U)$ with $a,a'\in \mathcal{H}^1(G,\Z_2)$ the gauge fields of first and second $\Z_2$ respectively. 
If we consider the action of the twist string on the $m_1$ magnetic particle, the attached Wilson line becomes ${\pi}_{(1,0)}[(j,k)] = \frac{k}{2}$ mod 1 for $(j,k)\in\Z_2\times\Z_2$. According to the action in Eq.~\eqref{eq:twiststringaction}, it means that the magnetic particle is acted by the twist string as
\begin{align}
    m_1\to m_1e_2.
\end{align}
Similarly, the action on the magnetic particle $m_2$ is given by
\begin{align}
    m_2\to m_2e_1.
\end{align}
These actions on the magnetic particle correspond to the twist string realized in the (2+1)D $\Z_2\times \Z_2$ toric code shown in Fig.~\ref{fig: string operator m1-m1e2 defect}. 
For (3+1)D $\Z_2\times\Z_2$ gauge theory, due to Eq.~\eqref{eq:twistactiongenericd} the action of the twist string $\mathcal{D}$ is expressed as
\begin{align}
    \mathcal{D}(S^1\times \gamma)\cdot \xi_{m_1}(\Sigma) = \eta_{e_2}(\gamma)\cdot \xi_{m_1}(\Sigma), \qquad \mathcal{D}(S^1\times \gamma)\cdot \xi_{m_2}(\Sigma) = \eta_{e_1}(\gamma)\cdot \xi_{m_2}(\Sigma)
\end{align}
where $\Sigma$ is a 2D surface that supports the magnetic surfaces $\xi_{m_1}, \xi_{m_2}$. This action corresponds to Fig.~\ref{fig: membrane operator m1-m1e2 defect} for the twist string in (3+1)D $\Z_2\times\Z_2$ gauge theory.

\subsection{Non-Abelian $G$}
\label{nonAbelianSPT}

The construction of the twist string based on gauging (1+1)D SPTs can be generalized to non-Abelian gauge groups. For a non-Abelian (untwisted) $G$-gauge theory in $(d+1)$D, we again consider a codimension-$(d-1)$ defect given by decorating a 2-submanifold with a (1+1)D bosonic $G$-SPT phase. 
In this case, the defect again has the effect of attaching the electric Wilson line to the magnetic surface operator labeled by a conjugacy class $[g]\subset G$ that contains group element $g\in G$, where now the electric charge is labeled by the elements in the centralizer of $g$. By repeating the discussion in the previous subsection, one can see that the action of the twist string on the magnetic excitations are expressed in the same fashion. That is, in (2+1)D $G$-gauge theory, the twist string $\mathcal{D}$ acts by permutation of the anyons in the quantum double $D(G)$
\begin{align}
    \rho_{\mathcal{D}}: ([g],\mathbf{1})\to ([g], \sigma_{[g]})
    \label{eq:fluxpi1transformSPT}
\end{align}
where $\sigma_{[g]}:=i_g\omega\in \mathcal{H}^1(C_g,\U)$ is the slant product of the (1+1)D SPT action $\omega\in \mathcal{H}^2(G,\U)$, which defines a 1d representation of the centralizer $C_g$ of $g\in G$. The slant product for generic non-Abelian group will be reviewed in Appendix~\ref{app:slant}. Here we emphasize that $\sigma_{[g]}\in \mathcal{H}^1(C_g,\U)$ is a representation of the centralizer group $C_g$ rather than the full gauge group $G$, and $\sigma_{[g]}$ itself does not correspond to the pure charge of the $G$-gauge theory. 
The action of the twist string $\mathcal{D}$ on dyons of $D(G)$ is also described in the same fashion,
\begin{align}
    \rho_{\mathcal{D}}: ([g],\pi_g)\to ([g], \pi_g\times \sigma_{[g]})
    \label{eq:fluxpigtransformSPT}
\end{align}
where $\pi_g\in \mathrm{Rep}(C_g)$.

Note that the action in Eq.~\eqref{eq:fluxpi1transformSPT},~\eqref{eq:fluxpigtransformSPT} only depend on the flux label $[g]$ and are independent of the irrep label. Therefore, this satisfies both conditions Eq.~\eqref{eq:fluxpreserving} and~\eqref{eq:twiststringtopological} discussed in Sec.~\ref{sec:twiststringnonabelianlayer}.

The permutation of anyons in Eq.~\eqref{eq:fluxpigtransformSPT} only gives the partial data to characterize the automorphism $\rho_{\mathcal{D}}$. In order to fully characterize the automorphism induced by the twist string $\mathcal{D}$, we also need to determine how $\rho_D$ acts on anyon fusion and splitting vertices~\cite{barkeshli2019}. Here we determine the complete data for characterizing the automorphism $\rho_{\mathcal{D}}$, and explicitly check that the twist string $\mathcal{D}$ indeed defines a 0-form global symmetry of the (2+1)D $G$-gauge theory $D(G)$ associated with the automorphism $\rho_{\mathcal{D}}$.

\begin{figure}[t]
    \centering
    \includegraphics[width=0.75\textwidth]{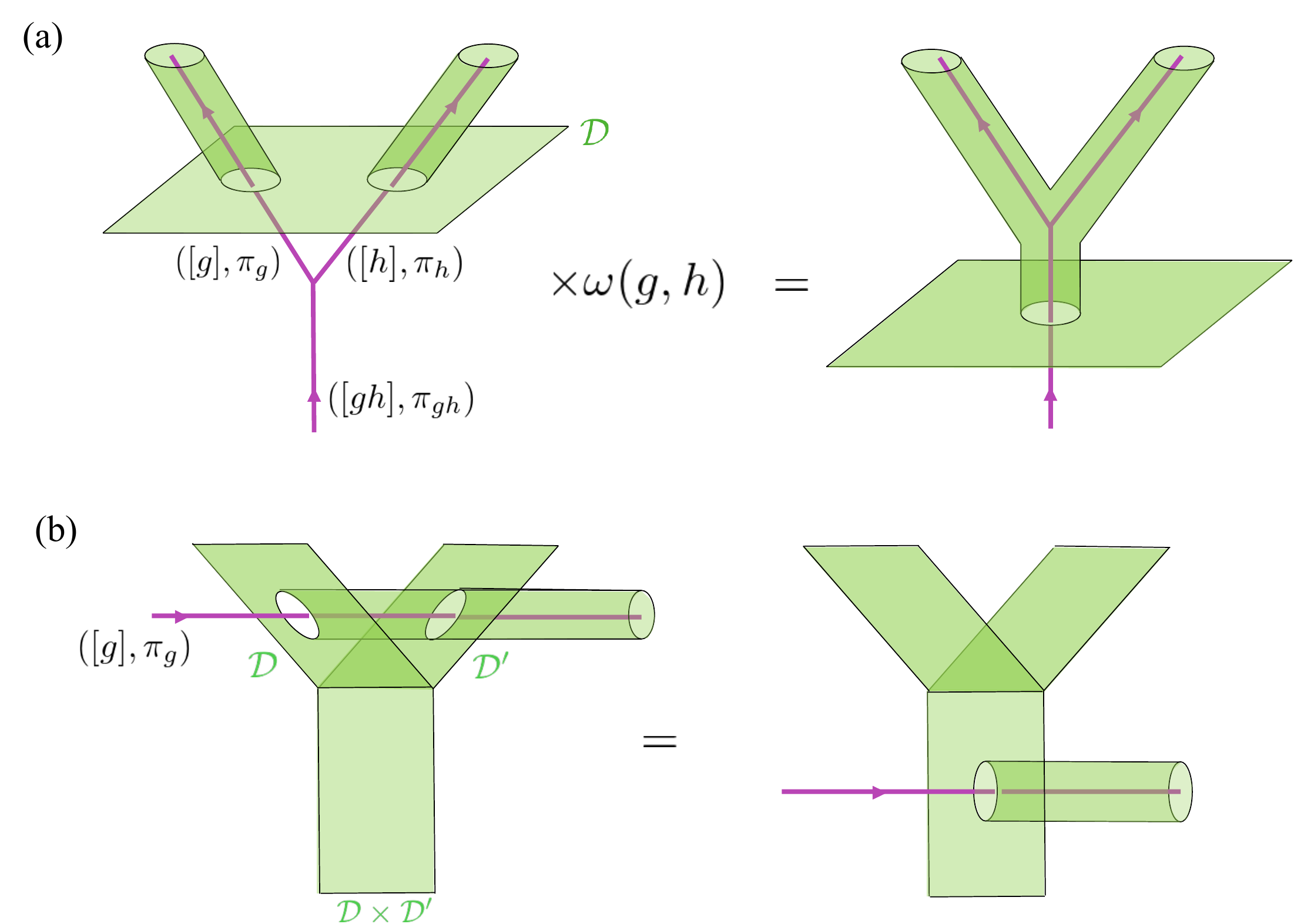}
    \label{fig:SPTdefectUeta}
\caption{The $U$ and $\eta$ symbols for the automorphism $\rho_{\mathcal{D}}$ associated with the (1+1)D $G$-SPT phase. (a): the symmetry action on the fusion vertex $U_{\mathcal{D}}$ is read by crossing the defect $\mathcal{D}$ through the junction of anyons. In the right figure after crossing, the junction of the $g,h\in G$ defects into $gh$ is introduced on the SPT phase encircling the junction of the anyons. The Boltzmann weight of the SPT phase emits a phase $\omega(g,h)$ locally at this junction, which is regarded as the symmetry action on the vertex gauge transformation $U_{\mathcal{D}}$. (b): $\eta_{([g],\pi_{g})}(\mathcal{D},\mathcal{D}')$ describes the symmetry fractionalization of the anyon $([g],\pi_{g})$ under the action of the defects $\mathcal{D},\mathcal{D}'$, which can be read by crossing the anyon $([g],\pi_{g})$ through the junction of $\mathcal{D}$ and $\mathcal{D}'$. Note that the whole process does not introduce the junction of gauge fields on the surface of SPT phase, so the SPT action does not produce any phase. We then conclude that $\eta_{([g],\pi_{g})}(\mathcal{D},\mathcal{D}')=1$. }
\end{figure}

In general, for given anyons $a,b,c\in D(G)$, the automorphism acts on the fusion/splitting space $V^{ab}_{c}$ as 
\begin{align}
\rho_{\mathcal{D}} : V_{ab}^c \rightarrow V_{\,^{\mathcal{D}}a \,^{\mathcal{D}}b}^{\,^{\mathcal{D}}c} .
\end{align}
We choose a basis $\ket{a,b;c,\mu}$ for $V_{ab}^c$, and the action of $\rho_{\mathcal{D}}$ on the basis states are expressed as
\begin{align}
\rho_{\mathcal{D}} |a,b;c, \mu\rangle = \sum_{\nu} [U_{\mathcal{D}}(\,^{\mathcal{D}}a ,
\,^{\mathcal{D}}b ; \,^{\mathcal{D}}c )]_{\mu\nu} \ket{\,^{\mathcal{D}} a, \,^{\mathcal{D}} b; \,^{\mathcal{D}}c,\nu},
  \label{eqn:rhoStates}
 \end{align}
where $^\mathcal{D}a := \rho_{\mathcal{D}}(a)$. The $N_{ab}^c \times N_{ab}^c$ matrix $U_{\mathcal{D}}$ generally determines the action on the fusion/splitting space. In our case where the defect $\mathcal{D}$ is given by the (1+1)D SPT action $\omega\in \mathcal{H}^2(G,\U)$, $U_{\mathcal{D}}$ is given by
 \begin{align}
     U_{\mathcal{D}}(([g],\pi_g),([h],\pi_h); ([gh],\pi_{gh})) = \omega(g,h).
     \label{eq:Usymbolstwiststring}
 \end{align}
Note that the $U$ symbol only depends on the flux labels of anyons, not on the irreps attached to them. The expression Eq.~\eqref{eq:Usymbolstwiststring} can be derived by considering  the SPT defect acting on the fusion vertex of the anyons carrying $[g]$ and $[h]$ fluxes, where it acts by encircling the fusion vertex by a junction of thin cylinders of the SPT phase, see Fig.~\ref{fig:SPTdefectUeta} (a). One can then see that the SPT action on the junction of cylinders weights as $\omega(g,h)$, since the junction of $G$-defects are induced on the SPT phase at the junction. This phase $\omega(g,h)$ is regarded as the vertex gauge transformation $U_{\mathcal{D}}$ acting on the fusion vertex.

To fully specify the automorphism, there is a further data written as $\eta_{([g],\pi_g)}(\mathcal{D},\mathcal{D}')$ for the pair of twist strings $\mathcal{D},\mathcal{D}'$~\cite{barkeshli2019}. This $\eta$ symbol characterizes the symmetry fractionalization of the anyon $([g],\pi_g)$ under the action of the twist string. That is, $\eta_{([g],\pi_g)}(\mathcal{D},\mathcal{D}')$ denotes a phase given by crossing the Wilson line of the anyon $([g],\pi_g)$ through the codimension-2 junction between $\mathcal{D}$ and $\mathcal{D}'$. We can see that $\eta_{([g],\pi_{g})}(\mathcal{D},\mathcal{D}') = 1$ for any pair of twist strings $\mathcal{D},\mathcal{D}'$; during the whole process the SPT action does not produce a phase because it does not involve the junction of $G$-gauge fields introduced on the SPT surface, see Fig.~\ref{fig:SPTdefectUeta} (b). 

One can check that the permutation action of $\rho_{\mathcal{D}}$ in Eq.~\eqref{eq:fluxpigtransformSPT},
$U_{\mathcal{D}}$ in Eq.~\eqref{eq:Usymbolstwiststring} together with $\eta_{([g],\pi_{g})} = 1$ give the consistent data that completely characterizes the automorphism $\rho_{\mathcal{D}}$, in the sense that they satisfies all the consistency conditions for the automorphism. For example, the consistency condition between $U$ and $\eta$ symbols are generally expressed as
\begin{align}
    {U}_{\mathcal{D}}({a},{b} ;{c}){U}_{\mathcal{D}'}(^{\mathcal{D}^{-1}}{a},^{\mathcal{D}^{-1}}{b} ;^{\mathcal{D}^{-1}}{c}) = {U}_{\mathcal{D}\times\mathcal{D}'}(a,b;c)\frac{\eta_c(\mathcal{D},\mathcal{D}')}{\eta_a(\mathcal{D},\mathcal{D}')\eta_b(\mathcal{D},\mathcal{D}')}
\end{align}
for $a,b,c\in D(G)$. This is satisfied with $\eta=1$ because $U_{\mathcal{D}}(([g],\pi_g),([h],\pi_h);([gh],\pi_{gh}))$ is given by the SPT action $\omega(g,h)$, and the equation $U_{\mathcal{D}}U_{\mathcal{D}'}=U_{\mathcal{D}\times\mathcal{D}'}$ simply follows from the stacking rule of the SPT phase where the action for the defect $\mathcal{D}\times\mathcal{D}'$ is given by the product of each SPT action $\mathcal{D}, \mathcal{D}'$.

In principle, it is possible that a nontrivial SPT phase $\omega\in \mathcal{H}^2(G,\U)$ leads to a trivial permutation of the anyons where $\sigma_{[g]}$ is trivial for all $g\in G$. Since $\sigma_{[g]}(h)$ is the same as the torus partition function $Z_{(g,h)}$ of the SPT phase with holonomy $(g,h)$, the above situation happens when a nontrivial element of $\mathcal{H}^2(G,\U)$ cannot be detected on a torus, though we are not aware of such an example. Even if one can find such an $\omega\in \mathcal{H}^2(G,\U)$ that leads to the trivial permutation of the anyons, the $U$ symbol for the automorphism $\rho_{\mathcal{D}}=\omega(g,h)$ is still nontrivial, and it is possible that it still gives a nontrivial class of the automorphism.

The automorphisms $\rho_1,\rho_2$ in the $A_6$ gauge theory from Sec.~\ref{subsubsec:A6} violate Eq.~\eqref{eq:twiststringtopological}, so do not fit into this approach of gauging (1+1)D SPT phases. 
Meanwhile, the composite automorphism given by $\rho_1\circ \rho_2$ satisfies both conditions Eq.~\eqref{eq:fluxpreserving} and~\eqref{eq:twiststringtopological}, and we conjecture that $\rho_1\circ \rho_2$ corresponds to the twist string given by the nontrivial (1+1)D $A_6$ SPT phase generating the $\Z_2$ classification, where the full classification is given by $\mathcal{H}^2(A_6,\U)=\Z_2\times \Z_3$. 

In generic $(d+1)$D $G$-gauge theory with $d>2$, the action of the twist string $\mathcal{D}$ is again described as
\begin{align}
    \mathcal{D}(S^1\times \gamma)\cdot \xi_{[g]}(M^{d-1}) = \eta_{\pi_g}(\gamma)\cdot \xi_{[g]}(M^{d-1}).
    \label{eq:twiststringnonabeliangenerald}
\end{align}

\subsection{Example: twist string in $D_{n}$ gauge theory}

Let us study a simple example given by the dihedral group $D_{n}$ (of order $2n$) where $n$ is even and group laws are described in Eq.~\eqref{eq: group law of Dn}.
We consider the twist string that corresponds to the non-trivial (1+1)D SPT given by the second cohomology group $\mathcal{H}^{2}(D_{n},\U) = \mathbb{Z}_{2}$. Here we write down the explicit form of the 2-cocycle that represents the nontrivial cohomology class. To do this, we note that the flat $D_n$ gauge field on a 2-simplex $\langle 012\rangle$ is realized by a pair of $\Z_n$-valued 1-cochain $r$ and $\Z_2$-valued 1-cochain $s$ that satisfies
\begin{align}
    \begin{split}
        r_{01} + r_{12}(1-2s_{01})  &=r_{02} \quad \mod n, \\
        s_{01} + s_{12} &= s_{02}\quad \mod 2, 
    \end{split}
\end{align}
due to the flatness condition $(r_{01},s_{01})\cdot (r_{12},s_{12}) = (r_{02},s_{02})$ on a 2-simplex. The above flatness conditions are neatly expressed by using cup product of cochains as
\begin{align}
     \begin{split}
        \delta r - 2 \hat{s}\cup r   &=0 \quad \mod n, \\
        \delta s &= 0\quad \mod 2, 
    \end{split}
\end{align}
where $\hat{s}$ denotes the lift of $s\in Z^1(M^2,\Z_2)$ to $\Z_n$, with $M^2$ a closed oriented 2-manifold.
We then consider a $\Z_n$-valued 2-cochain $\hat{s}\cup r\in Z^2(M^2,\Z_n)$. One can see that this cochain is closed, since
\begin{align}
    \begin{split}
        \delta (\hat{s}\cup r) &= \delta\hat{s}\cup r- \hat{s}\cup\delta r \\
        &= \delta\hat{s}\cup r-\hat{s}\cup (2\hat{s}\cup r)\\
        &= (\delta\hat{s}-2\hat{s}\cup\hat{s})\cup r\quad  \\
        &= 0 \quad
    \end{split}
\end{align}
where the equations are taken mod $n$. Hence, one can define a (1+1)D SPT action as
\begin{align}
    \exp\left(\frac{2\pi i}{n}\int \hat{s}\cup r \right).
    \label{eq:DnSPTaction}
\end{align}
One can see that the above action generates the nontrivial SPT phase, since the partition function on $T^2$ with the flat $D_n$ gauge field with holonomies for two fundamental 1-cycles $\{(\frac{n}{2},0), (0,1)\}$  becomes $-1$.

To explicitly write down the action of the twist string defect on magnetic excitations, we summarize the conjugacy classes of $D_n$ that label the magnetic strings, and their centralizers in Table~\ref{tab:conjcent}.

Let us consider a (2+1)D $D_n$ gauge theory $D(D_n)$, and we study a codimension-1 twist string defined as decoration of the SPT action in Eq.~\eqref{eq:DnSPTaction}. Suppose that a magnetic flux $[(r,s)]$ is acted on by the twist string, as shown in Fig.~\ref{fig:2dwrap}. The magnetic flux is then dressed by the Wilson line for the electric charge $\pi_{(r,s)}\in \mathrm{Rep}(C_{(r,s)})$, which is given by the slant product $i_{(r,s)}$ taken for the SPT action in Eq.~\eqref{eq:DnSPTaction}. The slant product $i_{(r,s)}$ is computed by the torus partition function on $T^2$ with holonomy on the compactifying cycle $(r,s)$, so the attached electric charge is evaluated as
\begin{align}
    \pi_{(r,s)}[(r',s')] = \exp\left(\frac{2\pi i}{n}(sr'-s'r)\right), \quad (r',s')\in C_{(r,s)}
\end{align}
which defines an 1D irreducible representation of the centralizer $C_{(r,s)}$.
According to this formula for the attached electric charge, one can see that only the magnetic fluxes labeled by the conjugacy classes $[(\frac{n}{2},0)],[(0,1)],[(1,1)]$ have an electric charge attached to them. For the magnetic particles labeled by $[(\frac{n}{2},0)]$, the attached electric charge is labeled by $\pi_{(\frac{n}{2},0)}(r',s') = (-1)^{s'}$,
\begin{equation}
   \left(\left[\left(\frac{n}{2},0\right)\right],\mathbf{1}\right)\to \left(\left[\left(\frac{n}{2},0\right)\right],\pi_{(\frac{n}{2},0)}\right).
\label{eq: r^2 J0 transform}
\end{equation}
For the non-Abelian magnetic string labeled by $[(0,1)]$, the attached Wilson line is given by ${\pi}_{(0,1)}[(r',s')] = e^{\frac{2\pi i}{n}r'}$, where we note that $r'\in {n}\Z/2$ since $(r',s')\in C_{(0,1)}\cong D_2$. So the attached electric charge is given by the irreducible representation of $D_2$,
\begin{equation}
([(0,1)], \mathbf{1})\to ([(0,1)], \pi_{(0,1)}).
\label{eq: s J0 transform}
\end{equation}
For the non-Abelian magnetic string labeled by $[(1,1)]$, we attach the Wilson line ${\pi}_{(1,1)}(r',s') = e^{\frac{2\pi i}{n}(r'-s')}$, where $r'-s'=0$ mod $n/2$ so that $(r',s')\in C_{(1,1)}\cong D_2$. So we have
\begin{equation}
  ([(1,1)], \mathbf{1})\to ([(1,1)], \pi_{(1,1)}).
    \label{eq: rs J0 transform}
\end{equation}
Eq.~\eqref{eq: r^2 J0 transform},~\eqref{eq: s J0 transform} and~\eqref{eq: rs J0 transform} completes all the nontrivial action of the twist string defect on magnetic strings, and realizes the automorphism described in Eq.~\eqref{eq:D4automorphism} for $D(D_4)$ in the case of $n=4$. 

For the $D_n$ gauge theory in generic $(d+1)$ dimensions with $d>2$, the nontrivial action of the twist string on the magnetic surfaces is given by Eq.~\eqref{eq:twiststringnonabeliangenerald} with $[g]=[(\frac{n}{2},0)],[(0,1)],[1,1]$. For $d=3$, the action corresponds to the twist string constructed via layer construction in Sec.~\ref{sec:twiststringnonabelianlayer}.

\section{Twist strings via gauging (1+1)D fermionic topological phases}
\label{sec:fermiongauge}

In this section, we study twist strings in (3+1)D topological orders that arise from gauging $\Z_2^f$ fermion parity symmetry of a (1+1)D fermionic invertible topological phase. That is, we start with a trivial fermionic invertible phase in (3+1)D, and then decorate a codimension-2 submanifold with a (1+1)D fermionic invertible phase with $\Z_2^f$ symmetry. We then gauge the $\Z_2^f$ symmetry. This process defines an invertible codimension-2 defect (twist string) of a $\Z_2$ gauge theory in (3+1)D with a fermionic point charge. We also study the more general situation where we gauge a fermionic symmetry group $G_f = G_b \times \Z_2^f$. 

The action of the twist string on the excitations of (3+1)D $\Z_2^f$ gauge theory is derived in a similar fashion to the bosonic case given in Sec.~\ref{sec:wrapping}. The twist string again acts on the magnetic string excitation by attaching the fermionic electric charge when the twist string crosses the magnetic string. 
The electric charge attached to the magnetic string is read off from the electric charge carried by the $\Z_2^f$ twisted sector of the (1+1)D fermionic invertible phase that corresponds to the twist string, since the magnetic string is regarded as a flux of the gauged $\Z_2^f$ symmetry. 

In the rest of the section, we derive the action of the defect on the magnetic excitation by studying the electric charge of the $\Z_2^f$-twisted sector in the fermionic topological phase. When we refer to the anti-periodic (resp.~periodic) boundary conditions with respect to $\Z_2^f$ symmetry, it is realized by the NS (resp.~R) spin structure around the cycle. In math terminology, these are the bounding and non-bounding spin structures, respectively. Throughout this section, we hence call the $\Z_2^f$-untwisted (resp.~$\Z_2^f$-twisted) sector the NS (resp.~R) sector.

\subsection{Warm-up: $e \leftrightarrow m$ twist string of (2+1)D $\Z_2$ toric code}
\label{subsec:kitaevfieldtheory}

Before discussing the (3+1)D $\Z_2$ gauge theory, we start with a (2+1)D gauge theory to illustrate the idea with a simple example.
We consider a trivial fermionic invertible phase in (2+1)D with $\Z_2^f$ symmetry, and then decorate a codimension-1 submanifold with the (1+1)D Kitaev chain. If we regard the decorated Kitaev chain as a symmetry defect of $\Z_2$ 0-form symmetry, this phase is regarded as a nontrivial fermionic SPT phase with $\Z_2\times \Z_2^f$ symmetry that generates the $\Z_8$ classification~\cite{Bhardwaj2017Statesum, Tarantino}.

We then obtain a (2+1)D bosonic $\Z_2$ gauge theory by gauging $\Z_2^f$ fermion parity symmetry of the theory. The Kitaev chain decoration of the original fermionic theory then defines an invertible codimension-1 defect of the $\Z_2$ gauge theory. We are interested in how this defect acts on anyons of the $\Z_2$ gauge theory.

Heuristically, one can apply the wrapping argument in Sec.~\ref{sec:wrapping}. Before we gauge the symmetry, we consider a background $\pi$-flux and bring it through the defect. The $\pi$-flux is dressed by the dimensional reduction of the Kitaev chain with a $\pi$-flux insertion. This Kitaev chain will have a periodic boundary condition and it is known that the ground state has an odd fermion parity \cite{kitaev2001}. Since the $\pi$-flux becomes the magnetic particle after we gauge the $\Z_2^f$ symmetry, we expect that the twist string acts on the magnetic particle as attaching a $\psi$ fermion. We provide a more detailed discussion at the level of the TQFT Hilbert space below.

In a fermionic QFT, gauging $\Z_2^f$ symmetry is performed by summing over all possible spin structures of the spacetime manifold, since the background gauge field of $\Z_2^f$ symmetry is expressed as the spin structure of the spacetime~\cite{Bhardwaj2017Statesum, aasen2019, tata2021anomalies}. 
After gauging $\Z_2^f$, the resulting bosonic $\Z_2$ gauge theory in (2+1)D has a dual $\Z_2$ 1-form symmetry generated by an Wilson line of a fermionic particle $\psi$.
Conversely, starting with this bosonic $\Z_2$ gauge theory, one can gauge the dual $\Z_2$ 1-form symmetry, by coupling the theory with spin structure. We then obtain the original fermionic phase with $\Z_2^f$ symmetry. 
Physically, this process corresponds to performing anyon condensation of a fermionic particle $\psi$ of $\Z_2$ gauge theory combined with a local fermion of a trivial fermionic gapped theory.
Hence, condensing $\psi$ of the $\Z_2$ gauge theory realizes the inverse process of gauging $\Z_2^f$ symmetry.

Since the original (2+1)D fermionic invertible phase can be obtained by condensation of a fermionic particle $\psi$ of the $\Z_2$ gauge theory, one can express the states of the original fermionic theory in terms of the states of the $\Z_2$ gauge theory~\cite{delmastro2021}. 
In particular, when the Hilbert space is defined on a torus $T^2$, the state of the bosonic $\Z_2$ gauge theory is labeled by the anyons of the $\Z_2$ gauge theory, $\ket{1},\ket{e},\ket{m},\ket{\psi}$. These states correspond to path integral of the bosonic $\Z_2$ gauge theory on a solid torus $D^2\times S^1$, with an insertion of the anyon lines along the longitude of the solid torus.
The state of the trivial fermionic invertible phase is expressed as a proper superposition of these states. We read the action of the defect on anyons by translating the effect of the Kitaev chain in the fermionic Hilbert space into a language of the bosonic states labeled by the anyons of the $\Z_2$ gauge theory.

To do this, let us consider a fermionic state on $T^2$ with the spin structure along the meridian of $T^2$ given by R spin structure. 
Inserting a $\psi$ Wilson line along the meridian on the state must act as a phase $-1$. 
So, the state on $T^2$ with R spin structure along the meridian is described by insertion of $m$ or $e$ Wilson line along the longitude, so that the $(-1)$ sign under insertion of $\psi$ is realized as the mutual braiding between $\psi$ and $e$ or $m$. The fermionic invertible phase with the given spin structure is expressed as
\begin{align}
\begin{split}
    \ket{T^2_{\mathrm{R},\mathrm{NS}}} &= \ket{e} +\ket{m}, \\
    \ket{T^2_{\mathrm{R},\mathrm{R}}} &= \ket{e} -\ket{m},
    \end{split}
\end{align}
where $T^2_{\mu,\lambda}$ denotes $T^2$ with spin structure $\mu$ along the meridian, and $\lambda$ along the longitude. 
A state $\ket{a}$ denotes a state on $T^2$ given by inserting an Wilson line $a$ along the longitude of a solid torus.
The rationale for the form of the state $\ket{e} \pm\ket{m}$ is that the $\psi$ line along the longitude must also act as $\pm 1$ phase, depending on spin structure along the longitude. Indeed, one can see that $\ket{e} \pm\ket{m}$ becomes an eigenvector under the action of fusing a $\psi$ line along the longitude.

\begin{figure}
    \centering
    \includegraphics[width=0.8\textwidth]{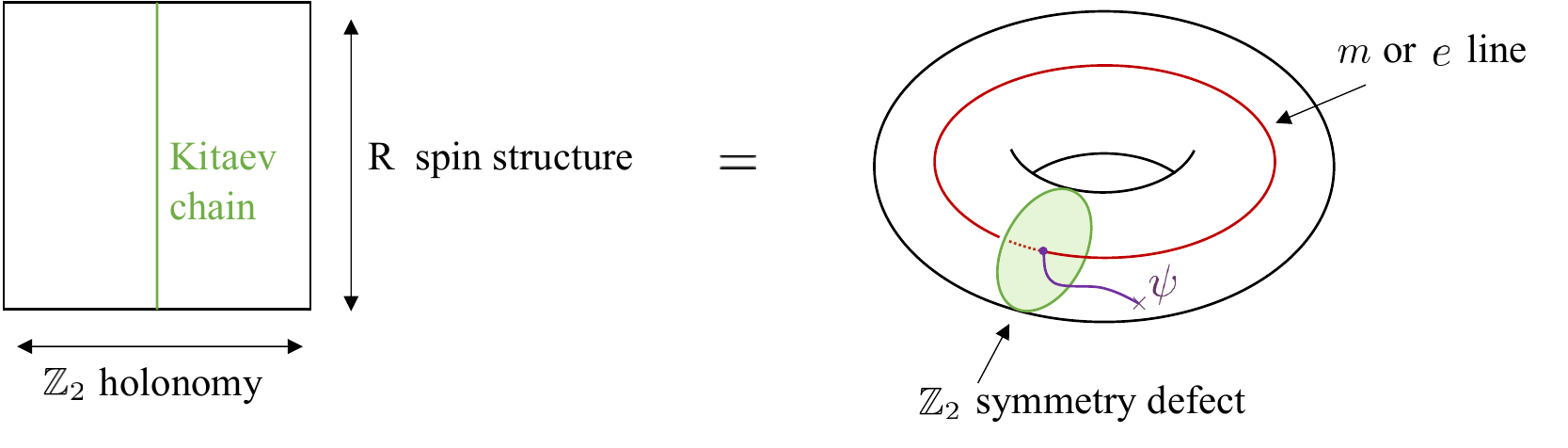}
\caption{A state of the fermionic SPT phase on a torus is described by a path integral of the bosonic shadow theory on the solid torus $D^2\times S^1$. When we have R spin structure along the meridian and the $\Z_2$ holonomy along the longitude, the state must carry odd fermion parity, so a $\psi$ line must be running along the time direction. This constrains the action of $\Z_2$ symmetry defect on anyons in the bosonic shadow theory.}
\label{fig:kitaevdefect}
\end{figure}

So far, we did not consider a state with an insertion of the Kitaev chain. Let us now consider a fermionic state with the codimension-1 defect on $D^2\times \{0\}$ in the solid torus, which amounts to putting a Kitaev chain along the meridian of $T^2$ with R spin structure, see Fig.~\ref{fig:kitaevdefect}.
Since a Kitaev chain put on a circle with R spin structure carries odd fermion parity, the state on $T^2$ with the codimension-1 defect must also carry odd fermion parity.

A state with odd fermion parity is expressed by a state of the bosonic $\Z_2$ gauge theory with a single $\psi$ Wilson lines extended along the time direction. So, in order to realize a state with odd fermion parity, a $\psi$ line must stem out of the $e$ or $m$ Wilson line along the longitude intersecting with the codimension-1 defect. This is possible only when the codimension-1 defect acts as an automorphism that permutes $e$ and $m$, where a $\psi$ line can stem out of the intersection between the codimension-1 defect and the Wilson line. This implies that the defect of the bosonic $\Z_2$ theory must exchange $e$ and $m$.
This fact was known in~\cite{Bhardwaj2017Statesum} in the context of fermionic SPT phase, where they found that the bosonic dual theory of the $\Z_2\times \Z_2^f$ fermionic SPT phase that generates the $\Z_8$ classification realizes the $e \leftrightarrow m$ exchange $\Z_2$ symmetry of the gauge theory.

\subsection{Twist string of (2+1)D $G_f$ gauge theory}\label{sec: Twist string of Gf gauge theory}
Here we generalize the above discussion to the (2+1)D $G_f$ gauge theory. That is, we start with the (2+1)D trivial fermionic invertible phase with $G_f$ symmetry that contains $\Z_2^f$ fermion parity symmetry, and then obtain a (2+1)D bosonic gauge theory with gauge group $G_f$ by gauging the whole symmetry including fermion parity. For simplicity, we restrict ourselves to the case where $G_f$ is expressed in the form of $G_f = G_b\times \Z_2^f$, i.e., direct product of a bosonic discrete group $G_b$ and fermion parity. We study the action of the codimension-1 twist string given by the decoration of (1+1)D fermionic invertible phase with $G_f = G_b\times\Z_2^f$ symmetry.

The (2+1)D $G_f$ gauge theory is obviously described by $D(G_b)\boxtimes D(\Z_2)$, where $D(\Z_2)$ is regarded as the $\Z_2$ gauge theory obtained by gauging $\Z_2^f$ symmetry. We hence take the electric particle in the $\Z_2$ gauge theory $D(\Z_2)$ as a fermion $\psi$, and $e, m$ particles are regarded as carrying the magnetic $\pi$-flux with respect to $\Z_2^f$ symmetry. So, we write the anyons of $D(G_b)\boxtimes D(\Z_2)$ naturally in the form of
\begin{align}
    ([g],\pi_g)\cdot(m^j, \psi^k), \quad \text{$g\in G_b,\  j,k\in \{0,1\}$}.
\end{align}
Analogously to what we have done in previous subsections, one can read the action of the twist string on each $G_b\times\Z_2$ flux sector, by studying the $G_b\times\Z_2^f$ charge of the (1+1)D SPT phase carried by the $([g],j)\in G_b\times\Z_2^f$ twisted sector.

The (1+1)D fermionic SPT phase with $G_b\times\Z_2$ symmetry is labeled by~\cite{Kapustin-Turzillo-You2018, kapustinThorngren2017FermionSPT}
\begin{align}
    (n_0, n_1, \nu_2)\in \Z_2\times \mathcal{H}^1(G_b,\Z_2)\times \mathcal{H}^2(G_b,\U),
\end{align}
where $n_0\in\Z_2$ represents the Kitaev chain, and $n_1\in  \mathcal{H}^1(G_b,\Z_2)$ represents the Gu-Wen SPT phase, and $\mathcal{H}^2(G_b,\U)$ represents the bosonic SPT phase. As discussed in Sec.~\ref{subsec:kitaevfieldtheory}, the Kitaev chain leads to the twist string with the action
\begin{align}
    ([g],\pi_g)\cdot (m^j, \psi^k) \to ([g],\pi_g)\cdot (m^j, \psi^{k+ n_0j} ),
\end{align}
that is, it attaches the fermion $\psi$ to the magnetic $\pi$-flux $e, m$, reflecting that the fermion parity of the Kitaev chain is odd in the R sector. Also, following the argument of Sec.~\ref{sec:wrapping}, the bosonic SPT phase $\nu_2$ leads to the action
\begin{align}
    ([g],\pi_g)\cdot (m^j, \psi^k) \to ([g],\pi_g\times i_g\nu_2)\cdot (m^j, \psi^{k} ),
\end{align}
where $i_g$ denotes the slant product that defines $i_g\nu_2\in \mathcal{H}^1(C_g,\U)$.

The remaining task is to evaluate the action of the Gu-Wen SPT phase labeled by $n_1\in \mathcal{H}^1(G_b,\Z_2)$. To do this, we utilize the path integral of the Gu-Wen phase given by the Gu-Wen Grassmann integral~\cite{Gaiotto:2015zta}. For our purpose, we express the path integral on a torus $T^2$ equipped with the spin structure $(s,s')$ on each fundamental cycle ($s,s'\in\{\mathrm{NS},\mathrm{R}\}$), and the $G_b$ flat background gauge field with the holonomy $(g,h)$ on each fundamental cycle. Then, the torus partition function is expressed in the form of~\cite{Gaiotto:2015zta, Kapustin-Turzillo-You2018} 
\begin{align}
    Z_{\mathrm{GW}}(T^2_{(s,s'),(g,h)}) = z_{(s,s')}(n_1(g),n_1(h)),
\end{align}
which is essentially the partition function of the fermionic SPT phase with $\Z_2\times\Z_2^f$ symmetry, coupled to the $\Z_2$ background gauge field on $T^2$ with holonomy $(n_1(g), n_1(h))$. $z_{s,s'}$ can be expressed in terms of the Gu-Wen Grassmann integral~\cite{Gaiotto:2015zta}. 

For convenience, let us label the spin structure by the $\Z_2$ number $s\in\Z_2$ as $s=0$ for NS spin structure, and $s=1$ for R. Then, the partition function $z_{(s,s')}$ can be computed as~\footnote{Here we use a couple of fundamental properties of the Grassmann integral. For the Grassmann integral $z_{\eta}(a)$ with $\eta$ the spin structure and $a\in Z^1(M^2,\Z_2)$ the $\Z_2$ gauge field, we firstly have the quadratic property that
\begin{align}
    z_{\eta}(a+b) = z_{\eta}(a)z_{\eta}(b)(-1)^{\int_{M^2}a\cup b},
\end{align}
from which one can write $z_{\eta}(n_1(g), n_1(h)) = z_{\eta}(n_1(g),0)z_{\eta}(0, n_1(h))(-1)^{n_1(g)n_1(h)}$. Also, the Grassmann integral has the property that it ``measures'' the spin structure, i.e., we have
\begin{align}
z_{\eta}(a)=\begin{cases}
1 & \text{the spin structure is NS around $C_a$ },\\
-1 & \text{the spin structure is R around $C_a$ }
\end{cases}
\end{align} 
where $C_a$ is a single, non-self-intersecting closed curve Poincar\'e dual to $a\in Z^1(M^2,\Z_2)$. This shows $z_{\eta}(n_1(g),0) = (-1)^{n_1(g)s'}, z_{\eta}(0, n_1(h)) = (-1)^{sn_1(h)}$, and we get Eq.~\eqref{eq:grassmannevaluation}.
}
\begin{align}
    Z_{\mathrm{GW}}(T^2_{(s,s'),(g,h)}) = z_{(s,s')}(n_1(g),n_1(h)) = (-1)^{n_1(g)\cdot s' + (s+ n_1(g))\cdot n_1(h)}.
    \label{eq:grassmannevaluation}
\end{align}
One can then read off the charge carried by $(g,s)\in G_b\times \Z_2^f$ twisted Hilbert space as~\cite{T17}
\begin{align}
    \{(-1)^{(s+n_1(g)) n_1}, n_1(g)\}\in \mathcal{H}^1(G_b,\U)\times \Z_2^f.
\end{align}
where we regard $(-1)^{(s+n_1(g)) n_1}$ as an element of $\mathcal{H}^1(G,\U)$ by a map $h\to (-1)^{(s+n_1(g)) n_1(h)}$ for $h\in G_b$.
Hence, the Gu-Wen phase leads to the action
\begin{align}
    ([g],\pi_g)\cdot (m^j, \psi^k) \to ([g],\pi_g\times (-1)^{(j+n_1(g)) n_1})\cdot (m^j, \psi^{k + n_1(g)} ).
\end{align}
Summarizing, the (1+1)D fermionic SPT phase labeled by $(n_0,n_1,\nu_2)$ defines the twist string with the action
\begin{align}
    ([g],\pi_g)\cdot (m^j, \psi^k) \to ([g],\pi_g\times (-1)^{(j+n_1(g)) n_1}\times i_g\nu_2)\cdot (m^j, \psi^{k + n_0 j + n_1(g)} ).
\end{align}

\subsection{Twist string of (3+1)D $\Z_2$ gauge theory with a fermionic particle}
\label{subsec:Z2fermion}

Next, we describe a codimension-2 defect of a bosonic $\Z_2$ gauge theory with a fermionic particle in (3+1)D, by generalizing the case of (2+1)D in the previous subsection. 
We again regard a (3+1)D bosonic $\Z_2$ gauge theory as a bosonic dual theory of a (3+1)D trivial fermionic invertible phase obtained by gauging $\Z_2^f$ fermion parity symmetry.

We again consider a codimension-2 defect of a trivial fermionic invertible phase in (3+1)D by decorating a codimension-2 submanifold with the Kitaev chain. We then gauge the $\Z_2^f$ symmetry of the whole system, and we are interested in how this codimension-2 defect is realized in the bosonic $\Z_2$ gauge theory with a fermionic particle.

To see this, we again note that the state of the fermionic invertible phase can be expressed by a superposition of the states of bosonic dual $\Z_2$ gauge theory, since the fermionic invertible phase can be obtained by condensation of a fermionic particle $\psi$ for the $\Z_2$ gauge theory that corresponds to gauging $\Z_2$ 2-form symmetry dual to $\Z_2^f$ symmetry.
Let us consider a Hilbert space of the fermionic invertible phase on a torus $T^3=S^1_x\times S^1_y\times S^1_z$, with R spin structure along the $x$ direction. We express this state in terms of path integral of the bosonic $\Z_2$ gauge theory on $D^2\times S^1_y\times S^1_z$, where $D^2$ is bounded by $S^1_x$. 

Since we have R spin structure along the $x$ direction, an insertion of a $\psi$ Wilson line along $x$ direction must act as a phase $-1$. These states are realized by inserting an $m$ surface operator along $S^1_y\times S^1_z$ of $D^2\times S^1_y\times S^1_z$, where $(-1)$ sign by inserting the $\psi$ Wilson line is understood as mutual braiding between $m$ and $\psi$.
In the absence of the codimension-2 symmetry defect for the Kitaev chain, the state with each spin structure along $y,z$ direction is given by
\begin{align}
\begin{split}
    \ket{T^3_{\mathrm{R},\mathrm{NS},\mathrm{NS}}} &= \ket{m_{yz}} +\ket{m_{yz}\times \psi_y} + \ket{m_{yz}\times \psi_z} +\ket{m_{yz}\times \psi_y\times \psi_z}, \\
    \ket{T^3_{\mathrm{R},\mathrm{NS},\mathrm{R}}} &= \ket{m_{yz}} +\ket{m_{yz}\times \psi_y} - \ket{m_{yz}\times \psi_z} -\ket{m_{yz}\times \psi_y\times \psi_z}, \\
    \ket{T^3_{\mathrm{R},\mathrm{R},\mathrm{NS}}} &= \ket{m_{yz}} -\ket{m_{yz}\times \psi_y} + \ket{m_{yz}\times \psi_z} -\ket{m_{yz}\times \psi_y\times \psi_z}, \\
    \ket{T^3_{\mathrm{R},\mathrm{R},\mathrm{R}}} &=  \ket{m_{yz}} -\ket{m_{yz}\times \psi_y} - \ket{m_{yz}\times \psi_z} +\ket{m_{yz}\times \psi_y\times \psi_z} 
    \end{split}
\end{align}
where $m_{yz}$ denotes a $m$ surface inserted along $S^1_y\times S^1_z$, and $\psi_y$ denotes a $\psi$ line inserted along $y$ direction.

Then, let us consider an insertion of the codimension-2 defect on $D^2\times \{0\}\times \{0\}$. On the spatial torus $T^3$, this amounts to putting a Kitaev chain running along $x$ direction. Since we have R spin structure in $x$ direction, the state with the codimension-2 symmetry defect must carry odd fermion parity. So, based on a similar logic to the case in (2+1)D, a $\psi$ Wilson line must be running along the time direction in order to realize a state with odd fermion parity. This is possible only when the intersection between the $m$ surface along $S^1_y\times S^1_z$ and the codimension-2 defect along $D^2$ can emit the $\psi$ Wilson line. This explains why the insertion of a Kitaev chain is realized as a codimension-2 twist string with the property described in  Fig.~\ref{fig: 3d defect membrane operator}. 

\section{Twist strings as condensation defects}
\label{sec:highergauge}
In this section, we provide another perspective of the codimension-2 defect studied in this paper, in terms of a condensation defect introduced in~\cite{Konstantinos2022Higher}. In general, a condensation defect in codimension-$p$ is defined as a defect obtained by summing over insertions of defects with codimension higher than $p$, supported on the codimension-$p$ manifold. For example, for a bosonic topological order in (2+1)D, all the codimension-1 topological defect can be realized as a condensation defect. This is because a generic codimension-1 defect of (2+1)D bosonic topological order given by a modular tensor category $\mathcal{C}$ is interpreted as a gapped boundary of $\mathcal{C}\boxtimes\overline{\mathcal{C}}$ by folding the theory along the defect, and then the gapped boundary is realized by condensation of the Lagrangian algebra anyon on the boundary. This means that the codimension-1 defect is generally realized as a condensation defect given by insertions of the Lagrangian algebra anyon of the folded theory $\mathcal{C}\boxtimes\overline{\mathcal{C}}$.

Here, we show that the codimension-2 twist strings of $\Z_2$ gauge theory and $\Z_2\times \Z_2$ gauge theory in (3+1)D can also be realized as a condensation defect, given by summing over insertions of Wilson line operators on the codimension-2 surface. This expression of twist strings as condensation defects is useful, since it allows us to easily compute the action of the twist strings on the magnetic string in terms of well-understood mutual braiding between the Wilson line and the magnetic string. In Sec.~\ref{sec:3groupfermion}, we utilize the expression of twist defects as condensation defects to derive the 3-group structure of global symmetries of the (3+1)D topological order.

\subsection{Twist string of two copies of standard $\Z_2$ toric code in (3+1)D}
\label{subsec:Z2Z2cond}
\begin{figure}
    \centering
    \includegraphics[width=0.8\textwidth]{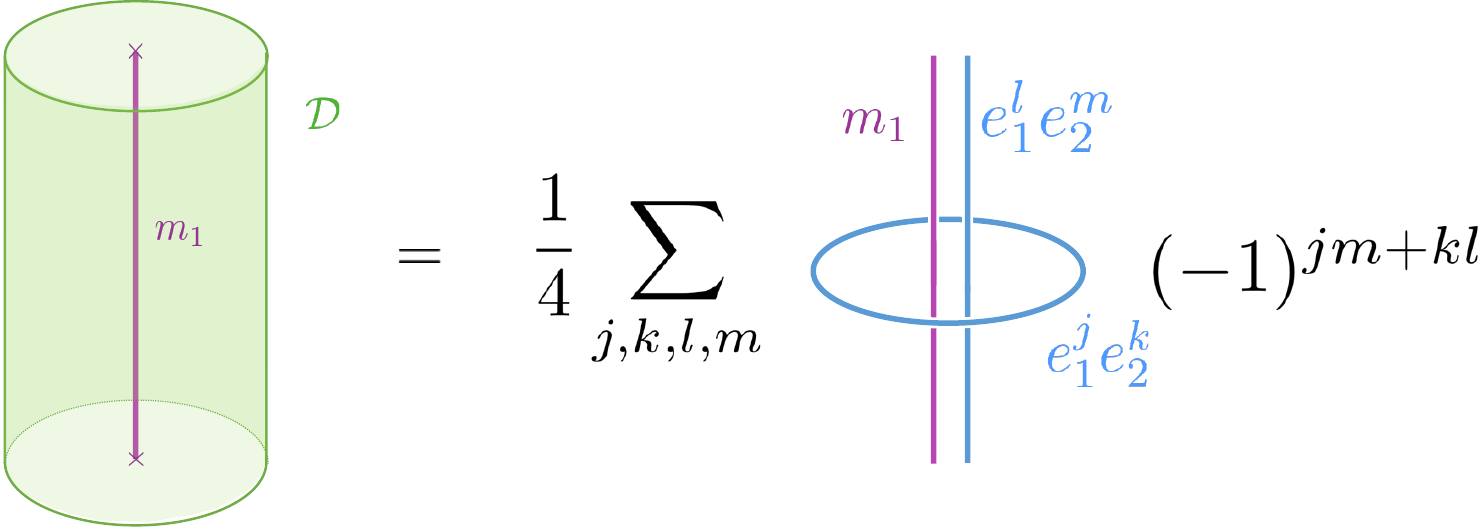}
\caption{The condensation defect acts on $m_1$ by enclosing the magnetic string by a cylinder in a 3D spatial slice.}
\label{fig:condensationZ2Z2}
\end{figure}

As we have seen in Sec.~\ref{subsec:3dZ2Z2exact}, the (3+1)D $\Z_2\times\Z_2$ gauge theory for two copies of toric codes has a codimension-2 twist string defect with the action on magnetic strings shown in Fig.~\ref{fig: membrane operator m1-m1e2 defect}. Here we describe this codimension-2 defect of (3+1)D $\Z_2\times\Z_2$ gauge theory in terms of the condensation defect.
The action for the (3+1)D $\Z_2\times\Z_2$ gauge theory is given by
\begin{align}
    \pi \int_{M^4} a^{e_1}\cup\delta b^{m_1} + a^{e_2}\cup \delta b^{m_2}
\end{align}
with the $\Z_2$ gauge fields $a^{e_1},a^{e_2}\in C^1(M^4,\Z_2), b^{m_1},b^{m_2}\in C^2(M^4,\Z_2)$.
We recall that the codimension-2 twist string is defined as a decoration of the (1+1)D $\Z_2\times\Z_2$ SPT action $a^{e_1}\cup a^{e_2}$ on it, which is expressed as
\begin{align}
    \mathcal{D}(M^2)=(-1)^{\int_{M^2} a^{e_1}\cup a^{e_2}}
\end{align}
This defect can be expressed as a condensation defect, i.e., sum over Wilson lines restricted to a 2-manifold $M^2$. To see this, we rewrite the above expression as
\begin{align}
    \mathcal{D}(M^2)=\frac{1}{\sqrt{|H_1(M^2,\Z_2\times\Z_2)|}}\sum_{\Gamma,\Gamma'\in H^1(M^2,\Z_2)}(-1)^{\int_{M^2}a^{e_1}\cup\Gamma}(-1)^{\int_{M^2}a^{e_2}\cup\Gamma'}(-1)^{\int_{M^2}\Gamma\cup\Gamma'}.
\end{align}
This expression is regarded as gauging the $\Z_2$ symmetries generated by the Wilson lines $(-1)^{\int a^{e_1}}, (-1)^{\int a^{e_2}}$ on the codimension-2 surface $M^2$, coupled with the additional discrete torsion term $(-1)^{\int \Gamma\cup\Gamma'}$. If we do not include the discrete torsion $(-1)^{\int \Gamma\cup\Gamma'}$, the above defect instead realizes a pair of the non-invertible Cheshire strings for each $\Z_2$ toric code, which is discussed in Appendix~\ref{app:cheshire}.
By writing the Poincar\'e dual of $\Gamma, \Gamma'$ as $\gamma,\gamma'$, the defect is expressed as a condensation defect given by
\begin{align}
    \mathcal{D}(M^2)=\frac{1}{\sqrt{|H_1(M^2,\Z_2\times\Z_2)|}}\sum_{\gamma,\gamma'\in H_1(M^2,\Z_2)}\eta_{e_1}(\gamma) \eta_{e_2}(\gamma') (-1)^{\sharp \mathrm{int}(\gamma,\gamma')}
\end{align}
where $\eta_{e_1}(\gamma)=\exp(\pi i\int_{\gamma} a^{e_1}), \eta_{e_2}(\gamma')=\exp(\pi i\int_{\gamma} a^{e_2})$ denote Wilson lines along the curves $\gamma, \gamma'$ respectively, and ${\sharp \mathrm{int}(\gamma,\gamma')}$ denotes the intersection number between $\gamma$ and $\gamma'$.
One can derive the action of the condensation defect  $\mathcal{D}(M^2)$ on the magnetic surface $m_1$ by considering a cylinder of the defect $\mathcal{D}(M^2)$ enclosing the magnetic string $m_1$ on a spatial 3-manifold, see Fig.~\ref{fig:condensationZ2Z2}.
The condensation defect on a cylinder is expressed as
\begin{align}
    \mathcal{D}=\frac{1}{4}\sum_{j,k,l,m\in\Z_2}\eta_{e_1}(\gamma_x)^j \eta_{e_1}(\gamma_y)^k\eta_{e_2}(\gamma_x)^l \eta_{e_2}(\gamma_y)^m (-1)^{jm+kl}
\end{align}
where $x,y$ denote the direction in the meridian and longitude respectively.
It acts on the $m_1$ magnetic string as
\begin{align}
\begin{split}
    \mathcal{D}\cdot m_1(\gamma_y) &= \frac{1}{4}\sum_{j,k,l,m}\eta_{e_1}(\gamma_y)^k \eta_{e_2}(\gamma_y)^m(-1)^{jm+kl+j}\cdot m_1(\gamma_y) \\
    &=\sum_{k,m}\delta(m+1)\delta(k)\eta_{e_1}(\gamma_y)^k \eta_{e_2}(\gamma_y)^m\cdot m_1(\gamma_y) \\
    &=\eta_{e_2}(\gamma_y)\cdot m_1(\gamma_y) \\
    \end{split}
\end{align}
where we use $\eta_{e_1}(\gamma_x)^j$ acts on the magnetic string $m_1$ as a phase $(-1)^j$ by unlinking and shrinking.
Hence, it turns out that the cylinder acts on a magnetic string by attaching an Wilson line $\eta_{e_2}$ parallel to $m_1$. This realizes the action of the twist string $\mathcal{D}(M^2)$ on $m_1$ explained in Sec.~\ref{subsec:wrap3dZ2Z2}. Using the same logic, the defect $\mathcal{D}$ acts on $m_2$ as
\begin{align}
     \mathcal{D}\cdot m_2(\gamma_y)=\eta_{e_1}(\gamma_y)\cdot m_2(\gamma_y)
\end{align}

\subsection{Twist string of (3+1)D $\Z_2$ gauge theory with a fermionic particle}
\label{subsec:fermionhighergauging}
Next, we describe a codimension-2 twist string defect of (3+1)D $\Z_2$ gauge theory with a fermionic particle in terms of the condensation defect. 
The action of the $\Z_2$ gauge theory is given by~\cite{kapustinThorngren2017FermionSPT, tata2021anomalies}
\begin{align}
    \pi \int_{M^4} a \cup \delta b + b\cup b + b\cup_1\delta b
\end{align}
with $a\in C^1(M^4,\Z_2), b\in C^2(M^4,\Z_2)$.
The Wilson line operator is defined as
\begin{align}
    \eta_{\psi}(\gamma) = \exp\left(i\pi \int_{\gamma} a\right).
\end{align}
This Wilson line generates a $\Z_2$ 2-form symmetry. If we couple the theory with the background gauge field of the 2-form symmetry $A_3\in Z^3(M^4,\Z_2)$, the theory has a 't Hooft anomaly characterized by the (4+1)D response action
\begin{align}
    (-1)^{\int\Sq^2(A_3)},
\end{align}
which represents the framing anomaly of the Wilson line with fermionic statistics.
For later convenience, we couple the $\Z_2$ gauge theory to an additional bosonic counterterm with $\Z_2$ 2-form symmetry. The partition function of the counterterm is denoted as $\sigma(A_3)$, which takes its value in $\{\pm 1\}$. This theory is called the Gu-Wen Grassmann integral~\cite{Gaiotto:2015zta}, and its phase shifts under background gauge transformations according to the (4+1)D response action
\begin{align}
    (-1)^{\int\Sq^2(A_3)+ w_2\cup A_3}.
\end{align}
Since the cocycle $\Sq^2(A_3)+ w_2\cup A_3$ is known to be exact on a closed oriented manifold due to the Wu relation~\cite{ManifoldAtlasWu}, the counterterm $\sigma(A_3)$ is regarded as a realization of a trivial gapped theory on the boundary of a trivial (4+1)D SPT phase.
After adding this the counterterm $\sigma(A_3)$, the 't Hooft anomaly of the $\Z_2$ 2-form symmetry of the Wilson line $\eta_{\psi}(\gamma)$ is given by 
\begin{align}
    (-1)^{\int w_2\cup A_3}.
    \label{eq:anomalyfermion}
\end{align}

 As we have seen in Sec.~\ref{subsec:Z2fermion}, the (3+1)D $\Z_2$ gauge theory with a fermionic particle has a codimension-2 defect $\mathcal{D}$ with the property that the intersection between $\mathcal{D}$ and the magnetic string emits a fermionic particle. Suppose that the codimension-2 closed oriented surface $M^2$ has the vanishing second Stiefel-Whitney class, i.e., $w_2(TM^4)=0$ when restricted to $M^2$. Then,
 we argue that the codimension-2 defect is expressed in terms of the condensation defect as
\begin{align}
    \mathcal{D}(M^2)=\frac{1}{\sqrt{|H_1(M^2,\Z_2)|}}\sum_{\gamma\in H_1(M^2,\Z_2)}\eta_{\psi}(\gamma)
    \label{eq:defecteta}
\end{align}
This operator is understood as gauging the $\Z_2$ symmetry generated by $\eta_{\psi}(\gamma)$ restricted to a codimension-2 defect $M^2$.
Since we assume $w_2=0$ on $M^2$, the 't Hooft anomaly in Eq.~\eqref{eq:anomalyfermion} is absent, so one can gauge the $\Z_2$ symmetry generated by the Wilson line restricted to $M^2$.

When a loop $\gamma$ is restricted to $M^2$, the insertion of the Wilson line along the loops in $M^2$ has a property that~\footnote{This relation comes from the quadratic property of the bosonic counterterm $\sigma$ given by~\cite{Gaiotto:2015zta}
\begin{align}
    \sigma(A_3)\sigma(A_3') = \sigma(A_3+A_3')\cdot  (-1)^{\int A_3\cup_2 A'_3}.
\end{align}
Since $A_3$ is the background gauge field of the 2-form symmetry generated by the Wilson line, $A_3, A'_3$ are regarded as the Poincar\'e dual of the inserted Wilson lines $\gamma, \gamma'$ respectively. When $\gamma, \gamma'$ are supported on $M^2$ and the normal bundle $NM^2$ of $M^2$ is trivial, $\int A_3\cup_2 A'_3$ is the mod 2 intersection number between $\gamma$ and $\gamma'$ evaluated on $M^2$. This is understood from the geometric interpretation of higher cup product $\cup_2$ developed in Ref.~\cite{tata2020}, which says that $A_3\cup_2 A'_3$ is regarded as the intersection between $\gamma$ and the 3D object obtained by ``thickening'' $\gamma'$ along the two directions determined by a framing of $NM^2$. This is the same as the intersection between $\gamma$ and $\gamma'$ on $M^2$, so we have $(-1)^{\int A_3\cup_2 A'_3}= (-1)^{\sharp\mathrm{int}(\gamma,\gamma')}$.
}
\begin{align}
    \begin{split}
         \eta_{\psi}(\gamma)\eta_{\psi}(\gamma') &=\eta_{\psi}(\gamma+\gamma')(-1)^{\sharp\mathrm{int}(\gamma,\gamma')}.
         \label{eq:etabraiding}
    \end{split}
\end{align}
As discussed in~\cite{Konstantinos2022Higher}, one can check that a 2D condensation defect Eq.~\eqref{eq:defecteta} for the Wilson line with the property Eq.~\eqref{eq:etabraiding}
gives an invertible defect with $\Z_2$ fusion rule.
Analogously to what we have done in the previous subsection, we derive the action of $\mathcal{D}$ on the magnetic string $m$ by enclosing a magnetic string by a cylinder. The defect $\mathcal{D}$ on a cylinder is expressed as
\begin{align}
    \mathcal{D}=\frac{1}{2}\sum_{j,k\in\Z_2}\eta_{\psi}(\gamma_x)^j \eta_{\psi}(\gamma_y)^k (-1)^{jk}
\end{align}
where we used Eq.~\eqref{eq:etabraiding}.
The action on the magnetic string in $y$ direction is then given by
\begin{align}
\begin{split}
    \mathcal{D}\cdot m(\gamma_y)&=\frac{1}{2}\sum_{j,k\in\Z_2} \eta_{\psi}(\gamma_y)^k (-1)^{jk+j}\cdot m(\gamma_y) \\
    &= \sum_{k}\delta(k+1)  \eta_{\psi}(\gamma_y)^k \cdot m(\gamma_y) = \eta_{\psi}(\gamma_y) \cdot m(\gamma_y),
    \end{split}
\end{align}
so it acts by attaching the fermionic Wilson line to the magnetic string. It realizes the action of the twist string on the magnetic string studied in Sec.~\ref{subsec:3dZ2exact}.

\section{3-group symmetry involving twist strings}
\label{sec:3group}

In this section, we argue that the codimension-2 defect of (3+1)D topological order discussed in this paper constitutes a 3-group structure of a global symmetry, together with 1-form and 2-form symmetry generated by Wilson lines and surfaces. We first describe the 3-group structure for each example of the codimension-2 defect studied in Sec.~\ref{sec:highergauge}, and also for the case of (3+1)D discrete $G$-gauge theory discussed in Sec.~\ref{sec:wrapping}. In Sec.~\ref{subsec:Z23toriccode3group}, we also consider the $\Z_2$ 0-form symmetry of the $\Z_2^3$ toric code in (3+1)D, and discuss the 3-group symmetry structure formed by 0, 1, and 2-form symmetries.

Let us first give a general overview for the equations of a flat background gauge field for a 3-group, formed by $K_0$ 0-form, $K_1$ 1-form, and $K_2$ 2-form symmetry. The 3-group has a hierarchical structure, where one first defines a 2-group $\Gamma_{0,1}$ as a nontrivial ``extension'' of $K_0$ by $K_1$, and then the 3-group is defined as a further extension of $\Gamma_{0,1}$ by $K_2$. Accordingly, the structure of the background gauge field is best explained in steps. 
First, let us consider the 2-group formed by $K_0$ and $K_1$.
In general, the 0-form symmetry can act on the generators of $K_1$ symmetry by a permutation of the generators. This means that the 0-form symmetry $K_0$ induces an automorphism of the 1-form symmetry,
\begin{align}
    \rho: K_0\to \mathrm{Aut}(K_1)
\end{align}
Let us denote the background gauge fields for $K_0, K_1$ symmetries as $B_1\in Z^1(M^4,K_0), B_2\in C^2(M^4,K_1)$ respectively. The flat background gauge field of the 2-group $\Gamma_{0,1}$ is then described by the equation
\begin{align}
    \delta_{\rho} B_2 =\gamma_0^* \Theta_3,
    \label{eq:2groupeqgeneral}
\end{align}
where $\gamma_0: M^4\to BK_0$ is the background gauge field for the 0-form symmetry $K_0$, and $\Theta_3\in H^3(BK_0,K_1)$ is called the Postnikov class or an ``$H^3$ obstruction'' that characterizes the nontrivial extension of $K_0$ by $K_1$. Summarizing, the 2-group $\Gamma_{0,1}$ is defined by a collection of data $(K_0,K_1,\rho,\Theta_3)$, and its background field satisfies the equation Eq.~\eqref{eq:2groupeqgeneral}.

Let us further consider the 3-group, which is obtained by a further extension of the 2-group structure $\Gamma_{0,1}$ by the 2-form symmetry $K_2$.
Again, the 0-form symmetry $K_0$ induces an automorphism of the 2-form symmetry,
\begin{align}
    \sigma: K_0\to \mathrm{Aut}(K_2)
\end{align}
Then let us denote the background gauge field of the $K_2$ 2-form symmetry as $B_3\in C^3(M^4,K_2)$. The equation for the 3-group is then given by
\begin{align}
    \delta_{\sigma} B_3 = \gamma_{0,1}^* \Theta_4,
    \label{eq:3groupeqgeneral}
\end{align}
where $\gamma_{0,1}$ denotes the background gauge field of the 2-group symmetry $\gamma_{0,1}: M^4\to B\Gamma_{0,1}$, and the Postnikov class $\Theta_4\in H^4(B\Gamma_{0,1},K_2)$ specifies the $H^4$ obstruction that characterizes the nontrivial mixture of the 2-form symmetry with 0-form and 1-form symmetries. The 3-group is then defined by a collection of data $(\Gamma_{0,1},K_2,\sigma,\Theta_4)$.
Summarizing, the background gauge fields $B_1, B_2, B_3$ satisfying Eq.~\eqref{eq:2groupeqgeneral},~\eqref{eq:3groupeqgeneral} amount to specifying a flat background gauge field for the 3-group, which is characterized by a map $\gamma_{0,1,2}: M^4 \to B \Gamma_{0,1,2}$ with $B \Gamma_{0,1,2}$ a classifying space for the 3-group $\Gamma_{0,1,2}$. We refer the reader to Appendix L of \cite{Lan2019fermion} for a description of the simplicial construction of classifying spaces of 2- and 3-groups aimed at physicists.

When one ignores the 0-form symmetry and the background gauge field of the 0-form symmetry $B_1$ is turned off, Eq.~\eqref{eq:3groupeqgeneral} reduces to a simpler equation
\begin{align}
    \delta B_3 = g_1^*\Theta_4,
    \label{eq:3groupeqsimpler}
\end{align}
where $g_1: M^4\to B^2K_1$ is the background gauge field for the $K_1$ 1-form symmetry, and $\Theta_4$ is regarded as an element $\Theta_4\in H^4(B^2K_1,K_2)$.

Below, we explicitly describe the 3-group structure for several examples of (3+1)D discrete gauge theories considered in the paper, by deriving the above equations for 3-groups Eq.~\eqref{eq:2groupeqgeneral},~\eqref{eq:3groupeqgeneral}.

\subsection{Twist string of $\Z_2\times\Z_2$ toric code in (3+1)D}
\label{sec:Z2Z23group}
For the defect of $\Z_2\times \Z_2$ toric code in (3+1)D, the twist string $\mathcal{D}$ has an effect of emitting a Wilson line operator $\eta_{e_1}, \eta_{e_2}$ from the intersection between $\mathcal{D}$ and magnetic Wilson surface operators $\xi_{m_1}$, $\xi_{m_2}$, as shown in Fig.~\ref{fig: membrane operator m1-m1e2 defect}. Let us denote background gauge fields of $\Z_2$ 2-form symmetries generated by $\eta_{e_1}, \eta_{e_2}$ as $A^{e_1}_3, A^{e_2}_3\in C^3(M^4,\Z_2)$ respectively, and also those of $\Z_2$ 1-form symmetries generated by $\xi_{m_1}, \xi_{m_2}$ as $B^{m_1}_2, B^{m_2}_2\in Z^2(M^4,\Z_2)$ respectively. By writing the background gauge field of $\Z_2$ 1-form symmetry generated by $\mathcal{D}$ as $C_2\in Z^2(M^4,\Z_2)$, the background gauge fields satisfy the relations
\begin{align}
    \delta A^{e_1}_3 = B^{m_2}_2 \cup C_2, \quad \delta A^{e_2}_3 = B^{m_1}_2 \cup C_2,
    \label{eq:3groupZ2Z2}
\end{align}
where each equation means that the generator of the 2-form symmetry is sourced from the intersection between $\mathcal{D}$ and $\xi_{m_i}$. These equations correspond to Eq.~\eqref{eq:3groupeqsimpler}, and imply that the 1-form $K_1=\Z_2\times\Z_2$ and 2-form $K_2=\Z_2\times\Z_2$ symmetry forms a nontrivial 3-group.

The non-trivial structure of the 3-group can actually be described in the language of quantum information in terms of the non-trivial commutation relations of the logical gates. Recall that the 2-form symmetry generated by $\eta_{e}$, and the 1-form symmetry generated by $\xi_{m}$ around non-trivial cycles correspond to the logical $\overline Z$ and $\overline X$ operators of the toric code, respectively. In addition, the 1-form symmetry generated by the twist string $\mathcal D$ corresponds to a logical $\overline{CZ}$ (in this particular case, the logical operators are moreover transversal). If the logical gates corresponds to a trivial 3-group symmetry, then we would have had that the group commutator of any pair of logical operators is a phase. Nevertheless, in this example, there exists a non-trivial relation. For concreteness, define the toric code on a 3-torus and consider the logical operators $\overline X^{m_1}_{xy}$, $\overline{Z}^{e_2}_x$ and $\overline{CZ}_{xz}$, where the subscript denotes the specific 2-cycles on which these logicals are defined. Visually, (see also Fig.~\ref{fig: membrane operator m1-m1e2 defect})
\begin{eqs}
    \overline{CZ}_{xz}=\raisebox{0.0cm}{\includegraphics[scale=0.5, valign=c ]{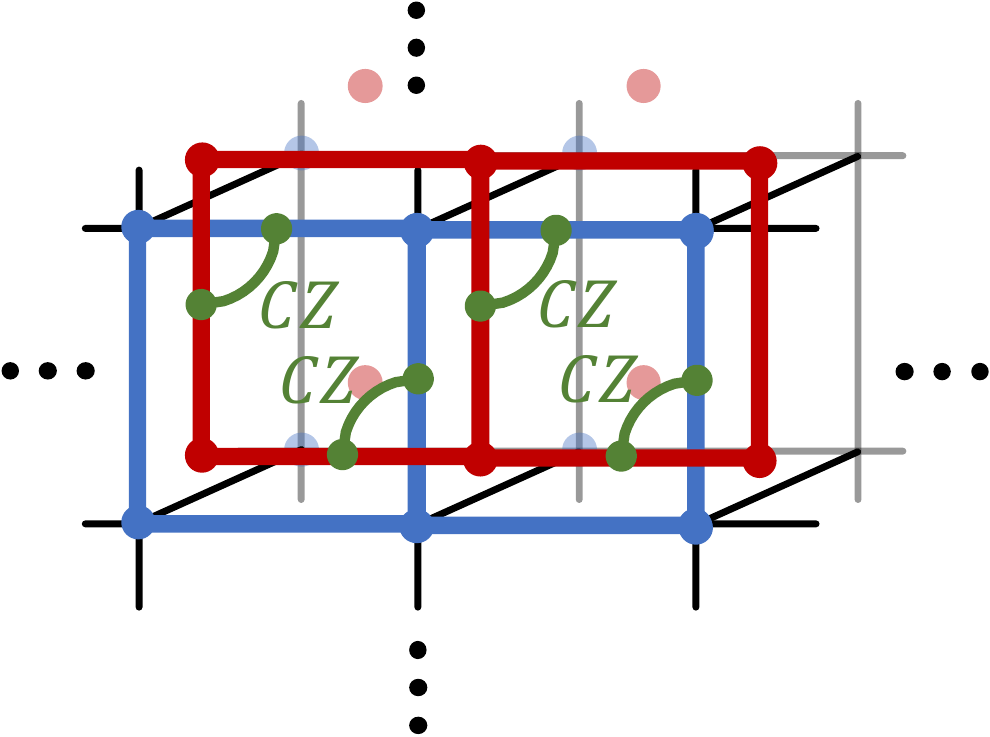}}
    \quad,
    \quad
    \overline{X}^{m_1}_{xy}=\raisebox{0.1cm}{\includegraphics[scale=0.5, valign=c ]{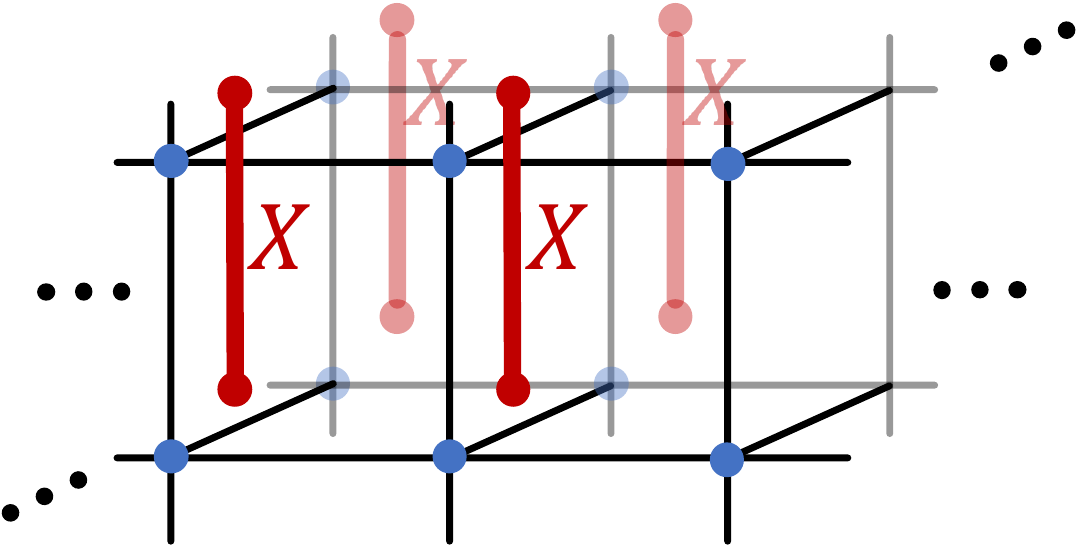}},
\end{eqs}
and
\begin{eqs}
    \overline{Z}^{e_2}_x=\raisebox{0.0cm}{\includegraphics[scale=0.5, valign=c ]{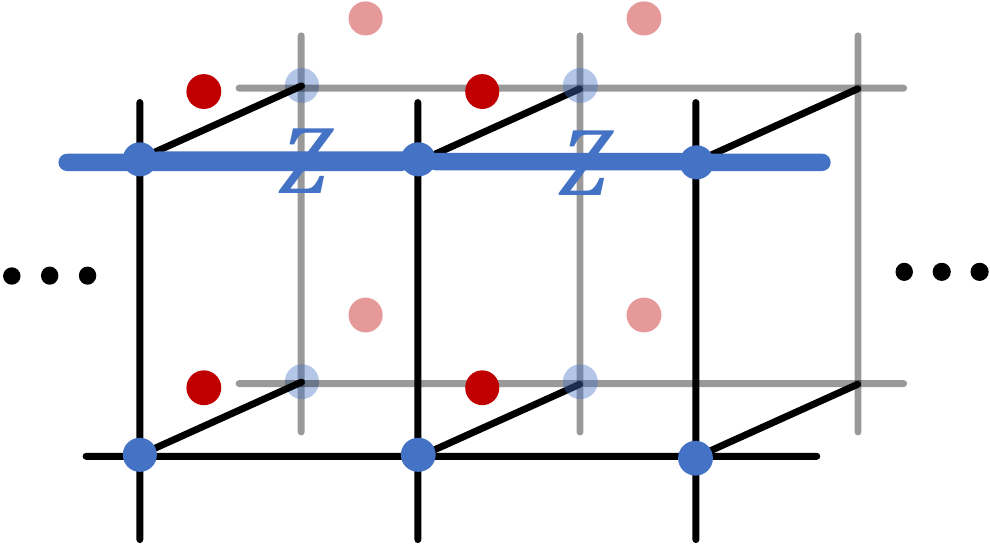}}
    \quad.
\end{eqs}
Then we find that the group commutator of $ \overline{CZ}_{xz}$ and $\overline X^{m_1}_{xy}$ gives
\begin{align}
    [\overline{CZ}_{xz},\overline X^{m_1}_{xy}] := \overline{CZ}_{xz} \overline X^{m_1}_{xy} {\overline{CZ}_{xz}}^{-1} {(\overline X^{m_1}_{xy})}^{-1} = \overline Z^{e_2}_x.
\end{align}
This is precisely the description of Eq.~\eqref{eq:3groupZ2Z2}. A logical $\overline Z^{e_1}$ which is a closed string of $e_1$ is present along the intersection of $\overline{CZ}$ and $\overline X^{m_2}$. The non-trivial relation of logical operators supported on cycles of different dimensions corresponds to the interplay between symmetries of different forms, emphasizing the higher group structure of the symmetries.

\subsection{Twist string of (3+1)D discrete $G$-gauge theory}
Here we describe the 3-group structure of the global symmetry of the (3+1)D discrete $G$-gauge theory, involving the invertible magnetic flux strings and the twist strings of the $G$-gauge theory.

The invertible flux string is labeled by $Z(G)$, which is the center of the gauge group $G$.
Also, as discussed in Sec.~\ref{sec:wrapping}, the twist string is given by a decoration of a (1+1)D bosonic $G$-SPT action on the codimension-2 defect. It generates a 1-form symmetry with the symmetry group $\mathcal{H}^2(G,\U)$, where the multiplication is given by the stacking rule of the (1+1)D SPT phase.
Let us write the background gauge fields of these 1-form symmetries as
\begin{align}
    B_2\in Z^2(M^4,Z(G)), \quad C_2\in Z^2(M^4,\mathcal{H}^2(G,\U))
\end{align}
where $B_2$ corresponds to the flux strings, and $C_2$ is for the twist strings respectively.
Analogously to the previous subsection, these 1-form symmetry again forms a non-trivial 3-group together with the 2-form symmetry generated by the invertible electric Wilson lines, due to the effect that the electric particle is emitted from the intersection between the twist string and the flux string. To describe the intersection effect, we define the coupling $\langle\rangle$: $Z(G)\times \mathcal{H}^2(G,\U)\to \mathcal{H}^1(G,\U)$ as
\begin{align}
    \langle g,\omega \rangle := i_{g}\omega \quad \text{for $g\in Z(G)$, $\omega\in \mathcal{H}^2(G,\U)$},
\end{align}
which tells the label of the electric particle sourced from the intersection between the invertible flux string $g\in Z(G)$ and twist string $\omega\in \mathcal{H}^2(G,\U)$. 

The 2-form symmetry generated by the electric Wilson line forms a group $\mathcal{H}^1(G,\U)$ that corresponds to the Abelian representation of $G$ carried by the Wilson line. By writing its background gauge field as $A_3\in C^3(M^4,\mathcal{H}^1(G,\U))$, the 3-group equation is expressed as
\begin{align}
    \delta A_3 = \langle B_2, \cup C_2 \rangle
\end{align}
where $ \langle B_2,\cup C_2 \rangle$ evaluates the coupling $\langle B_2(01,12), C_2(23,34)\rangle$ on each 4-simplex $\langle 01234\rangle$. The above equation specifies the $H^4$ Postnikov class for the 3-group structure that corresponds to $K_1 = Z(G)\times \mathcal{H}^2(G,\U), K_2 = \mathcal{H}^1(G,\U)$ in Eq.~\eqref{eq:3groupeqsimpler}.

\subsection{Twist string of (3+1)D $\Z_2$ gauge theory with a fermionic particle}
\label{sec:3groupfermion}
Next, we describe a 3-group structure for a codimension-2 twist string defect of (3+1)D $\Z_2$ gauge theory with a fermionic particle. 

Let us denote the background gauge field of the $\Z_2$ 2-form symmetry generated by the Wilson line $\eta_{\psi}(\gamma)$ as $A^\psi_3\in Z^3(M^4,\Z_2)$, and that of the $\Z_2$ 1-form symmetry generated by the twist string as $C_2\in Z^2(M^4,\Z_2)$. 
The 2-form symmetry then has a 't Hooft anomaly $(-1)^{\int w_2\cup A^\psi_3}$, as discussed in Sec.~\ref{subsec:fermionhighergauging}.
If we insert a codimension-2 defect $\mathcal{D}$ in the spacetime by turning on the background gauge field $C_2$, it turns out the symmetry has a 3-group structure
\begin{align}
    \delta A^\psi_3 = w_2\cup C_2,
    \label{eq:3groupfermion}
\end{align}
where $w_2$ is the second Stiefel-Whitney class of the spacetime manifold $M^4$. To see this, we enclose the codimension-2 Poincar\'e dual $W_2$ of $w_2$ by a cylinder of the codimension-2 defect $\mathcal{D}$, analogously to what we did in Fig.~\ref{fig:condensationZ2Z2} to study the action of $\mathcal{D}$ on magnetic Wilson surfaces. Then, the cylinder of $\mathcal{D}$ acts on $W_2$ extended along $y$ direction as
\begin{align}
\begin{split}
    \mathcal{D}\cdot W_2 &=\frac{1}{2}\sum_{j,k\in\Z_2} \eta_{\psi}(\gamma_y)^k (-1)^{jk+j}\cdot W_2(\gamma_y) \\
    &= \sum_{k}\delta(k+1)  \eta_{\psi}(\gamma_y)^k \cdot W_2(\gamma_y) = \eta_{\psi}(\gamma_y) \cdot W_2(\gamma_y),
    \end{split}
\end{align}
where we used $\eta_{\psi}(\gamma_x)$ acts by $(-1)$ sign on $W_2$ due to the 't Hooft anomaly $(-1)^{w_2\cup A_3}$ that manifests itself as the mutual braiding between $\eta_{\psi}(\gamma_x)$ and $W_2$.
So, $\mathcal{D}$ acts by attaching the fermionic Wilson line to the Poincar\'e dual of $w_2$. It means that when there is an intersection between the defect $\mathcal{D}$ and $w_2$ it emits a fermionic line operator, so it results in a 3-group shown in Eq.~\eqref{eq:3groupfermion}. If we further turn on the background gauge field $B_2^m$ of $\Z_2$ 1-form symmetry generated by $\xi_m$ 
the equation for 3-group becomes
\begin{align}
    \delta A^\psi_3 = (B^m_2+w_2)\cup C_2,
    \label{eq:3groupfermionA2added}
\end{align}
due to the intersection effect between the twist string $\mathcal{D}$ and the magnetic string $\xi_m$.~\footnote{
The 3-group structure in Eq.~\eqref{eq:3groupfermionA2added} is reminiscent of the Gu-Wen equation known in the context of the classification of (3+1)D fermionic SPT phase~\cite{kapustinThorngren2017FermionSPT, qingrui}. The fermionic SPT phase with 0-form symmetry $G$ is generally constructed by decoration of lower-dimensional fermionic invertible phases on the junction of $G$ symmetry defects, and the Gu-Wen equation gives the consistency condition for the decoration. The 3-group equation Eq.~\eqref{eq:3groupfermionA2added} corresponds to the consistency between the decoration of the Kitaev chain on the Poincar\'e dual of $B_2$, and the decoration of a complex fermion (i.e., (0+1)D fermionic invertible phase) on the Poincar\'e dual of $A_3$. From this perspective, our discussion is regarded as a derivation of the Gu-Wen consistency equation based on continuum field theory.}

\subsection{3-group structure of codimension-1,2, and 3 defects of $\Z_2^3$ toric code}
\label{subsec:Z23toriccode3group}
Finally, we describe the structure of symmetry groups of (3+1)D $(\Z_2)^3$ gauge theory given by three copies of $\Z_2$ toric codes in (3+1)D.
The action is given by
\begin{align}
    \pi \int a^{e_1}\cup\delta b^{m_1} + a^{e_2}\cup \delta b^{m_2} + a^{e_3}\cup \delta b^{m_3}
\end{align}
with $a^{e_1},a^{e_2},a^{e_3}\in C^1(M^4,\Z_2),b^{m_3}\in C^2(M^4,\Z_2)$. 
We first enumerate the global symmetries of the theory we want to consider. First, the theory has the $(\Z_2)^3$ 2-form symmetry generated by Wilson line operators 
\begin{align}
    \eta_{e_1}(\gamma) = \exp\left( \pi i \int_{\gamma} a^{e_1}\right), \quad  \eta_{e_2}(\gamma) = \exp\left( \pi i \int_{\gamma} a^{e_2}\right), \quad 
    \eta_{e_3}(\gamma) = \exp\left( \pi i \int_{\gamma} a^{e_3}\right)
\end{align}
we write the corresponding background gauge fields as $A^{e_i}_{3}\in C^3(M^4,\Z_2)$ for $i=1,2,3$.

Also, the theory has the $(\Z_2)^3$ 1-form symmetry generated by magnetic surface operators 
\begin{align}
    \xi_{m_1}(\Sigma) = \exp\left( \pi i \int_{\Sigma} b^{m_1}\right), \quad  \xi_{m_2}(\Sigma) = \exp\left( \pi i \int_{\Sigma} b^{m_2}\right), \quad 
    \xi_{m_3}(\Sigma) = \exp\left( \pi i \int_{\Sigma} b^{m_3}\right)
\end{align}
we write the corresponding background gauge fields as $B^{m_i}_{2}\in C^2(M^4,\Z_2)$ for $i=1,2,3$.

In addition, there is also the $(\Z_2)^3$ 1-form symmetry generated by the codimension-2 twist strings
\begin{align}
    \mathcal{D}_{e_1e_2}(\Sigma) = \exp\left( \pi i \int_{\Sigma} a^{e_1}\cup a^{e_2}\right), \quad  \mathcal{D}_{e_2e_3}(\Sigma) = \exp\left( \pi i \int_{\Sigma} a^{e_2}\cup a^{e_3}\right), \quad 
    \mathcal{D}_{e_3e_1}(\Sigma) = \exp\left( \pi i \int_{\Sigma} a^{e_3}\cup a^{e_1}\right)
\end{align}
we write these background gauge fields as $C_{2}^{e_ie_j}\in C^2(M^4,\Z_2)$.

Finally, we consider an interesting $\Z_2$ 0-form symmetry given by decorating the codimension-1 submanifold with the bosonic $(\Z_2)^3$ SPT phase in (2+1)D,
\begin{align}
    \mathcal{D}_{e_1e_2e_3}(M^3) = \exp\left( \pi i \int_{M^3} a^{e_1}\cup a^{e_2}\cup a^{e_3}\right)
\end{align}
whose background gauge field is written as $C_{1}\in Z^1(M^4,\Z_2)$.

This codimension-1 defect $ \mathcal{D}_{e_1e_2e_3}$ was studied in~\cite{Yoshida15}, and has an interesting action on the generators of 1-form symmetries. When the generators of 1-form symmetry crosses through $ \mathcal{D}_{e_1e_2e_3}$, the generators get permuted as
\begin{align}
    \xi_{m_1}\to \xi_{m_1} \times \mathcal{D}_{e_2e_3}, \quad
    \xi_{m_2}\to \xi_{m_2} \times \mathcal{D}_{e_3e_1}, \quad
    \xi_{m_3}\to \xi_{m_3} \times \mathcal{D}_{e_1e_2}
\label{eq: m1 to m1s23}
\end{align}
while leaving $\eta_{e_i}$ and $\mathcal{D}_{e_ie_j}$ invariant. 
Note that the action realizes the permutation of background gauge fields as
\begin{align}
    C_{2}^{e_1e_2} \to C_{2}^{e_1e_2} + B_{2}^{m_3}, \quad C_{2}^{e_2e_3} \to C_{2}^{e_2e_3} + B_{2}^{m_1}, \quad
    C_{2}^{e_3e_1} \to C_{2}^{e_3e_1} + B_{2}^{m_2}
\end{align}
while leaving $A^{e_i}_3$ and $B^{m_i}_{2}$ invariant. 
This permutation means that the $\Z_2$ 0-form symmetry realizes an automorphism of the $(\Z_2)^3\times (\Z_2)^3$ 1-form symmetry, which gives a group homomorphism $\rho: \Z_2\to\mathrm{Aut}((\Z_2)^3\times (\Z_2)^3)$.
Due to the permutation of background gauge fields, the gauge fields of 1-form symmetry satisfy the twisted cocycle conditions
\begin{align}
    \delta_{\rho} B_2^{m_i} = 0, \quad \delta_{\rho} C_2^{e_ie_j} = 0
    \label{eq:trivial2group}
\end{align}
which corresponds to the 2-group in Eq.~\eqref{eq:2groupeqgeneral} with the trivial $H^3$ obstruction class $\Theta_3=0$. The above equations are rewritten as
\begin{align}
    \begin{split}
        \delta B_2^{m_i} &= 0, \\
        \delta C_2^{e_1e_2} &= B_2^{m_3} \cup C_1, \\
        \delta C_2^{e_2e_3} &= B_2^{m_1} \cup C_1, \\
        \delta C_2^{e_3e_1} &= B_2^{m_2} \cup C_1.
        \label{eq:twistedcoboundary}
    \end{split}
\end{align}
Now we describe the 3-group structure of the global symmetry described above. First, when the background gauge field $C_1$ for the 0-form symmetry is turned off, the 3-group structure is a straightforward generalization of the case of $\Z_2\times\Z_2$ gauge theory obtained in Eq.~\eqref{eq:3groupZ2Z2}. That is, the background gauge fields $A^{e_i}_{3}$ of 2-form symmetry is not closed due to the crossing of twist strings with the magnetic strings, and these effects are written in the case of $C_1=0$ as
\begin{align}
    \begin{split}
        \delta A^{e_1}_3 &= B^{m_2}_2\cup C^{e_1e_2}_2 + B^{m_3}_2\cup C^{e_3e_1}_2, \\
    \delta A^{e_2}_3 &= B^{m_3}_2\cup C^{e_2e_3}_2 + B^{m_1}_2\cup C^{e_1e_2}_2,  \\
   \delta A^{e_3}_3 &= B^{m_1}_2\cup C^{e_3e_1}_2 + B^{m_2}_2\cup C^{e_2e_3}_2.
    \end{split}
    \label{eq:3groupZ2Z2Z2B0}
\end{align}
When we turn on the gauge field $C_1$, the RHS of Eq.~\eqref{eq:3groupZ2Z2Z2B0} is no longer closed due to the twisted cocycle condition in Eq.~\eqref{eq:twistedcoboundary}, hence the RHS must be modified so that the expression is closed. For the case $C_1$ is nonzero, the 3-group equations Eq.~\eqref{eq:3groupZ2Z2Z2B0} are generalized as
\begin{align}
    \begin{split}
       \delta A^{e_1}_3 &= B^{m_2}_2\cup C^{e_1e_2}_2 + B^{m_3}_2\cup C^{e_3e_1}_2 + (B^{m_2}_2\cup_1B^{m_3}_2)\cup C_1, \\
   \delta A^{e_2}_3 &= B^{m_3}_2\cup C^{e_2e_3}_2 + B^{m_1}_2\cup C^{e_1e_2}_2 + (B^{m_3}_2\cup_1B^{m_1}_2)\cup C_1, \\
   \delta A^{e_3}_3 &= B^{m_1}_2\cup C^{e_3e_1}_2 + B^{m_2}_2\cup C^{e_2e_3}_2 + (B^{m_1}_2\cup_1B^{m_2}_2)\cup C_1.
    \end{split}
    \label{eq:3groupZ2Z2Z2}
\end{align}
These three equations \eqref{eq:3groupZ2Z2Z2} correspond to the $H^4$ obstruction Eq.~\eqref{eq:3groupeqgeneral}, and together with Eq.~\eqref{eq:trivial2group} characterize the 3-group structure of the global symmetry.

\begin{figure}[t]
    \centering
    \includegraphics[width=0.95\textwidth]{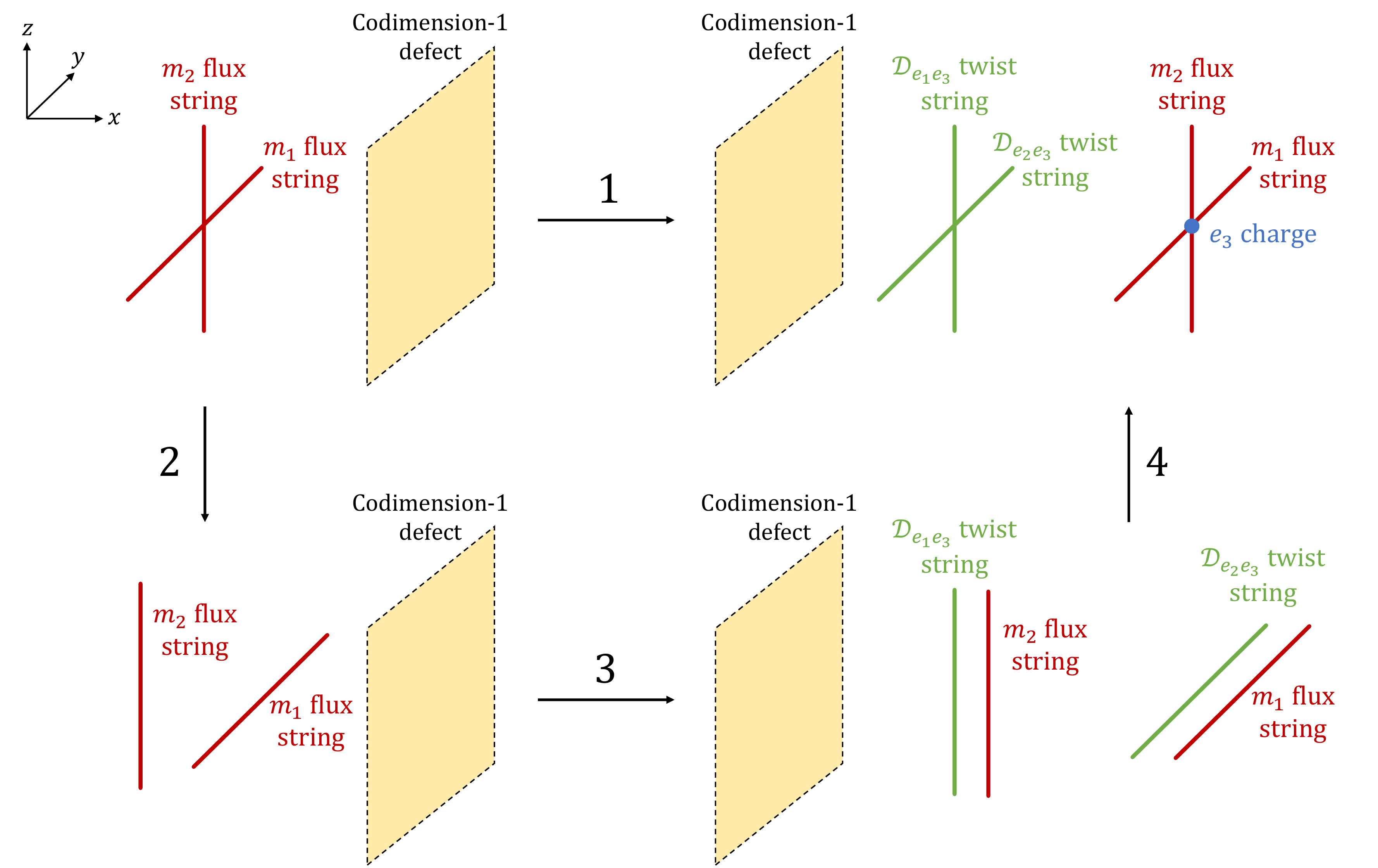}
\caption{The process $1$ corresponds to Eqs.~\eqref{eq: m1 to m1s23} and~\eqref{eq:3groupZ2Z2Z2}. This process can be decomposed into processes $2$, $3$, and $4$. We move the $m_1$ and $m_2$ flux strings across the codimension-1 defect separately as the $m_1 \mathcal{D}_{e_2 e_3}$ and $m_2 \mathcal{D}_{e_1 e_3}$ strings. In the process $4$, we re-arrange the $m_2$ flux string and the $\mathcal{D}_{e_2 e_3}$, and an $e_3$ charge appears. In terms of logical operators, the process $3$ can be written as: $\overline X^{m_2}_{xz} \overline X^{m_1}_{xy} \cdot\overline{CCZ} = \overline{CCZ}\cdot \overline{CZ}^{e_1 e_3}_{xz} \overline X^{m_2}_{xz} \cdot  \overline{CZ}^{e_2 e_3}_{xy} \overline X^{m_1}_{xy} $, which implies Eq.~\eqref{eq: logical CCZ and XX commutation}.}
\label{fig: codim-1 defect and m1m2}
\end{figure}

According to the geometrical interpretation of higher cup products reviewed in Appendix~\ref{sec: higher cup products}, the last term $(B^{m_1}_2\cup_1 B^{m_2}_2)\cup C_1$ in Eq.~\eqref{eq:3groupZ2Z2Z2} can be interpreted in the following way. If $m_1$ and $m_2$ strings are crossing at a point in 3d space and are moved across the codimension-1 defect together, an $e_3$ charge will appear at the crossing point (in addition to $\mathcal{D}_{e_2e_3}$ and $\mathcal{D}_{e_1 e_3}$ from $m_1$ and $m_2$ themselves). To see why a single $e_3$ charge would appear, we can regularize the process, as shown in Fig.~\ref{fig: codim-1 defect and m1m2}. We first move an $m_1$ string across the codimension-1 defect and it becomes an $m_1 \mathcal{D}_{e_2 e_3}$ string. Secondly, we move an $m_2$ string across the codimension-1 defect, which gives an $m_2 \mathcal{D}_{e_1 e_3}$ string. Next, we rearrange the ordering of $m_1$, $\mathcal{D}_{e_2 e_3}$, $m_2$, and $\mathcal{D}_{e_1 e_3}$ and separate the flux string part $m_1,~m_2$ and the twist string part $\mathcal{D}_{e_2 e_3},~\mathcal{D}_{e_1 e_3}$. This rearrangement requires moving one of the $m$ strings through a $\mathcal{D}$ string, which produces an $e_3$ charge.

In terms of logical operators for the theory defined on a spatial 3-torus, sweeping the defect $\mathcal D_{e_1e_2e_3}$ corresponds to a logical $\overline{CCZ}$ of the three toric codes. Sweeping an $m_1$ string across the $xy$-plane  corresponds to the logical operator $\overline{X}_{xy}^{m_1}$, while sweeping $m_2$ along the $xz$-plane corresponds to $\overline{X}_{xz}^{m_2}$. These planes intersect on a line along the $x$-direction. The emergence of the $e_3$ charge in the above discussion then corresponds to the following group commutator (see Fig.~\ref{fig: codim-1 defect and m1m2}):
\begin{align}
   [\overline{CCZ} ,\overline X^{m_1}_{xy} \overline X^{m_2}_{xz}] 
  := \overline{CCZ} (\overline X^{m_1}_{xy} \overline X^{m_2}_{xz}) {\overline{CCZ}}^{-1} {(\overline X^{m_1}_{xy} \overline X^{m_2}_{xz})}^{-1} = \overline{CZ}^{e_2e_3}_{xy}\overline{CZ}^{e_1e_3}_{xz} \overline Z^{e_3}_x.
\label{eq: logical CCZ and XX commutation}
\end{align}
On the right hand side, the first two contributions correspond to Eq.~\eqref{eq:twistedcoboundary} while the third term corresponds to the cup-1 contribution of Eq.~\eqref{eq:3groupZ2Z2Z2}.

\section{Codimension-3 defects at boundaries of twist strings: non-Abelian point defects}
\label{sec:codim3}

In contrast to the flux strings, twist strings $\mathcal{D}$ can exist on finite segments. The endpoints $\partial \mathcal{D}$ define non-Abelian codimension-3 defects that give rise to topological degeneracies, which hence  encode protected logical qubits. Moreover, the endpoints host ``zero-modes", in the sense that non-trivial electric point charges can be created or annihilated by local operators at these endpoints. This implies a non-conservation of electric charges in the vicinity of these point defects. 

In the following, we present the general layer construction of such codimension-3 defects as well as the lattice models for the (3+1)D $\ZZ_2 \times \ZZ_2$ toric code and the (3+1)D toric code with a fermionic charge in the presence of these codimension-3 defects. 

In the case of the (3+1)D $\Z_2$ toric code with fermionic charge, the endpoints of the twist strings localize unpaired Majorana zero modes (MZMs), which gives an interesting example of unpaired MZMs existing at isolated points in 3-dimensional space. This is reminiscent of discussions of non-Abelian statistics in 3 spatial dimensions in Refs.~\cite{teo2010,freedman2011,mcgreevy2011}. We note that Ref.~\cite{WB20} has also discussed some properties of codimension-3 defects in the context of the 3d Levin-Wen fermion model \cite{levin2003}, focusing mainly on the logical operators and encoding. Here we provide an explicit construction of an exactly solvable lattice model in the presence of these codimension-3 defects, which was not provided previously.

\subsection{General layer construction}

\begin{figure}[t]
    \centering
    \includegraphics[width=0.3\textwidth]{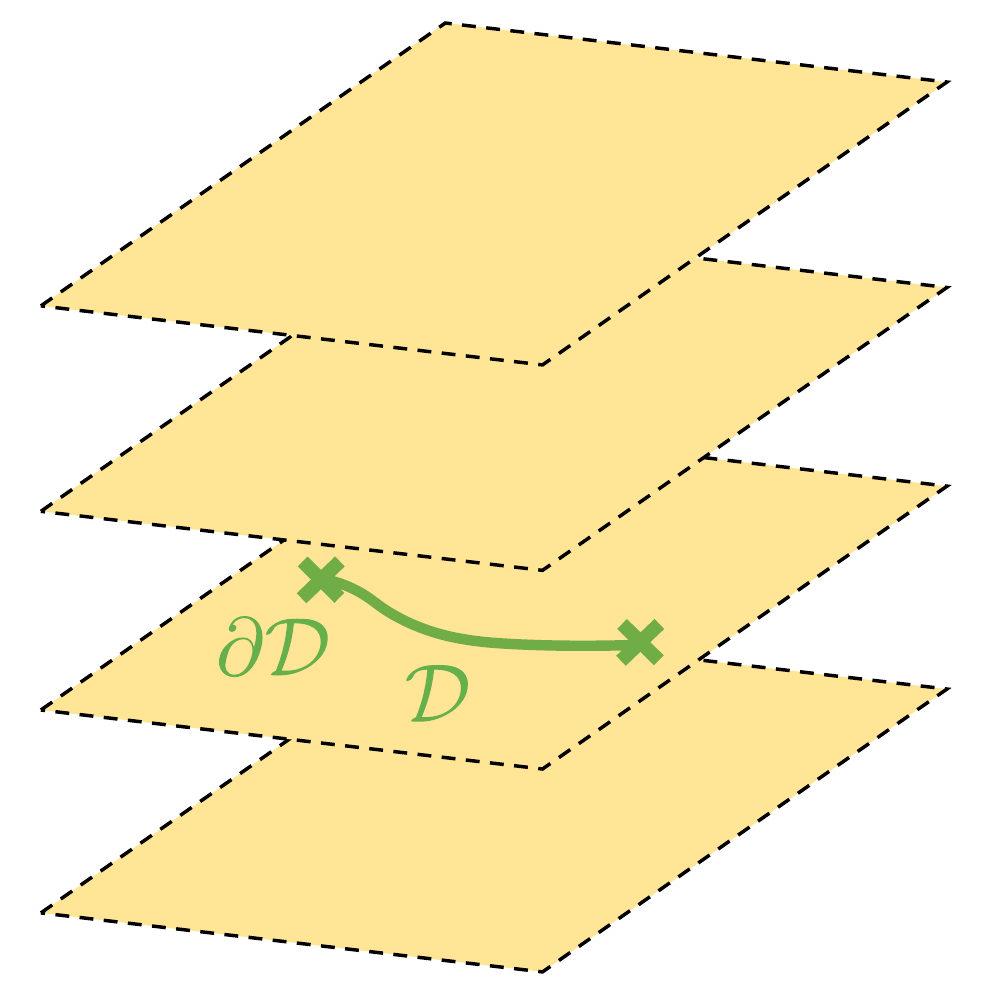}
\caption{Layer construction of the codimension-3 point defects in (3+1)D topological order.  The open twist string $\mathcal{D}$ from the (2+1)D topological order is embedded in the $j^\text{th}$ layer and becomes an open twist string in the (2+1)D topological order after condensing pair of pure charge and anti-charge in the neighboring layer. The endpoints of the open twist string, i.e., $\partial \mathcal{D}$ form the codimension-3 point defects in (3+1)D. }
\label{fig:layer_point_defect}
\end{figure}

We first discuss the general procedure of obtaining the codimension-3 defects from layer construction.
We start with twist defects in the (2+1)D topological order, which are the pair of point defects located on the boundary of the twist string (domain wall) $\mathcal{D}$, i.e., $\partial \mathcal{D}$. Similar to the discussion in Sec.~\ref{sec:layer}, we can place the twist defects along with the twist string in the $j^\text{th}$ layer, as illustrated in Fig.~\ref{fig:layer_point_defect}.  We can hence obtain 
the codimension-3 point defects $\partial \mathcal{D}$ at the boundaries of the twist string in (3+1)D topological order via the layer construction.

\subsection{Point defects in (3+1)D $\ZZ_2 \times \ZZ_2$ toric code}

\begin{figure}[H]
    \centering
    \includegraphics[width=0.8\textwidth]{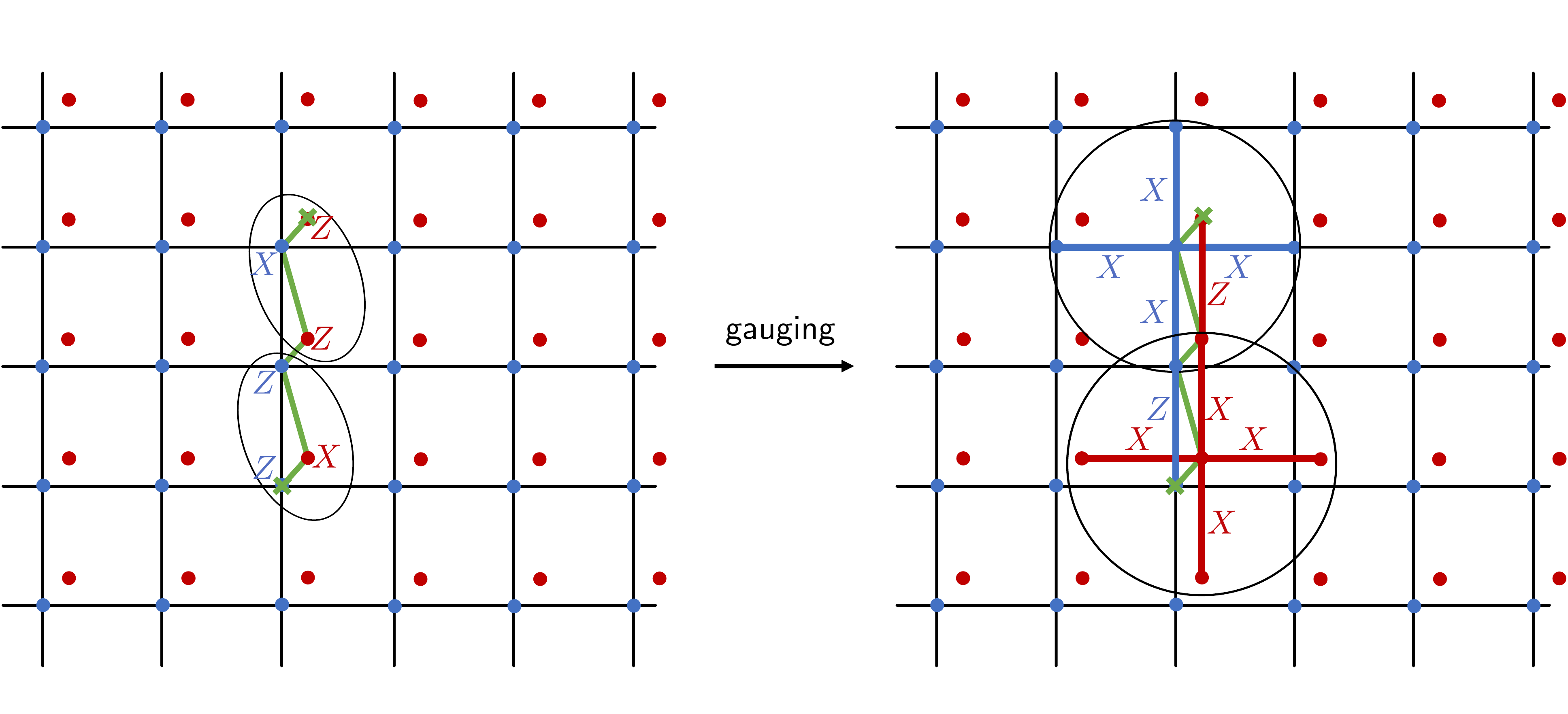}
\caption{Left: On the background of the 2d paramagnetic Ising model, a 1d cluster state is decorated on an open string (green line) terminated at two point defects (green crosses). The single-body paramagnetic Ising term $X_\nu$ is removed on the blue or red site coinciding with the green crosses. Right: After gauging the $\ZZ_2 \times \ZZ_2$ symmetry, the vertex $X$-stabilizer on the green line (between the endpoints) is dressed with an additional $Z$ of a different color. The 4-body vertex $X$-stabilizers centered on the blue or red vertices coinciding with the green crosses are removed.}
\label{fig:2d_point_defects_cluster}
\end{figure}

\begin{figure}[H]
    \centering
    \includegraphics[width=1.0\textwidth]{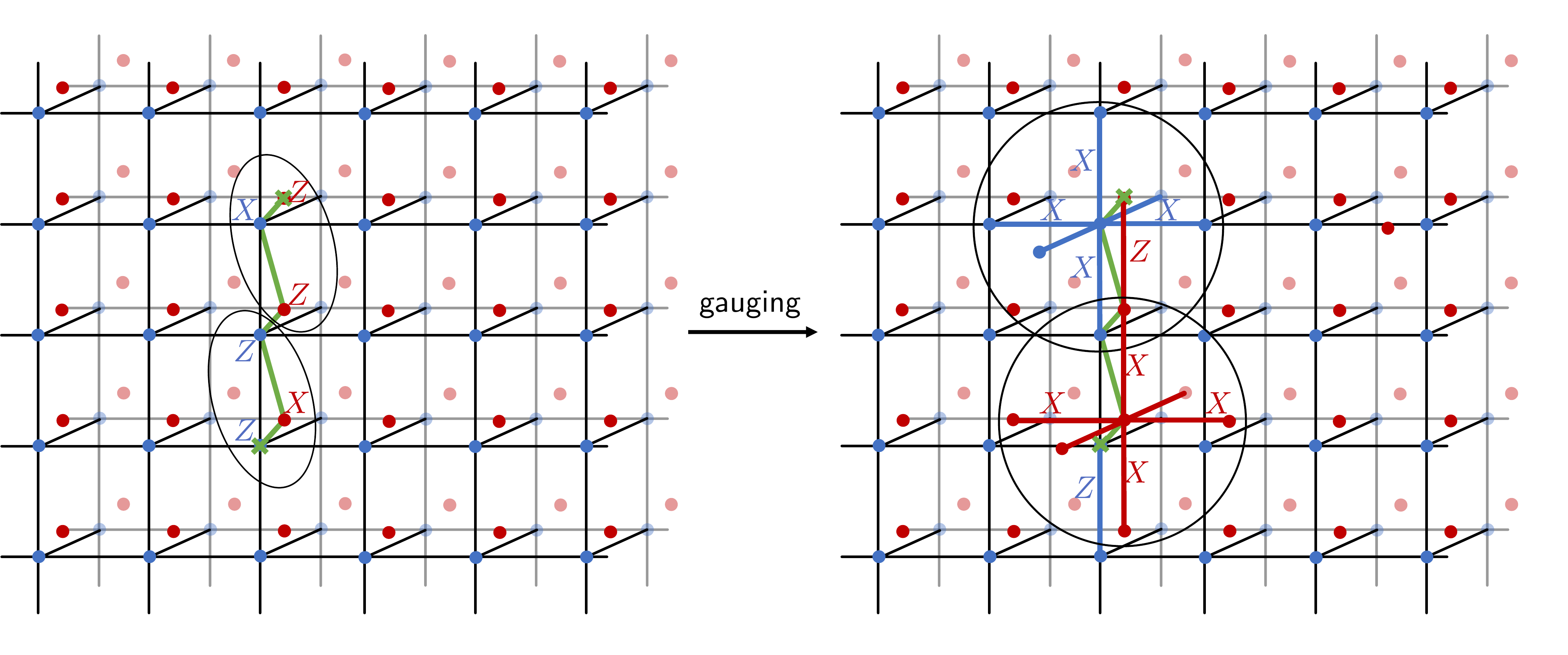}
\caption{Left: On the background of the 3d paramagnetic Ising model, a 1d cluster state is decorated on an open string (green line) terminated at two point defects (green crosses). The single-body paramagnetic Ising term $X_\nu$ is removed on the blue or red site coinciding with the green crosses. Right: After gauging the $\ZZ_2 \times \ZZ_2$ symmetry, the vertex $X$-stabilizer on the green line (between the endpoints) is dressed with an additional $Z$ of a different color. The 6-body vertex $X$-stabilizers centered on the blue or red vertices coinciding with the green crosses are removed.}
\label{fig:3d_point_defects}
\end{figure}

\begin{figure}[H]
    \centering
    \includegraphics[width=1\textwidth]{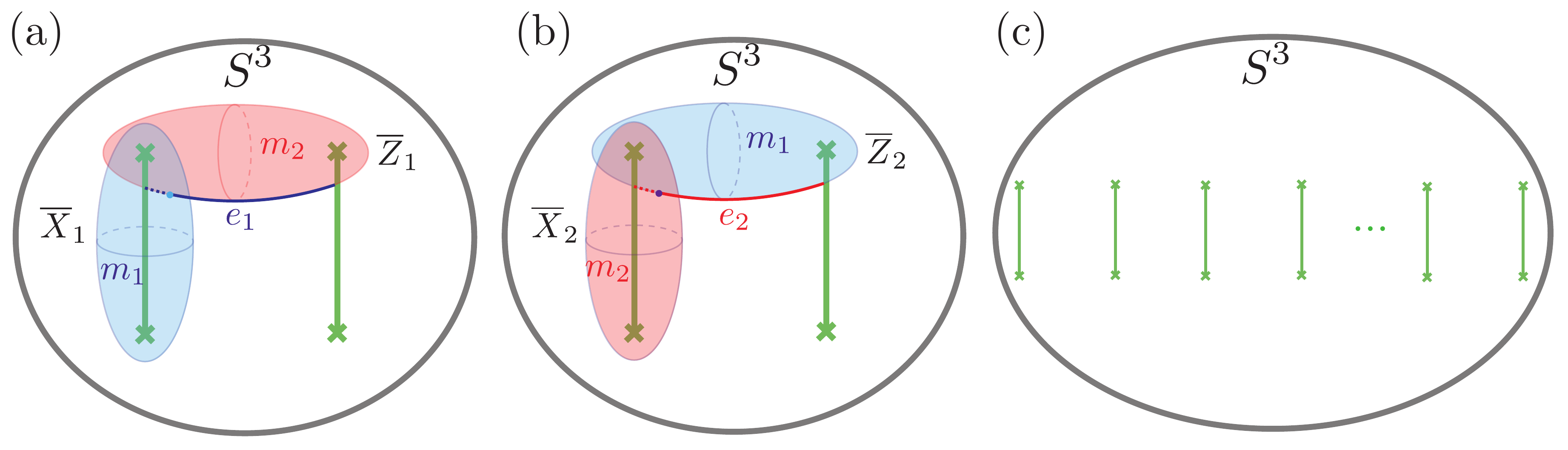}
\caption{(a,b) The anti-commuting pair of logical operators associated with each of the two logical qubits on a 3-sphere with a pair of open twist strings (4 point defects). In (a), the logical-$X$ operator $\overline{X}_1=\xi_{m_1}$ is a Wilson sheet operator (blue) of flux string $m_1$ supported on a 2-sphere enclosing a twist string. The logical-$Z$ operator $\overline{Z}_1=\xi_{m_2} \eta_{e_1}$ is a product of Wilson sheet operator $\xi_{m_2}$ (red) of flux string $m_2$ supported on a 2-sphere encircling two point defects belonging to the two different twist strings and a worldline operator $\eta_{e_1}$ (blue) of the charge $e_1$ connecting the two twist strings. The configuration of  the logical operators in (b) is the same as those in (a) up to a permutation of the copy label, i.e., $1\leftrightarrow 2$. (c) Placing $n$ open twist strings ($n$ pairs of point defects) on a 3-sphere leads to the ground-state degeneracy $4^{n-1}$ encoding $2(n-1)$ logical qubits. }
\label{fig:logical_Z2}
\end{figure}

We start with the construction of such point defects in the (2+1)D $\ZZ_2 \times \ZZ_2$ toric code, i.e., two copies of toric codes.  Such point defects are also called twist defects. 
As shown in Fig.~\ref{fig:2d_point_defects_cluster} (Left), we start with  two copies of the 2d paramagnetic Ising models $H_0 = -\sum_v X_v$ decorated with the 1d $\ZZ_2 \times \ZZ_2$ cluster state on an open string terminated at two point defects  (green crosses).  In particular, we remove the single-body paramagnetic Ising term $X_\nu$ at the two point defects (i.e., for $\nu \in \partial \mathcal{D}$), where the red (copy 1) or blue (copy 2) sites coincide with the blue crosses.  When gauging the $\ZZ_2 \times \ZZ_2$ symmetry, we obtain an open twist string, as illustrated in Fig.~\ref{fig:2d_point_defects_cluster} (Right).  Note that only between the two point defects $\partial \mathcal{D}$ (green crosses), there exist 5-body vertex stabilizers on the twist string coupling both copies of toric codes  (dressed with $Z$ operator with a different color). Moreover, two 4-body vertex $X$-stabilizers  centered on the two point defects, where the green crosses coincide with the blue or red sites, are removed to ensure commutativity between neighboring vertex stabilizers. 

When considering periodic boundary conditions in both directions on both copies, which corresponds to placing both copies on a torus, this removal of stabilizer constraints leads to an increase of topological degeneracy by a factor of four, thus encoding two additional logical qubits.  When considering a sphere topology $S^2$ for both copies of the toric codes, there is an additional redundancy of the vertex  stabilizers, $\prod_\nu A_\nu=1$, before introducing the twist string. The reduction of one stabilizer constraint per layer due to the introduction of the twist string hence does not increase the degree of freedom in the ground-state subspace, which hence does not contribute to the degeneracy. When introducing two open twist strings and hence two pairs of point defects on the sphere, one obtains a 4-fold degeneracy which encodes two logical qubits. More generally, when placing $n$ open twist strings and hence $n$ pairs of point defects on a sphere, the ground-state degeneracy becomes $\text{GSD}=4^{n-1}$, which encodes $2(n-1)$ logical qubits.

We now start constructing the point defects $\partial \mathcal{D}$ in the (3+1)D $\ZZ_2 \times \ZZ_2$ toric code, i.e., two copies of 3D toric codes.  Similar to the (2+1)D case, we start with two copies of the 3D paramagnetic Ising models and then  place the 1d $\ZZ_2 \times \ZZ_2$ cluster state on an open string terminated at two point defects (green crosses), as illustrated in Fig.~\ref{fig:3d_point_defects} (Left).  We also remove the single-body Ising term $X_\nu$ at the two point defects where the red or blue sites coincide with the green crosses.  After gauging the $\ZZ_2 \times \ZZ_2$ symmetry, we obtain the open twist string in the (3+1)D $\ZZ_2 \times \ZZ_2$ toric code, as shown in Fig.~\ref{fig:3d_point_defects} (Right).  Note that only between the two point defects $\partial \mathcal{D}$ (green crosses), there exist 7-body stabilizers on the twist string coupling both copies of toric codes (dressed with $Z$ operator with a different color). Moreover, two 6-body vertex $X$-stabilizers centered on the two point defects, where blue or red sites coincide with the green crosses, are removed to ensure commutativity.   When considering placing both copies on a 3-torus $T^3$, this removal of stabilizer constraints lead to the increase of topological degeneracy by a factor of four. When placing both copies on a 3-sphere $S^3$, the introduction of the open twist string does not increase the ground-state degeneracy due to the additional constraint of the vertex stabilizer in each copy before introducing the twist string, i.e., $\prod_\nu A_\nu=1$, in analogy to the (2+1)D case. More generally, when placing $n$ open twist strings and hence $n$ pairs of point defects on a 3-sphere as illustrated in Fig.~\ref{fig:logical_Z2}(c), the ground-state degeneracy becomes $\text{GSD}=4^{n-1}$, which encodes $2(n-1)$ logical qubits, i.e., same as the (2+1)D case.  

We now discuss the Wilson operator algebra which gives rise to the topological degeneracy in (3+1)D, with the background topology chosen as the 3-sphere $S^3$.  The pair of logical operators corresponding to the two logical operators on a 3-sphere with a pair of open twist strings are shown in Fig.~\ref{fig:logical_Z2}(a,b) respectively.  For the first logical qubit, the logical-$X$ operator is   $\overline{X}_1=\xi_{m_1}$, where $\xi_{m_1}$ is a  Wilson sheet operator of the flux $m_1$ supported on a 2-sphere $S^2$. On the lattice model, $\xi_{m_1}$ corresponds to a product of Pauli-$X$ operators supported on such a 2-sphere. The logical-$Z$ operator is $\overline{Z}_1=\xi_{m_2} \eta_{e_1}$, i.e., a product of the spherical Wilson sheet operator $\xi_{m_2}$ of flux $m_2$ (a product of Pauli-X) and a Wilson line operator $\eta_{e_1}$ of charge $e_1$ trapped between two open twist string. This Wilson line operator is emitted due to the crossing of flux $m_2$ through the twist string, and corresponds to a string of Pauli-$Z$ operators.  The second logical qubit has the same operator configuration up to an exchange of the copy label, $1\leftrightarrow 2$.  The four-fold topological degeneracy comes from the two anti-commutation relations $\overline{X}_i \overline{Z}_i = -\overline{Z}_i \overline{X}_i$ for $i=1,2$, where the minus sign is due to the intersection between the Wilson sheet $\xi_{m_i}$ trapped around one twist string and the Wilson line $\eta_{e_i}$ and hence the anti-commutation relation $\xi_{m_i}\eta_{e_i}=-\eta_{e_i}\xi_{m_i}$.

We note that on the 3-sphere, there can be logical-X operators encircling the right twist string, denoted by $\overline{X}'_{1,2}$, and logical-Z operators encircling the lower two point defects, denoted by $\overline{Z}'_{1,2}$.  However, these operators are equivalent to the logical operators discussed above, i.e., $\overline{X}'_{1,2}=\overline{X}_{1,2}$ and $\overline{Z}'_{1,2}=\overline{Z}_{1,2}$.  This is because when considering a spherical Wilson sheet operator on the $i^\text{th}$ copy of toric code  encircling both twist strings (i.e., the four point defects), it must equal to logical identity $1$ since it can be shrunk into a single point on the 3-sphere. Therefore, one has $\overline{X}'_{i}\overline{X}_{i}=1$ and  $\overline{Z}'_{i}\overline{Z}_{i}=1$.   Now consider general situation of placing $n$ open twist string ($n$ pairs of point defects) on the 3-sphere, we will have $n-1$ pairs of anti-commuting logical operators corresponding to $n-1$ logical qubits. This is because, although each open twist string can trap one logical-$X$ operators and hence lead to $n$ of such logical operators, only $n-1$ of them are independent due to the constraint that the product of them corresponds to the Wilson sheet operator encircling all twist strings and hence equaling the logical identity.

\subsection{Point defects in (3+1)D toric code with a fermionic particle: unpaired Majorana zero modes in (3+1)D}

\begin{figure}[H]
    \centering
    \includegraphics[width=0.8\textwidth]{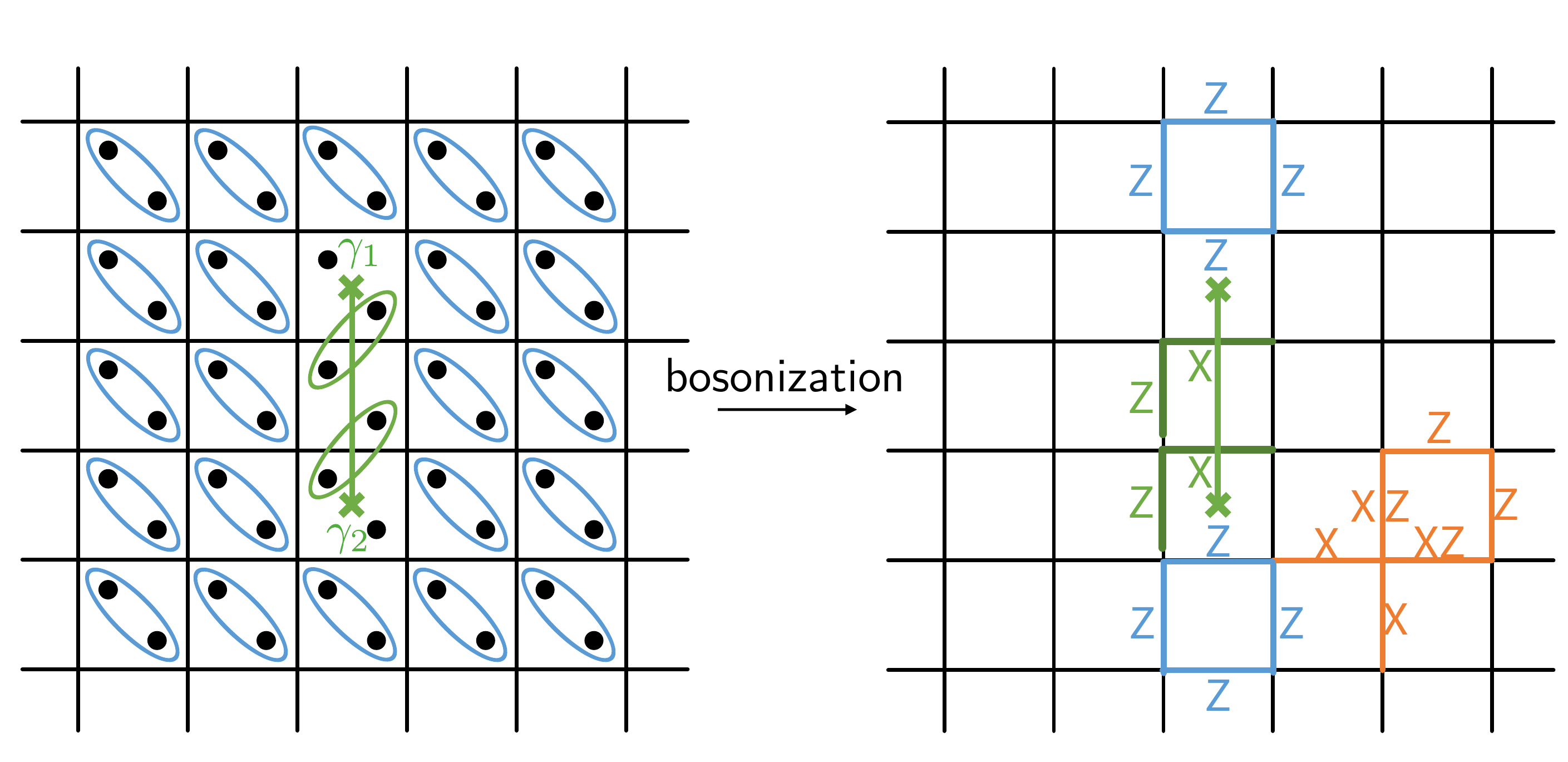}
\caption{Left: The Kitaev chain is decorated on the open green line terminated at the two endpoints corresponding to the point defects (green crosses),  which makes the Majorana fermion paired with the other one on its adjacent face; otherwise, Majorana fermions are paired within each face and form an atomic insulator. Two unpaired Majorana fermion mode $\gamma_1$ and $\gamma_2$ are left on the two point defects respectively.  Right: after bosonization, the pairing term on the Kitaev chain becomes the hopping operator $U_e$ on the open twist string, while the pairing term in each face not on the open twist becomes the 4-body plaquette $Z$-stabilizers. Moreover, the 4-body plaquette $Z$-stabilizers are removed from the two plaquettes where the point defects (green crosses) are located.}
\label{fig:majorana_2d}
\end{figure}

\begin{figure}[t]
    \centering
    \includegraphics[width=1.0\textwidth]{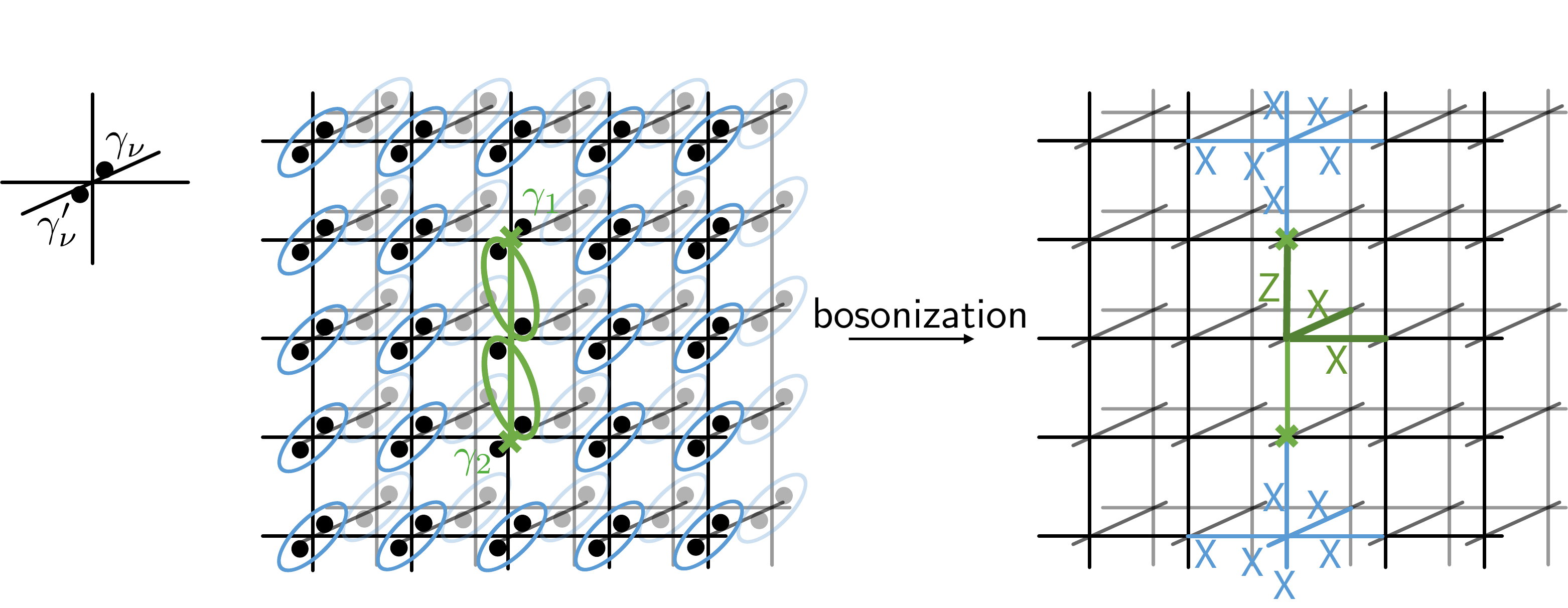}
\caption{Left: The Kitaev chain is decorated on the open green line terminated at the two endpoints corresponding
to the point defects (green crosses), which makes the Majorana fermion paired with the other one on its adjacent vertex; otherwise, Majorana fermions are paired within each vertex. Two unpaired Majorana fermion mode $\gamma_1$ and $\gamma_2$ are left on the two point defects respectively. Right: after bosonization, the pairing term on the Kitaev chain becomes the hopping operator $U_e$ on the open twist string, while the pairing term on each vertex not on the twist string  becomes the 6-body vertex $X$-stabilizers. Moreover, the 6-body vertex $X$-stabilizers are removed from the two vertices where the point defects (green crosses) are located. }
\label{fig:majorana_3d}
\end{figure}

\begin{figure}[t]
    \centering
    \includegraphics[width=0.4\textwidth]{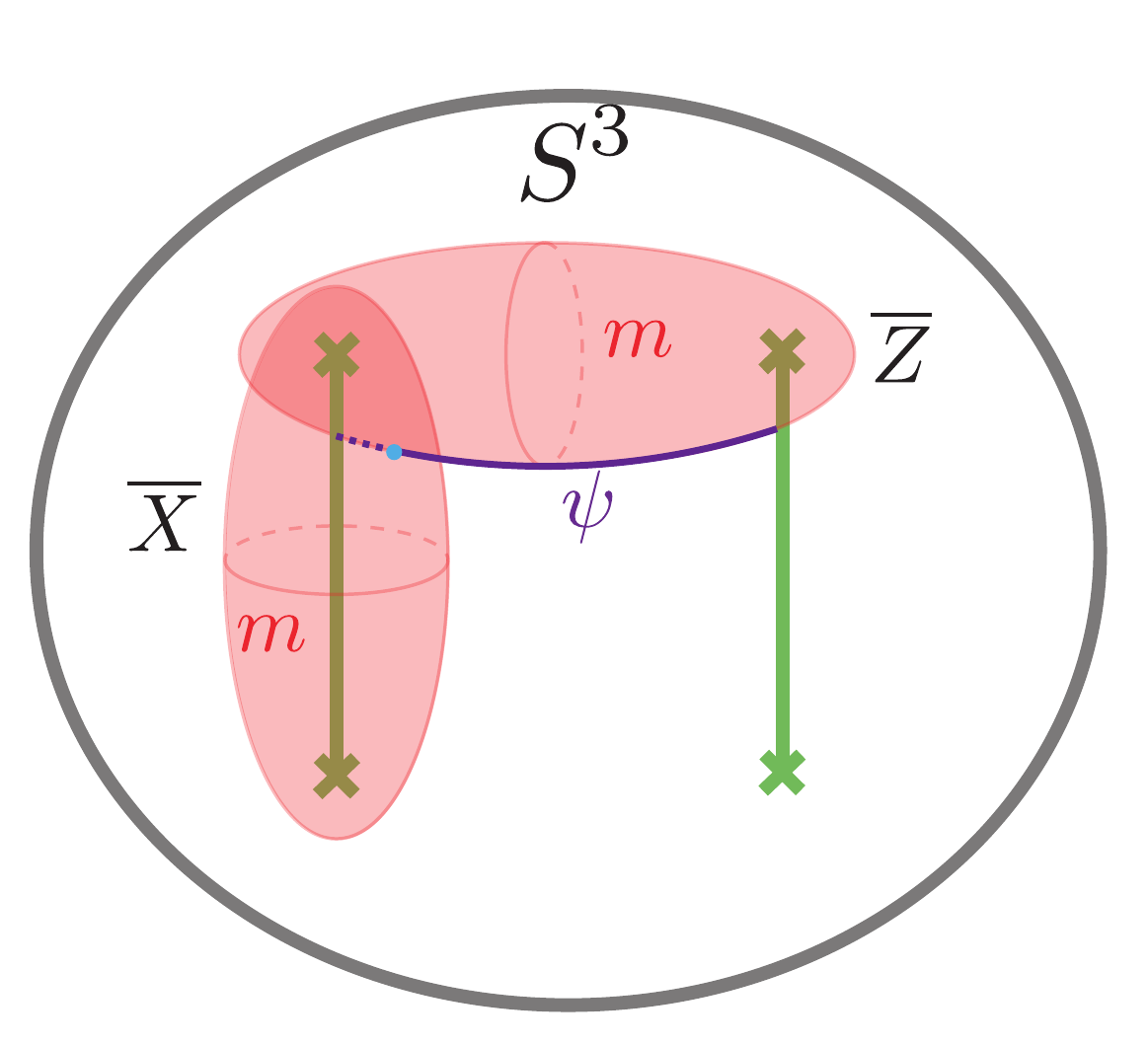}
\caption{The anti-commuting pair of logical operators on a 3-sphere with a pair of open twist strings (4 point defects). The logical-$X$ operator $\overline{X}=\xi_{m}$ is a Wilson sheet operator (red) of flux string $m$ supported on a 2-sphere enclosing an open twist string. The logical-$Z$ operator $\overline{Z}_1=\xi'_{m} \eta_{\psi}$ is a product of Wilson sheet operator $\xi_{m}$ (red) of flux string $m$ supported on a 2-sphere encircling two point defects belonging to the two different twist strings and a worldline operator $\eta_{\psi}$ (blue) of the fermion charge $\psi$.  The pair of open twist strings (four point defects or Majorana modes) encode a single logical qubit.}
\label{fig:logical_fermion}
\end{figure}

We first consider such point defects in (2+1)D toric code, which are sometimes referred to as $e$-$m$ twist defects. A construction of such open twist string and point defects (crosses) is shown in Fig.~\ref{fig:majorana_2d}. There are Majorana hopping operators located on the twist string between the two point defects on the boundaries.  Above and below the point defects, fermion parity operators live in the plaquettes.   For the two plaquettes containing the point defects, no fermion parity term exists, and hence one gets an unpaired Majorana zero mode located on each point defect, denoted by $\gamma_1$ and $\gamma_2$ respectively. After bosonization (right panel of Fig.~\ref{fig:majorana_2d}), we see that all the plaquettes above and below the point defects are just the 4-body $Z$-stabilizers $W_f$, while there are no such 4-body $Z$-stabilizers located on the plaquettes containing the two point defects on the boundary of the twist string.    On the other hand, Majorana hopping operators are still transformed to $U_e$ operators as shown in Fig.~\ref{fig:majorana_2d}.  When considering periodic boundary conditions in both directions, corresponding to the torus topology ($T^2$), the introduction of the twist defects reduces the stabilizer constraint by one, which leads to a 2-fold ground-state degeneracy or equivalently an additional logical qubit.  When considering a sphere topology $S^2$, there is an additional redundancy of the plaquette $Z$-stabilizers, i.e., $\prod_f W_f=1$, before introducing the twist string. The reduction of one stabilizer constraint hence does not increase the degree of freedom in the ground-state subspace, which hence does not contribute to the degeneracy. When introducing two open twist strings and hence two pairs of point defects on the sphere, one obtains a 2-fold degeneracy which encodes a single logical qubit. More generally, when putting $n$ open twist strings and hence $n$ pairs of point defects in a sphere, the ground-state degeneracy becomes $\text{GSD}=2^{n-1}$, which encodes $n-1$ logical qubits.

We remark that similar constructions have appeared in the context of dislocation defects in the Wen-plaquette model~\cite{bombin2010,you2012,KitaevKong12,you2013}. Here, the construction from gauging the Kitaev chain gives a natural way to realize such a defect, and is moreover readily generalizable to (3+1)D, as we now show.

We now investigate the case of the codimension-3 point defects in the (3+1)D toric code with a  fermionic charge. A construction of a codimension-2 open twist string and the corresponding point defects (crosses) on the boundaries is shown in Fig.~\ref{fig:majorana_3d}. Similar to the (2+1)D case, there are Majorana hopping operators located on the twist string between the two point defects on the boundaries.  Above and below the point defects, fermion parity operators live on the vertices.   For the two vertices containing the point defects, no fermion parity term exists, and hence one gets an unpaired Majorana mode located on each point defect, denoted by $\gamma_1$ and $\gamma_2$ respectively.  After bosonization (right panel of Fig.~\ref{fig:majorana_2d}), we see that all the vertices above and below the points defects are just the 6-body $X$-stabilizers $A_\nu$, while there are no such 6-body $X$-stabilizers located on the vertices containing the two point defects.    On the other hand, Majorana hopping operators are still transformed to the 3-body Pauli operators as shown in Fig.~\ref{fig:majorana_3d}.  When considering periodic condition in all directions, i.e., a 3-torus topology ($T^3$), the introduction of the twist defects reduce the number of stabilizer constraints by one, which leads to a 2-fold ground-state degeneracy or equivalently an additional logical qubit.  When considering a 3-sphere topology ($S^3$), there is an additional redundancy of the vertex $X$-stabilizers, i.e., $\prod_\nu A_\nu=1$, before introducing the twist string. The reduction of one stabilizer constraint hence does not contribute to the ground-state degeneracy.  When placing $n$ open twist strings and hence $n$ pairs of point defects on a 3-sphere, the ground-state degeneracy becomes $\text{GSD}=2^{n-1}$, which encodes $n-1$ logical qubits. 

We now discuss the Wilson operator algebra which gives rise to the topological degeneracy in (3+1)D, with the background topology chosen as the 3-sphere $S^3$.  The pair of logical operators on a 3-sphere with a pair of open twist strings are shown in Fig.~\ref{fig:logical_fermion}.  The logical-$X$ operator is   $\overline{X}_1=\xi_{m}$, where $\xi_{m}$ is a spherical Wilson sheet operator of the flux $m$.  The logical-$Z$ operator is $\overline{Z}=\xi'_{m} \eta_{\psi}$, i.e., a product of the spherical Wilson sheet operator $\xi'_{m}$ and a Wilson line operator $\eta_{\psi}$ of the fermion charge $\psi$ trapped between two open twist string. This Wilson line operator is emitted due to the crossing of flux $m$ through the twist string.  The two-fold topological degeneracy comes from the anti-commutation relation $\overline{X} \  \overline{Z} = -\overline{Z} \  \overline{X}$, where the minus sign is due to the intersection between the Wilson membrane $\xi_{m}$ trapped around one twist string and the Wilson line of fermion $\eta_{\psi}$ and hence the anti-commutation relation $\xi_{m}\eta_{\psi}=-\eta_{\psi}\xi_{m}$. 

Finally, we comment on the circuit complexity for creating open twist strings and moving their endpoints.  While a closed twist string along a loop $\gamma = \partial \mathcal{D}$ can always be created by a constant-depth local quantum circuit applied to a region $D$ corresponding to the higher-symmetry operator as explained in  Fig.~\ref{fig:domain_wall_sweeping}, any open twist string in (2+1)D and (3+1)D (including those in the $\ZZ_2 \times \ZZ_2$ and $\ZZ_2$ gauge theories discussed above) with the two endpoints separated by distance $l$ can only be created by a local quantum circuit of depth $O(l)$.  In other words, it takes $O(l)$ time to move one point defect away from the other by distance $l$.  Such a local quantum circuit is essentially composed of local Pachner moves, which re-triangulate or more generally re-cellulate the lattice \cite{Koenig:2010do, Zhu:2018CodeLong}.  We can also see that if one restricts the quantum circuit to act only on the support of a closed twist string, one needs a local quantum circuit with depth $O(l)$ to create a closed twist string with length $l$, which can be done by creating a pair of point defects connected by a short twist string with $O(1)$ length, moving them around $\gamma$, and reannihilating them.

\section{Discussion}

A major focus of this paper has been to understand through various constructions a certain class of invertible codimension-2 defects in (3+1)D topological phases, which we refer to as twist strings. Twist strings have the generic property that when they cross a flux string, a non-trivial electric charge is sourced, and this implies a non-trivial $H^4$ class for the 3-group symmetry of the theory. 

We have shown how twist strings can be understood through a layer construction of (2+1)D topological phases, where they simply descend from the known anyon permuting codimension-1 defects in (2+1)D. We have also seen how twist strings in a (3+1)D discrete $G$ gauge theory can be constructed by decorating a codimension-2 submanifold with a $G$ SPT or invertible topological phase, and then gauging $G$. 
We have provided an example of certain  geometric, not-fully topological twist strings in a (3+1)D non-Abelian gauge theory based on the group $A_6$ by utilizing the layer construction, which cannot be constructed from gauging a (1+1)D SPT. Meanwhile, we conjecture that the topological twist string in $A_6$ gauge theory is constructed from gauging (1+1)D SPT. We conjecture that the layer construction and gauged (1+1)D invertible state constructions for topological twist strings are equivalent, and we provided evidence for this in Sec.~\ref{subsec:sigmamodularinvariant} and~\ref{nonAbelianSPT}.

We note that the gauging perspective is not fully general in (2+1)D: not all anyon-permuting codimension-1 domain walls in (2+1)D arise by gauging lower dimensional bosonic or fermionic invertible phases. A simple example is the $e \leftrightarrow m$ twist string in $\ZZ_3$ toric code. Nevertheless, for describing topological twist strings in (3+1)D, it is possible that the gauging perspective is complete. Additionally, it is an open question whether the Gu-Wen fermionic SPT phases with $G_f=G_b\times\Z_2^f$ symmetry discussed in Sec.~\ref{sec: Twist string of Gf gauge theory} can generate a distinct invertible codimension-1 defect of (2+1)D $G_b\times\Z_2$ gauge theory from those obtained by bosonic SPT phases and Kitaev chains. In the case of $G = \ZZ_2 \times \ZZ_2$ symmetry, the cluster state and the Kitaev chains have already exhausted all automorphisms of $\Z_2\times\Z_2$ gauge theory, so the Gu-Wen SPT (characterized by non-trivial $n_1$) does not give rise to a distinct automorphism.

Further, it is also natural to ask whether all invertible pure codimension-2 topological defects correspond to either flux strings, twist strings, or some composite of the two. Assuming this is indeed the case, we can conjecture a complete characterization of invertible pure codimension-2 topological defects in (3+1)D discrete $G$ gauge theories. Assuming the case of untwisted $G$ gauge theory (i.e. trivial Dijkgraaf-Witten cocycle in $H^4(BG, \U)$), and with only bosonic point charges, we conjecture that invertible pure codimension-2 topological defects are fully characterized by $(g, \omega_2)$. Here $g \in Z(G)$, where $Z(G)$ is the center of $G$, labels the Abelian flux and $\omega_2$ labels the (1+1)D SPT decoration. That is, $K_1 = Z(G) \times \mathcal{H}^2(G, \U)$ in general. In the case where we allow the $G$ gauge theory to have fermionic charges, we expect that $\omega_2$ should be replaced by the data of a (1+1)D fermionic invertible phase, as discussed in Sec.~\ref{sec:fermiongauge}. 

Finally, we note that analogs of the codimension-2 twist string can also be found in continuous gauge theory, e.g., (3+1)D $\mathrm{PSU}(N)$ gauge theory, where it is identified as the magnetic $\Z_N$ 1-form symmetry of $\mathrm{PSU}(N)$ gauge theory. That is, $\mathrm{PSU}(N)$ gauge theory is obtained by gauging a center $\Z_N$ 1-form symmetry of the $\mathrm{SU}(N)$ gauge theory~\cite{Aharony2013reading}, and then the $\mathrm{PSU}(N)$ gauge theory has the dual symmetry of the center 1-form symmetry, generated by the Wilson surface of the dynamical $\Z_N$ 2-form gauge field of the center symmetry. This dual $\Z_N$ 1-form symmetry is called a magnetic 1-form symmetry. The 2-form gauge field for the Wilson surface is identified as the second Stiefel-Whitney class $w_2\in H^2(B\mathrm{PSU}(N),\Z_N)$ of the $\mathrm{PSU}(N)$ gauge field, and one can also regard $w_2$ as a response action of a (1+1)D bosonic SPT phase with $\mathrm{PSU}(N)$ symmetry. Therefore, the magnetic 1-form symmetry can also be understood as analogous to the twist string, obtained by a decoration of (1+1)D SPT phase on the codimension-2 defect. It would be interesting to see if the above perspective allows us to explicitly construct a symmetry defect of the magnetic 1-form symmetry in the lattice $\mathrm{PSU}(N)$ gauge theory~\cite{Sulejmanpasic}, e.g., by decorating the defect with a (1+1)D spin chain.

We now have a more complete understanding of invertible codimension-3 and codimension-2 topological defects, their interplay with each other, and how this defines a 3-group. The next step is to develop a comprehensive understanding of invertible codimension-1 defects and their interplay with invertible codimension-2, 3 defects, in order to develop a comprehensive understanding of the full categorical 3-group symmetry of (3+1)D topological phases. Eventually we hope to understand more systematically the full fusion 3-category of both invertible and non-invertible defects and higher symmetries in generic (3+1)D topological phases of matter. 
\section{Acknowledgments}
We thank David Aasen, Yichul Choi, Sahand Seifnashri, Shu-Heng Shao, and Dominic Williamson for discussions.
MB thanks David Penneys and Corey Jones for ongoing discussions and explanations of higher category theory. YC and NT thank Tyler Ellison for discussions of the $e\leftrightarrow m$ twist string in the (2+1)D toric code. YC also thanks Po-Shen Hsin and Anton Kapustin for discussions of defects and 3-groups.

MB acknowledges financial support from NSF CAREER DMR- 1753240. YC, SH, and RK are supported by the JQI postdoctoral fellowship at the University of Maryland. NT is supported by the Walter Burke Institute
for Theoretical Physics at Caltech. GZ is supported by the U.S. Department of Energy, Office of Science, National Quantum Information Science Research Centers, Co-design Center for Quantum Advantage (C2QA) under contract number DE-SC0012704.

\appendix

\section{Gauging fermion parity on lattices} \label{sec: gauging fermion parity}
\subsection{2d bosonization} \label{sec: review of 2d bosonization}

\begin{figure}
\centering
\resizebox{5.5cm}{!}{%
\begin{tikzpicture}
\draw[thick] (-3,0) -- (3,0);\draw[thick] (-3,-2) -- (3,-2);\draw[thick] (-3,2) -- (3,2);
\draw[thick] (0,-3) -- (0,3);\draw[thick] (-2,-3) -- (-2,3);\draw[thick] (2,-3) -- (2,3);
\draw[->] [thick](0,0) -- (1,0);\draw[->][thick] (0,2) -- (1,2);\draw[->][thick] (0,-2) -- (1,-2);
\draw[->][thick] (0,0) -- (0,1);\draw[->][thick] (2,0) -- (2,1);\draw[->][thick](-2,0) -- (-2,1);
\draw[->][thick] (-2,0) -- (-1,0);\draw[->][thick] (-2,2) -- (-1,2);\draw[->][thick](-2,-2) -- (-1,-2);
\draw[->][thick] (-2,-2) -- (-2,-1);\draw[->][thick] (0,-2) -- (0,-1);\draw[->] [thick](2,-2) -- (2,-1);
\filldraw [black] (-2,-2) circle (1.5pt) node[anchor=north east] {1};
\filldraw [black] (0,-2) circle (1.5pt) node[anchor=north east] {2};
\filldraw [black] (2,-2) circle (1.5pt) node[anchor=north east] {3};
\filldraw [black] (-2,-0) circle (1.5pt) node[anchor=north east] {4};
\filldraw [black] (0,0) circle (1.5pt) node[anchor=north east] {5};
\filldraw [black] (2,0) circle (1.5pt) node[anchor=north east] {6};
\filldraw [black] (-2,2) circle (1.5pt) node[anchor=north east] {7};
\filldraw [black] (0,2) circle (1.5pt) node[anchor=north east] {8};
\filldraw [black] (2,2) circle (1.5pt) node[anchor=north east] {9};
\draw (-1,-1) node{\large a};
\draw (1,-1) node{\large b};
\draw (1,1) node{\large c};
\draw (-1,1) node{\large d};
\end{tikzpicture}
}
\caption{For each face, there are two Majorana fermions $\g_f$, $\g_f^\prime$. For each edge, there is a qubit, corresponding to the Pauli matrices $X_e$, $Y_e$, $Z_e$.}
\label{fig:square}
\end{figure}
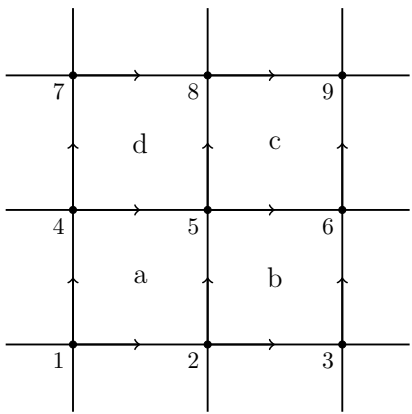

We begin by reviewing 2d bosonization on a square lattice following \cite{CKR18}. The elements of vertices, edges, and faces are denoted $v,e,f$. On each face $f$ of the lattice we place a single pair of fermionic creation-annihilation operators $c_f,c_f^\dagger$, or equivalently a pair of Majorana fermions $\gamma_f,\gamma'_f$. The even fermionic algebra consists of local observables with a trivial fermionic parity 
(i.e. those local observables which commute with the total fermion parity $(-1)^F \equiv \prod_f (-1)^{c^\dagger_f c_f}$).
The even algebra is generated by
\begin{equation}
    P_f=-i\gamma_f\gamma'_f
\end{equation}
and
\begin{equation}
    S_e=i\gamma_{L(e)}\gamma'_{R(e)}
\end{equation}
where $L(e)$ and $R(e)$ are faces to the left and right of $e$, with respect to some orientation of $e$.
Note that the even fermionic algebra (which is the same as the algebra of local observables containing an even number of Majorana operators)  is generated by $P_f$ and $S_e$. That is, any such local observable can be written as a linear combination of products of operators $P_f$ and $S_e$. In particular, this applies to any fermion bilinear involving two Majorana fermions on any two faces, and to interaction terms in an arbitrary finite-range Hamiltonian.

The bosonic dual of this system involves $\ZZ_2$-valued spins on the edges of the square lattice. For every edge $e$ we define a unitary operator $U_e$ which squares to $1$. Labeling the faces and vertices as in Fig.~\ref{fig:square}, we  define:
\begin{equation}
\begin{split}
U_{56} &= X_{56} Z_{25} \\
U_{58} &= X_{58} Z_{45} 
\end{split}
\label{eq: Ue definition}
\end{equation}
{ where $X$, $Z$ are Pauli matrices acting on a spin at each edge:
\begin{equation}
X_e=
\begin{bmatrix} 
0 & 1 \\
1 & 0 
\end{bmatrix},
~Z_e=
\begin{bmatrix} 
1 & 0 \\
0 & -1 
\end{bmatrix}.
\end{equation}
}
Operators $U_e$ for other edges are defined by using translation symmetry.  Pictorially, the operator $U_e$ is drawn as
\begin{equation}
    U_e=\begin{gathered}
    \xymatrix@=1cm{&{}\ar@{-}[d]|{\displaystyle X_e}\\
    {}\ar@{-}[r]|{\displaystyle Z}&}
    \end{gathered} \qquad \text{or} \qquad \begin{gathered}
    \xymatrix@=1cm{{}\ar@{-}[r]|{\displaystyle X_e}&\\
    {}\ar@{-}[u]|{\displaystyle Z}&}
    \end{gathered},
\label{eq: original Ue}
\end{equation}
corresponding to the vertical or horizontal edge $e$.
It has been shown in Ref.~\cite{CKR18}  that $U_e$ and $S_e$ satisfy the same commutation relations. We also identify fermionic parity $P_f$ at each face with the ``flux operator'' $W_f \equiv \prod_{e \subset f} Z_e$:
\begin{equation}
    W_f=\begin{gathered}
   \xymatrix@=1cm{%
    {}\ar@{-}[r]|{\displaystyle Z}\ar@{}[dr]|{\mathlarger f} & {}\ar@{-}[d]|{\displaystyle Z} \\
    {}\ar@{-}[u]|{\displaystyle Z} & {}\ar@{-}[l]|{\displaystyle Z}}
    \end{gathered}.
\label{eq: original Wf}
\end{equation}
The bosonization map is
\begin{equation}
\begin{split}
P_f = -i \g_f \g^\prime_f &\longleftrightarrow W_f \\
 S_e =  i \g_{L(e)} \g^\prime_{R(e)} &\longleftrightarrow U_e,
\end{split}
\label{eq: 2d bosonization map}
\end{equation}
or pictorially 
\begin{align}
    i\times
    \begin{gathered}
    \xymatrix@=1.2cm{
    &\\
    {}\ar@{}[ur]|{\displaystyle \gamma_{L(e)}} \ar@{}[dr]|{\displaystyle \gamma'_{R(e)}} \ar@{-}[r]^{ \mathlarger e} &  \\
    &
    }
    \end{gathered}
    &\begin{gathered}\xymatrix{
    {}\ar@{<->}[r] &{}}\end{gathered}
    \hspace{0.3cm}\begin{gathered}
    \xymatrix@=1cm{%
    {}\ar@{-}[r]|{\displaystyle X_e} &{} \\{}\ar@{-}[u]|{\displaystyle Z}& {}}
    \end{gathered} \quad ,
    \label{eq: Ue on horizontal edge}
    \\[-15pt]
    i\times
    \begin{gathered}
    \xymatrix@=1.2cm{
    &{}\ar@{-}[d]^{\mathlarger e}&\\
    \ar@{}[ur]|{\displaystyle \gamma_{L(e)}}&\ar@{}[ur]|{\quad \displaystyle \gamma'_{R(e)}}&}
    \end{gathered}
    &\begin{gathered}\xymatrix{
    {}\ar@{<->}[r] &{}}\end{gathered}
    \begin{gathered}
    \xymatrix@=1cm{%
    &{}\ar@{-}[d]|{\displaystyle X_e} \\{}\ar@{-}[r]|{\displaystyle Z}&
    }
    \end{gathered} \quad ,
    \\[10pt]
    -i\gamma_f\gamma_f'
    &\begin{gathered}\xymatrix{
    {}\ar@{<->}[r] &{}}\end{gathered}
    \hspace{0.3cm}
    \begin{gathered}
    \xymatrix@=1cm{%
    {}\ar@{-}[r]|{\displaystyle Z}\ar@{}[dr]|{\mathlarger f} & {}\ar@{-}[d]|{\displaystyle Z} \\
    {}\ar@{-}[u]|{\displaystyle Z} & {}\ar@{-}[l]|{\displaystyle Z}}
    \end{gathered} \quad .
\end{align}
The condition $ P_a P_c S_{{58}} S_{{56}} S_{{25}} S_{{45}} = 1$ on fermionic operators gives gauge constraints (stabilizer) $G_v=W_{f_c} \prod_{e \supset v_5} X_e  =1$ for bosonic operators, or generally for each vertex
\begin{equation}
    G_v =
    \begin{gathered}
   \xymatrix@=1cm{%
    &{}\ar@{-}[r]|{\displaystyle Z}\ar@{}[dr]|{\mathlarger f}& {}\ar@{-}[d]|{\displaystyle Z}\\
    {}\ar@{-}[r]|{\displaystyle X}&{v}\ar@{-}[u]|{\displaystyle xy} & {}\ar@{-}[l]|{\displaystyle XZ}\\
    &{}\ar@{-}[u]|{\displaystyle X}&}
\end{gathered}
= 1.
\label{eq:gauge constraint at vertex}
\end{equation}

\subsection{3d bosonization} \label{sec: review of 3d bosonization}

In the section, we review 3d bosonization in Ref.~\cite{CK19} (with a slightly different convention), which is a direct generalization of the 2d bosonization. This 3d bosonization originated from the Levin-Wen rotor model \cite{LW06}.
On the cubic lattice, we have two Majorana fermions at each vertex $\g_v, \g^\prime_v$ and one qubit at each edge $X_e, Y_e, Z_e$. The bosonization map can be summarized in Fig.~\ref{fig: 3d bosonization}, which will be explained in detail.

\begin{figure}[htb]
\centering
\includegraphics[width=0.4\textwidth]{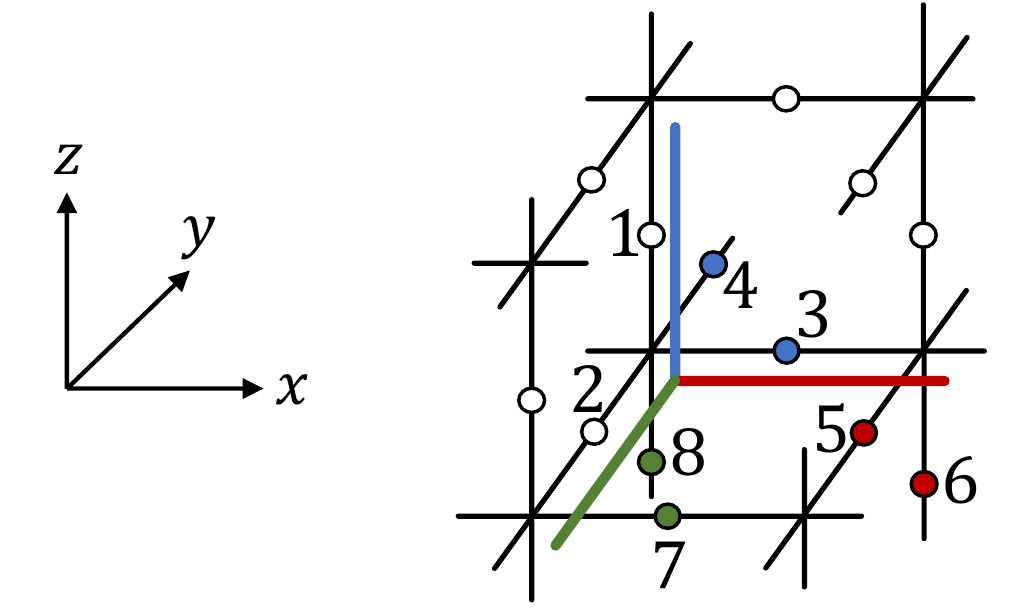}
\caption{For edges in the cubic lattice, the ``framing'' is defined by green, red, and blue edges, which is a small shift of the edges \cite{LW06}. Given an edge $e$, the hopping operator $U_e$ is defined as $Z_e$ times $X_{e'}$ for those $e'$ which intersect the framing of $e$ when projected to the plane (i.e. $U_{e_1}=Z_1 X_3 X_4,$ $U_{e_2}=Z_2 X_7 X_8,$ and $U_{e_3}=Z_3 X_5 X_6$  ).
\label{fig:cube1}}
\end{figure}

The on-site fermion parity operator on a vertex $v$ is $P_v = - i \gamma_v \gamma'_v$. It is a ``$\ZZ_2$ operator'' (i.e.\ it squares to $1$). All operators $P_v$ commute with each other. 
The even fermionic algebra is generated by the on-site fermion parity $P_v$ and the Majorana hopping operator $S_e = i \g_i \g^\prime_j$ along the edge $e=\lr{ij}$.\footnote{The direction of $e=\lr{ij}$ is assigned to be along $+x$, $+y$, and $+z$ directions (See Figs.~\ref{fig: 3d bosonization}~and~\ref{fig:cube1}).}
To illustrate the definition of these operators, in Fig.~\ref{fig:cube1}, fermions live on vertices and the orientation of each edge are taken to be along $+x$, $+y$, and $+z$ directions. The Majorana hopping operator is defined by $S_e= i \g_{L(e)} \g^\prime_{R(e)}$ where $L(e)$ and $R(e)$ are starting and ending points of the edge $e$ in the cubic lattice. $S_{e}$ and $S_{e'}$ anti-commute only when edges $e$ and $e'$ both start from the same point or both end at the same point.

\begin{figure}[htb]
\centering
\resizebox{8 cm}{!}{%
\includegraphics[width=0.8\textwidth]{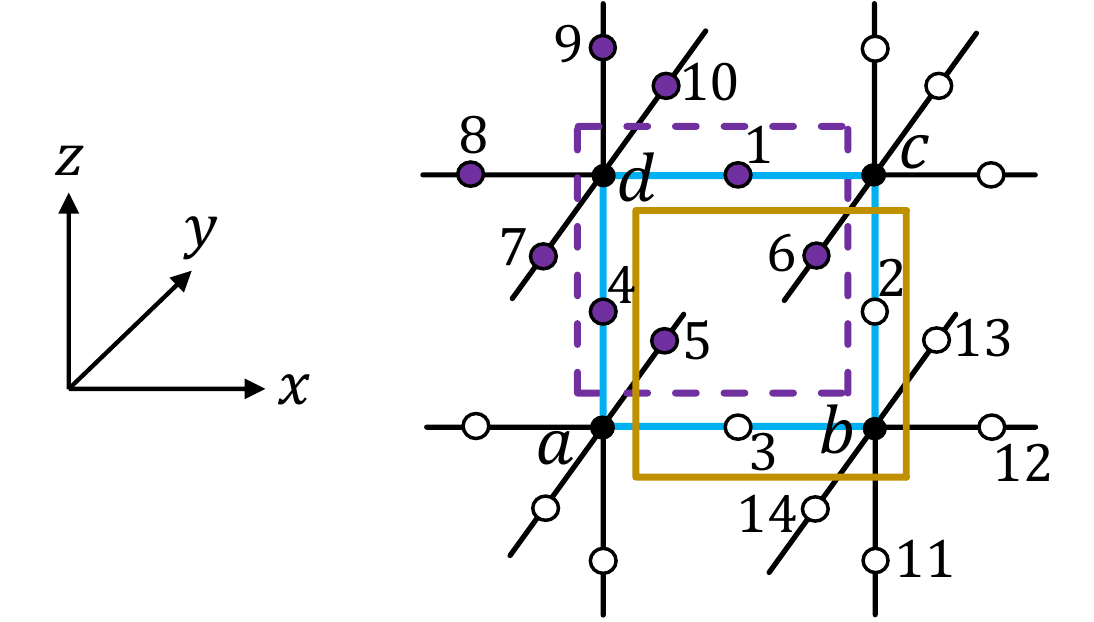}
}
\caption{The framing of the hopping term defined previously is indicated by the brown square, while the gauge constraint involves the $X$ operators in the opposite framing (purple dashed square).}
\label{fig:cube2}
\end{figure}

The dual bosonic system has qubits living on the edges of the cubic lattice. To define hopping operators $U_e$, we need to choose framing for each edge, i.e. a small shift of each edge along some orthogonal direction. We also assume that when projected on some generic plane (such as the plane of the page), a shifted edge intersects all edges transversally. For example, in Fig.~\ref{fig:cube1} such a framing is indicated by red, green and blue lines (for edges along $x$, $y$ and $z$ directions, respectively), and the shift of the edge $1$ intersects edges $3$ and $4$.\footnote{There are many choices of framing, and accordingly many versions of the bosonization map. By construction, they are related by automorphisms of the algebra of observables.} Now, we define $U_e$ as a product of $Z_e$ with all $X_{e^\prime}$ such that $e^\prime$ intersects the framing of $e$ when projected to the plane of the page. For example, the hopping operator for the edge $1$ is  $U_1=Z_1 X_3 X_4$. Notice that $U_1$, $U_3$, and $U_4$ anti-commute with each other and $U_3$, $U_5$, and $U_6$ anti-commute with each other, while $U_2$ and $U_3$ commute, and $U_1$ and $U_8$ commute. 
One can check that $S_f$ and $U_f$ have the same commutation relations. Therefore, the bosonization map in 3D can be defined as follows:
\begin{enumerate}
    \item
    For any vertex $v$, define the star term $A_v \equiv \prod_{e \supset v} X_e$. We identify the fermionic states $ |P_v = 1 \rangle$ and $ |P_v=-1 \rangle$  with bosonic states for which $A_v=1$ and $A_v=-1$, respectively. Thus
    \begin{equation}
    P_v= -i \g_v \g^\prime_v \longleftrightarrow A_v.
     \label{eq:bosonization assumption 1}
    \end{equation}
    \item
    The fermionic hopping operator $S_e$ is identified with $U_e$ defined above:
    \begin{equation}
     S_e =  i \g_{L(f)} \g^\prime_{R(f)} \longleftrightarrow U_e.
    \label{eq:bosonization assumption 2}
    \end{equation}
\end{enumerate}
As in the 2d bosonization, the bosonic operators satisfy some constraints. In Fig.~\ref{fig:cube2}, we calculate the product of $S_e$ around the a plaquette along with $P_v$ at vertices $b$ and $d$ on the lattice:
\begin{equation}
\begin{split}
 &-S_{e_1} S_{e_2} S_{e_3} S_{e_4} P_b P_d\\
 =& -(i \g_d \g^\prime_c)  (i \g_b \g^\prime_c)  (i \g_a \g^\prime_b)  (i \g_a \g^\prime_d) (-i \g_b \g^\prime_b) (-i \g_d \g^\prime_d) = 1.
\end{split}
\label{eq: S delta cube}
\end{equation}
Its bosonic dual defined by Eqs.~\eqref{eq:bosonization assumption 1}~and~\eqref{eq:bosonization assumption 2} is the product of the corresponding operators $U_f$ and $A_v$, which needs to be imposed as a gauge constraint:
\begin{equation}
\begin{split}
 1=& -U_{e_1} U_{e_2} U_{e_3} U_{e_4} A_b A_d\\
 =& -(Z_1 X_2 X_6) (Z_2 X_{12} X_{13}) (Z_3 X_{11} X_{14}) (Z_4 X_3 X_5) (X_2 X_3 X_{11} X_{12} X_{13} X_{14}) (X_1 X_4 X_7 X_8 X_9 X_{10})\\
=&  Z_1 Z_2 Z_3 Z_4 X_1 X_4 X_5 X_6  X_7 X_8 X_9 X_{10}.
\end{split}
\label{eq: U delta cube}
\end{equation}
The operators $X$'s are the edges crossed by the dashed square in Fig.~\ref{fig:cube2}. The framing for gauge constraints is opposite to the framing used to define hopping operators. We have a gauge constraint for each face of the lattice. The gauge constraints in other two directions are shown in Fig.~\ref{fig: 3d bosonization}. These constraints commute and thus define a $\ZZ_2$ 2-form gauge theory with an unusual Gauss law.

\section{Slant product for non-Abelian groups}
\label{app:slant}
Here we describe the slant product $i_g$ for group cohomology $\omega\in\mathcal{H}^2(G,\U)$ with non-Abelian group $G$. 
The slant product (corresponding to the element $g \in G$) of the 2-cocycle $\omega$ acting on a group element $h \in G$ is defined by \cite{WangWen2015nonAbelian}
\begin{eqs}
    i_g\omega(h) :=  \frac{\omega(g,h)}{\omega(h, h^{-1} g h)} \in\mathcal{H}^1(G,\U).
\end{eqs}
In this paper, we focus on the case that $h \in C_g$, i.e., the element $h$ is in the centralizer of the element $g$. In this case, the slant product $i_g\omega$ corresponds to the torus partition function of the (1+1)D SPT phase $\omega\in\mathcal{H}^2(G,\U)$, in the presence of holonomy $g\in G$ in one cycle of $T^2$:
\begin{align}
    Z_{\mathrm{SPT}}(T^2_{(g,h)})=i_g\omega(h) = \frac{\omega(g,h)}{\omega(h, g)},
\label{eq: definition of slant product}
\end{align}
where $T^2_{(g,h)}$ denotes a torus $T^2$ with the holonomy $(g,h)$ on each fundamental cycle, and we have used the condition $h\in C_g$ to assure that the holonomy $(g,h)$ defines a flat background gauge field on $T^2$.

We would like to show that 
\begin{align}
    i_g\omega(h) = i_{kgk^{-1}}\omega(k h k^{-1})
    \quad \text{for $k\in G$ and $h \in C_g$.}
\label{eq: two slant products}
\end{align}
The above equation implies that the charge attached to the flux after moving past a twist string only depends on the conjugacy class $[g]$.\footnote{For every $h \in C_g$, the centralizer of $g$, there exists an element $k h k^{-1} \in C_{k g k^{-1}}$, the centralizer of $k g k^{-1}$. Therefore, $C_{g}$ is isomorphic to $C_{k g k^{-1}}$. Eq.~\eqref{eq: two slant products} indicates that two irreps $i_g \omega$ and $i_{k g k^{-1}} \omega$ of $C_{g}$ and $C_{k g k^{-1}}$ are identical, corresponding to the same charge on the flux $[g]$.}

\begin{figure}[htb]
\centering
\resizebox{5cm}{!}{%
\includegraphics[width=0.8\textwidth]{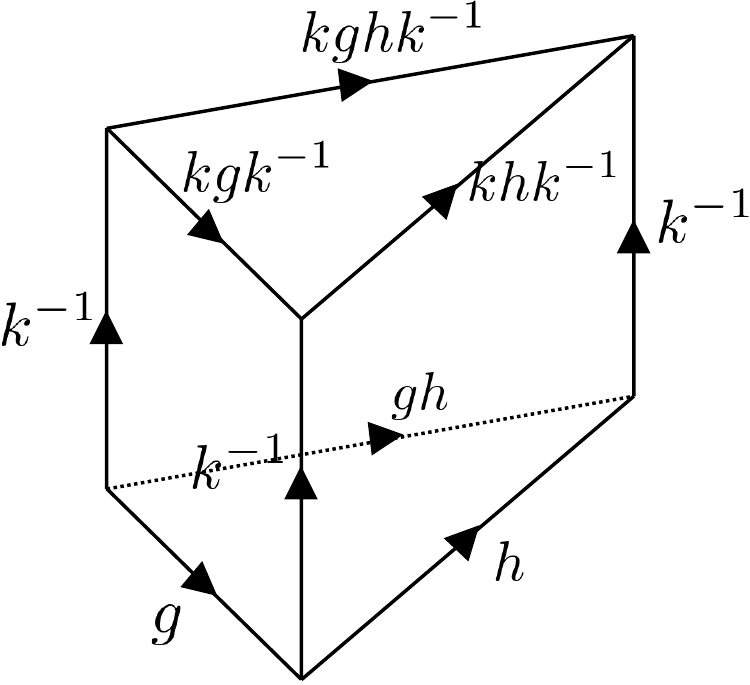}
}
\caption{The prism interpolating between two gauge equivalent 2-simplices.}
\label{fig:prism}
\end{figure}

To show this, let us first check that the cohomology class $[\omega]$ is invariant under the conjugation action. That is,
\begin{align}
    \omega(g,h) = \omega(kgk^{-1},khk^{-1})\cdot \delta\xi_k(g,h) \quad \text{for some $\xi_k \in C^1(G,\U)$}
    \label{eq:gaugetranscocycle}
\end{align}
This can be checked by considering a prism interpolating a pair of 2-simplices with the holonomy $(g,h)$ and $(kgk^{-1}, khk^{-1})$ respectively, see Fig.~\ref{fig:prism}.
Let us define $\xi_k(g):= \frac{\omega(g,k^{-1})}{\omega(k^{-1}, kgk^{-1})}$, which is regarded as a Boltzmann weight of the SPT phase on one rectangular face of the prism. Since one can regard the prism as a specific triangulation of a sphere $S^2$, we have\footnote{Eq.~\eqref{eq: Z on sphere equals 1} can also be derived from the cocycle condition of $\omega$.}
\begin{align}
    Z_{\mathrm{SPT}}(S^2) = \frac{\omega(kgk^{-1},khk^{-1})}{\omega(g,h)}\frac{\xi_k(g)\xi_k(h)}{\xi_k(gh)} =1.
\label{eq: Z on sphere equals 1}
\end{align}
which implies Eq.~\eqref{eq:gaugetranscocycle}. Applying this to the definition of the slant product, we find
\begin{align}
    i_g\omega(h) = i_{kgk^{-1}}\omega(khk^{-1})
    \quad \text{for $k\in G$},
\end{align}
which gives the desired result.

\section{Linear algebraic description of automorphisms} \label{sec: Z2 x Z2 automorphism}

Here we describe automorphisms of (2+1)D $\Z_2\times\Z_2$ gauge theory based on the $K$-matrix representation of the gauge theory. This gives a linear algebraic formulation of the automorphism which is useful for classifying them. We use this classification to show that all automorphisms of (2+1)D $\Z_2 \times \Z_2$ gauge theory can be implemented by twist strings arising from decorating $\Z_2 \times \Z_2$ SPTs and Kitaev chains. 

The $K$-matrix of the (2+1)D $\Z_2$ gauge theory without Dijkgraaf-Witten twist is given by
\begin{align}
    K= \begin{pmatrix}
    0 \ 0 \ 2 \ 0 \\
    0 \ 0 \ 0 \ 2 \\
    2 \ 0 \ 0 \ 0 \\
    0 \ 2 \ 0 \ 0 \\
    \end{pmatrix}
\end{align}
Each $\Z_2$-valued vector corresponds to an anyon generated by
\begin{align}
    e_1 = \begin{pmatrix}  1 \\ 0 \\ 0 \\ 0 \end{pmatrix} \quad 
    e_2 = \begin{pmatrix}  0 \\ 1 \\ 0 \\ 0 \end{pmatrix} \quad 
    m_1 = \begin{pmatrix}  0 \\ 0 \\ 1 \\ 0 \end{pmatrix} \quad 
    m_2 = \begin{pmatrix}  0 \\ 0 \\ 0 \\ 1 \end{pmatrix} 
\end{align}
For vectors $a,b$ that correspond to anyons, the mutual braiding and twist is expressed as
\begin{align}
    M_{a,b} = e^{2\pi i a^T K^{-1} b}.
\end{align}
\begin{align}
    \theta_a = e^{\pi i a^T K^{-1} a}.
\end{align}
Then, an automorphism is formulated as a linear transformation of integral vectors $a$ that correspond to anyons, $a\to A a$. $A$ is a $4\times 4$ $\Z_2$-valued matrix that preserves mutual braiding and topological twist. 
We are going to show that the automorphism of these anyons are generated by the following three matrices
\begin{align}
    H_1= \begin{pmatrix}
    0 \ 0 \ 1 \ 0 \\
    0 \ 1 \ 0 \ 0 \\
    1 \ 0 \ 0 \ 0 \\
    0 \ 0 \ 0 \ 1 \\
    \end{pmatrix},  \ 
    H_2=\begin{pmatrix}
    1 \ 0 \ 0 \ 0 \\
    0 \ 0 \ 0 \ 1 \\
    0 \ 0 \ 1 \ 0 \\
    0 \ 1 \ 0 \ 0 \\
    \end{pmatrix},  \ 
    CZ=\begin{pmatrix}
    1 \ 0 \ 0 \ 1 \\
    0 \ 1 \ 1 \ 0 \\
    0 \ 0 \ 1 \ 0 \\
    0 \ 0 \ 0 \ 1 \\
    \end{pmatrix}.
\end{align}
First, we define some useful matrices
\begin{align}
    CZ^T &\equiv H_1 \cdot H_2 \cdot CZ \cdot H_2 \cdot H_1 = \begin{pmatrix}
    1 \ 0 \ 0 \ 0 \\
    0 \ 1 \ 0 \ 0 \\
    0 \ 1 \ 1 \ 0 \\
    1 \ 0 \ 0 \ 1 \\
    \end{pmatrix}, \\
    \text{SWAP} &\equiv CZ \cdot H_1 \cdot H_2 \cdot CZ \cdot H_1 \cdot H_2 \cdot CZ \cdot H_1 \cdot H_2 = \begin{pmatrix}
    0 \ 1 \ 0 \ 0 \\
    1 \ 0 \ 0 \ 0 \\
    0 \ 0 \ 0 \ 1 \\
    0 \ 0 \ 1 \ 0 \\
    \end{pmatrix}, \\
    \text{Col}_{12}&\equiv H_2 \cdot CZ \cdot H_2 =
    \begin{pmatrix}
    1 \ 1 \ 0 \ 0 \\
    0 \ 1 \ 0 \ 0 \\
    0 \ 0 \ 1 \ 0 \\
    0 \ 0 \ 1 \ 1 \\
    \end{pmatrix}, \\
    \text{Row}_{12}&\equiv H_2 \cdot H_1 \cdot H_2 \cdot CZ \cdot H_2 \cdot H_1 \cdot H_2=
    \begin{pmatrix}
    1 \ 0 \ 0 \ 0 \\
    1 \ 1 \ 0 \ 0 \\
    0 \ 0 \ 1 \ 1 \\
    0 \ 0 \ 0 \ 1 \\
    \end{pmatrix}. 
\end{align}
All matrices above are order two.

Given an automorphism $A$, the first column must contain at least one 1. We can apply $H_1$, $H_2$, and $\text{SWAP}$ gates to make its first entry 1:
\begin{eqs}
    A \sim
    \left(
    \begin{array}{cc|cc}
    1 & * & * & * \\
    * & * & * & * \\
    \hline
    * & * & * & * \\
    * & * & * & * \\
    \end{array}
    \right),
\end{eqs}
where $\sim$ represents that the matrices are equal up to multiplying $H_1$, $H_2$, and $CZ$ on its left and right sides. Next, we apply $CZ$, $CZ^T$, $\text{Col}_{12}$ and $\text{Row}_{12}$ on its left or right side to get
\begin{eqs}
    A \sim
    \left(
    \begin{array}{cc|cc}
    1 & 0 & * & 0 \\
    0 & * & * & * \\
    \hline
    * & * & * & * \\
    0 & * & * & * \\
    \end{array}
    \right).
\end{eqs}
Since $A e_1$ is a boson, the $*$ in the first column must be $0$. By the braiding property of $A e_1$ with $A e_2$, $A m_1$, and $A m_2$, the third row is $0$ $0$ $1$ $0$:
\begin{eqs}
    A \sim
    \left(
    \begin{array}{cc|cc}
    1 & 0 & * & 0 \\
    0 & * & * & * \\
    \hline
    0 & 0 & 1 & 0 \\
    0 & * & * & * \\
    \end{array}
    \right).
\end{eqs}
The entries $A_{22}$ and $A_{24}$ contains at least one 1 (otherwise the second and third rows do not have the full rank), so we can apply $H_2$ to make $A_{22}=1$. Since $A e_2$ is a boson, $A_{42}$ must be 0:
\begin{eqs}
    A \sim
    \left(
    \begin{array}{cc|cc}
    1 & 0 & * & 0 \\
    0 & 1 & * & * \\
    \hline
    0 & 0 & 1 & 0 \\
    0 & 0 & * & * \\
    \end{array}
    \right).
\end{eqs}
Next, by the braiding property of $A e_2$ with $A m_1$ and $A m_2$, the last row is $0$ $0$ $0$ $1$:
\begin{eqs}
    A \sim
    \left(
    \begin{array}{cc|cc}
    1 & 0 & * & 0 \\
    0 & 1 & * & * \\
    \hline
    0 & 0 & 1 & 0 \\
    0 & 0 & 0 & 1 \\
    \end{array}
    \right).
\end{eqs}
$A m_1$ being a boson implies $A_{13}=0$. Similarly, $A m_2$ being a boson gives $A_{24}=0$:
\begin{eqs}
    A \sim
    \left(
    \begin{array}{cc|cc}
    1 & 0 & 0 & 0 \\
    0 & 1 & * & 0 \\
    \hline
    0 & 0 & 1 & 0 \\
    0 & 0 & 0 & 1 \\
    \end{array}
    \right).
\end{eqs}
By the braiding property of $A m_2$, $A_{23}$ is 0 and we show that $A$ is the identity matrix by multiplying $H_1$, $H_2$, and $CZ$ on its left and right sides.

\section{Cheshire string in (3+1)D $\Z_2$ toric code}
\label{app:cheshire}
Here we describe Cheshire string in the (3+1)D $\Z_2$ toric code with a bosonic electric particle. The Cheshire string is a non-invertible codimension-2 defect of the (3+1)D toric code, which is distinct from $m$ magnetic surface operator. 
Roughly speaking, the Cheshire string is obtained by condensing $e$ particle on a region restricted to the codimension-2 surface. 
Here, we construct the Cheshire string in the (3+1)D $\Z_2$ toric code by gauging the (3+1)D bosonic trivial invertible phase with $\Z_2$ 0-form symmetry, with a decoration of the 
$\Z_2$ spontaneously-symmetry-broken (SSB) Ising theory on the (1+1)D defect.
We show that the $\Z_2$ SSB phase becomes the non-invertible Cheshire string in (3+1)D $\Z_2$ gauge theory after gauging the $\Z_2$ symmetry.
We provide an exactly solvable model of the (3+1)D toric code with the Cheshire string in Sec.~\ref{subsec:cheshireexact}, and then give the field theoretical description of the Cheshire string in Sec.~\ref{subsec:cheshirecondense} in terms of the condensation defect.
We note that the Cheshire string in the lattice model of the $\Z_2$ toric code is also described in Refs.~\cite{else2017, Kong2020defects}. 

\subsection{Exactly solvable model for the Cheshire string in (3+1)D $\Z_2$ toric code}
\label{subsec:cheshireexact}

\begin{figure}
    \centering
    \includegraphics[width=0.9\textwidth]{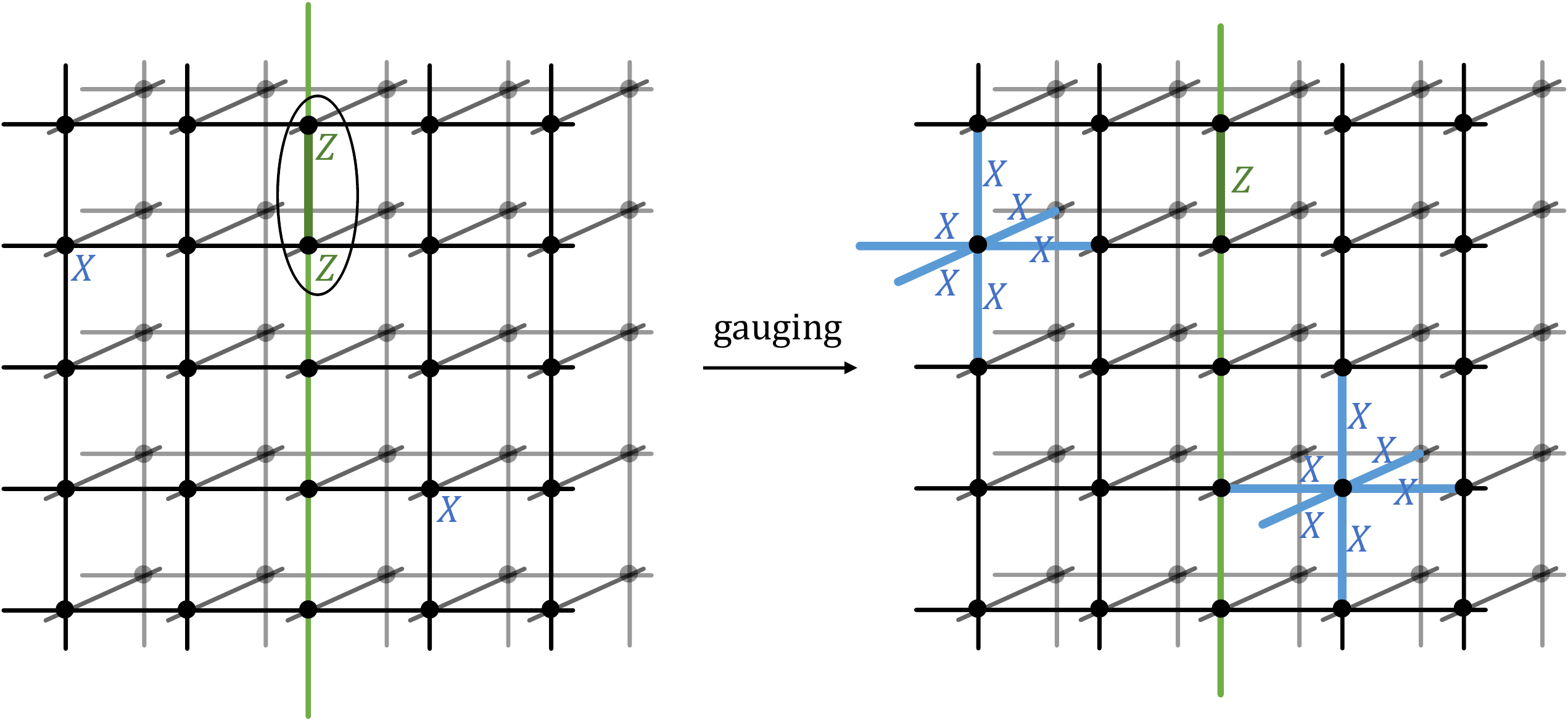}
\caption{The Cheshire string is obtained by gauging the (3+1)D trivial invertible phase with $\Z_2$ symmetry, with a decoration of the $\Z_2$ symmetry-broken Ising theory on the 1d defect line (green line).}
\label{fig: gaugingSSB}
\end{figure}

Starting from a 3d cubic lattice with a qubit at each vertex, we prepare a trivial Hamiltonian 
\begin{eqs}
    H_0 = -\sum_v X_v,
\end{eqs}
which has a global $\ZZ_2$ symmetry $\prod_v X_v$.
As shown in the left of Fig.~\ref{fig: gaugingSSB}, we decorate a 1d defect line by replacing $X_v$ with $Z_v Z_{v+1}$, where $v$ and $v+1$ are adjacent vertices on the defect line. This new Hamiltonian still respects the $\ZZ_2$ symmetry, but this 1d defect line hosts symmetry-broken ground states.

Next, we gauge the $\ZZ_2$ symmetry, and the resulting Hamiltonian is shown in the right of Fig.~\ref{fig: gaugingSSB}. The $X_v$ term becomes an $X$-star term, while the $Z_v Z_{v+1}$ term on the defect becomes $Z_e$ on the edge $e=\langle v, v+1 \rangle$. This is exactly the original model of the Cheshire string constructed in Ref.~\cite{else2017}. The $e$-line ($Z$ string) can end on the defect without violating any term in the Hamiltonian. We are going to provide the field-theoretical understanding of this defect in the next section.

\subsection{Cheshire string as a condensation defect}
\label{subsec:cheshirecondense}
Here we derive the expression of the Cheshire string in terms of the condensation defect of the (3+1)D $\Z_2$ gauge theory, and then explain its relation to (1+1)D $\Z_2$ symmetry-broken Ising theory. 
The $\Z_2$ toric code in (3+1)D is given by the action
\begin{align}
    S[a, b] = \pi \int_{M^4} a \cup \delta b
\end{align}
with $a\in C^1(M^4,\Z_2), b\in C^2(M^4,\Z_2)$.
The Cheshire string is realized as condensing the $e$ particle on the codimenison-2 defect, which is realized by gauging the symmetry generated by the $e$ Wilson line restricted to the codimension-2 surface. 
To describe this gauging process, we consider a region in the form of $D^2\times M^2$ in a spacetime 4-manifold $M^4$, which is regarded as a vicinity of the codimension-2 submanifold $M^2$ with a small disc $D^2$. We gauge the $\Z_2$ 2-form symmetry generated by the line operator $\exp(\pi i\int a)$ restricted to this region. This is performed by adding a term
\begin{align}
    \pi \int_{D^2\times M^2} a \cup B_3,
\end{align}
to the action, where a dynamical 3-form gauge field $B_3$ satisfies the Dirichlet boundary condition $B_3=0$ on $\partial (D^2\times M^2)$. Because of the boundary condition, $B_3$ is valued in $[B_3]\in H^3(D^2\times M^2; \partial(D^2\times M^2), \Z_2)$. Gauging $B_3$ amounts to condensing $e$ particle at $D^2\times M^2$. By shrinking $D^2$ to a point, we effectively obtain a codimension-2 defect supported on $M^2$.

One can express Cheshire string as a sum of Wilson line operators $\exp(\pi i\int a)$ supported on $M^2$, i.e., a condensation defect supported on $M^2$. That is, an insertion of a Cheshire string on $M^2$ is performed by gauging the 2-form symmetry of the toric code on a region $X=D^2\times M^2$ as 
\begin{align}
 \frac{|H^1(X;\partial X,\Z_2)|}{|H^2(X;\partial X,\Z_2)||H^0(X;\partial X,\Z_2)|}\sum_{B_3\in H^3(X; \partial X,\Z_2)}\exp\left( iS[a,b]\right)\exp\left(\pi i\int_{X} a \cup B_3\right)
\label{eq: action with Cheshire string defect}
\end{align}
where we follow a standard normalization of $\Z_2^{[2]}$ gauge theory provided in Ref.~\cite{Choi2021noninvertible}.\footnote{In the original $\ZZ_2$ toric code, the partition function is obtained from $\exp (i S[a,b])$ by summing over $a\in C^1(M^4,\Z_2)$ and $b\in C^2(M^4,\Z_2)$ with a proper normalization factor. After we insert the Cheshire string, the original summand $\exp (i S[a, b])$ is replaced by Eq.~\eqref{eq: action with Cheshire string defect}. }
This is interpreted as summing the Wilson line $\exp(\pi i\int a)$ supported on the Poincar\'e dual of $B_3$. 
When $M^2$ is oriented, one can use Lefschetz duality to obtain $H^k(X;\partial X,\Z_2)=H_{4-k}(X,\Z_2)=H_{4-k}(M^2,\Z_2)$, which is a generalized version of Poincar\'e duality. Then, the above process is equivalent to inserting an operator given by the sum
\begin{align}
    \mathcal{D}_{\mathrm{ch}}=\frac{1}{|H_2(M^2,\Z_2)|}\sum_{\gamma\in H_1(M^2,\Z_2)}\exp(\pi i \int_\gamma a)
    \label{eq:cheshiresum}
\end{align}
where $\gamma\in H_1(D_2\times M^2,\Z_2)=H_1(M^2,\Z_2)$ is the Lefschetz dual of $B_3$.
For example, if we consider $M^2=T^2=S_x^1\times S_y^1$, the Cheshire charge becomes
\begin{align}
    \mathcal{D}_{\mathrm{ch}}=\frac{1}{2}\left( 1+ \exp(\pi i \int_{C_x} a) + \exp(\pi i \int_{C_y} a)+ \exp(\pi i \int_{C_x+C_y} a)\right)
\end{align}
By using the expression Eq.~\eqref{eq:cheshiresum}, one can see the fusion rule as
\begin{align}
    \begin{split}
        \mathcal{D}_{\mathrm{ch}}\times \mathcal{D}_{\mathrm{ch}} = \frac{|H_1(M^2,\Z_2)|}{|H_2(M^2,\Z_2)|}\mathcal{D}_{\mathrm{ch}} = \frac{|H_1(M^2,\Z_2)|}{|H_2(M^2,\Z_2)||H_0(M^2,\Z_2)|}(\mathcal{D}_{\mathrm{ch}}+\mathcal{D}_{\mathrm{ch}})=2^{-\chi(M^2)}\cdot (\mathcal{D}_\mathrm{ch} + \mathcal{D}_\mathrm{ch}),
    \end{split}
\end{align}
so the fusion rule is given by
$\mathcal{D}_\mathrm{ch}\times \mathcal{D}_\mathrm{ch}=2^{-\chi(M^2)}\cdot (\mathcal{D}_\mathrm{ch} + \mathcal{D}_\mathrm{ch})$,
where we used $M^2$ is connected, and $\chi(M^2)$ is the Euler number of $M^2$. 
By redefining the defect by $\mathcal{D}'_\mathrm{ch}:= 2^{\chi(M^2)}\mathcal{D}_\mathrm{ch}$, the fusion rule becomes $\mathcal{D}'_\mathrm{ch}\times \mathcal{D}'_\mathrm{ch} = \mathcal{D}'_\mathrm{ch} +  \mathcal{D}'_\mathrm{ch}$, which is consistent with the fusion rule in Ref.~\cite{Kong2020defects}.\footnote{In general, a topological defect can be redefined by adding Euler term $\lambda^{\chi(M^2)}$ for real positive number $\lambda$, since the Euler term is regarded as a partition function of a trivial invertible TQFT~\cite{Yonekura2019Cobordism}.}

As we have seen in Sec.~\ref{subsec:cheshirecondense}, the (3+1)D $\Z_2$ gauge theory with a Cheshire string $\mathcal{D}_\mathrm{ch}$ can be constructed by starting with an invertible phase in (3+1)D with $\Z_2$ symmetry, then decorating the codimension-2 submanifold of the (3+1)D spacetime with a (1+1)D $\Z_2$ SSB phase, and gauging its global $\Z_2$ symmetry.
The decorated (1+1)D phase in the (3+1)D invertible theory results in a defect $\mathcal{D}_{\mathrm{ch}}$ in the $\Z_2$ gauge theory after gauging.
The partition function of the decorated (1+1)D $\Z_2$ SSB phase is given by
\begin{align}
    Z_{\mathrm{SSB}}(M^2,a) = 2^{-\chi(M^2)}\cdot 2 \delta(a),
\end{align}
where $\delta(a)$ gives 1 if $[a]\in H^1(M^2,\Z_2)$ is trivial, otherwise zero. Note that $2 \delta(a)$ is regarded as a partition function of a $\Z_2$ broken Ising theory. The delta function $\delta(a)$ arises because the $\Z_2$ symmetry defect in the SSB phase gives a symmetry domain wall that costs energy, so nontrivial gauge field $a$ is projected out from the spectrum. The factor $2$ denotes the super-selection sector, and $2^{-\chi(M^2)}$ is a normalization by Euler term. Inserting a theory $Z_{\mathrm{SSB}}$ on a codimension-2 surface $M^2$ in a trivial (3+1)D $\Z_2$ SPT phase and then gauging $\Z_2$ symmetry is equivalent to inserting the Cheshire string $\mathcal{D}_{\mathrm{ch}}$ in $M^2$. This is because an insertion of $\mathcal{D}_{\mathrm{ch}}$ is regarded as turning off $a$ on $M^2$, and it is rewritten as
\begin{align}
    \frac{|H_1(M^2,\Z_2)|}{|H_2(M^2,\Z_2)|}\delta(a)=\frac{|H_1(M^2,\Z_2)|}{|H_2(M^2,\Z_2)||H_0(M^2,\Z_2)|}2\delta(a)=Z_{\mathrm{SSB}}(M^2,a).
\end{align}

\section{Modular S and T matrices for $D(D_{4})$}
\label{app:D4ST}
In this section, we present the modular $S$ and $T$ matrices for the quantum double $D(D_{4})$. 

\begin{equation}
	S= \frac{1}{8}
	\resizebox{0.88\hsize}{!}{
	$\left(
	\begin{array}{cccccccccccccccccccccc}
	    1 & 1 & 1 & 1 & 1 & 1 & 1 & 1 & 2 & 2 & 2 & 2 & 2 & 2 & 2 & 2  & 2 & 2 & 2 & 2 & 2 & 2   \\
		1 & 1 & 1 & 1 & 1 & 1 & 1 & 1 & 2 & -2 & -2 & -2 & -2 & 2 & 2 & 2  & -2 & -2 & 2 & 2 & -2 & -2  \\
		1 & 1 & 1 & 1 & 1 & 1 & 1 & 1 & -2 & 2 & -2 & -2 & 2 & -2 & 2 & -2  & 2 & -2 & 2 & -2 & 2 & -2 \\
		1 & 1 & 1 & 1 & 1 & 1 & 1 & 1 & -2 & -2 & 2 & 2 & -2 & -2 & 2 & -2  & -2 & 2 & 2 & -2 & -2 & 2  \\
		1 & 1 & 1 & 1 & 1 & 1 & 1 & 1 & 2 & 2 & 2 & 2 & 2 & 2 & -2 & -2  & -2 & -2 & -2 & -2 & -2 & -2 \\
		1 & 1 & 1 & 1 & 1 & 1 & 1 & 1 & 2 & -2 & -2 & -2 & -2 & 2 & -2 & -2  & 2 & 2 & -2 & -2 & 2 & 2 \\
		1 & 1 & 1 & 1 & 1 & 1 & 1 & 1 & -2 & 2 & -2 & -2 & 2 & -2 & -2 & 2  & -2 & 2 & -2 & 2 & -2 & 2  \\
    	1 & 1 & 1 & 1 & 1 & 1 & 1 & 1 & -2 & -2 & 2 & 2 & -2 & -2 & -2 & 2  & 2 & -2 & -2 & 2 & 2 & -2 \\
		2 & 2 & -2 & -2 & 2 & 2 & -2 & -2 & 4 & 0 & 0 & 0 & 0 & -4 & 0 & 0 & 0 & 0 & 0 & 0 & 0 & 0 \\
		2 & -2 & 2 & -2 & 2 & -2 & 2 & -2 & 0 & 4 & 0 & 0 & -4 & 0 & 0 & 0 & 0 & 0 & 0 & 0 & 0 & 0 \\
		2 & -2 & -2 & 2 & 2 & -2 & -2 & 2 & 0 & 0 & 4 & -4 & 0 & 0 & 0 & 0 & 0 & 0 & 0 & 0 & 0 & 0 \\
		2 & -2 & -2 & 2 & 2 & -2 & -2 & 2 & 0 & 0 & -4 & 4 & 0 & 0 & 0 & 0 & 0 & 0 & 0 & 0 & 0 & 0 \\
		2 & -2 & 2 & -2 & 2 & -2 & 2 & -2 & 0 & -4 & 0 & 0 & 4 & 0 & 0 & 0 & 0 & 0 & 0 & 0 & 0 & 0 \\
		2 & 2 & -2 & -2 & 2 & 2 & -2 & -2 & -4 & 0 & 0 & 0 & 0 & 4 & 0 & 0 & 0 & 0 & 0 & 0 & 0 & 0 \\
		2 & 2 & 2 & 2 & -2 & -2 & -2 & -2 & 0 & 0 & 0 & 0 & 0 & 0  & 4  & 0 & 0 & 0 & -4 & 0 & 0 & 0 \\
		2 & 2 & -2 & -2 &  -2 & -2 & 2 & 2 & 0 & 0 & 0 & 0 & 0 & 0 & 0  & 4 & 0 & 0 & 0 & -4 & 0 & 0  \\
		2 & -2 & 2 & -2 & -2 & 2 & -2 & 2 & 0 & 0 & 0 & 0 & 0 & 0 & 0  & 0 & 4 & 0 & 0 & 0 & -4 & 0  \\
		2 & -2 & -2 & 2 & -2 & 2 & 2 & -2 & 0 & 0 & 0 & 0 & 0 & 0 & 0  & 0 & 0 & -4 & 0 & 0 & 0 & 4  \\
		2 & 2 & 2 & 2 & -2 & -2 & -2 & -2 & 0 & 0 & 0 & 0 & 0 & 0 & -4 & 0 & 0 & 0 & 4 & 0 & 0 & 0 \\
		2 & 2 & -2 & -2 & -2 & -2 & 2 & 2 & 0 & 0 & 0 & 0 & 0 & 0 & 0  & -4 & 0 & 0 & 0 & 4 & 0 & 0 \\
		2 & -2 & 2 & -2 & -2 & 2 & -2 & 2 & 0 & 0 & 0 & 0 & 0 & 0 & 0  & 0 & -4 & 0 & 0 & 0 &  4 & 0  \\
		2 & -2 & -2 & 2 & -2 & 2 & 2 & -2 & 0 & 0 & 0 & 0 & 0 & 0 & 0  & 0 & 0 & 4 & 0 & 0 & 0 & -4  \\
	\end{array}
    \right)$
    }
\end{equation}

\begin{align}
	T= \text{diag}
	\left(
	\begin{array}{cccccccccccccccccccccc}
		1 & 1 & 1 & 1 & 1 & 1 & 1 & 1 & 1 & 1 & 1 & -1 & -1 & -1 & 1 & 1  & 1 & i & -1 & -1 & -1 & -i   
	\end{array}
	\right)
\end{align}

We use the following basis for the modular matrices:
\begin{align}
		&([(0,0),J_{0}]), ([(0,0),J_{3}]), ([(2,0),J_{0}]), ([(2,0),J_{3}]), ([(0,0),J_{2}]), ([(0,0),J_{1}]), ([(2,0),J_{2}]), 
     	\\ \nonumber
		&	([(2,0),J_{1}]), ([(0,0),\alpha]), ([(1,1),A_{0}]), ([(1,1),A_{3}]), ([(1,1),A_{1}]), ([(1,1),A_{2}]), ([(2,0),\alpha]), 
		\\ \nonumber
		& ([(0,1),A_{0}]), ([(0,1),A_{2}]), ([(1,0),\beta_{0}]), ([(1,0),\beta_{1}]), ([(0,1),A_{1}]), ([(0,1),A_{3}]), ([(1,0),\beta_{2}]), ([(1,0),\beta_{3}]).
\end{align}
One can explicitly check that the flux-preserving automorphism Eq.~\eqref{eq:D4automorphism} leaves the above $S$ and $T$ matrices invariant. We have checked this by a computer.

\section{Geometrical interpretation of higher cup products}\label{sec: higher cup products}

\begin{figure}[H]
    \centering
    \includegraphics[width=0.37\textwidth]{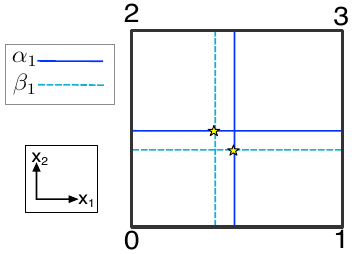}
    \caption{The cup product of $\alpha_1$ and $\beta_1$ is determined by the intersections (labeled by stars) of their dual edges. More precisely, the right star corresponds to $\alpha_1 ( \lr{01} ) \beta_1 (\lr{13})$ and the left star gives $\alpha_1 ( \lr{02} ) \beta_1 (\lr{23})$ in the definition of the cup product Eq.~\eqref{eq: cup-0 definition}.}
    \label{fig: square cup}
\end{figure}

In this section, we review geometrical properties of higher cup products in Refs.~\cite{tata2020, CT21}. The (higher) cup products are defined for both arbitrary triangulations and hypercubic lattices. For simplicity, we will demonstrate the geometrical properties on square and cubic lattices.

First consider 2d square lattices $M_2$ with two closed 1-cochains $\alpha_1, \beta_1 \in C^1(M_2, \ZZ_2)$, i.e., $\delta \alpha_1 = \delta \beta_1 = 0$.\footnote{The definition of cup products works for all cochains, not necessary closed one.} We draw the dual edges for $\alpha_1$ and $\beta_1$ separately in Fig.~\ref{fig: square cup}. Note that the dual edges of $\alpha_1$ and $\beta_1$ are differed by a small shift. The edges with $+1$ values of $\alpha_1$ and $\beta_1$ correspond to two closed loops in the dual lattice. The cup product is defined by
\begin{eqs}
    \alpha_1 \cup \beta_1 (\Box_{0123}) = \alpha_1 ( \lr{01} ) \beta_1 (\lr{13}) + \alpha_1 ( \lr{02} ) \beta_1 (\lr{23}),
\label{eq: cup-0 definition}
\end{eqs}
which counts the number (mod 2) of intersections of $\alpha_1$ loop and $\beta_1$ loop in a square, shown in Fig.~\ref{fig: square cup}.

\begin{figure}[H]
    \centering
    \includegraphics[width=0.4\textwidth]{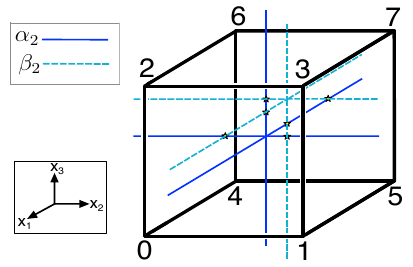}
    \caption{The cup-1 product of $\alpha_2$ and $\beta_2$ is determined by the intersections (labeled by stars) of their dual edges, after the projection from a certain angle. There are six intersections of the dual edges, which gives the six terms in the definition of the cup-1 product Eq.~\eqref{eq: cup-1 definition}.}
    \label{fig: cubic cup}
\end{figure}

Next, the 3d cubic lattice $M_3$ with two closed 2-cochains $\alpha_2, \beta_2 \in C^2(M_3, \ZZ_2)$ ($\delta \alpha_2 = \delta \beta_2 = 0$) can be analyzed in a similar way. We draw the dual edges of $\alpha_2$ and $\beta_2$ with a small shift in Fig.~\ref{fig: cubic cup}. The $+1$ values of $\alpha_2$ and $\beta_2$ form two closed loops of dual edges.
The definition of the cup-1 product is
\begin{eqs}
    \alpha_2 \cup_1 \beta_2 (\text{cube}) =~ & \alpha_2 (\Box_{1357}) \beta_2(\Box_{0145}) + \alpha_2 (\Box_{0246}) \beta_2(\Box_{0123}) \\
    +& \alpha_2 (\Box_{2367}) \beta_2(\Box_{0123}) + \alpha_2 (\Box_{2367}) \beta_2(\Box_{0246}) \\
    +& \alpha_2 (\Box_{4567}) \beta_2(\Box_{1357}) + \alpha_2 (\Box_{4567}) \beta_2(\Box_{0145}),
\label{eq: cup-1 definition}
\end{eqs}
which counts the number (mod 2) of intersections of $\alpha_2$ loop and $\beta_2$ loop according to the convention in Fig.~\ref{fig: cubic cup}. Note that the intersections depend on the direction of projection. Although we only demonstrate the cubic lattice, the same picture holds for arbitrary triangulations \cite{tata2020}.

Finally, we analyze the physical meaning of the $(B^{m_1}_2\cup_1 B^{m_2}_2)\cup C_1$ term in Eq.~\eqref{eq:3groupZ2Z2Z2}. We decompose it as
\begin{eqs}
    (B^{m_1}_2\cup_1 B^{m_2}_2)\cup C_1 (\lr{01234}) = (B^{m_1}_2\cup_1 B^{m_2}_2)(\lr{0123}) C_1(\lr{34}).
\end{eqs}
If we treat $\lr{34}$ as the temporal direction and consider $(B^{m_1}_2\cup_1 B^{m_2}_2)(0123)$ as the intersection of the $m_1$ and $m_2$ strings in the spatial manifold (a time slice), the whole term can be interpreted as the intersection of the $m_1$ and $m_2$ strings crossing the codimension-1 defect (where $C_1 = 1$).

\bibliographystyle{utphys}
\bibliography{TI,bibliography}

\end{document}